\documentstyle[10pt,epsfig,amssymb,amstex,titlepage]{article}
\newcommand{\be}{\begin{equation}}
\newcommand{\ee}{\end{equation}}
\newcommand{\beqn}{\begin{eqnarray}}
\newcommand{\eeqn}{\end{eqnarray}}

\newcommand{\frho}{{\bf \rho}}
\newcommand{\qf}{{\bf q}}

\newcommand{\kf}{{\bf k}}
\newcommand{\lf}{{\bf l}}
\newcommand{\ef}{{\bf e}}

\newcommand{\Pam}{I \!\! P}
\newcommand{\disc}{\mbox{disc}}
\newcommand{\fez}{\frac{1}{2}}
\newcommand{\ftz}{\frac{3}{2}}
\newcommand{\fnz}{\frac{n}{2}}
\newcommand{\om}{\omega}
\newcommand{\alp}{\frac{N_c \alpha_s}{\pi}}
\newcommand{\al}{\alpha_s}
\newcommand{\bs}{\boldsymbol}

\newcounter{savefig}

\begin{document}
\begin{titlepage}                                                              
\hfill
\hspace*{\fill}
\begin{minipage}[t]{4cm}
DESY--96--262\\
hep-ph/9705288\\
\end{minipage}
\vspace*{2.cm}                                                                 
\begin{center}                                                                 
\begin{LARGE}                                                                  
{\bf
Phenomenology of the BFKL Pomeron 
 \\
and Unitarity Corrections at low x
}\\
\end{LARGE}                                                                    
\vspace{2.5cm}                                                  
\begin{Large}
{ 
H.\ Lotter
}
\\
\end{Large}
\end{center}
\vspace{1.5cm}
\begin{center}
II.\ Institut f.\ Theoretische Physik, 
Universit\"at Hamburg, \\Luruper Chaussee 149, 
D-22761 Hamburg
\end{center}                                                   
\vspace*{2.cm}                          
\begin{quotation}                                                              
\noindent
The low $x$ limit of deep inelastic electron proton scattering is 
considered using methods of perturbative QCD. 
In the first part we investigate the phenomenological consequences 
of the resummation of leading logarithms in $1/x$ given by the 
BFKL pomeron. 
We apply the BFKL pomeron to the inclusive structure function
$F_2$, to the diffractive production of vector mesons at large momentum
transfer, to inclusive photon diffractive dissociation in DIS and to 
quark-antiquark production with large transverse momenta in DIS 
diffractive dissociation. 
For the last process we perform 
extensive numerical calculations based on the double
logarithmic approximation. \\
The BFKL pomeron is known to violate unitarity.
In the second part the first next-to-leading corrections which 
have to be taken into account to restore unitarity 
of the scattering amplitude are investigated.
A compact configuration space representation of the two to four 
gluon transition vertex is derived. Conformal symmetry of the vertex 
is proven and its relation to a conformal covariant three point 
function is established.
The important role of the spectral function $\chi_4$ of the four gluon
state is pointed out. We relate this function to the twist expansion of
the four gluon amplitude. Motivated by this relation we develop a method
to perform the twist expansion of the amplitude.
Based upon first results of our analysis we draw conclusions concerning 
the singularity structure of the function $\chi_4$.
\end{quotation}                                                                
\vfill
\vspace{1cm}
\end{titlepage} 
\newpage
\thispagestyle{empty}
\vspace*{\fill}
\newpage
\setcounter{page}{1}
\tableofcontents
\newpage
\section{Introduction}
The physics of hadrons is the regime of the theory 
of the strong interaction. A realization of a theory 
of the strong interaction satisfying the requirements
of general quantum theory and relativity is Quantum
Chromodynamics (QCD). The objective of QCD is to describe 
the variety of strong interaction phenomena ranging from 
the spectrum of light meson states to high energetic hadronic
collisions. Unfortunately, the mathematical structure of QCD
is quite complicated so that a solution of QCD, e.\ g.\ 
the calculation of all correlation functions, appears impossible.
Consequently, to make predictions in QCD, 
one has to make use of an approximation scheme. 
In a very coarse-grained distinction these approximation methods
can be divided into two classes. One can distinguish
between perturbative\footnote{The term 'perturbative' is used in 
a narrow sense here denoting methods which are based on the 
expansion in powers of the coupling constant.} 
methods on the one hand and nonperturbative methods 
on the other hand.
\\
Perturbative calculations start from the elementary degrees 
of freedom of QCD, namely quarks and gluons.
Perturbation theory begins with the observation that the coupling constant 
of QCD which specifies the strength of the interactions of quarks and gluons 
decreases when the energy of these particles increases (asymptotic freedom).
The perturbative approach is therefore appropriate when the specific 
phenomenon under consideration can be shown to be describable in terms
of the elementary interactions of highenergetic quarks and gluons. 
Nonperturbative methods have to be applied when such a description 
is not possible. Since the interaction of quarks and gluons is 
strong then, nonperturbative considerations often start from 
composite objects or collective excitations of quarks and 
gluons as e.\ g.\ condensates, mesons or instantons. 
\\
It is a major challenge in QCD to find a connection between these 
nonperturbative composite objects and the perturbative quark and 
gluon degrees of freedom. This is the place where this thesis is 
settled. Using perturbative methods it attempts to approach a composite,
probably nonperturbative, object of QCD, the pomeron. 
\\
The pomeron is a concept which originates from Regge theory.
It has a mathematical definition as the rightmost singularity
with vacuum quantum numbers of the proton-proton scattering amplitude 
in the complex angular momentum plane. Phenomenologically it admits the
interpretation of an exchanged object
 mediating the proton 
proton interaction at very high center of mass energies.
It leads to a slow rise of the total proton-proton cross section
at high energies. From the theoretical point of view 
this rise can at most be logarithmic since a power rise would 
ultimately lead to a violation of unitarity bounds. 
Since in the bulk of the total proton-proton cross section processes which
are determined by an underlying high energetic quark or gluon interaction
cannot be identified, the pomeron is commonly considered to be of 
nonperturbative origin.
Stated differently, perturbation theory is not the correct approximation 
scheme to describe the total proton-proton cross section.
It should be emphasized that up to now
the behavior which is associated with the 
pomeron has not been derived ab initio from QCD.
\\
There are, on the other hand, processes which are describable in terms
of elementary quark and gluon interactions, in which perturbation theory 
is thus applicable. It is then perfectly legitimate to ask for the behavior 
of these processes at high center of mass energy, i.\ e.\ in the regime 
of Regge theory.
\\
An example of such a process is onium-onium scattering, i.\ e.\ 
the scattering of two colorless heavy quark - heavy antiquark states.
When processes of this type are studied at high center of mass energy
in the framework of QCD perturbation theory a problem arises.  
When perturbative contributions of higher order are considered
they are found to be proportional to powers of the logarithm of the 
center of mass energy. Since one is interested in high energies, this 
logarithm is a large parameter and it is eventually 
sufficiently large to compensate the smallness of the coupling constant. 
Because of this observation it is of course senseless to apply fixed order 
perturbation theory to study the high energy behavior of the process.
In this situation a possible escape is resummation.
It is clearly not possible to resum all perturbative contributions
but a sensible starting point is to isolate in each order the 
contribution with the highest power of the logarithm and 
to resum these leading terms. The result of this procedure is termed
the leading logarithmic approximation. 
\\
This resummation was performed, in perturbative QCD, for the first 
time by Balitskii and Lipatov \cite{blip}.
They made use of results which were obtained by Lipatov and coworkers
in the framework of the massive gauge theory
some time before \cite{lip76,klf1,klf2}.
The infinitely many Feynman diagrams they resummed have the topology
of ladder diagrams with reggeized gluons in the $t$-channel.
The ladders are summed by an integral equation, the BFKL equation. 
The reggeized gluon itself also represents an infinite number of 
Feynman diagrams. In this sense it can be regarded as a collective
excitation within perturbative QCD. 
The resummed amplitude turns out to have a cut in the complex angular 
momentum plane. It has been termed the perturbative or BFKL pomeron.
\\
We now give an introduction to the investigations pursued
in this thesis where we refer also to previous and related work.
The original contributions contained in the present work 
are indicated in the end.
\\
The BFKL pomeron has a number of remarkable properties which 
we summarize in the second chapter of this thesis. 
First of all it is an
infrared finite quantity, i.\ e.\ all singularities which arise individually 
from real and virtual corrections cancel in the sum.
A quite interesting property 
of the BFKL equation is its invariance w.\ r.\ t.\ 
global conformal transformations in two-dimensional transverse configuration
space. Due to this property the equation can easily be solved in 
configuration space by a conformal partial wave expansion \cite{lip86}.
Furthermore the resummed scattering amplitude can be interpreted in 
the framework of an effective conformal field theory. In this interpretation 
the reggeized gluon appears as a primary field of conformal dimension 
and spin zero. The BFKL amplitude which depends only on the 
anharmonic ratios of the coordinates can consistently be interpreted 
as the four point function of this field.
\\
With these interesting results in mind one can ask two questions
\begin{itemize}
\item
Is the BFKL pomeron phenomenologically relevant, or, stated differently,
is there a possibility to observe the effect of the resummation of 
perturbative leading logarithms in a physical process?
\item
What is the relation between the BFKL pomeron and the true QCD pomeron?
\end{itemize} 
The first of these questions is addressed in chapter 3 
of this work whereas chapter
4 discusses the present status of the second question.
\\
Since onium-onium scattering is not experimentally feasible 
presently
, a different physical process has to be used to study the 
phenomenological applications of the BFKL resummation.
A class of processes which is for many reasons
appropriate for this undertaking 
are deep inelastic electron-proton scattering (DIS) processes
at low values of the Bjorken scaling variable $x$.
In DIS perturbative QCD has been very successful in describing 
observable quantities at intermediate values of $x$ 
due to the large virtuality $Q^2$ of the 
photon which in many processes sets the energy scale for the
strong coupling constant.   
The perturbative QCD approach to inclusive DIS, 
i.\ e.\ structure function calculations,
is based on collinear factorization \cite{collfac}
which allows to separate perturbative and 
nonperturbative parts of the scattering amplitude, 
and on the DGLAP \cite{dglap}
evolution equations. 
The latter serve to calculate the scale dependence 
of the structure functions. This scale dependence is controlled by the 
purely perturbatively calculable evolution kernels while the nonperturbative
part has been factored off in the initial conditions of the equations.
DIS is of interest for our purposes since when $x$ becomes small one 
is entering the Regge limit due to the fact that $x$ is inversely
proportional to the center of mass energy of the virtual photon-proton
system. Keeping $Q^2$ large one is entering this limit from a side where 
perturbation theory is applicable. 
\\
Repeating the reasoning sketched above one comes to the conclusion that 
logarithms of $1/x$ are important in this region and eventually have to 
be summed up.
For this reason the BFKL equation has been considered as an alternative 
to the DGLAP equations for the description of the scale and $x$ dependence
of the structure function at low $x$. 
The BFKL and DGLAP equations have their origin in quite different 
approximations of the phase space integration.
The latter is an 
evolution equation in $Q^2$ and takes into account the region of phase space 
in which the transverse momenta are strongly ordered, whereas the former
is an evolution equation in $1/x$ and takes into account
the phase space with strongly
ordered longitudinal momenta with the transverse momenta 
being disordered. There is an overlap region of both strongly ordered 
transverse and longitudinal momenta where these equations coincide, 
the so-called
double logarithmic region. 
\\
We show in section \ref{sec21} that
by extending the framework of collinear factorization to high-energy 
(or $\kf$-) factorization \cite{ktfac} the BFKL equation can be used to 
calculate DIS structure functions. 
We perform a calculation of the structure function $F_2$ based on a 
numerical evaluation of the equation.
It predicts a strong increase 
of the structure function with decreasing $x$. This is qualitatively 
in good agreement with HERA data \cite{h1,zeus} in a region of $x$ of 
$10^{-4} \cdots 10^{-2}$. There is still an intense debate on the 
theoretical interpretation of the data. 
Section \ref{sec21} gives an interpretation
from the point of view of the BFKL equation. 
\\
The application of the BFKL equation to the inclusive structure function 
faces a serious problem, namely the separation of perturbative
and nonperturbative scales which is in some sense perfect in the 
collinear factorization framework is limited here.
A characteristic feature of the BFKL evolution is the diffusion
in transverse momentum space which we display explicitly by calculating 
transverse momentum distributions. 
Starting from some characteristic
momentum scale of an initial condition the transverse momenta diffuse 
during evolution into both the infrared 
and the ultraviolet region. This means
that the evolution receives a contribution from a part of phase space 
which is certainly not treated correctly in the perturbative
approach. This is a serious problem for the structure function calculation
because on the proton side of the evolution one starts already deep 
in the nonperturbative regime and the diffusion tends to make it worse.
As a potential way out of this problem we consider the possibility
of imposing a modification of the infrared region 
to suppress the incorrectly treated region of 
phase space. With this procedure one can quite successfully describe
the data, but it is not really conclusive since there is some 
arbitrariness inherent in it. This is demonstrated by the rather
strong dependence of the result on the precise choice of the 
parameters of the modification.
One should state, however, that the HERA data demonstrate the importance
of the logarithms in $1/x$. This is emphasized by the success of the 
double logarithmic approximation (or the double asymptotic scaling
approach) in which only terms are taken into account which contain both 
a logarithm of $Q^2$ and of $1/x$. But it could not be unambiguously 
demonstrated that one needs the BFKL logarithms (the ones without the $Q^2$)
to describe the small $x$ data presently available. 
This includes also studies \cite{anom} in which these logarithms were
implemented into the evolution in form of higher order 
contributions to the gluon anomalous dimension.      
\\
To overcome the theoretical problem arising from diffusion in the 
BFKL equation one has too look for a process in which the contribution
of infrared momenta is suppressed due to the kinematical characteristics 
of the process. One example for such a process is inclusive DIS with an 
associated jet in the final state \cite{jet}. 
In this case the large transverse momentum of the 
additional jet lifts the lower side of the evolution out of the infrared
region \cite{balo}. 
Indeed first data on this process indicate a better 
agreement with BFKL based predictions than with fixed-order matrix element 
calculations, indicating the importance of higher 
order logarithms in $1/x$ \cite{jet2}. 
A related possibility is the study of the total cross section of 
virtual photon-photon scattering at $e^+-e^-$ colliders. 
Here the virtuality of the second photon takes the role of the jet transverse
momentum in the example above.
A discussion and estimate of event rates can be found in \cite{albert}.
\\
Another possibility to suppress the region of low momenta is to 
require a nonzero momentum transfer along the BFKL ladder.
Such a finite $t$ acts as an effective infrared momentum cut-off.
If the momentum transfer is increased the cross section  
decreases rapidly due to the decrease of the proton formfactor. 
At small $x$ one could still expect a sizeable
effect in a region of moderate $t$ due to the strong rise of the
cross section as a function of $x$ which could compensate the 
$t$-suppression. 
The study of the $t$-dependence
of the BFKL pomeron is also interesting since it allows the investigation of 
the slope of the BFKL pomeron trajectory.
As a consequence of the conformal invariance of the BFKL kernel its 
eigenvalues which control the energy dependence of the amplitude 
are momentum transfer independent. From this one would expect the 
trajectory associated with the BFKL pomeron to be fixed, 
i.\ e.\ to have no slope. In a physical process, however, conformal 
invariance is broken by characteristic mass scales of external particles.
Due to this breaking of conformal symmetry one can indeed derive 
a small effective slope of physical amplitudes which contain the 
BFKL pomeron.
\\
As a case study of a process of this kind we investigate 
in section \ref{sec22} diffractive vector meson production at large 
momentum transfer in DIS which was proposed by Forshaw and Ryskin
\cite{jeff1} (The study of large $t$ diffractive processes was also 
considered earlier by Frankfurt and Strikman \cite{frastrik}). 
For these calculation it is required to have at hand
the momentum space
expression for the nonforward BFKL amplitude. It is derived from 
the configuration space representation in chapter 2 of this work. 
\\
A class of processes where Regge theory traditionally is 
successfully applied 
is diffractive dissociation.
Single diffractive dissociation in hadron-hadron collisions denotes 
a process in which of two colliding hadrons one stays intact and
is only weakly deflected after the scattering whereas the other one 
dissociates into a many particle hadronic system with vacuum 
quantum numbers.
One can consider a related process in DIS by replacing the hadron 
which dissociates by the virtual photon.
This type of process has received much attention after the HERA
experiments have reported an excess of 'rapidity-gap' events compared with 
predictions from standard DIS Monte-Carlo generators. 
These rapidity gap events at low $x$ can be interpreted
as diffractive dissociation of the photon. 
The 'gap' refers to a region in the detector in which no final state 
particles are observed.
In terms of Regge theory this process is interpreted by assuming the exchange 
of a pomeron\footnote{We disregard secondary trajectories for the moment.}
between the proton and the photon. In the region of large invariant
masses of the produced hadronic system also more general exchange
processes have to be taken into account in which the photon 
does not couple to the pomeron directly but through another pomeron 
and a triple pomeron interaction vertex.
\\
Since for large $Q^2$ one is in a region where perturbation theory  
in principle is applicable one can raise the question to which extent
perturbative QCD can describe photon diffractive dissociation in DIS.
The simplest possibility to start with is to consider two gluon exchange
between the photon and the proton. The gluons couple to the two quarks 
in which the photon dissociates in this picture.
At low $x$ one finds again that two gluon exchange is not enough since
additional loops generate logarithms in $1/x$.
The resummation of this logarithms finally leads to the BFKL pomeron as the 
object which is exchanged between the photon and the proton.
By using the results of Bartels \cite{bartels} and Bartels and W\"usthoff
\cite{bartels-wue} on the triple Regge limit in perturbative QCD one can 
also extend this approach to the triple Regge region of large invariant 
masses of the hadronic system.
This approach is pursued  in section \ref{sec23} of this thesis where 
the inclusive DIS diffractive cross section is calculated for finite and zero 
momentum transfer. As a result we finds in this model that the 
$x$-dependence of the cross section is rather steep, it rises
at low $x$ with a power of about 1. Although the data are not yet 
conclusive one can say that this $x$-dependence is certainly too 
strong compared with the experimental observation \cite{h1diff,zdiff}.
The applicability of the BFKL pomeron to integrated DIS 
diffractive dissociation therefore 
has to be questioned. 
\\
It is possible to understand this drawback by again considering 
the contribution of different regions of phase space to the evolution
in this process. Numerical studies of the BFKL equation in diffractive
dissociation have demonstrated \cite{mvogt} that almost the whole evolution 
takes place in the infrared region where the perturbative approach cannot 
be consistently applied. In particular it was shown that in inclusive 
diffractive DIS the virtuality $Q^2$ of the photon does not act as 
a hard scale in the effective photon pomeron interaction. On the 
contrary, the coupling of the gluons to the quarks is dominated by
very low scales
\footnote{
The dominance of soft scales is emphasized in other models of 
diffractive dissociation. In the aligned jet model \cite{ajm}
one of the produced quarks is very soft and inetracts with the proton
like a hadronic constituent quark. In the model of \cite{buchheb}
the proton is treated as a soft gluonic background field.
The fluctuation of the photon into the $q\bar{q}$-pair is then 
calculated in this soft background.}.
Consequently, this process appears not to be appropriate to 
phenomenologically identify
the effect of the large perturbative logarithms of $1/x$. 
\\
Since, on the other hand, the data on DIS diffractive dissociation 
show a rise of the cross section at low $x$, similar to the inclusive
structure function, one can still ask if there is a subclass of 
events in which the effect of the large logarithms is visible.
\\    
In \cite{mvogt} it was shown that the scale of the effective photon pomeron
vertex becomes hard if restrictions are imposed on the diffractively 
produced final state.
The authors considered a large transverse momentum of the 
outgoing quarks but one could also think of heavy quarks or the 
production of a heavy vector meson. With such an effective large
scale perturbation theory has a much better foundation 
compared to the inclusive case dominated by low scales.
\\
As an example of such a process we consider the diffractive production 
of a quark-antiquark pair with either a large 
quark mass or a large transverse
momentum of the (anti)quark \cite{baj}. Starting again from BFKL 
pomeron exchange we analyze the process in detail in the 
double logarithmic approximation in which the coupling of the two 
gluons to the proton can be expressed in terms of the gluon density.
Within this approximation we perform an extensive numerical evaluation of 
the cross section.
As a remarkable property of the two-gluon exchange model
(see also \cite{diehl-phd}) we consider the azimuthal dependence of the
cross section. This shows a pattern quite different from the one 
obtained in the photon-gluon fusion process.
We propose to use this signal to reveal the two gluon exchange 
nature of the interaction underlying the process.
For the case of heavy quarks (we consider charm) the transverse momentum 
integration can be performed without entering the infrared region.
This allows the calculation of the charm contribution to the 
diffractive structure function in our model.    
\newline
\newline
So far we have specified the energy dependence of the BFKL pomeron
only qualitatively. The cut in the complex angular momentum plane transforms
into a power behavior (in $1/x$) of the amplitude
with the power being given by $4 \log 2 \,N_c\alpha_s/\pi$.
For reasonable values of the strong coupling constants the numerical
value of this power is 0.5. In numerical simulations in which the infrared 
region is suppressed by a cut-off the effective power turns out to be 
somewhat smaller. The common feature of all calculations of DIS processes
based on BFKL pomeron exchange is accordingly the 
power rise of the cross section at low values of $x$.
This behavior is in principle conflict with the unitarity 
bound which states 
that the cross section can at most increase logarithmically.
The leading logarithmic approximation thus violates a fundamental
principle of quantum theory.
The same problem is observed when only double logarithmic terms are
taken into account although the asymptotic rise turns out to be 
somewhat smaller then. Unitarity is therefore also violated in the DGLAP
formalism for asymptotically low $x$.
\\
The key problem of low $x$ physics is therefore to identify the corrections
which restore unitarity and to calculate them. The attractive feature
of DIS is without doubt that one can use perturbation theory as a starting 
point here since this approximation is highly successful at 
intermediate $x$.   
It is therefore natural to ask first for the perturbative corrections
which render the amplitude unitary.
If one succeeds in constructing a unitary amplitude in perturbation
theory one can expect that from such a highly constrained result
one will get indications of potential nonperturbative corrections.
Ultimately these considerations could then lead to the true QCD 
pomeron which applies also in hadron-hadron scattering.
As an example of nonperturbative corrections which are related to perturbation
theory, namely to ambiguities in the asymptotic QCD perturbation 
series, we mention
renormalons \cite{renormalon} which have been intensively studied
in the last few years.
\\
Let us shortly sketch a physically intuitive picture to motivate 
the necessity of corrections at low $x$.
At fixed $x$ and $Q^2$ the proton as seen by the virtual photon can be 
regarded as a dilute gas of partons of transverse size $1/Q^2$.
The partons are nearly noninteracting since the interaction strength
is small. If, at fixed $Q^2$, $x$ is decreased the density of the gas
starts to rise. 
The interaction strength, specified by the coupling constant 
evaluated at the scale $Q^2$ is still weak but interactions occur
frequently since the density is high. The state of the partons can now be 
regarded as a liquid. The frequent interactions now diminish the rise
of the density eventually until a steady state is reached (saturation). 
This would then 
correspond to the unitarity limit of the scattering amplitude.   
\\
How is this picture realized in terms of higher order perturbative 
corrections? The key step is to extend the single parton evolution equations
to take into account density reducing
parton-parton interactions.
\\
The first approach in this direction is due to Gribov, Levin and Ryskin
\cite{glr}. They considered freely evolving parton cascades which 
merge at a triple ladder vertex (fan diagrams).
The corresponding diagrams were resummed by a nonlinear evolution 
equation where the nonlinearity represents the parton screening.
This approach already emphasizes the necessity to consider diagrams
with more than two partons in the $t$-channel of the amplitude.
In fact the relevant diagrams at small $x$ contain only gluons in the 
$t$-channel due to the vector nature of the gluon. The 
contribution of spin-$\frac{1}{2}$ particles is suppressed at high energy.
\\
A systematic program towards the calculation of higher order corrections
of this kind has been initiated by 
Bartels \cite{bart-veryold,bart-old,bartels}.
In this program the BFKL amplitude represents the starting point and 
the guideline to calculate higher order corrections is unitarity.
This implies that in each order only those corrections are taken
into account which are absolutely necessary to restore unitarity.
The corrections are calculated using unitarity integrals and 
dispersion relations.  
The outcome of this formalism are sets of coupled integral equations
for amplitudes with $n$ gluons in the $t$-channel, where $n=2$ 
corresponds to the BFKL equation. It should be strongly emphasized 
that the gluons are reggeized in this approach, i.\ e.\ the elementary 
degree of freedom is a composite object which resums infinitely many 
Feynman diagrams. In this way the reggeized gluon is the first example of 
what is called a reggeon. The ultimate aim of the program proposed here
is to combine as many elementary interactions as possible into collective
excitations, called reggeons, and to encode all higher order contributions
in an effective reggeon field theory \cite{gribov,abarbanel,white,bart-old}.
A prerequisite for the inductive construction of such an 
effective field theory is the thorough understanding of the basic elements
which arise in low orders.
\\
Bartels \cite{bartels} and Bartels and W\"usthoff \cite{bartels-wue}   
have considered the system of coupled equations up to $n=4$.
They succeeded in solving this system partially by reducing 
the  $n=3$ system and parts of the $n=4$
system to the BFKL amplitude and they decomposed the remaining part of the 
$n=4$ system into two factors.
The first factor is an effective vertex which mediates the transition
from the two (reggeized) gluon system to the four gluon system.
The second part is the amplitude associated with the four gluon state.
\\
Chapter 4 of this thesis is devoted to the investigation of these two 
elements. The transition vertex has originally been derived in momentum
space but with regard to an effective field theory it is sensible to 
represent it in configuration space. This is the case since - as already 
indicated - the starting point of the formalism, the BFKL amplitude,
has in configuration space an interesting interpretation in terms
of a conformal field theory. Since conformal symmetry,
especially in two dimensions, is known to be a powerful property,
it is an essential question if this property persists in higher orders.
If this turns out to be the case, one could think of conformal 
field theory as a potential 
effective field theory encoding all unitarity corrections.
As an encouraging first step in this direction the 
proof of the conformal symmetry of the transition vertex has been 
given in \cite{balipwue}. In the present work we formulate an alternative
and simpler proof which is based on the symbolic operator representation
introduced by Lipatov \cite{holsep} 
to prove holomorphic separability of the BFKL
kernel.  
Unfortunately it turns out that holomorphic separability cannot be shown
for the transition vertex. Using conformal symmetry we then elaborate
on the interpretation of the vertex in terms of elements of a conformal 
field theory.  
It is demonstrated that in this framework the vertex admits 
the simple representation as a conformal three point function.
The field entering this correlation function is a field which can 
be associated with the BFKL pomeron. A virtue of the configuration 
space representation as a three point function is that the number 
of terms contributing to the vertex function is drastically reduced. 
\\
Since the understanding of the transition vertex is rather satisfactory
the calculation of the four (reggeized) gluon amplitude is an urgent task.
The four gluon amplitude is defined as the solution of the four 
particle BKP \cite{bart-veryold,bkp-k} equation which is of the form of a 
Bethe Salpeter equation with summation over all pairwise interactions of
two gluons. These pairwise interactions are given by the BFKL kernel.
The solution of this equation would complete the analysis of the first 
nontrivial unitarity corrections.
It is of great phenomenological significance since it would allow
to investigate in which region of $x$ the corrections which reduce
the rise of the cross section become sizeable.
It would be highly interesting to see whether these terms, added to 
the BFKL amplitude, can describe experimental data, both for the 
structure function and for diffractive events. 
From the theoretical point of view the analysis of the four gluon state     
constitutes only the first step. Asymptotically this contribution will 
probably also generate a power behavior which violates unitarity.
Only after resummation of all $n$-gluon contributions a unitary 
amplitude can be expected.
\\
The $n$-gluon system is currently under intense study in the large 
$N_c$-approximation. In this limit the BKP equations 
in configuration space have been shown to be holomorphic 
factorizable \cite{holsep,quant}.
The holomorphic and the antiholomorphic part have been demonstrated 
\cite{fad-kor}
to be equivalent to the integrable model of the XXX Heisenberg 
spin chain for spin $s=0$ \cite{takh}. 
Accordingly one tries to solve the model using quantum inverse 
scattering methods \cite{kor,wall}.
\\
In this work we restrict ourselves to $n=4$ and we do not 
consider the large $N_c$-approximation. The aim is  
to derive
the operator algebra \cite{bpz} of the fields which we associate with 
reggeons constructed from the coupling of two gluons.
The idea behind is that from the knowledge of the operator
algebra further information can be obtained regarding a potential effective 
conformal field theory.  
To derive the operator algebra we consider the short-distance 
or twist expansion of the four gluon amplitude.
There is another interesting information which one could
obtain from this expansion. Namely, for the BFKL equation a very close 
relation between the anomalous dimensions appearing in the twist expansion
and the spectrum of the BFKL kernel can be demonstrated. The anomalous 
dimensions correspond to the residues of the poles of the eigenvalue
at integer points in the $\nu$-plane where $\nu$ corresponds 
to the quantum number associated with the conformal symmetry. 
Postulating a similar relationship for the four gluon 
spectrum one could think of reproducing the 
spectrum from the anomalous dimensions
with the help of
a dispersion relation. The knowledge of the spectrum, i.\ e.\ the 
eigenvalues of the four-particle interaction operator
would already allow important conclusions concerning the 
high energy behavior of 
the amplitude. It remains then of course to calculate the eigenfunctions.
\\
Section \ref{sec33} of this thesis concentrates on the twist expansion
of the four gluon amplitude. We use the Faddeev reordering \cite{faddeev}
and try to make use of conformal invariance of the two particle 
interaction kernels by resumming pairwise interactions with the BFKL
amplitude. In this way the problem is formulated as an effective
two reggeon problem. Our analysis of the twist expansion starts 
from identifying singularities of the reggeon propagators
and the effective reggeon interaction vertices.  
For the simplest examples it is then shown how the anomalous dimensions
can be obtained from the iteration of these singularities.
\\
\\
Let us briefly recapitulate the organization of this thesis.
\\
Chapter 2 collects some background material on the BFKL pomeron
with the emphasis 
on the conformal properties of the amplitude. 
The sections 2.2 and 2.3 mainly review known results. 
The momentum space expressions discussed in section 2.3
constitute an original contribution of this thesis 
partly published in \cite{blotwue}.
\\
Chapter 3 is devoted to the phenomenological applications of the 
BFKL pomeron in DIS at low $x$. 
The numerical calculations on the structure function $F_2$
in sections 3.1.2 and 3.1.3 are 
results of this work. Results similar to the ones in 3.1.2 have been reported 
in \cite{akms}. 
The calculation of the cross section of diffractive vector meson
production at large $t$ 
in section 3.2.1 is a new contribution of this 
thesis. The following sections 3.2.2 and 3.2.3 contain some corollaries
and numerical evaluations. 
Results similar to the ones in 3.2.2 and 3.2.3 have been obtained in the joint
publication \cite{jeff2}. 
Section 3.3 on inclusive diffractive dissociation
is based on the joint publication \cite{blotwue}.
The main contribution of the present work is the explicit momentum
space calculation based on the results of section 2.3. The results of 
section 3.3.3 entered into the joint publication \cite{baliplofo}.
Section 3.4 on quark-antiquark production in DIS diffractive dissociation
almost exclusively contains results which have been obtained
independently by the author. They partly entered the joint 
publications \cite{baj}. The generalization to finite quark mass and the
results of section 3.4.5 have not been published so far.
\\
Chapter 4 focusses on the 
unitarity corrections. The section 4.1 serves as an introduction 
and review of the background upon which the investigation is based.
Apart from section 4.1.2 the material presented here is well-known.
Section 4.2 is devoted to the analysis of the transition vertex.
The content of the sections 4.2.1 and 4.2.2 has been obtained
in the collaboration \cite{rhein}. Section 4.2.3 is based on the 
author's independent research.
Section 4.3 discusses the significance of the four-gluon state. 
The material contained therein has not been presented in this form before.
In section 4.4 the four gluon state is considered in detail.
Apart from a short review of the basic results of \cite{bartels}
the approach developed here constitutes an original contribution of this 
thesis which has not been published so far.
\\
Some technical supplements are collected in the appendix.
\newpage
\section{Properties of the BFKL Pomeron}
\label{chap1}
In 1976 a program was initiated \cite{lip76} to investigate the ideas 
of Regge theory in the framework of QCD perturbation theory.
The aim was to understand, in this framework, the nature of the leading 
singularity with vacuum quantum numbers in the complex angular
momentum plane, the so-called Pomeranchuk singularity (Pomeron).
In the asymptotic region of large center of mass 
energy $s$ large logarithms
of $s$ compensate the small value of the perturbative coupling 
$\alpha_s$ and have to be resummed. In \cite{lip76}-\cite{blip} 
the partial wave amplitude of gluon-gluon scattering
was calculated in the leading-log($s$) approximation, which takes 
into account contributions of the form $s(\alpha_s \log s)^n \phi_n(Q^2)$
with a hard momentum scale $Q^2$.
In the original papers the calculations were performed in a 
massive gauge theory in which the natural scale is the mass of the 
vector boson. In the massless theory the hard scale is given by
the characteristic scale of perturbatively calculable impact factors 
with which the gluon-gluon scattering amplitude is convoluted. 
This partial wave amplitude 
satisfies a Bethe-Salpeter equation which generates a ladder-like structure
with gluons being produced in the $s$-channel and reggeized gluons being
exchanged in the $t$-channel.
In the vacuum channel (color singlet exchange) the equation generates 
a fixed cut in the angular momentum plane.
In the color octet channel (the channel with gluon quantum
numbers) the equation is solved by a Regge-pole ansatz. 
This bootstrap property demonstrates the reggeization of 
the gluon in perturbative QCD.  
The BFKL equation with zero momentum transfer has been studied
extensively in the last few years.
Relatively little attention has been devoted to the generalization to 
finite momentum transfer (nonforward direction).
This generalized equation is important for phenomenological
applications and in particular for theoretical reasons since it 
appears as an essential building block in a systematic study
of subleading corrections to the BFKL equation. 
The analysis of the generalized equation is highly 
facilitated by its symmetry 
w.\ r.\ t.\ two-dimensional conformal transformations, due to which 
it can be diagonalized by conformal partial waves \cite{lip86}. 
The conformal symmetry reveals itself in the configuration space 
representation which is particularly well suited to study the conformal
properties of the BFKL amplitude.
In this representation the results of 
the leading-log approximation can be interpreted
in the general framework of conformal field theory \cite{bpz}. 
The reggeized gluon can be interpreted as an elementary field with 
conformal dimension zero and the BFKL amplitude has a natural interpretation 
as the four point function of that field.
An interpretation along these lines might in the long run serve as 
a guideline to obtain an effective theory of QCD in the Regge limit
as a conformal field theory.
Further evidence for this conjecture will be given 
in the last chapter of this thesis.
\\
The first part of this 
section serves as an introduction to these aspects of the 
BFKL theory. We will introduce
the equation and derive its configuration space 
representation. The proof of conformal symmetry will be 
given and the interpretation in terms of a conformal field theory 
will be sketched.
\\
The configuration space representation is elegant and allows for 
promising interpretations but using it one looses the connection
to perturbation theory since QCD perturbation theory is usually 
performed in momentum space. 
The amplitudes of the scattering processes which are studied in chapter 
\ref{chap2} and  
the subleading corrections to the BFKL theory 
which are the objective in 
chapter \ref{chap3} are calculated within perturbative
QCD in momentum space. 
To apply the BFKL amplitude to the calculation of these scattering 
amplitudes and the analysis of unitarity corrections 
a back transformation of the configuration space results to 
momentum space is desirable.
This is the aim of the second part of this section. 
We will formulate the solution of the nonforward BFKL equation 
in momentum space and study some of its properties. These results 
will be used in forthcoming sections of this thesis.
\\
The formulation of the BFKL equation as a two particle Schr\"odinger 
equation in
two dimensions with a non-trivial kinetic term \cite{holsep}
which is a further development of the results of section 
\ref{secconf} will be discussed in conjunction with the unitarity
corrections in the last chapter of this work.   
\subsection{The Bethe Salpeter equation in momentum space}
\label{sec11}
The results of the leading-log($s$) approximation are the following.
The amplitude for the scattering of colorless objects can be written
in factorized form
\beqn
A(s,t) &=& i s \int_{\cal C} \frac{d \omega}{2 \pi i}
s^{\omega}\Phi_{\omega}(\qf^2) \;,\; t=-\qf^2
\label{amp}
\\
\Phi_{\om}(\qf^2)&=&\int
 \frac{d^2 \kf}{(2 \pi)^3}\frac{d^2 \kf'}{(2 \pi)^3}
\Phi_{\om}(\kf,\kf';\qf)
\phi_1(\kf,\qf)\phi_2(\kf',\qf)
\label{pwamp}
\eeqn
The integration contour ${\cal C}$ is located to the right of all
singularities of $\Phi_{\om}$.
The function $\Phi_{\om}(\kf,\kf';\qf)$
can be interpreted as the $t$-channel partial wave amplitude
for the scattering of virtual gluons with virtualities
 $-\kf^2,-(\qf-\kf)^2,-\kf{'}^2$ and $-(\qf-\kf')^2$ respectively.
The functions $\phi_{1,2}$ are the wave functions of the scattered 
colorless states. The color neutrality condition manifests itself 
in the following property of the wave functions
\beqn
\phi_{1,2}(\kf=0,\qf)=\phi_{1,2}(\kf=\qf,\qf)=0
\label{ward}
\eeqn
which is essential for the infrared finiteness of the amplitude.
The partial wave amplitude $\Phi_{\om}$ is given in 
the leading-log approximation as the solution of a 
Bethe-Salpeter type of equation in the transverse momentum space 
\beqn
\om \Phi_{\om}(\kf,\kf';\qf)
=\frac{\delta^{(2)}(\kf-\kf')}{\kf^2\kf{'}^2} 
&+&({\cal K}_{\mbox{\tiny BFKL}}\otimes\Phi_{\om})
(\kf,\kf';\qf)  
\nonumber \\ &-&
[\beta(\kf^2)+\beta((\qf-\kf)^2)]
\Phi_{\om}(\kf,\kf';\qf)
\label{bfklmom}
\eeqn
This equation resums the radiative corrections to the Born level two
gluon exchange which is represented by the inhomogeneous term. The integral
kernel ${\cal K}_{\mbox{\tiny BFKL}}$ is given by the square of an
effective real gluon production vertex
and reads
\beqn
({\cal K}_{\mbox{\tiny BFKL}} \otimes \Phi_{\om})
(\kf,\kf';\qf) =
\frac{1}{\kf^2(\qf-\kf)^2}
\frac{N_c \alpha_s}{2\pi^2}\int d^2 \lf
\left[-\qf^2+\frac{\lf^2(\qf-\kf)^2+\kf^2(\qf-\lf)^2}{(\kf-\lf)^2} 
\right]\Phi_{\om}(\lf,\kf';\qf)
\label{kernmom}
\eeqn
The function $\beta(\kf^2)$ is the gluon trajectory function
\footnote{Loosely speaking
we call $\beta$ the trajectory function of the gluon. The
trajectory of the gluon in the sense of Regge theory is 
$\alpha(\kf^2)=1-\beta(\kf^2)$. It passes through the physical spin 1
for zero momentum transfer $\alpha(0)=1$.}    
which resums the 
virtual corrections 
\beqn
\beta(\kf^2)=
\frac{1}{2}
\frac{N_c \alpha_s}{2\pi^2}\int d^2 \lf
\frac{\kf^2}{\lf^2(\kf-\lf)^2}
\label{traj}
\eeqn
The trajectory function as well as the production vertex contain
infrared singularities. The sum of real and virtual corrections,
however, gives a finite result in the case of color singlet exchange.
This cancellation of singular contributions relies on the 
property (\ref{ward}) of the wavefunctions.
It should be remarked that on the level of the leading-log
accuracy $\alpha_s$ is fixed and the scale is, strictly speaking,
not determined. For consistency one should use the hard scale 
of the process as the argument of $\alpha_s$.
Logarithms of a relevant scale
$M^2$ which are absorbed by introducing 
the running coupling $\alpha_s(M^2)$ only appear in next-to-leading 
order. They are expected to be found in the complete NLO corrections
to the BFKL equation which are studied presently by Lipatov and Fadin 
\cite{fadlip}.
Iteration of the Bethe-Salpeter equation 
leads to a ladder like structure with 
reggeized gluons being exchanged in the $t$-channel and real gluons
being produced in the $s$-channel.
Setting the momentum transfer $t$ equal to zero one obtains the
BFKL equation in forward direction which can be diagonalized
by power functions 
\beqn
e^{(\nu,n)}(\kf) = 
2\pi \sqrt{2} \,
(\kf^2)^{-\frac{3}{2}
-i\nu} e^{-i n \phi}\phantom{xx};\;\nu \in {\Bbb{R}} , n\in {\Bbb{Z}}
\label{eigenzero}
\eeqn
with eigenvalues 
\beqn
\chi(\nu,n)=\alp\left[2\psi(1)-\psi(\frac{1+|n|}{2}+i\nu)
                               -\psi(\frac{1+|n|}{2}-i\nu)\right]
\label{chi}
\eeqn
The function 
$\psi$ is defined as the logarithmic derivative of the 
$\Gamma$-function: $\psi(x)=\Gamma'(x)/\Gamma(x)$.
Some important properties of the function $\chi(\nu,n)$
are given in appendix \ref{app3}.
The solution of the BFKL equation consequently reads
\beqn
\Phi_{\om}(\kf,\kf')=
 \sum_{n=-\infty}^{+\infty}
\int_{-\infty}^{\+\infty} \frac{d \nu}{2 \pi}   
\frac{1}{\omega-\chi(\nu,n)} e^{(\nu,n)}(\kf)e^{(\nu,n)\,\ast}(\kf')
\label{solzero}
\eeqn
For the case of non-zero momentum transfer simple eigenfunctions
are not known in momentum space. In that case it proves 
to be useful to
discuss the equation in configuration space.
\subsection{The configuration space representation}
\label{secconf}
We define the Fourier transformation of the 
off-shell partial wave amplitude 
$\Phi$
\beqn
\delta^{(2)}(\qf-\qf')\Phi_{\om}(\kf,\kf';\qf')
=\int \prod_{i=1}^2 d^2\rho_i\,
\prod_{i=1}^2 d^2\rho_{i'}\,
\phantom{xxxxxxxxxxxxxxxxxxxxxxxxxxxxx}
\nonumber \\
\cdot
\Phi_{\om}(\rho_1,\rho_2;
\rho_{1'},\rho_{2'})
e^{i\kf\rho_1+i(\qf-\kf)\rho_2-i\kf'\rho_{1'}-i(\qf'-\kf')
\rho_{2'}}
\label{fourier}
\eeqn
Consequently we can express the partial wave amplitude as
\beqn
\delta^{(2)}(\qf-\qf')\Phi_{\om}(\qf^2)
=
\frac{1}{(2\pi)^2} \int \prod_{i=1}^2d^2\rho_i\,
\prod_{i=1}^2d^2\rho_{i'}\, 
\Phi_{\om}(\rho_1,\rho_2;
\rho_{1'},\rho_{2'})
\phi_1(\rho_1,\rho_2;\qf)\phi_2(\rho_{1'},\rho_{2'};\qf)
\label{pwampconf}
\eeqn
with the configuration space representation 
of the wavefunctions
\beqn
\phi_1(\rho_1,\rho_2;\qf)=
\frac{1}{(2\pi)^2} \int d^2\kf \;\phi_1(\kf,\qf)e^{i\kf\rho_1+
i(\qf-\kf)\rho_2}
\label{waveconf}
\eeqn
In configuration space the color neutrality condition 
is equivalent to the vanishing of the space integral over 
the wavefunction
\beqn
\int d^2 \rho_1 \;\phi_1(\rho_1,\rho_2;\qf)=
\int d^2 \rho_2 \;\phi_1(\rho_1,\rho_2;\qf)=0
\label{wardconf}
\eeqn
To proceed with the configuration space representation
it will be convenient to introduce complex coordinates in the 
two-dimensional transverse momentum and configuration space
\footnote
{In the following, boldtype letters 
are used for 
vectors in two-dimensional transverse 
momentum 
space with positive euclidean metric.
In configuration space normal type letters are used both
for the two-dimensional vector 
$\rho=(\rho_x,\rho_y)$
and the complex number $\rho=\rho_x+i\rho_y$.
Which object is meant should be clear from the context in which  
the expression appears.
}
\beqn
\kf &=& (k_x,k_y) \nonumber \\
k=k_x+ik_y &,& k^{\ast}=k_x-ik_y \nonumber \\
\rho_i=\rho_{i \,x}+i\rho_{i\, y} &,&
 \rho_i^{\ast}=\rho_{i \,x}-i\rho_{i \,y} \nonumber \\
\partial_i=\frac{\partial}{\partial \rho_i} &,&
\partial_i^{\ast}=\frac{\partial}{\partial \rho_i^{\ast}} \nonumber \\
\kf \cdot \rho &=&\frac{1}{2}(k\rho^{\ast}+k^{\ast}\rho)
\label{coord}
\eeqn
Rewriting the Bethe-Salpeter
equation in terms of these coordinates and performing the Fourier 
transformation we obtain (cf. appendix {\ref{appft})
\beqn
\om |\partial_1|^2|\partial_2|^2 \Phi_{\om}(\rho_1,\rho_2;\rho_{1'},
\rho_{2'}) &=& \frac{1}{(2\pi)^4}
\delta^{(2)}(\rho_{11'})\delta^{(2)}(\rho_{22'})
\hspace{2cm}
\nonumber \\
&+& \alp \left[  \partial_1\partial_2^{\ast} \frac{1}{2}\int
\frac{d^2\rho_0}{|\rho_{10}|^2}
\theta\left(\frac{|\rho_{10}|}{|\rho_{12}|}
-\epsilon\right)
\partial_0^{\ast}\partial_2 \Phi_{\om}(\rho_0,\rho_2;\rho_{1'},
\rho_{2'})  
\right. \nonumber \\  
&+& \left.
 \pi \log \epsilon  |\partial_1|^2|\partial_2|^2
 \Phi_{\om}(\rho_1,\rho_2;\rho_{1'},\rho_{2'})
\right. \nonumber \\ 
&+& \left. 
(\mbox{h.c.}) + (1 \leftrightarrow 2)  \right] 
\label{bfklconf}
\eeqn
We have introduced the shorthand notation $\rho_{ij}=\rho_i-\rho_j$.
The variable $\epsilon$ is a fictious parameter which regularizes 
the ultraviolet singularity of the integral arising from 
the region near $\rho_{10}=0$.
The kernel of the $\rho_0$-integral encodes in a compact form the
contribution of the gluon production vertex and the gluon 
trajectory function.
\subsubsection{Conformal invariance of the equation}
In two dimensions the global conformal transformations
can be represented by the mapping
\beqn
\rho \rightarrow \frac{a \rho+b}{c \rho+d} \,,\,a,b,c,d \in {\Bbb{C}} 
\,,\, ad -bc \neq 0 
\label{transform}
\eeqn
i.\ e.\ a linear fractional transformation.
To be precise, global conformal transformations are 
well-defined on the compactified complex plane, i.e. the
Riemannian sphere $S^2$.
These transformations form the pseudoorthogonal group
$SO(3,1)$ which is isomorphic to $SL(2,{\Bbb{C}})/\{\pm1\}$. 
Each of the 
elements
of this group is obtained by a superposition 
of transformations of the following types
\beqn
&\mbox{Translations:}&\rho \rightarrow \rho+b 
\nonumber \\
&\mbox{Rotations\phantom{xx}:}&\rho \rightarrow a\rho , |a|=1
\nonumber \\
&\mbox{Dilatations\phantom{x}:}&\rho \rightarrow \lambda \rho, \lambda \in \Bbb{R}_+
\nonumber \\
&\mbox{Inversions\phantom{xx}:}&\rho \rightarrow \frac{1}{\rho} 
\nonumber
\eeqn                                                      
We want to show that the group of linear fractional 
transformations is the invariance group of the 
Bethe-Salpeter equation.
Conformal invariance implies restrictions on the form
of the correlation
functions of a theory. These restrictions and the interpretation 
of the amplitudes $\Phi_{\om}$ in the framework of a conformal 
invariant theory will be the subject of the next section. 
\\
First we show that the solution
of the Born level equation ($\alpha_s=0$) can be written in 
conformally invariant form, then we show that the integral kernel
is invariant under conformal transformations. From this we 
conclude that the general solution of the Bethe-Salpeter 
equation can be represented in conformally invariant form 
and that the solution can be obtained by conformal partial
wave diagonalization. \\ 
First we note that the Bethe-Salpeter
equation (\ref{bfklconf})
reduces for $\alpha_s=0$ to a Poisson type equation
in two dimensions the solution of which is given by
\beqn
\Phi_{\om}^{(0)}(
\rho_1,\rho_2;\rho_{1'},\rho_{2'})
=\frac{1}{\om}\frac{1}{(2\pi)^6} \log |\rho_{11'}\rho_{22'}|
\label{born1}
\eeqn
By adding logarithmic terms this solution can be 
transformed into an expression which depends only on the 
two independent conformally invariant anharmonic ratios which 
can be constructed from four coordinates
\beqn
\Phi_{\om}^{(0)}(
\rho_1,\rho_2;\rho_{1'},\rho_{2'})
=\frac{1}{2\om}
\frac{1}{(2\pi)^6}
\log\frac{|\rho_{11'}\rho_{22'}|}{|\rho_{12}\rho_{1'2'}|}
\log\frac{|\rho_{11'}\rho_{22'}|}{|\rho_{12'}\rho_{1'2}|}
\label{born2}
\eeqn
When integrated with wave functions of color neutral systems
due to the property eq. (\ref{wardconf})  
the additional terms give no contribution.
Hence we have shown that the zeroth order solution 
can be represented in
conformally invariant form.
Note that eq.\ (\ref{born1}) corresponds to two-gluon exchange 
whereas the terms which where added to obtain 
(\ref{born2}) do not have a counterpart in terms of 
Feynman diagrams.
\\
Now we want to verify that the integral kernel of the 
configuration space representation is invariant under 
conformal transformations. It is obvious that the kernel
in (\ref{bfklconf}) remains invariant under rotations, translations
and dilatations. It remains to show the
invariance with respect to inversions. Let us perform simultaneously
the transformations
\beqn
\rho_i & \rightarrow & \frac{1}{\rho_i}  
\label{trans} \\
\partial_i & \rightarrow & -\rho_i^2\partial_i
\\
d^2 \rho_0 & \rightarrow & \frac{d^2 \rho_{0}}
{|\rho_{0}|^4} \\
|\rho_{i0}|^2 & \rightarrow &  \frac{|\rho_{i0}|^2}
{|\rho_{i}|^2|\rho_{0}|^2}
\eeqn
Under these transformations the lhs and the inhomogeneous 
part of the eq.\ (\ref{bfklconf}) are multiplied by the factor
$|\rho_1|^4|\rho_2|^4$. Applying these transformations
to the integral kernel, i.\ e.\ the terms in brackets 
in (\ref{bfklconf}), we obtain
\beqn
& &\frac{1}{2} \rho_1^2\rho_2^{\ast\,2}\partial_1\partial_2^{\ast}
\int \frac{d^2 \rho_0}{|\rho_{10}|^2}
\frac{|\rho_1|^2|\rho_0|^2}{|\rho_0|^4}
\theta\left(\frac{|\rho_{10}||\rho_2|}{|\rho_{12}||\rho_0|}-
\epsilon\right)\rho_0^{\ast\,2}\rho_2^2 
\partial_0^{\ast}\partial_2 \Phi_{\om}(\rho_0,\rho_2) 
\nonumber \\
& & \phantom{xxxxxxxxxx}
+\pi \log \epsilon |\rho_1|^4|\rho_2|^4 
|\partial_1|^2|\partial_2|^2 \Phi_{\om}(\rho_1,\rho_2)
+[\mbox{h.c.}] + [1 \leftrightarrow 2]
\nonumber \\
&=&|\rho_1|^4|\rho_2|^4\left[\frac{1}{2} 
\partial_1\partial_2^{\ast}
\int \frac{d^2 \rho_0}{|\rho_{10}|^2}
\frac{\rho_1\rho_0^{\ast}}{\rho_0\rho_1^{\ast}}
\theta\left(\frac{|\rho_{10}||\rho_2|}{|\rho_{12}||\rho_0|}-
\epsilon\right) 
\partial_0^{\ast}\partial_2 \Phi_{\om}(\rho_0,\rho_2)
\right. \nonumber \\ & & 
\left. \phantom{xxxxxxxxxx}
+\pi \log \epsilon |\partial_1|^2|\partial_2|^2
\Phi_{\om}(\rho_1,\rho_2) 
+[\mbox{h.c.}] + [1 \leftrightarrow 2] \right]
\label{proof1}
\eeqn
We rewrite the numerator in front of the $\theta$-function as
\beqn
\rho_1\rho_0^{\ast}&=&(\rho_{01}^{\ast}+\rho_1^{\ast})
(\rho_{10}+\rho_0)\nonumber \\
&=& \rho_{01}^{\ast}\rho_{10}+\rho_{01}^{\ast}\rho_0+\rho_1^{\ast}
\rho_{10}+\rho_1^{\ast}\rho_0
\label{decomp}
\eeqn
and consider the four terms seperately. For the first three terms the 
singularity in $\rho_0=\rho_1$ becomes integrable and the $\theta$-function
can be removed. We obtain for the first term
\beqn
\partial_1\partial_2^{\ast} 
\int d^2 \rho_0 \frac{(-1)}{\rho_1^{\ast}\rho_0}
\partial_0^{\ast}\partial_2 \Phi_{\om}(\rho_0,\rho_2)
=\pi \delta^{(2)}(\rho_1)|\partial_2|^2 
\Phi_{\om}(\rho_1,\rho_2)
\label{cancel}
\eeqn
where the rhs emerges after partial integration with 
respect to $\rho_0$ and the use of the following identities
($\Delta$ is the conventional twodimensional Laplace operator) 
\beqn
& &\Delta \log |\rho| = 4 \partial \partial^{\ast} \log 
|\rho|
= 2 \pi \delta^{(2)}(\rho) \\ 
& &\partial \frac{1}{\rho^{\ast}}=
\partial^{\ast} \frac{1}{\rho}=\pi \delta^{(2)}(\rho)
\label{ident}
\eeqn
The second and third term of (\ref{decomp}) are calculated in a 
similar way
\beqn
& &
\partial_1\partial_2^{\ast} 
\int d^2 \rho_0 
\left[\frac{1}{\rho_1^{\ast}\rho_{01}}+\frac{1}{\rho_0\rho_{10}^{\ast}}
\right] \partial_0^{\ast}\partial_2 \Phi_{\om}(\rho_0,\rho_2)
\nonumber \\
&=& \left[-\partial_1\frac{1}{\rho_1^{\ast}}+\frac{1}{\rho_1}
\partial_1^{\ast} \right]|\partial_2|^2 \Phi_{\om}(\rho_1,\rho_2)
\label{deco1}
\eeqn
The conjugate expression gives
\beqn
\left[-\partial_1^{\ast}\frac{1}{\rho_1}+\frac{1}{\rho_1^{\ast}}
\partial_1 \right]|\partial_2|^2 \Phi_{\om}(\rho_1,\rho_2)
\label{deco2}
\eeqn
After commutation of derivatives we obtain terms proportional to
$\delta^{(2)}(\rho_1)$-functions which cancel against (\ref{cancel});
the remaining terms in (\ref{deco1}) and (\ref{deco2}) add up to zero.
Consequently the only contribution comes from the fourth term
of (\ref{decomp}), i.\ e.\ after inversion the kernel reads
\beqn
|\rho_1|^4|\rho_2|^4\left[\frac{1}{2}
\partial_1\partial_2^{\ast}
\int \frac{d^2 \rho_0}{|\rho_{10}|^2}
\theta\left(\frac{|\rho_{10}||\rho_2|}{|\rho_{12}||\rho_0|}-
\epsilon\right)
\partial_0^{\ast}\partial_2 \Phi_{\om}(\rho_0,\rho_2)
\right. \nonumber \\
\left.
+\pi \log \epsilon |\partial_1|^2|\partial_2|^2
\Phi_{\om}(\rho_1,\rho_2)
+[\mbox{h.c.}] + [1 \leftrightarrow 2] \right]
\label{proof2}
\eeqn
Now we use the identity
\beqn
\int \frac{d^2 \rho_0}{|\rho_{10}|^2}
\theta\left(\frac{|\rho_{10}||\rho_2|}{|\rho_{12}||\rho_0|}
-\epsilon\right) \psi(\rho_0)=
\int \frac{d^2 \rho_0}{|\rho_{10}|^2}
\theta\left(\frac{|\rho_{10}|}{|\rho_{12}|}
-\epsilon\right) \psi(\rho_0)-2 \pi \log \frac{|\rho_2|}{|\rho_1|} 
\psi(\rho_1)
\label{theta}
\eeqn
Where $\psi$ is some testfunction
to eliminate the factor $|\rho_2|/|\rho_0|$ from the argument of the 
$\theta$-function. The extra term $\log |\rho_2|/|\rho_1|$ 
cancels due to its antisymmetry with respect to the exchange 
of $\rho_1$ and $\rho_2$. \\
After removing the overall factor $|\rho_1|^4|\rho_2|^4$ we recover the 
original expression (\ref{bfklconf}). This completes the proof of
conformal invariance of the Bethe-Salpeter equation in 
configuration space.
\subsubsection{Conformal partial wave expansion}
In the last section it was shown that the amplitude 
$\Phi_{\om}(\rho_1,\rho_2;\rho_{1'},\rho_{2'})$ can be written
in a conformally invariant way. It follows that the solutions 
of the Bethe-Salpeter equation can be classified according
to the irreducible representations of the conformal group
$SO(3,1)$. The general solution is obtained as an expansion
with respect to the representation functions of the 
corresponding irreducible representation, the conformal
partial wave expansion
\beqn
\Phi_{\om}(\rho_1,\rho_2;\rho_{1'},\rho_{2'})=
\sum_{n=-\infty}^{\infty} \int \frac{d\nu}{2 \pi}
\int d^2 \rho_0 \;\; 
E^{(\nu,n)}(\rho_{10},\rho_{20})
\Gamma^{(\nu,n)}(\rho_{1'0},\rho_{2'0};\om)
\label{pwexp}
\eeqn
The representation functions $E^{(\nu,n)}(\rho_{10},\rho_{20})$
are the (up to a constant) uniquely defined conformally
invariant three-point functions and will be given explicitly below.
The representation is labeled by the parameters 
$\nu \in \Bbb{R}$ and $n \in \Bbb{Z}$
which enumerate the unitary irreducible 
principal series representation of the group $SO(3,1)$.
The coordinate $\rho_0$ represents an additional quantum number.
For a detailed mathematical discussion we refer to \cite{mack} 
and references given therein.
The explicit form of the representation functions is
\beqn
E^{(\nu,n)}(\rho_{10},\rho_{20})=
c(\nu,n)
\left(\frac{\rho_{12}}{\rho_{10}\rho_{20}}\right)^{
\frac{1+n}{2}-i\nu}
\left(\frac{\rho_{12}^{\ast}}
{\rho_{10}^{\ast}\rho_{20}^{\ast}}\right)^{
\frac{1-n}{2}-i\nu}
\label{repfunc}
\eeqn
The constant $c(\nu,n)$ is undetermined and for the moment
set equal to unity. The functions $E^{(\nu,n)}(\rho_{10},\rho_{20})$
are the eigenfunctions of the two Casimir operators
$L^2=-\rho_{12}^2\partial_1\partial_2$,
$L^{\ast\,2}=-\rho_{12}^{\ast\,2}
\partial_1^{\ast}\partial_2^{\ast}$ of the global conformal 
group
\beqn
L^2E^{(\nu,n)}&=&h(h-1)E^{(\nu,n)} \nonumber \\
L^{\ast\,2}E^{(\nu,n)}&=&\bar{h}(\bar{h}-1)E^{(\nu,n)}\nonumber \\
h=\frac{1-n}{2}+i\nu &,& \bar{h}=\frac{1+n}{2}+i\nu 
\label{casimir}
\eeqn
where $h$ and $\bar{h}$ are the conformal weights of the 
representation.
For the representation 
functions $E^{(\nu,n)}(\rho_{10},\rho_{20})$ the 
following orthonormalization condition holds
\beqn
\int \frac{d^2\rho_1d^2\rho_2}{|\rho_{12}|^4}
E^{(\nu,n)}(\rho_{10},\rho_{20})
E^{(\mu,m)}(\rho_{10'},\rho_{20'})=
a_{\nu,n}\delta_{n,m}\delta(\nu-\mu)\delta^{(2)}(\rho_{00'})
\hspace{1.4cm}
\nonumber \\ \hspace{1cm}
+b_{\nu,n}\delta_{n,-m}\delta(\nu+\mu)
\frac{1}{|\rho_{00'}|^{2+4i\nu}}\left(
\frac{\rho_{00'}}{\rho_{00'}^{\ast}}\right)^n
\label{ortho}
\eeqn
The proof of this relation can be found in the 
appendix of \cite{lip86}. The coefficients $a_{\nu,n}$ and 
$b_{\nu,n}$ are determined as 
\beqn
a_{\nu,n}&=&\frac{2 \pi^2}{4 \nu^2+n^2}
\\
b_{\nu,n}&=& \pi^3 4^{2 i \nu}
\frac{\Gamma(-i\nu+\frac{|n|}{2})
      \Gamma(i\nu+\frac{|n|}{2})}
     {\Gamma(i\nu+\frac{|n|}{2})
      \Gamma(1-i\nu+\frac{|n|}{2})}
\eeqn
The appearance of the second term on the rhs of the 
orthonormalization relation is due to the fact that
the representations corresponding to the quantum 
numbers $[\nu,n]$ and $[-\nu,-n]$ are equivalent.
We have the following intertwining relation
\beqn
E^{(-\nu,-n)}(\rho_{10},\rho_{20})=
\frac{b_{n,-\nu}}{a_{n,\nu}}
\int d^2\rho_{0'}E^{(\nu,n)}(\rho_{10'},\rho_{20'})
|\rho_{00'}|^{-2+4i\nu}\left(\frac{
\rho_{00'}^{\ast}}{\rho_{00'}} \right)^n
\label{inter}
\eeqn
This relation can be proven with the methods used in 
appendix \ref{appeigen}.
By insertion of the conformal partial wave expansion 
into the Bethe-Salpeter equation and use of eqs. (\ref{ortho})
and (\ref{inter}) one can prove that the partial
wave $\Gamma(\rho_{1'0},\rho_{2'0};\omega)$ is proportional
to the conjugate representation function 
$E^{(\nu,n)\, \ast}(\rho_{1'0},\rho_{2'0})$.
It follows that the explicit solution of the Bethe-Salpeter 
equation can be written in the following form
\beqn
\Phi_{\om}(\rho_1,\rho_2;\rho_{1'},\rho_{2'}) =
\sum_{n=-\infty}^{\infty} \int_{-\infty}^{+\infty}
 \frac{d \nu}{2 \pi} 
\frac{16\nu^2 + 4 n^2}
{[4 \nu^2+(n-1)^2][4 \nu^2+(n+1)^2]}
\hspace{1cm} \nonumber \\
\cdot
\frac{1}{\omega-\chi(\nu,n)} \int d^2 \rho_0
E^{(\nu,n)}(\rho_{10},\rho_{20})
E^{(\nu,n)\,\ast}(\rho_{1'0},\rho_{2'0})
\label{solnonzero}
\eeqn
Here $\chi(\nu,n)$ is given by eq.\ (\ref{chi}) and corresponds to the
eigenvalue of the integral kernel acting on the representation
function $E^{(\nu,n)}$. Due to conformal invariance the 
eigenvalue has to be independent of the momentum transfer and hence 
has to coincide with the eigenvalue of the zero momentum transfer
equation.
An explicit calculation of this eigenvalue in configuration
space is performed in appendix \ref{appeigen}.
Note that for $n=\pm1$ the $\nu$-integral in the expression
above is understood in the sense of the principal value.
\subsubsection{Conformally invariant $n$-point functions}
In a general conformally invariant field theory there exists 
a set of primary fields $\phi_{h,\bar{h}}$ which transform under a conformal
transformation $\rho\rightarrow f(\rho),\rho^{\ast}\rightarrow
f^{\ast}(\rho^{\ast})$ according to
\beqn
\phi_{h,\bar{h}}(\rho,\rho^{\ast})
\rightarrow 
\left(\frac{\partial f}{\partial \rho}\right)^h
\left(\frac{\partial f^{\ast}}{\partial \rho^{\ast}}\right)^{\bar{h}}
\phi_{h,\bar{h}}(f(\rho),f^{\ast}(\rho^{\ast}))
\eeqn
where $h$ and $\bar{h}$ are the conformal weights of the field.
Conformal covariance then implies that the correlation functions
of the theory satisfy
\beqn
<\phi_{h_1,\bar{h}_1}(\rho_1,\rho_1^{\ast}) \cdots
\phi_{h_n,\bar{h}_n}(\rho_n,\rho_n^{\ast})> =
\prod_{i=1}^n\left(\frac{\partial f}{\partial \rho_i}\right)^{h_i}_{
|\rho=\rho_i}
\left(\frac{\partial f^{\ast}}{\partial \rho_i^{\ast}}
\right)^{\bar{h}_i}_{
|\rho^{\ast}=\rho_i^{\ast}} \nonumber 
\hspace{1cm} \\
<\phi_{h_1,\bar{h}_1}(f(\rho_1),f^{\ast}(\rho_1^{\ast})) \cdots
\phi_{h_n,\bar{h}_n}(f(\rho_n),f^{\ast}(\rho_n^{\ast}))>
\label{corr}
\eeqn
This covariance property imposes restrictions on the form of
the correlation functions \cite{polyakov}. The two-point functions 
and the three-point functions of the primary fields are fixed 
up to a constant
\beqn
<\phi_{h_1,\bar{h}_1}(\rho_1,\rho_1^{\ast}) 
\phi_{h_2,\bar{h}_2}(\rho_2,\rho_2^{\ast})> &=& 
\delta_{h_1,h_2}\delta_{\bar{h}_1,\bar{h}_2}
\frac{c_{h_1,\bar{h}_1}}{\rho_{12}^{2 h_1}\rho_{12}^{\ast\,2 \bar{h}_1}}
            \label{two}
\\
<\phi_{h_1,\bar{h}_1}(\rho_1,\rho_1^{\ast})
\phi_{h_2,\bar{h}_2}(\rho_2,\rho_2^{\ast})
\phi_{h_3,\bar{h}_3}(\rho_3,\rho_3^{\ast})> &=&
     \nonumber 
\eeqn
\begin{equation}
c_{h_1 h_2 h_3,\bar{h}_1\bar{h}_2\bar{h}_3}
\frac{1}{\rho_{12}^{h_1+h_2-h_3}\rho_{13}^{h_1+h_3-h_2}
\rho_{23}^{h_2+h_3-h_1}}
\frac{1}
{\rho_{12}^{\ast\,\bar{h}_1+\bar{h}_2-\bar{h}_3}
\rho_{13}^{\ast\,\bar{h}_1+\bar{h}_3-\bar{h}_2}
\rho_{23}^{\ast\,\bar{h}_2+\bar{h}_3-\bar{h}_1}}
\label{three}
\end{equation}
The higher $n$-point functions are not completely fixed by 
conformal covariance but have to obey certain constraints. 
The four-point function takes the form
\beqn
<\phi_{h_1,\bar{h}_1}(\rho_1,\rho_1^{\ast}) \cdots
\phi_{h_4,\bar{h}_4}(\rho_4,\rho_4^{\ast})> =
\Psi(x,x^{\ast})\prod_{i<j} \rho_{12}^{-h_i-h_j+\frac{1}{3}
\sum_i h_i} \rho_{12}^{\ast\, -\bar{h}_i-\bar{h}_j
+\frac{1}{3}\sum_i \bar{h}_i}
\label{four}
\eeqn
where $\Psi(x,x^{\ast})$ is an undetermined function 
of the anharmonic ratio $x=\frac{\rho_{12}\rho_{34}}
{\rho_{13}\rho_{24}}$ and its conjugate.
In general the $n$-point function
$(n \geq 4)$ contains an a priori undetermined 
function which depends on the $n(n-3)/4$ anharmonic ratios
which can be constructed from n coordinates and their conjugate.
\\
Let us now interpret the amplitudes which were calculated
in the preceeding sections in terms of the general
conformal n-point function. By comparing the conformal 
representation functions $E^{(\nu,n)}$ (\ref{repfunc}) with the general
three-point function (\ref{three}) we note that the former can be 
interpreted as a conformal three-point function build up 
from two fields with conformal weights $h=0,\bar{h}=0$ at the 
points $\rho_1,\rho_2$ and a field with conformal weights
$h=(1+n)/2-i\nu,\bar{h}=(1-n)/2-i\nu$ 
at the point $\rho_0$
\beqn
E^{(\nu,n)}(\rho_{10},\rho_{20})  = <\phi_{0,0}(\rho_1)\phi_{0,0}(\rho_2)
O_{h,\bar{h}}(\rho_0)>_{ {h=(1-n)/2+i\nu \choose \bar{h}=(1+n)/2+i\nu} }
\label{desy1}
\eeqn
We interpret $\phi_{0,0}$ as an elementary field representing
the reggeized gluon and $O_{}(\rho_0)$ as a field which represents
a composite state of two reggeized gluons which emerges from 
the solution of the dynamical equations of the theory.  
After having introduced an elementary field representing the
reggeized gluon it is natural to interpret the off-shell
scattering amplitude $\Phi_{\om}(\rho_1,\rho_2;\rho_{1'},\rho_{2'})$
as the four-point function of the field $\phi_{0,0}$. It then follows
from the general theory outlined above that this function can 
depend on the four points only as a function of the anharmonic
ratio. 
This was already shown for the Born-level approximation
of this fuction (\ref{born2}) and can be demonstrated for the 
exact four-point function (\ref{solnonzero}) by integrating over the 
coordinate $\rho_0$. For this 
calculation we limit ourselves to the case $n = 0$. 
Since $\Phi_{\omega}(\rho_1,\rho_2;\rho_{1'},\rho_{2'})$
is conformally invariant the $\rho_0$-integration can be simplified 
by choosing e.\ g.\ $\rho_{2'}=\infty$. Then we use the methods 
and results of \cite{dot} to perform the integration and 
restore the $\rho_{2'}$-dependence afterwards by requiring 
the correct transformation properties. In this way we obtain
\beqn
\Phi_{\om}(\rho_1,\rho_2;\rho_{1'},\rho_{2'}) = 
\int_{-\infty}^{+\infty}
\frac{d \nu}{2 \pi} \frac{16 \nu^2}{[4 \nu^2+1]^2}
\frac{\pi}{\omega-\chi(\nu,0)}
\phantom{xxxxxxxxxxxxxxxxxxxxxxxx}
\phantom{xxxxxxxxxxxxxxxxxxxxxxxx}
\!\!\!\!\!\!\!\!\!\!\!\!\!\!\!\!\!\!
\!\!\!\!\!\!\!\!\!\!\!\!\!\!\!\!\!\!
\!\!\!\!\!\!\!\!\!\!\!\!\!\!\!\!\!\!
\!\!\!\!\!\!\!\!\!\!\!\!\!\!\!\!\!\!
\\
\left[
\eta^{\fez+i\nu}\eta^{*\,\fez+i\nu}
\frac{\Gamma(2i\nu\!-\!1)\Gamma(2\!-\!2i\nu)}
     {\Gamma^2(1+2i\nu)}
\frac{\Gamma^2(\fez\!+\!i\nu)}{\Gamma^2(\fez\!-\!i\nu)}\,
_2F_1\left(\!\fez\!+\!i\nu\!,\!\fez\!+\!i\nu\!,\!1\!+\!2i\nu\!,
\!\eta\!\right)\,
_2F_1\left(\!\fez\!+\!i\nu\!,\!\fez\!+\!i\nu\!,\!1\!+\!2i\nu\!,
\!\eta^*\!\right)
\right. \nonumber \\ \left. 
+
\eta^{\fez-i\nu}\eta^{*\,\fez-i\nu}            
\frac{\Gamma(\!-\!2i\nu\!-\!1)\Gamma(2\!+\!2i\nu)}
     {\Gamma^2(1-2i\nu)}
\frac{\Gamma^2(\fez\!-\!i\nu)}{\Gamma^2(\fez\!+\!i\nu)}\,
_2F_1\left(\!\fez\!-\!i\nu\!,\!\fez\!-\!i\nu\!,\!1\!-\!2i\nu\!,\eta\right)\,
_2F_1\left(\!\fez\!-\!i\nu\!,\!\fez\!-\!i\nu\!,\!1\!-\!2i\nu\!,\eta^*\right)
\right]
\nonumber 
\label{new}
\eeqn
with the anharmonic ratios
\beqn
\eta = \frac{\rho_{12}\rho_{1'2'}}{\rho_{11'}\rho_{22'}}
\,,\,\eta^*= \frac{\rho_{12}^*\rho_{1'2'}^*}{\rho_{11'}^*\rho_{22'}^*}
\eeqn 
Eq.\ (\ref{new}) confirms that the four-point function 
of four fields associated with the reggeized gluon depends on 
the coordinates only through the two anharmonic ratios 
in agreement with the general theory.
The representation above could be useful 
in the study of the short-distance 
limits of the BFKL amplitude in configuration space.  
\subsection{The momentum space representation
of the conformal eigenfunctions}
\label{sec13}
In the preceding sections the non-forward 
Bethe-Salpeter equation was solved in configuration
space by conformal partial wave expansion. An interpretation in terms
of a conformal field theory in two dimensions was sketched.
Although the resulting expression looks elegant and compact 
it seems preferable for many reasons to work with a 
momentum space representation. \\
To this end we apply the inverse Fourier transformation
to the 
configuration 
space solution (\ref{solnonzero}) 
of the Bethe Salpeter equation.
The $\rho_0$-integration can be performed easily to give the
$\delta^{(2)}(\qf-\qf')$-function. The resulting expression
factorizes in two terms, one depending only
on $\kf$ and the other one depending only on $\kf'$.
The $\kf$-dependent term reads
\beqn
E^{(\nu,n)}(\kf,\qf-\kf)
=c(\nu,n) \int d^2 \rho_1 d^2 \rho_2 
e^{i \kf \rho_1+i(\qf-\kf)\rho_2}\left(
\frac{\rho_{12}}{\rho_1\rho_2}\right)^{\frac{1+n}{2}-i\nu}
\left(\frac{\rho_{12}^{\ast}}{\rho_1^{\ast}
\rho_2^{\ast}}\right)^{\frac{1-n}{2}-i\nu}
\label{mom1}
\eeqn
and corresponds to the momentum space representation of the 
conformal three-point function.
We reintroduced the constant $c(\nu,n)$ which will be adjusted
in the end to match the eigenfunction 
(\ref{eigenzero}) in the forward direction.
The $\kf$-factorization is of course broken by the $\nu$-integration
as will be seen in a moment.
In the following we restrict ourselves to the case $n=0$.
As we will show later, results for nonzero $n$ can be obtained in a 
very similar fashion.
Introducing center of mass 
coordinates $\rho=\rho_{12},R=\rho_1+\rho_2$,
we first obtain the mixed representation for the three-point function
\beqn
E^{(\nu,0)}(\rho,\qf) 
&=& c(\nu,0)
(|\rho|^2)^{\frac{1}{2}-i\nu} 
\int d^2 R \;e^{i\qf R}
\left(\frac{1}{|\rho+R|^2|\rho-R|^2}\right)^{\frac{1}{2}-i\nu}
\\
&=& c(\nu,0)
2 \pi \frac{4^{i\nu}}{\Gamma^2(\frac{1}{2}-i\nu)}
|\rho| (\qf^2)^{-i \nu}
\nonumber \\
& &\int_0^1 d x  [x(1-x)]^{-\frac{1}{2}} 
e^{-i \qf \rho(1-x)}K_{-2i \nu}(|\qf||\rho| 
\sqrt{x(1-x)})
\label{mix}
\eeqn
where $K$ denotes the modified Bessel function of the second kind.
Now we perform the $\rho$-integration using polar coordinates
which results in
\beqn
E^{(\nu,0)}(\kf,\qf-\kf) 
&=& c(\nu,0)
 (\qf^2)^{-i \nu}
    \int_0^\infty d |\rho|^2
|\rho|^{} 
J_0(|\rho||\kf-(1-x)\qf|) K_{-2 i \nu}(|\qf||\rho|\sqrt{x(1-x)})
\\ 
&=& c(\nu,0)
8 \pi^2 4^{i\nu}
\frac{\Gamma(\frac{3}{2}+i\nu)\Gamma(\frac{3}{2}-i\nu)}
{\Gamma^2(\frac{1}{2}-i\nu)}
(\qf^2)^{-\frac{3}{2}-i\nu}
\int_0^1 d x [x(1-x)]^{-2}
\,
\nonumber \\ 
& & 
\hspace{1cm} \cdot_2F_1\left(
\frac{3}{2}+i\nu,\frac{3}{2}-i\nu,1;
-\frac{x(1-x)\qf^2}{(\kf-x\qf)^2}
\right)
\\
&=& c(\nu,0)
8 \pi^2 4^{i\nu}
\frac{\Gamma(\frac{3}{2}+i\nu)\Gamma(\frac{3}{2}-i\nu)}
{\Gamma^2(\frac{1}{2}-i\nu)}
\int_0^1 d x [x(1-x)]^{-\frac{1}{2}+i\nu}
\left[ x(1-x)\qf^2+(\kf-x\qf)^2\right]^{-\frac{3}{2}-i\nu} 
\,
\nonumber \\
& & \hspace{1cm}
\cdot_2F_1\left( 
\frac{3}{2}+i\nu,-\frac{1}{2}+i\nu,1;
\frac{ (\kf-x\qf)^2}
{x(1-x)\qf^2+(\kf-x\qf)^2}
\right)
\label{momspacef}
\eeqn
where in the last line a transformation of the hypergeometric function
was used which leading to a result in which the argument of the 
$_2F_1$-function is bounded by 1 from above for every value of $\kf,\qf$
and $x$. By performing the transformation $x \to 1-x$ we 
establish the symmetry w.\ r.\ t.\ the exchange of $\kf$ and $\qf-\kf$.
Now we want to recover the result for the forward direction 
($\qf=0$) from this expression. In the limit $\qf=0$ the argument
of the hypergeometric function moves on the unit circle
where the convergence properties of the hypergeometric
series depend on the values of the first three arguments. 
To make this more 
transparent we use an analytic transformation \cite{bateman} 
for the 
hypergeometric function which gives in the limit $\qf \to 0$
\beqn
_2F_1\left( 
\frac{3}{2}+i\nu,-\frac{1}{2}+i\nu,1;
\frac{ (\kf-x\qf)^2}
{x(1-x)\qf^2+(\kf-x\qf)^2}
\right) 
\phantom{xxxxxxxxxxxxxxxxxxxxxxxxxxxxxx}
\nonumber \\
=
\left[
\frac{\Gamma(-2 i \nu)}
     {\Gamma(-\frac{1}{2}-i\nu)\Gamma(\frac{3}{2}-i\nu)}
+ 
\left(
x(1-x)
\frac{\qf^2}{\kf^2}
\right)^{-2 i \nu} 
\cdot
\frac{\Gamma(2i\nu)}
     {\Gamma(\ftz+i\nu)\Gamma(-\fez+i\nu)}
\right]
\left[1 + O(\qf^2)\right]
\eeqn
Inserting this into eq.\ (\ref{momspacef}) we find 
\beqn
E^{(\nu,0)}(\kf,\qf-\kf) 
= c(\nu,n) 4 \pi^2 
(\kf^2)^{-\ftz-i\nu}
\left[
4^{-i\nu}\frac{\Gamma(-i\nu)}{\Gamma(1+i\nu)}
\frac{\Gamma(\ftz+i\nu)\Gamma(\fez+i\nu)}
     {\Gamma(\fez-i\nu)\Gamma(-\fez-i\nu)}
\right. \nonumber \\ \left.
+\left(\frac{\qf^2}{\kf^2}\right)^{-2 i\nu}
4^{i\nu}\frac{\Gamma(i\nu)}{\Gamma(1-i\nu)}
\frac{\Gamma(\ftz-i\nu)\Gamma(\fez+i\nu)}
     {\Gamma(\fez-
i\nu)\Gamma(-\fez+i\nu)}
\right]
\left[
1+O(\qf^2)
\right]
\label{limzero}
\eeqn
We choose the normalization $c(\nu,n)$ to cancel 
the first term in square brackets
\beqn
E^{(\nu,0)}(\kf,\qf-\kf)
=  4 \pi^2
(\kf^2)^{-\ftz-i\nu}
\left[
1+ 
4^{2 i\nu}\left(\frac{\qf^2}{\kf^2}\right)^{-2 i\nu} 
\frac{\Gamma(i\nu)\Gamma(1+i\nu)}
     {\Gamma(-i\nu)\Gamma(1-i\nu)}
\frac{\Gamma(\ftz-i\nu)\Gamma(-\fez-i\nu)}
     {\Gamma(\ftz+i\nu)\Gamma(-\fez+i\nu)}
\right]
\label{limzero2}
\eeqn
From this representation we conclude that the 
the momentum space eigenfunction $E^{(\nu,0)}$ 
has a well 
defined limit in the forward direction only if 
$Im(\nu) > 0$. In the opposite case, 
$Im(\nu) < 0$, the function $E^{(\nu,0)}$ becomes singular
in the forward direction. For the conjugate eigenfunction
$E^{(\nu,0) \, \ast}$ the converse is true. 
Thus the $\qf=0$-limit of the product $E^{(\nu,0)}E^{(\nu,0)\, \ast}$
which enters into the momentum space expression of the BFKL amplitude
seems to be ill-defined in the whole strip $ -1 < Im(\nu) < 1 $
in which the coefficient of the second term 
in eq. (\ref{limzero}) and its conjugate
are analytic functions of $\nu$.
However, according to this equation  
in the limit $\qf \to 0$ the product 
$E^{(\nu,0)}E^{(\nu,0)\, \ast}$
can be decomposed
into a sum of four terms
and we can deform the contour of the $\nu$-integration for each 
term seperately. Explicitly we have
\beqn
E^{(\nu,0)}(\kf,\qf-\kf)E^{(\nu,0)\, \ast}(\kf',\qf-\kf')
=
\frac{16 \pi^4}{\nu^2} 
\left(\nu^2+\frac{1}{4}\right)^2 
(\kf^2)^{-\ftz-i\nu}(\kf{'}^2)^{-\ftz+i\nu}
\nonumber \\
\left[ 1 + \left(\frac{\kf^2}{\kf{'}^2}\right)^{2i\nu}
+
4^{ 2 i\nu}
\left(\frac{\qf^2}{\kf^2}\right)^{-2 i \nu}
C(\nu)
+
4^{- 2 i\nu}
\left(\frac{\qf^2}{\kf{'}^2}\right)^{2 i \nu}
C^*(\nu)
\right]
\left[1 + O(\qf^2)
\right]
\eeqn
with $C(\nu)$ representing the combination of $\Gamma$-functions
in eq.\ (\ref{limzero2}).
For the third and the fourth term
we can shift the integration contour to the lower, resp.\ upper half 
of the complex $\nu$-plane 
This shift is legitimate
due to the analyticity properties 
of $C(\nu)$.
After this deformation the limit
$\qf \to 0$ of these terms can be performed and yields zero.
The first and the second term are identical since in the second 
term we can shift from $\nu$ to $-\nu$. 
With this prescription we end up with
\beqn
E^{(\nu,0)}(\kf,-\kf)E^{(\nu,0)\, \ast}(\kf',-\kf')
=
2 \frac{16 \pi^4}{\nu^2} 
\left(\nu^2+\frac{1}{4}\right)^2 
(\kf^2)^{-\ftz-i\nu}(\kf{'}^2)^{-\ftz+i\nu}
\eeqn
The $\nu$-dependent prefactors cancel after inserting this into 
eq.\ (\ref{solnonzero}). To match the solution in the forward 
direction we finally have to divide by $4\pi^2$. This is done by absorbing 
an additional factor $\sqrt{2}/(2\pi)$ in $c(\nu,0)$
Consequently the normalization factor $c(\nu,0)$ is determined as    
\beqn
c(\nu,0) = \frac{\sqrt{2}}{2 \pi}4^{i\nu}\frac{\Gamma(1+i\nu)}{\Gamma(-i\nu)}
\frac{\Gamma(\fez-i\nu)\Gamma(-\fez-i\nu)}
{\Gamma(\ftz+i\nu)\Gamma(\fez+i\nu)}
\label{bali}
\eeqn
For the $\qf=0$-limit of the eigenfunctions we prescribe
to neglect the $(\qf^2)^{-2 i \nu}$-term 
in eq.\ (\ref{limzero})
by choosing $Im(\nu)>0$.
With this effective prescription we reproduce the correct result.
When this normalization is used for the eigenfunctions in momentum space
the factor $\nu^2/(\nu^2+1/4)^2$ which appears in the solution of the 
configuration space equation has to be omitted. It is absorbed into 
the eigenfunctions.
\\
In eq.\ (\ref{momspacef}) we have realized a rather compact form 
of the momentum space expression of the conformal three-point
function for $n=0$. We were not able to give a similar representation
for $n \neq 0$.
Of course for $n \neq 0$ the momentum space expressions can
be obtained by linear combination of 
functions generated from
expressions similar to
eq.\ (\ref{momspacef}) by partial differentiation.
Switching from complex to cartesian coordinates
in configuration space, for $n = \pm 1$ we can construct
the linear combinations
\beqn
\frac{1}{2}\left[
E^{(\nu,+1)}\pm
E^{(\nu,-1)}
\right](\rho_1,\rho_2)
=
\frac{(|\rho_{12}|)^{-i\nu}}{(|\rho_1|^2|\rho_2|^2)^{1-i\nu}}
\left[
{\rho_2^x \choose -i\rho_2^y}|\rho_1|^2 +
{-\rho_1^x \choose i \rho_1^y}|\rho_2|^2
\right]
\eeqn 
Changing from cartesian components to partial derivatives 
we find the momentum space representation of the 
above expression
\beqn
\frac{1}{2} \!
\left[
E^{(\nu,+1)} \!\pm \!E^{(\nu,-1)}
\right] \!\!
(\kf,\qf-\kf)
\!=\!
-2i\, 
c(\nu,1)\!
\left[\!
{\partial_{\qf_x} \choose -i\partial_{\qf_y} } 
I^{(\nu,1)}_0\!(\kf,\qf-\kf)
\!+\!
{-\partial_{\kf_x}-\partial_{\qf_x} \choose i 
 \partial_{\kf_y}+i\partial_{\qf_y} }
I^{(\nu,1)}_1\!(\kf,\qf-\kf)\!
\right]
\label{neqone}
\eeqn
where we have introduced the function
\beqn
I^{(\nu,1)}_m(\kf,\qf-\kf)=
4 \pi^2 4 ^{i\nu} \frac{\Gamma(1+i\nu)}{\Gamma(-i\nu)}
\int_0^1 d x x^{i\nu-1+m}(1-x)^{i\nu-m}
[\qf^2x(1-x)+(\kf-x\qf)^2]^{-1-i\nu}
\nonumber \\
_2F_1\left(1+i\nu,i\nu,1;\frac{(\kf-x\qf)}{\qf^2x(1-x)+(\kf-x\qf)^2}
\right)
\eeqn
It is clear that representations of this type can also
be generated for higher $n$ but the explicit expressions 
become rather complicated. 
For use at a later stage of this work it will still be useful
to know explicitly the normalization factors 
$c(\nu,n)$ for arbitrary $n$. For $n=1$ this factor can be found 
by performing first the differentiations in eq.\ (\ref{neqone})
and following the steps described above for $n=0$. 
\subsubsection{The expansion in powers of the momentum transfer}
There is an alternative approach which allows to obtain a closed
formula for the factor $c(\nu,n)$. 
It turns out that it is possible to compute the complete
expansion of $E^{(\nu,n)}(\kf,\qf-\kf)$ in $|\qf|$ 
for arbitrary $n$. From this expansion $c(\nu,n)$ is obtained
by normalizing the zero-order contribution.
\\
To construct the expansion we go back to the definition of the momentum
space representation in eq.\ (\ref{mom1})
\beqn
E^{(\nu,n)}(\kf,\qf-\kf)
= c(\nu,n)
\int d^2 \rho \;e^{i\kf \rho}(\rho^2)^{\fez-i\nu}
\left(\frac{\rho}{\rho^*}\right)^{\frac{n}{2}}
\int d^2 \rho_2 \frac{e^{i\qf\rho_2}}
{(\rho_2^2(\rho+\rho_2)^2)^{\fez-i\nu}}
\left[\frac{\rho_2^*(\rho+\rho_2)^*}{\rho_2(\rho_2+\rho)}
\right]^{\frac{n}{2}}
\label{socke}
\eeqn
It is essential to realize that there are 
two regions in the $\rho_2$-integration which give 
contributions of the same order in the limit $\qf \to 0$.
The first region is $|\rho_2| > |\rho|$ and the coefficients
are obtained by first expanding the integrand of the $\rho_2$ integral in 
powers of $|\rho|/|\rho_2|$ and then integrating the coefficients
over $\rho_2$ and $\rho$. The second region is 
$|\qf| < |\rho_2|$ and the coefficients are obtained by first 
expanding the exponential in powers of $|\qf|$
and integrating then again the coefficients over $\rho_2$ and $\rho$.
As to the first region we obtain from the $\rho_2$-integration
\beqn
c(\nu,n)\int d^2 \rho \;e^{i\kf\rho}(|\rho^2|)^{\fez-i\nu}
\left(
\frac{\rho}{\rho^*}
\right)^{\frac{n}{2}}
\sum_{M=0}^{\infty}(|\qf||\rho|)^M 
\sum_{m=0}^M
\left(\frac{q}{q^*}\right)^{\frac{M}{2}-m-n}
\left(\frac{\rho}{\rho^*}\right)^{m-\frac{M}{2}}
\beta_{M,m}^{(\nu,n)}
(\frac{q}{q^*},\frac{\rho}{\rho^*})
\eeqn
with
\beqn
\beta_{M,m}^{(\nu,n)}
(\frac{q}{q^*},\frac{\rho}{\rho^*})
=
\pi 4^{2 i \nu-\frac{M}{2}}
\frac{(-1)^M}{\Gamma(1+M)}\,
i^{|2n+2m-M|}
{M \choose m}
\phantom{xxxxxxxxxxxxxxxxxxxxxxxx}
\nonumber \\
\cdot
\frac{\Gamma(\fez-i\nu+\frac{n}{2}+m)\Gamma(\fez-i\nu-\frac{n}{2}+M-m)}
     {\Gamma(\fez-i\nu+\frac{n}{2})\Gamma(\fez-i\nu-\frac{n}{2})}
\frac{\Gamma(2 i\nu+|n+m-\frac{M}{2}|-\frac{M}{2})}
     {\Gamma(1- 2 i\nu+|n+m-\frac{M}{2}|+\frac{M}{2})}
\eeqn
As to the second integration region we expand the exponential 
function in the $\rho_2$-integral, use a complex represenation for
the scalar product $\qf\rho=1/2(q\rho^* + q^*\rho)$ and expand the higher
orders of this expression using the binomial theorem.
This gives us in each order of $(|\qf|^2)^M$ a sum over $M+1$
contributions of the form
\beqn
\int d^2 \rho_2
\left(\frac{1}{\rho_2}\right)^{\fez-i\nu+\frac{n}{2}-(M-m)}
\left(\frac{1}{\rho_2^*}\right)^{\fez-i\nu-\frac{n}{2}-m}  
\left(\frac{1}{\rho_2+\rho}\right)^{\fez-i\nu+\frac{n}{2}}
\left(\frac{1}{\rho_2^*+\rho^*}\right)^{\fez-i\nu-\frac{n}{2}}
\label{dotint}
\eeqn
This integral can be calculated by reducing it to a product of two
contour integrals, a method which is also used in appendix \ref{appeigen}
to calculate the eigenvalues of the BFKL-kernel and is described in detail
in \cite{dot}.
After introducing the rescaled variables $\sigma=\rho_2/\rho,\sigma^*=
\rho_2^*/\rho^*$ one performs a Wick rotation
$\sigma_y \to i\exp(-2i\epsilon)\sigma_y \simeq i(1-2i\epsilon)\sigma_y$
and turns over to light-cone coordinates $\sigma_+=\sigma_x+\sigma_y,
\sigma_-=\sigma_x-\sigma_y$. The integrals over 
$\sigma_+$ and $\sigma_-$ then factorize up to terms of order 
$\epsilon$ which define the way in which the contours of
integration by-pass the singular points. 
Deformation of the contours then leads to the product 
of two integrals which reduce to $\Gamma$-functions.
Eq.\ ({\ref{dotint}) can be expressed as
\beqn
\pi (|\rho|^2)^{2 i \nu} \rho^M
\left(\frac{\rho^*}{\rho}\right)^{n+m}
(-1)^{M+n}
\frac{\Gamma(-2i\nu-n-m)}{\Gamma(1+2i\nu-n+M-m)}
\frac{\Gamma(\fez+i\nu-\fnz)\Gamma(\fez+i\nu-\fnz+M-m)}
     {\Gamma(\fez-i\nu-\fnz)\Gamma(\fez-i\nu-\fnz-m)}
\eeqn
This leads to the following contribution to the expansion 
of eq.\ (\ref{socke})
in $|\qf|$
\beqn
c(\nu,n)\int d^2 \rho \,
e^{i\kf\rho}(|\rho^2|)^{\fez+i\nu}
\left(\frac{\rho}{\rho^*}\right)^{\fnz}
\sum_{M=0}^{\infty}(|\qf||\rho|)^M
\sum_{m=0}^{M}
\left(\frac{q}{q^*}\right)^{m-\frac{M}{2}}
\left(\frac{\rho}{\rho^*}\right)^{-m-n+\frac{M}{2}}
\alpha_{M,m}^{(\nu,n)}(\frac{q}{q^*},\frac{\rho}{\rho^*})
\eeqn
with coefficients
\beqn
\alpha_{M,m}^{(\nu,n)}(\frac{q}{q^*},\frac{\rho}{\rho^*})
=\frac{\pi}{\Gamma(1+M)}\left(\frac{i}{2}\right)^M (-1)^{M+n}
{ M \choose m}
\phantom{xxxxxxxxxxxxxxxxxxxxxxxx}
\nonumber \\
\cdot
\frac{\Gamma(-2i\nu-n-m)}{\Gamma(1+2i\nu-n+M-m)}
\frac{\Gamma(\fez+i\nu-\fnz)\Gamma(\fez+i\nu-\fnz+M-m)}
     {\Gamma(\fez-i\nu-\fnz)\Gamma(\fez-i\nu-\fnz-m)}
\eeqn
It remains to perform the $\rho$-integration. This is straightforward
and we obtain the complete expansion of the momentum space eigenfunction
in powers of $|\qf|/|\kf|$ for arbitrary $n$.
\beqn
E^{(\nu,n)}=c(\nu,n)(|\kf|^2)^{-\ftz-i\nu}
\left(\frac{k}{k^*}\right)^{-\fnz}
\sum_{M=0}^{\infty}
\left(\frac{|\qf|}{|\kf|}\right)^M
\sum_{m=0}^M
\left(\frac{k}{k^*}\right)^{-m+\frac{M}{2}}
\left(\frac{q}{q^*}\right)^{m-\frac{M}{2}}
\phantom{xxxx}
\nonumber \\
\cdot \left[ 
a_{M,m}^{(\nu,n)}
+
\left(\frac{\qf^2}{\kf^2}\right)^{-2i\nu}
\left(\frac{q^* k}{q k^*}\right)^{n}
b_{M,m}^{(\nu,n)}
\right] 
\eeqn
with coefficients
\beqn
a_{M,m}^{(\nu,n)}&=&
2 \pi 4^{1+i\nu+\frac{M}{2}}
(-1)^{|\fnz-m+\frac{M}{2}|}
\frac{\Gamma(\ftz+i\nu+\frac{M}{2}+|\fnz-m+\frac{M}{2}|)}
     {\Gamma(-\fez-i\nu-\frac{M}{2}+|\fnz-m+\frac{M}{2}|)}
\alpha_{M,m}^{(\nu,n)}
\nonumber \\
b_{M,m}^{(\nu,n)}&=&
2 \pi 4^{1-i\nu+\frac{M}{2}}
(-1)^{|\fnz+m-\frac{M}{2}|}
\frac{\Gamma(\ftz-i\nu+\frac{M}{2}+|\fnz+m-\frac{M}{2}|)}
     {\Gamma(-\fez+i\nu-\frac{M}{2}+|\fnz+m-\frac{M}{2}|)}
\beta_{M,m}^{(\nu,n)}
\label{guhl}
\eeqn
For $n=0$ and low orders in $M$ this can be compared with the 
expansion which can be worked out starting from 
eq.\ (\ref{momspacef}).
\\
A comment is in order concerning the existence of the integrals 
which determine the coefficients of the expansion.
These integrals converge only in a strip in the complex $\nu$-plane
the location of which depends on the order $M$. The value  
outside this strip is obtained from the result of the integration
by analytical continuation.
\\ 
By repeating the same line of arguments as given for the case $n=0$ we 
determine the normalization factor $c(\nu,n)$ for general $n$ to be
\beqn
c(\nu,n)=i^n\frac{\sqrt{2}}{2 \pi}4^{i \nu}
\frac{\Gamma(1+i\nu+\frac{|n|}{2})}{\Gamma(-i\nu+\frac{|n|}{2})}
\frac{\Gamma(-\fez-i\nu+\frac{|n|}{2})\Gamma(\fez-i\nu+\frac{|n|}{2})}
{\Gamma(\fez+i\nu+\frac{|n|}{2})\Gamma(\ftz+i\nu+\frac{|n|}{2})}
\eeqn
For the $\qf=0$-limit of the eigenfunction we use the same prescription 
as before, namely to give $\nu$ a small imaginary part.
The second term in brackets in eq.\ (\ref{guhl}) then vanishes in the limit 
$\qf=0$.
%
\newpage
\setcounter{equation}{0}
\setcounter{figure}{0}
\setcounter{table}{0}
\section{Phenomenology of the BFKL Pomeron
in Deep Inelastic Scattering}
\label{chap2}
The past few years have seen a renewed interest in the BFKL 
theory from the phenomenological point of view.
In deep inelastic scattering (DIS) QCD perturbation theory,
in particular the resummation of large logarithms
of $Q^2$, has been successfully applied for a long time.
It was then realized that with DIS entering the regime 
of small Bjorken-$x$ there might be the need to resum also 
logarithms of $1/x$ which - in leading order approximation -
is done by the BFKL equation.
Processes at small $x$ - which corresponds to large 
photon-proton cms energy - are traditionally described within
Regge theory. In Regge theory scattering processes are mediated by 
Regge trajectories - singularities of the partial wave amplitude
in the complex angular momentum plane. 
The leading singularity at high energies - the pomeron - 
has been regarded as an object of nonperturbative origin in QCD.
Since, on the other hand,
DIS is successfully described within perturbation 
theory, the question arises if in DIS at small $x$ a novel object,
a 'hard' pomeron appears, which emerges from the resummation of 
large logarithms in $1/x$ within perturbation theory.
To leading order this 'hard' pomeron has to be identified with the
BFKL pomeron. 
In this chapter the idea that in DIS at small $x$
large logarithms of $1/x$ have to be resummed is pursued.
\\ 
The BFKL resummation is first used to determine the inclusive 
DIS structure function $F_2$.
The conventional approach towards hard scattering processes in DIS
is based on the collinear factorization theorem which allows to seperate 
perturbative and nonperturbative contributions.
In order to resum the logarithms in 
$1/x$ this framework has to be extended to the 
more general formalism of high energy factorization.
The key difference is that in the latter the gluon 
\footnote{The quark distribution has not been considered in this 
framework up to now.}
distribution of
the proton becomes transverse momentum dependent and  
a convolution in transverse momentum space has to be carried out.
In section \ref{sec21} we 
give a short introduction to collinear and high energy factorization
and show how the BFKL equation can be used to calculate the 
unintegrated (transverse momentum dependent) gluon density.
\\
In the transverse momentum convolution the integration is performed 
also over nonperturbative scales. This raises the question 
of the consistency of this approach since it uses perturbative
results in a region where perturbation theory is not applicable. 
Even worse, it can be shown that in principle one has to expect a 
finite contribution from this region. The logarithms in $1/x$ are built
up in the multi-Regge region of phase space to which hard and low
scales contribute equally. The evolution which resums these
contributions can be regarded as a diffusion process in transverse 
momentum space.
The center of the diffusion is set by the characteristic external scales 
of the process. 
In inclusive DIS there are two scales, the virtuality $Q^2$ 
of the photon and the inverse size $1/R_P^2$ of the proton. 
The diffusion process which starts from the proton side
is centered around $Q_0^2= 1/R_P^2$ in the nonperturbative region
and consequently the part of the phase space which is treated incorrectly
in the BFKL theory gives a substantial contribution.
This dependence on the nonperturbative region is investigated in detail 
in section \ref{sec21}.  
We will furthermore discuss modifications of the infrared region of the 
BFKL equation and show their effect on the evolution 
and the results for $F_2$.
An additional part of  section \ref{sec21} is devoted to the transverse 
energy distribution of the gluons which evolve according to the BFKL equation.
The measurement of the transverse energy distribution has been advocated
as a possible 'footprint' of BFKL dynamics, especially of the 
characteristic gaussian
distribution of transverse momenta.
\\ \\
The sections \ref{sec22} - \ref{sec24} deal with exclusive 
processes in deep inelastic scattering.
In section \ref{sec22} diffractive production of vector mesons
at large momentum transfer $t$ and small $x$ is investigated.
This process deserves interest for two reasons.
First, for large $t$ the physical picture of the BFKL evolution changes
drastically. The scale $\qf^2=-t$ acts as a lower cutoff for the 
diffusion in transverse momentum space, i.\ e.\ we have diffusion
with a boundary. This means that the serious infrared problem which the 
application of BFKL to inclusive DIS ($F_2$) faces is absent here and the 
BFKL prediction has a much better theoretical foundation.
Second, this process allows to investigate the $t$-dependence of 
the hard pomeron trajectory. It was stated in the first chapter of this 
work that due to conformal symmetry the BFKL singularity is a fixed cut,
i.\ e.\ there is no $t$-dependence.
The conformal symmetry is broken when the BFKL amplitude is 
convoluted with the impact factors of external particles 
and in principle a $t$-dependence of the trajectory can be observed.
We parametrize such a dependence in terms of the slope of the trajectory
and derive an effective slope for the amplitude under consideration.
\\
In section \ref{sec23} we turn to diffractive dissociation of the 
photon in DIS. Diffractive dissociation is well-known from 
hadron-hadron scattering and is traditionally described in terms of Regge
singularity exchange. Given the presence of the large scale $Q^2$ 
one can ask again to which extent QCD perturbation theory 
can describe these processes. The perturbative analysis of 
diffractive dissociation is fairly complicated especially if 
one turns to large masses of the diffractively produced system and 
consistent approximation schemes exist only for specific limiting 
cases. In the simplest case the virtual photon dissociates into a 
quark-antiquark pair which scatters off the proton. We will use 
the BFKL pomeron to describe
this scattering at large and zero momentum transfer.
The result for this simplest case shows some specific properties 
which are expected to hold also when more complicated final states are 
taken into account. From the phenomenological point of view
the results for zero momentum transfer are not so interesting 
since it has been shown that in the diffractive case the infrared 
problem of the BFKL equation is even more severe than in inclusive DIS.
It turns out that the effective scale at the photon side is much lower
than the virtuality $Q^2$.
Therefore 
the BFKL evolution is driven deeply into 
the nonperturbative domain. From the theoretical point of view
these calculations are nevertheless important since they represent the 
first steps towards the perturbative unitarization of the BFKL pomeron.
For finite momentum transfer the theoretical foundation of the 
perturbative results is better. For this case we discuss some subtleties
concerning the coupling of the BFKL pomeron to quarks.
\\
We will then generalize the calculation to include also  
more complicated final
states. Use will be made of results which were obtained for the triple
Regge limit in which the final state mass $M^2$ is very large.
We will show that the $q\bar{q}$-calculation can easily be extended
and we derive interesting results for the 
energy dependence of the cross section
in the zero momentum transfer limit. These results can be shown to follow 
from a conservation law for the conformal dimension of the BFKL pomerons 
at an effective triple pomeron vertex.
In this last part the analysis will, however, remain 
incomplete in so far as we will 
not consider a certain group of 
contributions which is beyond theoretical reach 
at the moment.     
\\
In section \ref{sec24} we turn to exclusive photon diffractive dissociation.
It has been shown that a large momentum scale in the 
diffractively produced 
final state acts as a hard scale at the effective photon pomeron vertex
in diffraction. In this case the diffusion scenario of the BFKL evolution
is comparable to inclusive DIS with the proton scale $Q_0^2$ at the 
lower end and the hard scale at the upper end.
Examples for such hard scales are a heavy vector meson mass, a large
transverse momentum of the produced quark-antiquark pair or the mass 
of a produced charm quark. We will concentrate on the last two cases,
namely we will consider the production of $q\bar{q}$-pairs with 
either large transverse momentum or large mass. Our starting point 
will be the exchange of the BFKL pomeron and we will show how 
to apply high energy factorization in this context.
It is known that in the double logarithmic limit in which the large 
logarithms in $Q^2$ and $1/x$ are resummed the collinear factorization 
and the high-energy factorization coincide. This limit provides a 
very accurate approximation to the data on inclusive $F_2$ at small $x$.
Based upon this observation we perform the double logarithmic limit
of the leading-log$(1/x)$ results. This allows us to express the 
cross section for the processes under consideration in terms of the 
conventional gluon density and in turn to obtain a parameter free 
prediction. Several properties of the result are studied in detail
numerically and predictions for event rates are given. 
The azimuthal dependence which is obtained within our two-gluon model
might be of special interest since a remarkable difference to the 
azimuthal dependence of the photon-gluon fusion process is found.
\\
It has been argued that the process of diffractive heavy vector
meson production might serve as a tool to constrain the gluon density
since the square of
this quantity enters the cross section. 
The same is true for diffractive $q\bar{q}$-production 
and this process could  
even be better suited for that purpose since ambiguities associated 
with the undetermined vector meson wave function are absent.
\newpage
\subsection{Inclusive Scattering at small $x$:
Collinear and High Energy Factorization}
\label{sec21}
\subsubsection{Collinear factorization}
The well-established framework to study deep inelastic 
scattering or more general hard processes
in perturbative QCD is the formalism of collinear factorization
\cite{collfac}.
This formalism applies whenever there is a strong interaction process
with one large momentum scale $Q^2$ at which the coupling 
constant $\alpha_s$ of QCD is small.
Collinear factorization then allows a clean separation of the 
contribution of large (hard) scales, for which perturbation theory is
applicable from the contribution of low (soft) scales which are determined
by nonperturbative dynamics of QCD.
This separation is made explicit by representing the cross section
of the process as a convolution of a hard quark- or gluon-subprocess
cross section with a universal probability distribution of 
finding a quark (gluon) in the hadron which enters the process.
The master formula of collinear factorization for DIS processes 
with hard scale $Q^2$ reads
\footnote{Here we have set the factorization scale $\mu^2$ 
equal to the
hard scale $Q^2$ for simplicity.}
\beqn
F(x,Q^2) = \sum_{i} \int_x^1 d z\; 
C_i(\alpha_s(Q^2),\frac{x}{z},Q^2)
f_i(z,Q^2)
\label{facttheorem}
\eeqn
In this expression the function $C_i$ represents the hard subprocess
cross section for an incoming parton of type $i$.
These coefficient functions are calculable in perturbative QCD as a 
power series in $\alpha_s$ and have been calculated for the 
important processes at least to next-to-leading order.
The quantity $f_i(z,Q^2)$ represents the probability to pick the parton of 
type $i$ with longitudinal momentum fraction $z$ and virtuality 
$\leq Q^2$ out of the hadron.
In the following we prefer the moment space representation of the
factorization formula which unfolds the $z$-convolution
\beqn
F(N,Q^2) = \sum_{i} C_i(\alpha_s(Q^2),N,Q^2) f_i (N,Q^2) \;\;,
\mbox{with}\;\;
F(N,Q^2) = \int_0^1 dx x^{N-1} F(x,Q^2)
\label{factthmom}
\eeqn
The scale dependence of the parton distribution functions is
governed by the renormalization group equations, the so-called 
DGLAP evolution equations \cite{dglap}, which read in the moment 
representation
\beqn
\frac{d}{d \log Q^2} f_i(N,Q^2) = \sum_j \gamma_{i j}(\alpha_s(Q^2),N)
f_j(N,Q^2)
\label{dglap}
\eeqn
The important quantity entering here is the anomalous dimension 
matrix $\gamma_{ij}(N)$ which controls the scaling violations 
of the process. The scale dependence of the parton densities which follows
from eq.\ (\ref{dglap}) is
\footnote{For simplicity the solution with fixed $\alpha_s$ is given.}
\beqn
f_k(N,Q^2)= f_k(N,Q_0^2)\left(\frac{Q^2}{Q_0^2}\right)^
{\hat{\gamma}_k(\alpha_s,N)}
\label{scalingviolations}
\eeqn
with $\hat{\gamma}_k(\alpha_s,N)$ being the eigenvalues 
of the anomalous dimension matrix
and the index $k$ referring to the corresponding eigenvectors.
The elements of the anomalous dimension matrix are also computable 
perturbatively as a power series in $\alpha_s$ and are known
explicitly up to ${\cal O}(\alpha_s^2)$.
The nonperturbative contribution enters into the formalism as the inital 
condition of the coupled evolution equations (\ref{dglap})
at a lower momentum scale $Q_0^2$. These input distributions have to
be obtained from experimental data
by means of global analysis.
With coefficient functions and anomalous dimensions evaluated in 
leading order the above formalism can be shown to resum the leading 
logarithms in $Q^2$ (the hard scale), i.\ e.\ all perturbative 
contributions of the form $\left(\alpha_s \log (Q^2/Q_0^2)\right)^n$ 
are taken into account. This is due to the fact that the evolution 
equations (\ref{dglap}) resum logarithmic contributions 
(collinear logarithms) of an infinite number of Feynman diagrams.
In a physical gauge these diagrams correspond to ladder diagrams
with strong ordering of parton transverse momenta along the ladder.
\\
If, for fixed $Q^2$, the variable $x$ is decreased one approaches the 
Regge limit of DIS. Using the inversion formula for the 
moments of parton densities one notices that the leading contribution 
in the limit of small $x$ is determined through the part of the 
gluon anomalous dimension which is singular
\footnote{The effect of the singularity of the anomalous dimension
matrix element $\gamma$ at $N=0$ is of higher order 
in $\log 1/x$
since the gluon 
does not couple directly to the photon.} at $N=0$.
This leads to the following behavior of the gluon distribution
at small $x$
\beqn
x f_G(x,Q^2) = 
\sqrt{\frac{\pi}{4}} 
\left[
\frac{3 \alpha_s}{\pi}
\frac{\log Q^2/Q_0^2}{\log 1/x}
\right]^{\frac{1}{4}}
\exp{\sqrt{4 \frac{N_c \alpha_s}{\pi}\log\frac{1}{x}\log
\frac{Q^2}{Q_0^2}}}\,
\cdot \mbox{const.}
\label{dla-gluon}
\eeqn
Rewritten in the variables $\rho=\sqrt{\log(1/x) / \log (Q^2/Q_0^2)}$ and 
$\sigma=\sqrt{\log (1/x) \cdot \log (Q^2/Q_0^2)}$ 
this behavior has been 
termed double asymptotic scaling 
\cite{das,bf} and has been shown to be in good agreement 
with small-$x$ data from the HERA collider \cite{bf}.
\\The presence of the singularity at $N=0$ in the gluon anomalous dimension 
is due to the fact that the diagrams which are resummed by the anomalous 
dimension do not only contain the collinear logarithm (in $Q^2$) 
in each order of $\alpha_s$ but also a logarithm in $1/x$ (soft logarithm).
In eq.\ (\ref{dla-gluon}) all contributions of the form 
$\left(\alpha_s \log (Q^2/Q_0^2) \log 1/x \right)^n$ are summed , consequently
the corresponding approximation is referred to as the 
double logarithmic approximation. 
\subsubsection{High energy factorization and the BFKL equation for $F_2$}
Given the presence of the singularity at $N=0$
one may ask for the importance of higher order contributions
to the anomalous dimension which also contain a singularity at $N=0$.
The presence of these singularities might spoil the convergence
of the perturbative expansion of the anomalous dimension 
at small $x$. This is the reason why the BFKL equation was 
considered to describe inclusive deep inelastic scattering in        
the small-$x$ region. The BFKL equation effectively resums
all singularities at $N=0$, or, formulated in $x$-space,
it sums the perturbative contributions of the form 
$\left(\alpha_s \log 1/x \right)^n$. Transformed to $x$-space the
BFKL equation (\ref{bfklmom}) in the forward direction 
has the form of an evolution equation 
\footnote{This equation is obtained from eq. (\ref{bfklmom}) by inverse 
mellin transform w.\ r.\ t.\ $N=\omega-1$ and removing one propagator
from the function $\Phi_{\omega}(\kf)$.
The function that results from this operation is then termed
${\cal F}_G$.}
with
the logarithm of $x$ being the evolution parameter
\beqn
-\frac{d}{d \log x} {\cal F}_G(x,\kf) =
\frac{N_c \alpha_s}{2 \pi^2} 
\int d^2 \kf'
\left[
\frac{1}{(\kf-\kf')^2} - \delta^{(2)}(\kf-\kf')
\int \frac{d^2 \lf}{\lf^2}\frac{\kf^2}{[\lf^2+(\lf-\kf)^2]}
\right]{\cal F}_G(x,\kf')
\label{bfklevo}
\eeqn
The function ${\cal F}_G(z,\kf)$ has to be interpreted as the 
unintegrated (w.\ r.\ t.\ to $\kf$) gluon density.
In order to obtain the hadronic cross section (the structure function
in DIS) the unintegrated gluon density has to be convoluted
with a $\kf$-dependent coefficient function.
\beqn
F(x,Q^2)= 
\int d^2 \kf
\int_x^1 \frac{d z}{z} \hat{C}(\alpha_s,z,\kf,Q^2)
{\cal F}_G(z,\kf)
\label{ktfact}
\eeqn
This equation constitutes the high-energy, or $\kf$-factorization theorem 
\cite{ktfac}.
It is a generalization of the collinear factorization theorem in
eq.\ (\ref{facttheorem}) which allows for the inclusion of leading-log$(1/x)$ 
contributions in the structure function.
The double logarithmic result is recovered by assuming dominance of the 
region $\kf^2 \ll Q^2$ in the $\kf$-integration by means of which one 
obtains
\beqn
F(x,Q^2)=\int_x^1 d z \;C(\alpha_s,\frac{x}{z},Q^2)
f_G(z,Q^2) \;\;,
\mbox{with} \;\; z f_G(z,Q^2) =\int_{\kf^2 \leq Q^2}d^2 \kf\; 
{\cal F}_G(z,\kf)  
\label{kt-dla}
\eeqn
i.\ e.\ the collinear factorization.
\\
The main theoretical problem which one faces when using the BFKL equation 
to predict the behavior of the structure function at small $x$ is the fact 
that the BFKL evolution does not seperate perturbative and 
nonperturbative scales. In the DGLAP evolution the partons transverse 
momenta $\kf^2$ increase starting from a lower but still perturbative scale 
$Q_0^2$ to the hard scale
$Q^2$, i.\ e.\ the whole evolution process takes place in the 
perturbative domain.
In the BFKL evolution there is no $\kf^2$-ordering, since the leading 
logarithms in $x$ are built up in the multi-Regge region of
strongly ordered longitudinal momenta with the transverse 
momenta being 
disordered. To be more spcific we recall that the rescaled
(with $\sqrt{\kf^2}$) solution of the BFKL equation for 
asymptotically small values of $x$ reads
\beqn
\psi(z,\kf)
=\sqrt{\kf^2}{\cal F}_G(x,\kf)
= 
\frac{a}{\sqrt{\Lambda_0^2}}
\left(\frac{1}{x}\right)^{\frac{N_c\alpha_s}{\pi} \; 4 \log 2} 
\frac{1}{\sqrt{14 N_c \alpha_s \zeta(3)\log 1/x}} \;
e^{- \pi \frac{\log \kf^2/\Lambda_0^2}{N_c \alpha_s 56 \zeta(3)\log 1/x}} 
\label{saddle}
\eeqn
with $a$ and $\Lambda_0^2$ being related to the normalization and mass scale 
of an inital condition at large $x$. From this solution one concludes first
that the BFKL equation predicts a power rise of the unintegrated 
gluon structure function at small $x$ with the famous 
exponent $4 \log 2 N_c \alpha_s/\pi $, which has to be compared with the 
exponential rise in eq.\ (\ref{dla-gluon}).
In the second place one realizes that the transverse momenta  
are distributed according to a normal distribution in $\log \kf^2$
with mean value $\log \Lambda_0^2$ and width 
$\sqrt{N_c \alpha_s/\pi \cdot 28 \zeta(3) \log 1/x}$. 
This normal distribution
is the outcome of the diffusion mechanism underlying the BFKL evolution.
The partons perform a random walk in $\log \kf^2$-space.
As a result there will inevitably be a contribution of nonperturbative
momentum scales to the evolution. In this region the perturbative BFKL results
are of course not valid and the application of the evolution 
equation becomes inconsistent
\footnote{We do not discuss here the additional problem
of violation of energy conservation due to the contribution
of very large scales which has been studied in \cite{coll-la} and 
\cite{fhs}.}.
\\
In order to overcome this infrared consistency problem it was 
proposed \cite{akms,ross} to modify the infrared sector of the 
BFKL equation. Several {\em ad hoc} -prescpriptions are described in 
the literature which were designed to exclude nonperturbative 
effects from the BFKL evolution, or, more ambitious, to model
contributions from the nonperturbative region.
The common feature of these prescriptions is that they 
introduce additional parameters the precise value of which 
is not under theoretical control. The results obtained with these
modified BFKL evolution equations of course depend on the 
values chosen for the parameters. With this freedom one gets 
on the one hand more room to describe succesfully the 
experimental data, but, on the other hand, the theoretical
foundation of this description becomes somewhat obscure.  
The uncertainty associated with this model- and parameter-dependence 
is related to the magnitude of the contribution which the BFKL evolution
receives from the infrared region in a given application.
Therefore, a first estimate of the reliability of a BFKL prediction 
requires an investigation of this quantity. 
\\
To illustrate this we perform a calculation of the structure 
function $F_2$ at small $x$ based on the $\kf$-factorization formalism
with the unintegrated gluon density being calculated from the
BFKL equation. We use eqs.\ (\ref{bfklevo}) and (\ref{ktfact}) 
with the well-known coefficient function $\hat{C}$ \cite{coeff}
representing the coupling of a gluon to the virtual photon.
To solve the BFKL equation we have to specify  
an inital condition at a starting point $x_0$.
In order to be consistent with the $F_2$-calculations for 
$x > x_0$ we express the unintegrated gluon distribution at
$x_0$ as the logarithmic derivative of the gluon density 
(cf.\ eq.\ (\ref{kt-dla})) and take the latter from one of the 
current parameterizations in the literature.
These parameterizations give the gluon density only above 
a boundary value $Q_0^2$ and for our purposes we have to 
continue it down to $Q^2=0$. There is a freedom in this continuation
and here we follow \cite{akms} by taking 
${\cal F}_G(x_0,\kf)=c /(\kf^2+\kf_a^2)$ for $\kf^2 \leq \kf_0^2$
and
${\cal F}_G(x_0,\kf) = 1/(\kf^2+\kf_a^2) \partial/(\partial
\log \kf^2) x_0 f_G(x_0,\kf^2)$ for $\kf^2 \geq k_0^2$
with $c$ determined from continuity and $\kf_0^2$, $\kf_a^2$ two momentum
scales to be specified later.
This ansatz guarantees consistency with the gluon density for 
large $\kf^2$ and respects the gauge invariance constraint 
$\lim_{\kf^2 \to \infty} \kf^2 {\cal F}_G(x_0,\kf)=0$.
Now we solve the BFKL equation with a particular modification of the 
infrared region. We evolve the inital condition down from $x=x_0$
but after each evolution step cut off the infrared tail of the 
distribution ${\cal F}_G(x,\kf)$ with $\kf^2 \leq \kf_0^2$
and replace it with the model function $c/(\kf^2+\kf_a^2)$
with $c$ again determined by continuity and $\kf_0^2,\kf_a^2$ as above. 
\\
In fig.\ \ref{fig21a} we show the results of this calculation for different
values of $Q^2$ where we have taken 
$\kf_0^2 = 1 \, \mbox{GeV}^2$ and two different
values for $\kf_a^2$, $\kf_a^2=1 \, \mbox{GeV}^2$ 
and $\kf_a^2=2 \, \mbox{GeV}^2$.
In addition the result is displayed which is obtained without the 
modification of the infrared region.\\
\begin{figure}[!h]
\begin{center}
\input{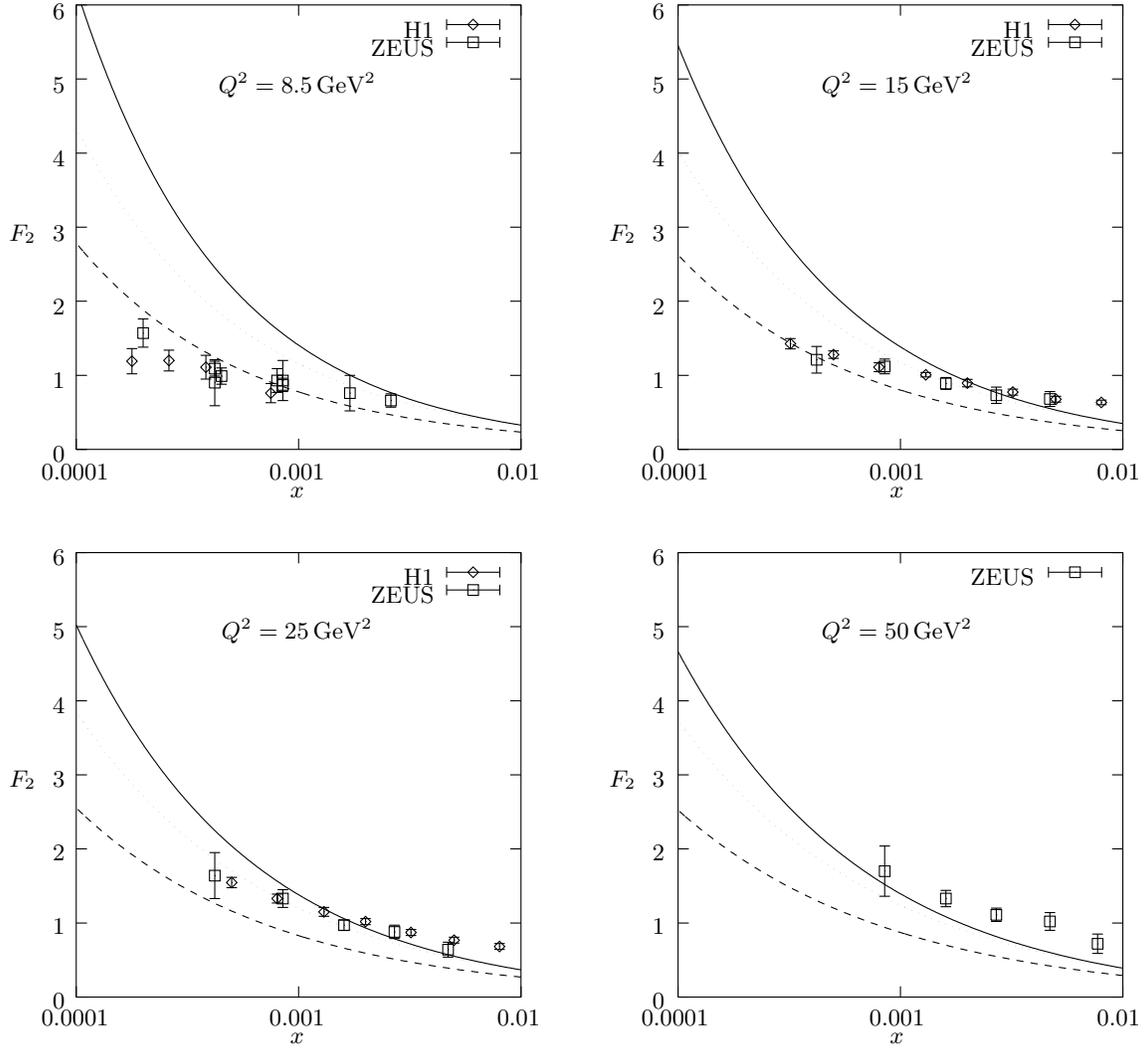}
\end{center}
\caption{
Results of the BFKL-based $F_2$-calculation for $Q^2 = 8.5, 15, 25$
and $50 \,\mbox{GeV}^2$. Displayed are the non-modified solution
(solid line) and the modified solution with $\kf_a^2 = 1 \,\mbox{GeV}^2$
(dotted line) and $\kf_a^2 = 2 \,\mbox{GeV}^2$ (dashed line)
in comparison with data from the H1 
\protect\cite{h1}, ZEUS \protect\cite{zeus} 
and E665 \protect\cite{e665} collaborations.
\label{fig21a}
}
\end{figure}
One recognizes that the normalization of the result depends strongly
on the value of the infrared parameter $\kf_a^2$ whereas the slope 
is rather stable.
This dependence becomes weaker with increasing $Q^2$. It should be remarked 
that these results were obtained from a calculation with 
fixed $\alpha_s=\alpha_s(Q^2)$. Corrections which introduce a scale
dependence of $\alpha_s$ are subleading in $\log(1/x)$
and hence beyond the scope of the BFKL formalism.
We choose here $\alpha_s=\alpha_s(Q^2)$ because $Q^2$ is the only hard
scale in the present context. The value of $\alpha_s$ enters in the 
global normalization and in the $x$-slope in the calculation.
One sees that this produces a wrong scaling behaviour in $Q^2$ 
for very small $x$. It was shown \cite{akms} that after 
introduction of running $\alpha_s$ a more realistic $Q^2$-scaling is 
obtained. One should however keep in mind that all attempts to implement 
running $\alpha_s$ into the BFKL evolution remain preliminary 
unless the analysis of the complete NLO-corrections (see \cite{fadlip}
and references therein) is completed. 
\\
In the following we
concentrate on the interpretation of the 
observed strong sensitivity of the
results to the infrared parameters. To this end 
we notice \cite{balo} that we can represent $F_2$
as
\beqn
F_2(x,Q^2)= \sqrt{Q^2\Lambda_0^2}\int_{-\infty}^{+\infty}
d \xi \psi_1(Q^2,\xi;\frac{x}{z})\psi_2(\Lambda_0^2,\xi,\frac{z}{x_0})
\:\:,\;\;\xi=\log \frac{\kf^2}{\kf_0^2},\;\;x \leq z \leq x_0
\label{cigar}
\eeqn
and calculate, for every $z$ the function $\psi_1\psi_2$ 
which determines the transverse momentum distribution inside 
the BFKL evolution. For these distributions we calculate
the mean value $<\xi>$ and the width $\Delta(\xi)
= (<\xi^2>-<\xi>^2)^{\frac{1}{2}}$. 
In figs.\ \ref{fig21b}, \ref{fig21c} 
we display for the extreme values of $Q^2$
the results for $<\xi>$ and $\xi_{\pm}=<\xi>\pm\Delta(\xi)$ translated back
to the variable $\kf^2$ as a function of z.  
\begin{figure}[!h]
\begin{center}
\input{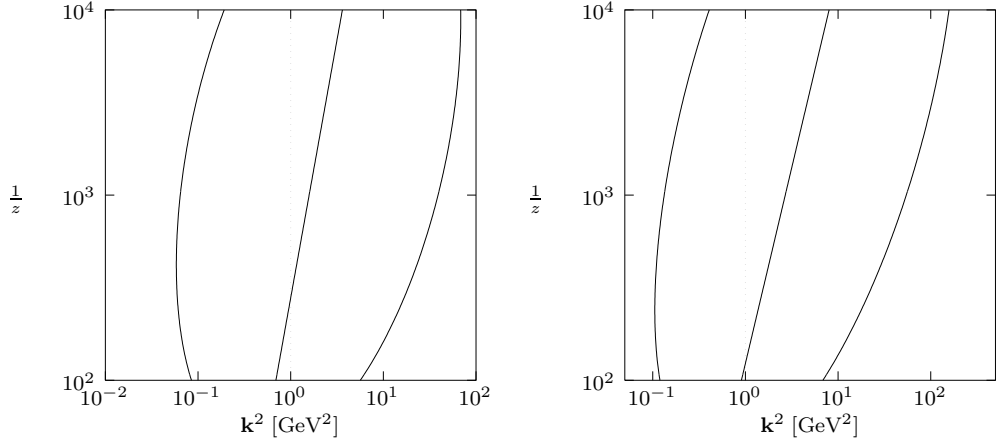}
\end{center}
\caption{
Mean value $<\xi>$ and width $\Delta(\xi)$ of the distribution 
of $\xi=\log \kf^2/\kf_0^2$ inside the 
non-modified BFKL evolution for $F_2$ with
$x=10^{-4},x_0=10^{-2}$ and $Q^2=8.5\,\mbox{GeV}^2$
(left diagram) resp.\ $Q^2=50 \,\mbox{GeV}^2$ (right diagram).
\label{fig21b}
}
\end{figure}
\begin{figure}[!h]
\begin{center}
\input{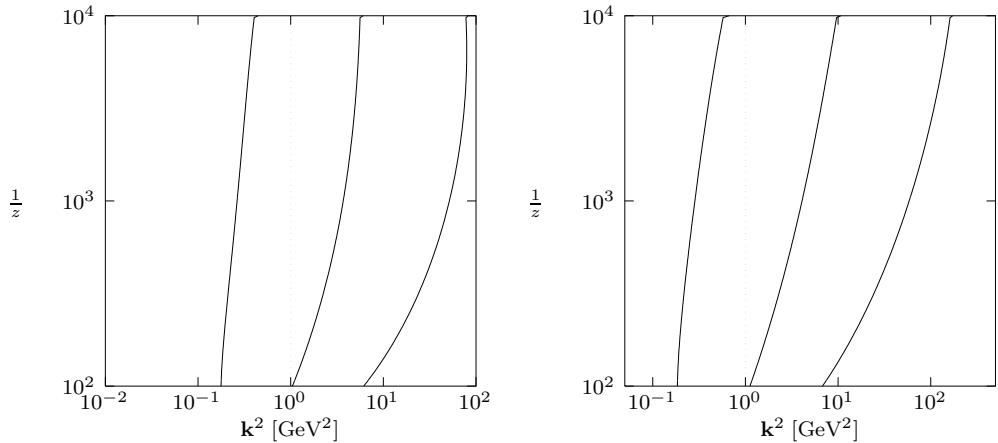}
\end{center}
\caption{The same as in 
fig.\ \protect\ref{fig21b} but with the BFKL evolution
modified in the infrared region with $\kf_a^2 = 1\,\mbox{GeV}^2$.
\label{fig21c}
}
\end{figure}
Fig.\ \ref{fig21b} illustrates the basic 
diffusion mechanism of the BFKL evolution.
Starting from the inital distributions at $z=x_0$ and $z=x$
the transverse momenta diffuse into the infrared and ultraviolet
region. Coming from below the center of diffusion drifts to the 
right since the hard scale at the upper end pulls the evolution 
out of the infrared domain which is to the left of the dotted vertical 
line. The contribution which comes from this nonperturbative region
is clearly increasing with decreasing $Q^2$.
Assuming as an approximation 
gaussian distributions for $\psi_1$ and $\psi_2$
the pattern of fig.\ \ref{fig21b} can be reproduced analytically.
The evolution picture changes drastically if the modification
of the infrared region is applied. 
Fig.\ \ref{fig21c} shows that in this case
the diffusion into the infrared is cut off
whereas there is still diffusion into the ultraviolet.
Given that in the application to $F_2$ the contribution from the 
infrared region is large as observed in fig.\ \ref{fig21b} 
it is clear that the result of the evolution is rather sensitive
to the details of the infrared modifications.
It is also easy to realize that this sensitivity
becomes stronger if $Q^2$ is decreased
in agreement with the behavior observed in fig.\ \ref{fig21a}. 
\\
To summarize one can say that by using the BFKL equation and 
$\kf$-factorization one obtains predictions for $F_2$ at small $x$ 
which are seriously affected by the way in which
one treats the infrared region leading to a large theoretical 
uncertainty. In the preasymptotic region $x \simeq 10^{-3}$ good
agreement with data can be achieved but for smaller $x$ the slope
generated by the BFKL evolution becomes to steep. 
It should be recalled that we are calculating only the gluonic 
contribution to $F_2$. Agreement with data in the preasymptotic region
requires of course the addition of a non-gluonic background contribution. 
\subsubsection{BFKL equation and transverse energy distribution}
In the context of the $F_2$-measurements at HERA it has been advocated
\cite{transenergyflow} to use the transverse energy flow in the final 
state as a window to examine the nature of the parton evolution
in the deep inelastic scattering process. In particular one would
expect an enhanced energy flow in the central rapidity region for a BFKL 
type of evolution compared to conventional evolution
due to disordered transverse momenta. 
In order to study the transverse energy flow, the 
transverse energy weighted single parton inclusive cross section
has been calculated in the BFKL formalism \cite{transenergyflow}.
It should be remarked that the single parton inclusive cross section
is not a well-defined quantity within the BFKL approximation
since the collinear singularities present in the real and virtual 
corrections cancel only at the fully inclusive level.
For the one-parton inclusive case an infrared finite result is 
obtained only after energy weighting. A more consistent
definition of the transverse energy flow requires the analysis
of associated distributions at small $x$ based on the color 
coherence approach \cite{marchesini}. 
\\
As a preparatory study we examine in the following the transverse 
energy distribution of the gluons which are exchanged in the $t$-channel
in the BFKL evolution. From this one should obtain at least
a qualitative impression of the energy distribution of the final 
state particles.
We focus on the center of rapidity in the photon proton cms frame
which is defined by the relation $z=(x_B\cdot \kf^2/Q^2)^{\frac{1}{2}}$.
For $\kf^2$ we take a mean value of the transverse momentum which can be read
off from fig.\ \ref{fig21b}. The results for the transverse energy 
distribution for $Q^2=50\, \mbox{GeV}^2$ and three different
values of $x_B$ are shown in fig.\ \ref{fig21d}.
\\
\begin{figure}[!h]
\begin{center}
\input{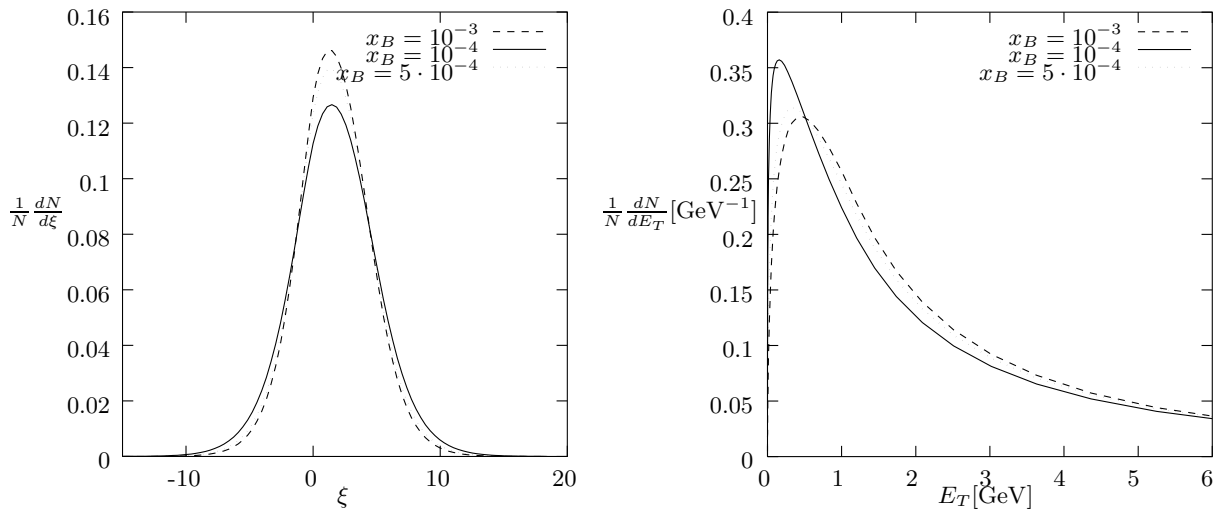}
\end{center}
\caption{ 
The $\xi$-distributions in the central rapidity region for different
values of $x_B$ and the corresponding transverse energy distributions.
\label{fig21d}
}
\end{figure}
On the left hand side the product distribution $\psi_1(\xi)\psi_2(\xi)$
is shown as a function of $\xi$. On the right hand side we display the 
transverse energy distribution calculated from the product distributions.
The characteristic feature of the $\xi$-distributions is the 
increase of the width with incrasing rapidity interval which is
a manifestation of the diffusion mechanism. The width of these distributions
is shown as a function of $x_B$ 
for two values of $Q^2$
on the left hand side of fig.\ \ref{fig21e}.
The increase of the width transforms into an increase of the mean transverse
energy which is shown on the right hand side of fig.\ \ref{fig21e}.
\\
\begin{figure}[!h]
\begin{center}
\input{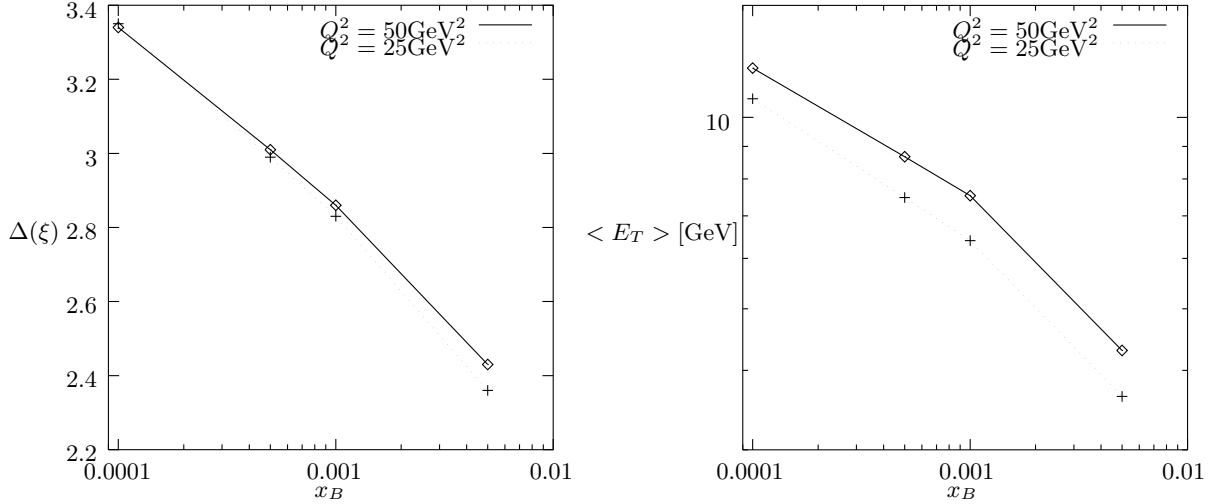}
\end{center}
\caption{
Width of the $\xi$-distributions and mean value of the $E_T$ distributions 
for $Q^2=25 \,\mbox{GeV}^2$ and $Q^2=50 \,\mbox{GeV}^2$.
\label{fig21e}
}
\end{figure}
Based on the asymptotic solution (\ref{saddle})
of the BFKL equation 
simple analytic estimates for the observables studied above can be
obtained. 
Characterizing the inital distributions
$\psi_i(\xi)$ at $x_B$ and $x_0$ by their mean value $<\xi_i>$ and width
$\delta_i(\xi)$ we find for the mean value of the product distribution
in the central rapidity region
\beqn
\xi = \frac{<\xi_1>(\delta_1(\xi)
                    +\frac{1}{2}\log \frac{1}{x_B}\frac{\kf^2}{Q^2})
           +<\xi_2>(\delta_2(\xi)
                    +\frac{1}{2}\log \frac{x_0^2}{x_B}\frac{Q^2}{\kf^2})}
           { \delta_1(\xi)+\delta_2(\xi)
            + \frac{1}{2}\log \frac{1}{x_B}\frac{\kf^2}{Q^2}
            + \frac{1}{2}\log \frac{x_0^2}{x_B}\frac{Q^2}{\kf^2}}
\eeqn
For small $x_B$ it becomes constant as observed in fig.\ \ref{fig21d}.
For the width we find 
\beqn
\Delta(\xi)= \left[
             \frac{N_c\alpha_s}{\pi} 28 \zeta(3)
             \frac{(\delta_1(\xi)
                    +\frac{1}{2}\log \frac{1}{x_B}\frac{\kf^2}{Q^2})
                   (\delta_2(\xi)
                    +\frac{1}{2}\log \frac{x_0^2}{x_B}\frac{Q^2}{\kf^2})}
                  { \delta_1(\xi)+\delta_2(\xi)
            + \frac{1}{2}\log \frac{1}{x_B}\frac{\kf^2}{Q^2}
            + \frac{1}{2}\log \frac{x_0^2}{x_B}\frac{Q^2}{\kf^2}}
              \right]^{\frac{1}{2}} 
\eeqn
The width becomes independent of the details of the boundary conditions
for small $x_B$ and thus represents a characteristic BFKL signal.
Asymptotically it rises with the square root of the rapidity. 
The latter can thus be interpreted as the time coordinate 
of the gluon's random walk. 
The analytic prediction for the asymptotic
mean transverse energy $<E_T>$ is
\beqn
<E_T>= \exp\left[\frac{1}{8}\Delta^2(\xi)+\frac{1}{2}<\xi>\right]
\eeqn
Since $<\xi_1>$ increases with increasing $Q^2$ the mean $E_T$ 
increases with $Q^2$ in accordance with the behavior observed in
fig.\ \ref{fig21e}.
\\ 
To conclude, we want to stress that it is the square root increase of 
the width of the $\log \kf^2$-distribution with rapidity
which is the fundamental manifestation of the diffusion mechanism
underlying the
BFKL equation. An experimental study of this observable might 
provide valuable insight into the parton evolution at small $x$.
The problem with this quantity, however, could be its large 
sensitivity to hadronization effects which wash out the 
characteristics of the parton level distributions.       
\newpage
\subsection{Diffractive Production of Vector Mesons at large $t$}
\label{sec22}
In the preceeding section it was shown that the diffusion of gluon 
momenta in the BFKL evolution leads to a large theoretical
uncertainty associated with the contribution of nonperturbative
momentum scales.
This is not the case if there is a large momentum scale present 
all along the ladder which keeps the diffusion in the perturbative
domain. An immediate example for such a large scale is the momentum 
transfer $-t$. The authors of \cite{jeff1,jeff2}
introduced a reaction in which colorless exchange at large 
momentum transfer applies. They studied the process 
$\gamma^*(Q) + P(p) \to V(Q+q)+ X(p-q)$
where $V$ is a vector meson with mass $M_V$
, e.\ g.\ a $J/\Psi$, which is produced 
with large momentum transfer $t = q^2$ and $X$ is a final state 
resulting from the dissociation of the proton.
If the hadronic energy $W^2=(p+Q)^2$ becomes large logarithmic 
corrections become important which are resummed by the BFKL amplitude
(represented by the blob in fig.\ \ref{fig221}).
\begin{figure}[!h]
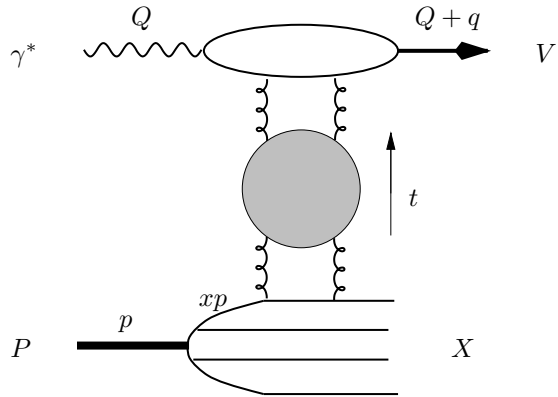

\begin{center}
\input vmatlarget.pstex_t
\end{center}
\caption{
Graphical representation of the process of diffractive vector meson
production in DIS.
\label{fig221}
}
\end{figure}
\subsubsection{The result for the cross section}
The cross section for the process can be written in the 
following factorized form \cite{jeff1}
\beqn
x \frac{d \sigma^{\gamma^*P}}{d x d t}
= \frac{d \sigma^{\gamma^* q}}{d t}(x,t)
\left[ \sum_f(x q_f(x,-t)+ x \bar{q}_f(x,-t)) + 
\frac{81}{16} x G(x,-t)\right]
\label{eq22init}
\eeqn
This factorization means that the photon scatters off a single
parton with longitudinal momentum $x p$ and virtuality $\leq t$.
It is expected to hold as long as $-t$ is larger than the inverse 
size $Q_0^2$ of the proton. \\
The photon-parton subprocess energy is $s=x W^2$ and the subprocess 
cross section reads
\beqn
\frac{d \sigma^{\gamma^* q}}{d t} (x,t)
= \pi \left(\frac{4}{9}\right)^2 
\frac{\alpha_s^4}{t^4}
\left|\int \frac{d \omega}{2 \pi i}
\left(\frac{s}{Q^2+M_V^2-t}\right)^{\omega} 
\Phi_{\omega}(\qf^2) 
\right|^2 \phantom{xxx}; \phantom{x}\qf^2 =-t
\label{carl}
\eeqn
with $\Phi_{\omega}$ being given as the convolution of 
the BFKL amplitude with the impact factors of the incoming 
particles
\beqn
\Phi_{\omega}(\qf^2) = 
|\qf|^4 
\int \frac{d^2 \kf}{(2 \pi)^3} \frac{d^2 \kf'}{(2 \pi)^3} 
\; \Phi_{\omega}(\kf,\kf';\qf)
\; \phi_V(\kf,\qf) \; \phi_q(\kf',\qf)
\label{vmfact}
\eeqn   
Here  $\phi_V$ is the impact factor for the photon producing 
a vector meson and $\phi_q$ is the impact factor for the 
incoming parton.
The photon-vector meson transition can be described in terms
of a non-relativistic form factor \cite{ryskinvm}
\beqn
\phi_V(\kf,\qf)=-\frac{{\cal C}}{2}
\left[\frac{1}{\Delta^2+(\kf-\qf/2)^2}
- \frac{1}{\Delta^2+\qf^2/4}
\right]
\label{phot-vecmes}
\eeqn
where $4 \Delta^2=Q^2+M_V^2$ 
and the coefficient is determined through the electromagnetic 
width of the vector meson
\beqn
{\cal C}^2 = 3 \Gamma_V^{ee}\frac{M_V^3}{\alpha_{\mbox{\tiny em}}}
\eeqn
This transition is helicity conserving and it is given here
for transverse polarization of the photon. For the longitudinal
polarization one has to multiply the righ hand side of eq.\ (\ref{carl})
with $Q^2/M_V^2$. 
\\
The parton formfactor deserves more discussion. An incoming 
parton is not a colorless state and hence eq.\ \ref{vmfact}
seems to be in contradiction with the results of section
2.1. .
The description in terms of a parton formfactor constitutes an
effective prescription, valid for $-t \gg Q_0^2$,
which requires a particular modification of the BFKL 
amplitude $\phi_{\omega}(\kf,\kf';\qf)$.
For the conformal eigenfunction $E^{(\nu,0)}$ 
which 
couples to the parton line the Mueller-Tang subtraction
\cite{muellertang} has to be performed
\beqn
E^{(\nu,0)} (\rho_1,\rho_2)=
\left( \frac{\rho_{12}^2}{\rho_1^2\rho_2^2}
\right)^{\frac{1}{2}-i\nu}
 \longrightarrow 
E^{(\nu,0)}_{MT} (\rho_1,\rho_2) =
E^{(\nu,0)} (\rho_1,\rho_2)
-\left(\frac{1}{\rho_1^2}\right)^{\frac{1}{2}-i\nu}
-\left(\frac{1}{\rho_2^2}\right)^{\frac{1}{2}-i\nu}
\eeqn 
This modification subtracts $\delta$-function like contributions
from the conformal eigenfunction which are not present in perturbation 
theory. These $\delta$-function terms have implicitly been added to the 
BFKL equation to ensure conformal invariance.
They are not present in the physical amplitude since they give zero 
after convolution with color neutral impact factors.
In the effective prescription in which the BFKL pomeron couples to
a colored state, however, they have to be subtracted.
It should be emphasized here that the Mueller-Tang prescription
does not apply in general \cite{baliplofo}.
We postpone a more detailed discussion to the following section.
With the above prescription the quark formfactor becomes unity
and insertion of the general solution of the nonforward BFKL
equation with the Mueller-Tang subtraction performed gives
\beqn
\Phi_{\omega}(\qf^2) = |\qf|^4 \!\!
\int_{-\infty}^{+\infty} \frac{d \nu}{2 \pi} 
\frac{\nu^2}{(\nu^2+1/4)^2}
\frac{1}{\omega-\chi(\nu)}\!
\int \!\frac{d^2 \kf}{(2 \pi)^3} E^{(\nu,0)}(\kf,\qf-\kf) \phi_V(\kf,\qf)
\!
\int \!\frac{d^2 \kf'}{(2 \pi)^3} E_{MT}^{(\nu,0)\,\ast}(\kf',\qf-\kf')
\label{eq22}
\eeqn
We consider only conformal spin $n=0$ since this gives the dominant 
contribution in the high-energy limit.
The integration of the Mueller-Tang eigenfunction gives the 
simple result
\beqn
\int \frac{d^2 \kf'}{(2 \pi)^3} \int d^2 \rho_1 d^2 \rho_2 
e^{i \kf' \rho_1+i(\qf-\kf')\rho_2}
\left[
\left(\frac{\rho_{12}^2}{\rho_1^2\rho_2^2}\right)^{\frac{1}{2}+i\nu}
-\left(\frac{1}{\rho_1^2}\right)^{\frac{1}{2}+i\nu}
-\left(\frac{1}{\rho_2^2}\right)^{\frac{1}{2}+i\nu}
\right]
\\ 
=
- 2 \;4^{-i\nu} (\qf^2)^{-\frac{1}{2}+i\nu}
\frac{\Gamma(\frac{1}{2}-i\nu)}
     {\Gamma(\frac{1}{2}+i\nu)}
\eeqn
The evaluation of the $\kf$-integration is more complicated.
Instead of using the explicit representation for the momentum
space eigenfunctions we prefer to work with the
mixed representation (\ref{mix}). 
The $\kf$-integration then gives a 
modified Bessel-function
\beqn
\int \frac{d^2 \kf}{(2 \pi)^3} \phi_V(\kf,\qf) e^{i\kf \rho}
=-\frac{{\cal C}}{2} \frac{1}{(2 \pi)^2}
\left[ K_0(|\rho|\Delta) 
e^{\frac{i}{2}\qf \rho} 
- 2\pi \;\delta^{(2)}(\rho)\frac{1}{\Delta^2+\qf^2/4}
\right]
\label{besselk0}
\eeqn
If we assume $\mbox{Re}(i\nu) < 1/2$ the only non-zero contribution comes 
from the first term and for the $\kf$-integral in eq.\ (\ref{eq22}) we find
\beqn 
-\frac{{\cal C}}{2} \frac{1}{2 \pi} 4^{i \nu} 
\frac{(\qf^2)^{-\frac{3}{2}-i\nu}}{\Gamma^2(\frac{1}{2}-i\nu)}
\int_0^{\infty} d \xi \xi^2 K_0(\xi \frac{\Delta}{|\qf|})
\int_0^1 d x[x(1-x)]^{-\frac{1}{2}} J_0(\xi|x-\frac{1}{2}|)
K_{2 i \nu}(\xi \sqrt{x(1-x)})
\eeqn
where the the integral over the angle of $\rho=\rho_{12}$ 
was performed and the 
dimensionless variable $\xi$ was introduced.
The next step is the insertion of the following representation of a
$\delta$-function
\beqn
\delta(\xi-\xi')= \int_{-\infty}^{+\infty} 
\frac{d \lambda}{\pi}
\frac{1}{\xi}\left(\frac{\xi'}{\xi}\right)^{2 i \lambda}
\eeqn
which allows to separate the $K_0$ function from the $x$-dependent
factors
\beqn
-\frac{{\cal C}}{2} \frac{1}{2 \pi} 4^{i \nu} 
\frac{(\qf^2)^{-\frac{3}{2}-i\nu}}{\Gamma^2(\frac{1}{2}-i\nu)}
\int \frac{d \lambda}{\pi}
\left(\frac{|\qf|}{\Delta}\right)^{3+2 i \lambda}
\int_0^{\infty}
d \xi' \xi{'}^{2+2i\lambda} K_0(\xi')
\phantom{xxxxxxxxxxx}
\nonumber \\
\cdot
\int_0^{\infty} d \xi \xi^{-1-2i\lambda} 
\int_0^1 d x[x(1-x)]^{-\frac{1}{2}} J_0(\xi|x-\frac{1}{2}|)
K_{2 i \nu}(\xi \sqrt{x(1-x)})
\\
=
-\frac{{\cal C}}{2} \frac{1}{2 \pi} 4^{i \nu} 
\frac{(\qf^2)^{-\frac{3}{2}-i\nu}}{\Gamma^2(\frac{1}{2}-i\nu)}
\int \frac{d \lambda}{\pi}
\left(\frac{|\qf|}{\Delta}\right)^{3+2 i \lambda}
2^{1+2 i \lambda} \Gamma^2(\frac{3}{2}+i\lambda)
\phantom{xxxxxxxxxxx}
\nonumber \\
\cdot
\int_0^{\infty} d \xi \xi^{-1-2i\lambda} 
\int_0^1 d x[x(1-x)]^{-\frac{1}{2}} J_0(\xi|x-\frac{1}{2}|)
K_{2 i \nu}(\xi \sqrt{x(1-x)})
\eeqn
Here the contour of the $\lambda$-integration has been shifted in the
complex plane such that the condition 
$-3/2<\mbox{Re}(i\lambda)<\pm \mbox{Re}(i\nu)$ is fulfilled.
Now the remaining $\xi$ and $x$ integrations can be performed.
The $\xi$-integral gives 
\beqn
2^{-2-2 i \lambda} 
\Gamma(-i\lambda+i\nu)\Gamma(-i\lambda-i\nu)
\int_0^1 dx[x(1-x)]^{-\frac{1}{2}+i\lambda} 
\,_2F_1(-i\lambda-i\nu,-i\lambda+i\nu,1;-\frac{(1-2 x)^2}{4x(1-x)})
\eeqn
After changing the integration variable to $z = (1-2x)^2$ and using 
an analytic continuation of the hypergeometric function the remaining 
integral can be performed to give
\beqn
4^{-1-2 i \lambda} 
\Gamma(-i\lambda+i\nu)\Gamma(-i\lambda-i\nu)
\frac{\Gamma(\frac{1}{2})\Gamma(\frac{1}{2}-i\nu)}{\Gamma(1-i\nu)}
\,_3F_2(-i\lambda-i\nu,1+i\lambda-i\nu,\frac{1}{2};1,1-i\nu;1)
\\
=
4^{-1-2 i \lambda} 
\pi \frac{\Gamma(\frac{1}{2}+i\nu)\Gamma(\frac{1}{2}-i\nu)
\Gamma(-i\lambda+i\nu)\Gamma(-i\lambda-i\nu)}
{\Gamma(\fez-\fez(i\lambda+i\nu))\Gamma(\fez-\fez(i\lambda-i\nu))
\Gamma(1+\fez(i\lambda+i\nu))\Gamma(1+\fez(i\lambda-i\nu))} 
\eeqn
where the last identity follows from Watson's theorem \cite{bateman}.
Collecting everything together we arrive at the following expression 
for the partial wave amplitude 
\beqn
\Phi_{\omega}(\qf)= 4 \pi {\cal C} 
\int_{-\infty}^{+\infty} 
\frac{d \nu}{2 \pi}\frac{\nu^2}{(\nu^2+1/4)^2}
\frac{1}{\omega-\chi(\nu)} 
\phantom{xxxxxxxxxxxxxxxxxxxxxxxxxxxxxxxxxxxxxxxxxx}
\nonumber \\
\cdot
\int \frac{d \lambda}{\pi}
\left(\frac{\qf^2}{4\Delta^2}\right)^{\frac{3}{2}+i \lambda}
\frac{\Gamma^2(\frac{3}{2}+i\lambda)
\Gamma(-i\lambda+i\nu)\Gamma(-i\lambda-i\nu)}
{\Gamma(\fez-\fez(i\lambda+i\nu))\Gamma(\fez-\fez(i\lambda-i\nu))
\Gamma(1+\fez(i\lambda+i\nu))\Gamma(1+\fez(i\lambda-i\nu))}
\label{eq22fin}
\eeqn
\subsubsection{Comparison of the exact formula with asymptotic expressions} 
The remaining contour integral $I(\nu)$ 
(the second line in eq.\ (\ref{eq22fin})
over $\lambda$ can either be 
performed numerically or
evaluated as a power series in in the ratio 
$|\qf|/(2\Delta)$ (resp.\ $2\Delta/|\qf|$).
In the limit of large momentum transfer, when 
$|\qf|/(2\Delta) \gg 1$, the dominant contribution is obtained by shifting
the contour past the singularity at $i\lambda = -3/2$. The residue 
of the double pole leads to the leading behavior
\beqn
I(\nu)=8 \frac{\Gamma(\fez+i\nu)\Gamma(\fez-i\nu)}
              {\Gamma^2(\frac{1}{4}+\fez i\nu)
               \Gamma^2(\frac{1}{4}-\fez i\nu)}
\left[ 
\log \frac{\qf^2}{4 \Delta^2} -2 \gamma_E 
 - 2 \mbox{Re}\;\psi(\frac{1}{2}+i\nu)
\right]
\label{asyhigh}
\eeqn
In the opposite limit we have two poles at $i\lambda = \pm i\nu$
which contribute equally as long as the imaginary part of $\nu$ is small
which is true in the high-energy limit. For $|\qf|/(2\Delta) \ll 1$
we thus have
\beqn
I(\nu)=
\frac{4}{\sqrt{\pi}}\left(\frac{\qf^2}{4 \Delta^2}\right)^{
\frac{3}{2}} \mbox{Re}\; \left[
\left(\frac{\qf^2}{4 \Delta^2}\right)^{i\nu}
\frac{\Gamma^2(\frac{3}{2}+i\nu)\Gamma(-2i\nu)}
{\Gamma(\fez-i\nu)\Gamma(1+i\nu)}
\right]
\label{asylow}
\eeqn
\begin{figure}[!h]
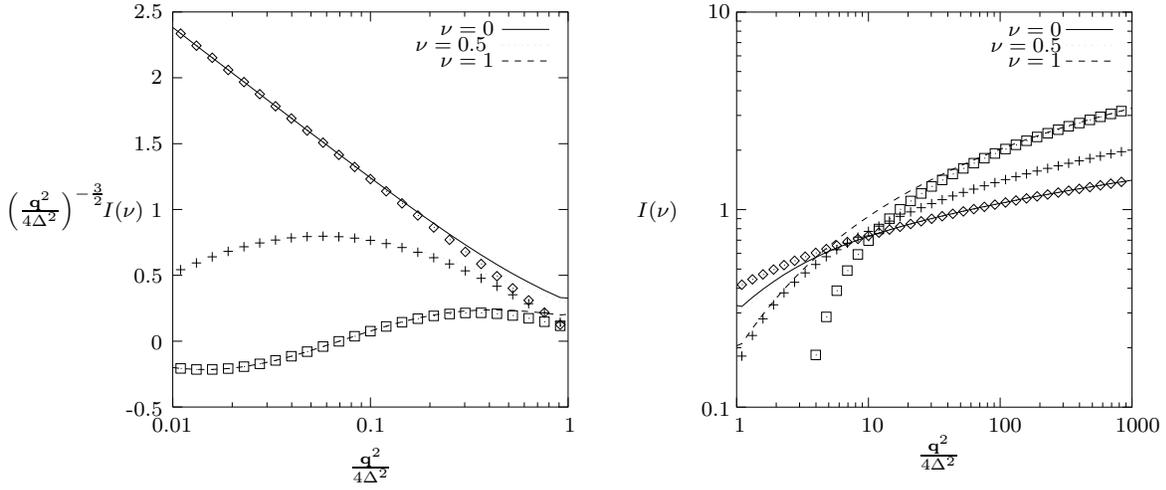

\begin{center}
\input comp.pstex_t
\end{center}
\caption{
Comparison of a numerical evaluation of $I(\nu)$ (lines)
with the asymptotic analytic predictions (symbols) for
different values of $\nu$.
\label{fig222}
}
\end{figure}
In fig.\ \ref{fig222} the result of a numerical calculation of $I(\nu)$
is compared with the asymptotic formulas eqs.\ 
(\ref{asyhigh}) and (\ref{asylow}) for different values of $\nu$.
\\
It remains to find the energy dependence of the cross section
by performing the $\nu$ and $\omega$ integration.
The $\omega$ integration leads to the familiar exponential factor
$\exp[y \chi(\nu)]$ ($y=\log s/(Q^2+M_V^2-t)$) and we end with
\beqn
\frac{d \sigma^{\gamma^* q}}{d t} (x,t)
= \pi \left(\frac{4}{9}\right)^2 
\frac{\alpha_s^4}{t^4}
\; \left| \,4 \pi {\cal C} \int \frac{d \nu}{2 \pi} 
\frac{\nu^2}{(\nu^2+1/4)^2}
\frac{1}{\omega-\chi(\nu)} I(\nu) 
\, \right|^2
\eeqn 
where $I(\nu)$ is given by the second line
in eq.\ (\ref{eq22fin}).
Analytic results can again be obtained in certain kinematical limits.
First we conclude from eq.\ (\ref{asyhigh})
that for large $|\qf|/(2 \Delta)$ the $|\qf|$ and the $y$ dependence
factorize. For large $y$ the $\nu$ integral is then dominated by the 
exponential $\exp[y \cdot \chi(\nu)]$ and the asymptotic behavior is 
obtained by expansion around the saddle point which is located at 
$\nu = 0$
\beqn
\chi(\nu)=\frac{N_c \alpha_s}{\pi}\left[ 4 \log 2 - 
28 \zeta(3) \, \nu^2 \right] +O(\nu^4)
\eeqn
The saddle point approximation then gives the following result in the 
limit of $y \gg 1$ for fixed $|\qf|/(2 \Delta) \gg 1$
\beqn 
\frac{d \sigma^{\gamma^* q}}{d t} (x,t)
= \pi \left(\frac{4}{9}\right)^2 
\frac{\alpha_s^4}{t^4}\;
\left| \,
{\cal C}\,
128 \frac{\pi^3}{\Gamma^4(1/4)}\left[\log
\left(\frac{|\qf|^2}{4 \Delta^2}\right)+4 \log 2\right]
\frac{e^{\frac{N_c\alpha_s}{\pi}4 \log 2 \cdot y}}
{[N_c \alpha_s 14 \zeta(3) y]^{\frac{3}{2}}}
\, \right|^2
\label{approx1}
\eeqn
Note the $y^{\frac{3}{2}}$ in the denominator which has to be compared 
with the $(\log 1/x)^{\fez}$ in the denominator of eq.\ (\ref{saddle}).
This difference is a manifestation of the different types of 
leading $\omega$-plane singularities of the $t=0$ and 
$t \neq 0$ solutions of the BFKL equation. \\
For the case $|\qf|/(2 \Delta) \ll 1$ the analysis is more complicated 
since we have from eq.\ (\ref{asylow}) an additional factor 
$[|\qf|/(2 \Delta)]^{2 i\nu}$ in the $\nu$ integral which leads to 
a correlation between the $|\qf|$ and the $y$ dependence. 
Now we have to find the saddle point of the function
$y \cdot \chi(\nu) \pm i\nu \log |\qf|^2/(4 \Delta^2) $. 
If we assume $y \gg \log (4 \Delta^2)/|\qf|^2$ the saddle
point is found at $\nu=\pm i \log |\qf|^2/(4 \Delta^2)
\cdot 1/(y\,|\chi''(0)|)$ and the corresponding result for
the cross section reads
\beqn 
\frac{d \sigma^{\gamma^* q}}{d t} (x,t) = 
\pi \left(\frac{4}{9}\right)^2 
\frac{\alpha_s^4}{t^4}\;
\left| \,
{\cal C}
\, 16 \pi^2 \left(\frac{|\qf|^2}{4 \Delta^2}
\right)^{\frac{3}{2}}
(-2+3 \log 2)
\frac{e^{\frac{N_c\alpha_s}{\pi}4 \log 2 \cdot y}}
{[N_c \alpha_s 14 \zeta(3) y]^{\frac{3}{2}}}
e^{-\pi \frac{ \log^2 \frac{|\qf|^2}{4 \Delta^2} }
{N_c \alpha_s 56 \zeta(3) y}} \,
\right|^2
\label{approx2}
\eeqn
The last exponential factor induces a
(rather weak) correlation between $y$ and $|\qf|$, i.\ e.\
the cms energy and the momentum transfer of the process. 
The above approximation can be valid only in a limited range.
In the formal limit of $y \to \infty$ one has to perform
the integrations over $\lambda$ and $\nu$ in reverse order.
Integrating over $\nu$ first one obtains the saddle point 
at $\nu=0$ and evaluating the integral in the corresponding 
approximation one finds that the singularity structure of the 
integrand of the $\lambda$ integration has changed.
Instead of two separated poles at $\lambda=\pm \nu$ one 
double pole at $\lambda=0$ is found. In the case $|\qf|/(2 \Delta) \ll 1$
the residue of this pole dominates and one gets the result
\beqn
\frac{d \sigma^{\gamma^* q}}{d t} (x,t) = 
\pi \left(\frac{4}{9}\right)^2 
\frac{\alpha_s^4}{t^4}\;
\left| \,
{\cal C} \, 16 
\pi^2 
\left(\frac{|\qf|^2}{4 \Delta^2}
\right)^{\frac{3}{2}}
(3\log 2-2 - \fez \log \frac{|\qf|^2}{4 \Delta^2})
\frac{e^{y \frac{ N_c\alpha_s}{\pi}4 \log 2}}
{[N_c \alpha_s 14 \zeta(3) y]^{\frac{3}{2}}}
\, \right|^2
\label{approx3}
\eeqn
In this limit the $y$ and $|\qf|$ dependence again factorize,
i.\ e.\ there is no energy-momentum transfer correlation.
\\
The last interesting case to be studied is the limit of 
small momentum transfer at fixed $y$, i.\ e.\ we start from the 
approximation eq.\ (\ref{asylow}) and assume 
$\log (4 \Delta^2)/|\qf|^2 \gg y$. The saddle point 
of the $\nu$ integration is then found 
at $i\nu=\pm 1/2 \mp 
[N_c \alpha_s/\pi \cdot y \, \log^{-1}(4 \Delta^2)/|\qf|^2]^{\fez}$  
and the cross section is expressed through the amplitude in 
the double-leading-log limit
\beqn
\frac{d \sigma^{\gamma^* q}}{d t} (x,t) = 
\pi \left(\frac{4}{9}\right)^2 
\frac{\alpha_s^4}{t^4}\;
\left| \,
{\cal C} \,
\frac{2}{\sqrt{\pi}}
\left(\frac{|\qf|^2}{4 \Delta^2}\right)^2 
\frac{e^{\sqrt{4 \frac{N_c \alpha_s}{\pi} y 
\log \frac{4\Delta^2}{|\qf|^2}}}}
{[N_c \alpha_s/\pi \, y \log 4 \Delta^2/|\qf|^2]^{\frac{3}{4}}} 
\log \frac{4\Delta^2}{|\qf|^2}
\, \right|^2
\label{approx4}
\eeqn
In this expression $|\qf|^2$ no longer has the meaning of the
momentum transfer. In the limit which is considered here
$|\qf|$ can be neglected along the gluon ladder and the result for the 
amplitude becomes identical to the $t=0$ result for the gluon density
in the double logarithmic (DL) limit
with the upper large scale $4 \Delta^2$ and the virtuality
of the incoming parton being of the order of $t$. Thus $|\qf|^2$ becomes 
the lower scale of the gluon evolution and it is clear that the
amplitude diverges in the formal limit $|\qf| \to 0$ since we are
considering a colored initial state. 
To obtain a finite $t=0$ cross section one has to go beyond the simple
factorization formula (\ref{eq22init}) 
and has to couple the two gluon ladder 
to the proton in a gauge invariant way. 
An example for such a coupling will be presented 
and discussed in the next section.
The above expression gives a higher twist contribution to DIS 
since it scales as $1/Q^8$. 
\begin{figure}[!h]
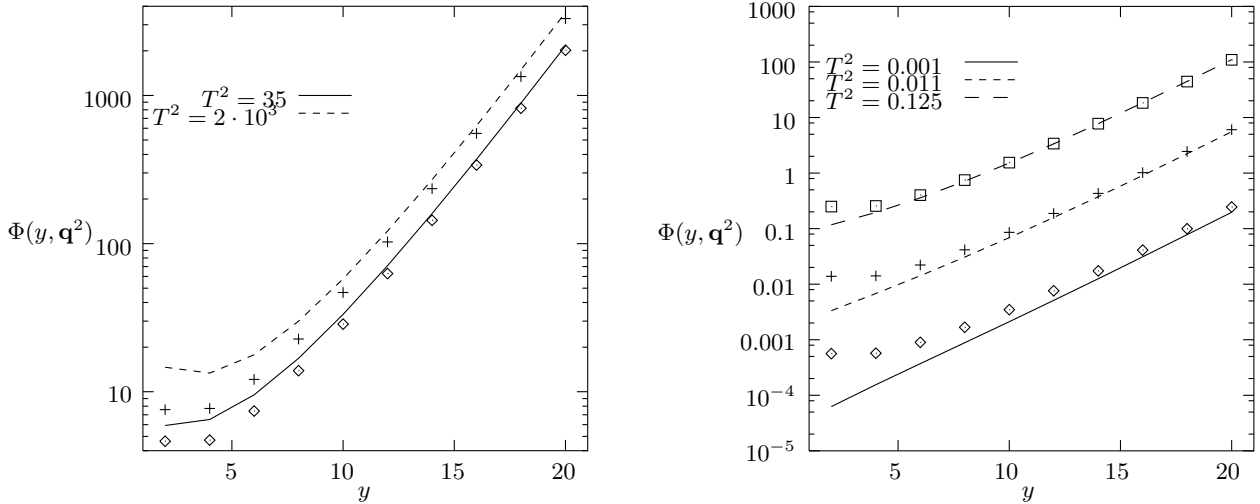

\begin{center}
\input vmfigydep.pstex_t
\end{center}
\caption{
Comparison of the analytic results (symbols) contained in 
eq.\ (\ref{approx1}) (left hand side) and eq.\ (\ref{approx2})
(right hand side) with the exact numerical evaluation (lines) of 
$\Phi(y,\qf^2)$ 
based on eq.\ (\ref{eq22fin})
for different value of $T^2= |\qf|^2/(4 \Delta^2)$.
\label{fig223}
}
\end{figure}
The accuracy of the large $y$ approximations can be read off from fig.\
\ref{fig223}. We display the inverse Mellin transformed partial wave amplitude
$\Phi(y,\qf^2)$ as a function of $y$ for different values of 
$|\qf|/(2 \Delta)$ and compare with the analytic results eqs.\
(\ref{approx1}) and (\ref{approx3}). 
One finds very good agreement for $y \gtrsim 10$.
From the right hand figure one can conclude that the approximation 
contained in eq.\ (\ref{approx2}) is rather bad in the kinematic range 
considered here. This approximation misses the $\log |\qf|/(2 \Delta)$
contribution which is very important since $3 \log 2 -2 \simeq 0.079 $
is a very small coefficient. Hence this approximation largely underestimates
the true result.
\begin{figure}[!h]
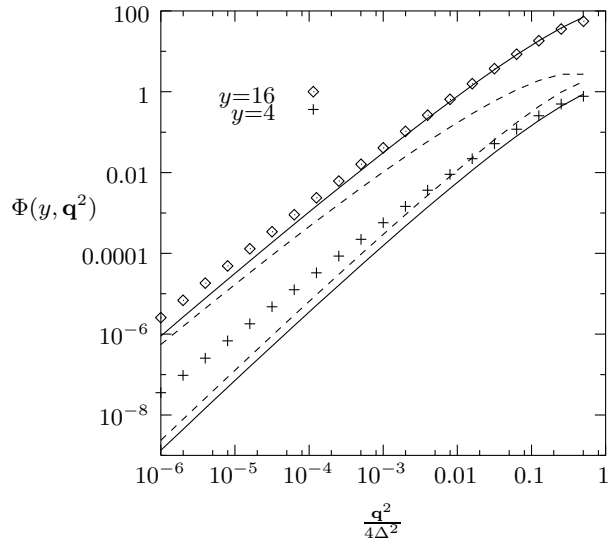

\begin{center}
\input vmfig5.pstex_t
\end{center}
\caption{
Comparison of the $|\qf|$-dependence of the analytical results in 
eq.\ (\ref{approx3}) (symbols) and eq.\ (\ref{approx4}) (dashed line)
with the numerical result (solid line).
\label{fig224}
}
\end{figure}
In fig.\ \ref{fig224} we compare the approximations contained in
eqs.\ (\ref{approx3}) and (\ref{approx4}) with the exact numerical
result. One sees that for very small $|\qf|/(2\Delta)$ the double
logarithmic approximation indeed becomes valid, whereas in the range of not so 
small $|\qf|/(2\Delta)$ the BFKL type approximation is better, 
especially for large $y$. 
\subsubsection{The slope of the BFKL amplitude}
The fact that for very small $|\qf|/(2\Delta)$ the double logarithmic
approximation provides a good description
is remarkable since it 
implies a correlation of the 
momentum transfer 
and the energy.   
Such a behavior is usually related to the slope of the Regge trajectory
which is exchanged in the process. The slope $\alpha'$
is defined as the coefficient of the term linear in $t$ in the small-$t$
expansion of the trajectory function $\alpha(t)=\alpha_0+\alpha' \cdot t$.
Assuming dominance of this trajectory the amplitude
behaves as $(s/s_0)^{\alpha(t)} \;= \;
(s/s_0)^{\alpha_0}\exp(\alpha' y t)$ ($y =\log s/s_0$)
and $\alpha'$ determines the shrinkage of the diffraction peak, i.\ e.\
the increase of the slope of the $t$-dependence with increasing energy.
Using the above expression the slope 
\footnote{The term 'slope' appears here with two different meanings.
On the one hand we have the slope parameter $\alpha'$ of the 
Regge trajectory function. The slope of the diffraction peak on the 
one hand is a measure for the steepness of the decrease of the 
diffractive cross section as a function of $t$ in the small-$t$ region.
} 
$\delta$ of the diffraction peak 
equals
\beqn
\delta=\alpha' \log s/s_0 +\delta_0
\eeqn
where $\delta_0$ is determined through the $t$-dependent residue function.
Due to conformal invariance the BFKL-singularity is a fixed 
($t$-independent) cut and 
{\em prima facie} there is no shrinkage.
An effective slope is found if the conformal invariance is broken by
introducing running $\alpha_s$ and additional scale parameters to 
modify the infrared region as discussed 
in \cite{levrysk} and \cite{niko1}. 
In our approach conformal symmetry is maintained but there are scale
parameters which enter through the convolution with the impact factors
of the scattered particles. The correlation with energy
is mediated through the $\nu$ integration. For large energy the
saddle point of this integration becomes constant and the $s$ and $t$
dependence factorize as in eq. (\ref{approx3}), i.\ e.\ 
there is indeed no 
shrinkage. In the subasymptotic region the saddle point is a function of 
$s$ and $t$ and the behavior of eq.\ (\ref{approx4}) is found which implies
an effective slope. Using the general formulae given above we can define
the effective slope parameter $\alpha'_{\mbox{\tiny{eff}}}$ 
through the relation
\beqn
 \alpha'_{\mbox{\tiny{eff}}} = - \frac{\partial}{\partial y}
\frac{\partial}{\partial \qf^2}  \log \Phi(y,\qf^2)
\eeqn
and find from eq.\ (\ref{approx4})
\beqn
\alpha'_{\mbox{\tiny{eff}}} =  \frac{1}{|\qf^2|}\frac{N_c\alpha_s}{2 \pi}
\frac{1}{\sqrt{N_c\alpha_s/\pi \; y \; \log 4 \Delta^2/ |\qf|^2}} \, ,
\label{slope}
\eeqn 
i.\ e.\ the slope is a function of $t$ and $s$.
The same energy dependence of the effective slope  
was found in \cite{levrysk} by using the diffusion approximation 
in impact parameter space for an infrared modified BFKL equation.
It seems reasonable to assume that this effective slope is universal,
i.\ e.\ it does not depend on the specific process. The process
enters only through the scale which is $4\Delta^2$ here.
It should be stressed, however, that the result in eq.\ (\ref{slope})
is obtained in a limit which does not correspond to the diffusion 
limit of the BFKL equation.  
Using eq.\ (\ref{slope}) as an estimate for the slope and inserting the
realistic values $-t = 1 \,\mbox{GeV}^2$, 
$y=8$, $4\Delta^2=10 \,\mbox{GeV}^2$
we find $\alpha'_{\mbox{\tiny{eff}}} \simeq 0.05 \,\mbox{GeV}^{-2}$
which has to be regarded as very small in comparison with the 
slope $\alpha_{\mbox{\tiny{soft}}}=0.25 \,\mbox{GeV}^{-2}$ of the 
soft (Donnachie-Landshoff) pomeron \cite{donnla}.
A more accurate calculation \cite{jeff2}
based on the exact numerical evaluation
of $\Phi(y,\qf^2)$ in eq.\ (\ref{eq22fin})
leads to a similar result. It should be stressed here again 
that in the small-$t$ limit which is considered here one should go
beyond the Mueller-Tang prescription. We will come back to this 
point in the 
next section. 
\\
Finally in fig.\ \ref{fig225} we present the full differential partonic
cross section normalized to its value at $-t = 1 \,\mbox{GeV}^2$
for small values of $-t$.
\begin{figure}[!h]
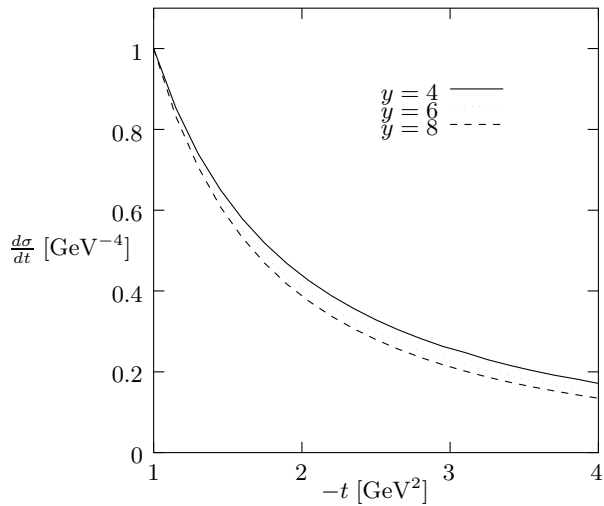

\begin{center}
\input vmfig6.pstex_t
\end{center}
\caption{
The normalized photon-parton cross section as a function of 
$t$ for different values of $y$.
\label{fig225}
}
\end{figure}
These curves illustrate the shrinkage phenomenon.
For large energy the slope of the $t$-dependence 
increases.
Based on the starting formula (\ref{eq22init})
an absolute prediction for the cross section is possible.
To obtain an estimate for the total cross section we perform a 
fourfold numerical integration over the auxiliary parameter $\lambda$,
the conformal dimension $\nu$ and the phase space variables $x$ and $t$.
We have chosen to integrate $\lambda$ and $\nu$ by iteratively using 
conventional methods based on the interpolation of the integrand 
between a discrete set of points. The phase space integration is done 
with a Monte-Carlo algorithm \cite{vegas}. 
For the photoproduction case $Q^2=0$ and 
the vector meson being the $J/\Psi$ with mass $M_{J/\Psi}=3.10\, 
\mbox{GeV}$ we integrate over $x$ from 0.2 to 0.6 and over $|t|$ from
$1 \,\mbox{GeV}^2$ to $20 \,\mbox{GeV}^2$. We find for the 
total photon proton cross section values 
of 480\,nb, 99\,nb and 24\,nb for $W = 200 \,\mbox{GeV}, 100 \,\mbox{GeV}$
an $50  \,\mbox{GeV}$, respectively.
\newpage
\subsection{The BFKL pomeron in DIS diffractive dissociation}
\label{sec23}
The subject of this section is a study of deep inelastic diffractive 
dissociation at zero and finite momentum transfer based 
on the BFKL pomeron, or, more general, the leading log($1/x$)
resummation of perturbative QCD.
We investigate the process 
$\gamma^*(q)+P(p) \to X(q+\xi)+ P(p-\xi)$.
Experimentally, e.\ g.\ in electron-proton collisions at HERA, 
these processes show the 
characteristic signature of a rapidity gap. There is a region between the 
outgoing proton and the system $X$ in which no particles are observed.
In the first step we will consider the system $X$ as being made up 
of a quark-antiquark pair, afterwards we will generalize to 
$q\bar{q} + n \; \mbox{gluon}$ final states. The analysis presented here
is to a large extent based on the results on the triple Regge limit 
in perturbative QCD contained in \cite{bartels} and 
\cite{bartels-wue}.  
\\
The interest in this type of process stems from the fact that it 
is closely related to the unitarization problem.
Both the leading-log ($Q^2$) (eq.\ (\ref{dla-gluon})) 
and the
leading-log ($1/x$) (eq.\ (\ref{saddle}))
asymptotics at small $x$ violate unitarity,
i.\ e.\  they grow to fast with decreasing $x$.
From unitarity, at most a logarithmic increase at small $x$ is 
expected
\footnote{
A rigorous proof of this statement has not been given up to 
now. It is however commonly accepted, based on the proof for 
hadron-hadron scattering and the use of vector meson dominance.
For recent attempts towards a proof see \cite{buchhaidt} and 
\cite{levuni}.
\label{foot}
}.   
The first subleading corrections which serve to restore unitarity
have been classified and analyzed in \cite{bartels} and 
\cite{bartels-wue}.
It turns out that a large part of these corrections contributes to 
diffractive dissociation. In other words, DIS diffractive dissociation
is the appropriate process to study, both theoretically and experimentally,
properties of unitarity corrections at small $x$. 
It will be shown that the BFKL four-gluon amplitude constitutes a 
very important element of these corrections.
\\
From the start we will assume that the large scale $Q^2$ of 
deep inelastic scattering justifies the perturbative approach.
A critical discussion of this assumption will be given  
after the results have been stated.
\subsubsection{The production of $q\bar{q}$-pairs}
We begin with the discussion of diffractive $q\bar{q}$-production.
Photon diffractive dissociation means that the virtual photon which is 
radiated off a fast electron dissociates into a $q\bar{q}$-pair 
which in turn scatters diffractively off the proton, i.\ e.\ the photon
stays intact or is weakly excited after the collision.
No color is exchanged between the proton and the $q\bar{q}$-pair
and consequently in the framework of perturbative QCD
the amplitude in lowest order contains a two-gluon exchange. 
At high photon-proton cms (or small $x$ equivalently)
the dominant higher order corrections to the amplitude are resummed by the 
BFKL equation.
The general structure of the squared amplitude is represented graphically
in fig.\ \ref{figincdiff1}. 
\begin{figure}[!h]
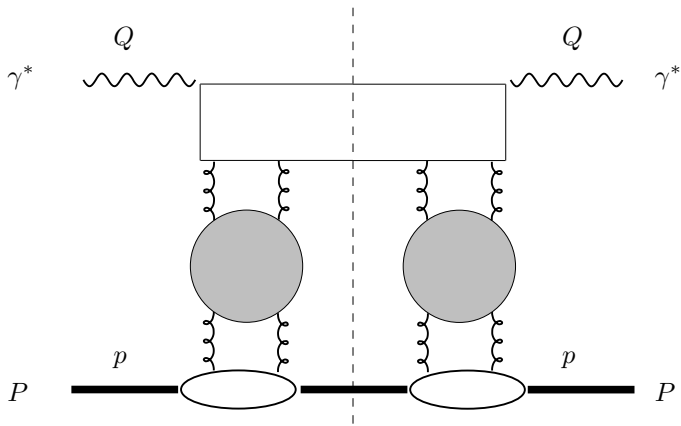

\begin{center}
\input incdiff.pstex_t
\end{center}
\caption{
Photon
diffractive dissociation into a $q\bar{q}$-pair.
\label{figincdiff1}
}
\end{figure}
The shaded blobs represent the BFKL amplitude 
which is coupled to the proton through an impact factor to be 
specified below.
Mathematically, we use the partial wave representation for the cross section
\beqn
\frac{d \sigma^{\gamma^*P}}{d t}(x,t)
= 
\frac{1}{Q^4}
\int \frac{d \omega_1}{2 \pi i} \int \frac{d \omega_2}{2 \pi i}
\left(\frac{1}{x}\right)^{\omega_1+\omega_2}
\Phi_{\omega_1\omega_2}(\qf^2)
\;\;\;; \;\;\;\qf^2=-t
\eeqn
where the partial wave amplitude $\Phi_{\omega_1\omega_2}$ has the 
following form
\beqn
\frac{1}{Q^4}\Phi_{\omega_1\omega_2}(\qf^2) = 
\int \frac{d^2\kf_1}{(2 \pi)^3}
\int \frac{d^2\kf_2}{(2 \pi)^3}
D_{(4,0)}(\kf_1,\qf-\kf_1,\kf_2,-\qf-\kf_2) \phantom{xxxxxxxxxx}
\nonumber \\
\int \frac{d^2\lf_1}{(2 \pi)^3} 
 \Phi_{\omega_1}(\kf_1,\lf_1;\qf) \,\phi_P(\lf_1,\qf)
\int \frac{d^2\lf_2}{(2 \pi)^3}
\Phi_{\omega_2}(\kf_2,\lf_2;-\qf) \,\phi_P(\lf_2,-\qf) 
\label{start1}
\eeqn
Here $\Phi_{\omega_i}$ is the BFKL amplitude,
$D_{(4,0)}$ represents the quark-loop diagram with four gluons 
attached and $\phi_P$ is the coupling of the BFKL amplitude to the 
proton. This coupling is not determined theoretically and one has to 
use a phenomenological model. A convenient choice is the 
rescaled photon-meson
formfactor $\Delta^2 \phi_V$ 
from the previous section (eq.\ (\ref{phot-vecmes}))
with $\Delta^2$
a characteristic proton scale of the order of $1 \,\mbox{GeV}^2$ and 
${\cal C}$ an undetermined dimensionless constant.  
As to the four gluon amplitude $D_{(4,0)}$, one has to consider 16 diagrams
corresponding to the different ways to couple four gluons to a 
quark-antiquark pair
\cite{muelld,nikd,levwued}.
In the inclusive case which is under study here
the momentum structure of these diagrams simplifies considerably
and can be reduced to the two gluon amplitude $D_{(2,0)}$.
This is due to the fact that two or three gluons which 
couple to the same quark line collapse into one reggeon.
The reggeon is a composite object consisting of 
two or more gluons, but it behaves
effectively like a single gluon.
The expression for the four gluon amplitude then reads
\beqn
D_{(4,0)}(\kf_1,\kf_2,\kf_3,\kf_4)
=
4 \pi \alpha_s \frac{\sqrt{2}}{3}
\left[ \sum_{i=1}^4 D_{(2,0)}(\kf_i,-\kf_i)
-\sum_{i=2}^4 D_{(2,0)}(\kf_1+\kf_i,-\kf_1-\kf_i)
\right]
,\left( \sum_{i=1}^4 \kf_i =0\right)
\label{de4}
\eeqn
The coefficient in front results from the projection of the color structure
of the four gluon amplitude
on color zero in the subsystems of the gluons with momenta 
$(\kf_1,\kf_2)$ and $(\kf_3,\kf_4)$. For further details see 
\cite{bartels-wue}.
The function $D_{(2,0)}(\kf,-\kf)$ is well known \cite{coeff}.
It is identical to the coefficient function in DIS in the 
$\kf$-factorization formalism as discussed in section \ref{sec21}.
We consider transverse and longitudinal photons separately
and use the Mellin representation
\beqn
D_{(2,0)}(\kf,-\kf) &=& \int \frac{d \nu}{2 \pi} 
\left(\frac{\kf^2}{Q^2}\right)^{\frac{1}{2}+i\nu}D_{(2,0)}(\nu)
\\
D_{(2,0)\,T}(\nu) &=& \sum_f e_f^2 
\alpha_{\mbox{\tiny em}}\alpha_s \frac{1}{\sqrt{8}}
\frac{\Gamma(5/2-i\nu)}{\Gamma(2-i\nu)}
\frac{\Gamma(1/2-i\nu)}{1/2-i\nu}
\frac{\Gamma(1/2+i\nu)}{1/2+i\nu}
\frac{\Gamma(5/2+i\nu)}{\Gamma(2+i\nu)}
\\
D_{(2,0)\,L}(\nu) &=& \sum_f e_f^2 
\alpha_{\mbox{\tiny em}}\alpha_s \frac{1}{\sqrt{8}}
\frac{\Gamma(5/2-i\nu)}{\Gamma(2-i\nu)}
\frac{\Gamma(1/2-i\nu)}{3/2-i\nu}
\frac{\Gamma(1/2+i\nu)}{3/2+i\nu}
\frac{\Gamma(5/2+i\nu)}{\Gamma(2+i\nu)}
\eeqn
The key difference between the transverse and the longitudinal 
part is the double pole of $D_{(2,0)\,T}(\nu)$ at $i\nu=1/2$
where $D_{(2,0)\,L}(\nu)$ has a simple pole. It means that 
at large $Q^2$ the transverse part is logarithmically enhanced relative to
the longitudinal part. This is well known in standard DIS where the 
longitudinal cross section is an order $\alpha_s$ contribution
(it is zero in the naive parton model).
It should be mentioned that a compact expression for the four gluon 
amplitude $D_{(4,0)}$ is obtained if the configuration space 
representation is used \cite{muelld,nikd}. 
This corresponds to Fourier transformation of eq.\ (\ref{de4})
\beqn
D_{(4,0)}(\kf_1,\kf_2,\kf_3,\kf_4)
=
4 \pi \alpha_s \frac{\sqrt{2}}{3}
\int d^2 \rho \; |\Psi(\rho)|^2 \prod_{i=1}^4(1-e^{i\kf_i\rho})
\eeqn
Here $\Psi(\rho)$ is the light cone wave function of the photon.
The explicit expression for $|\Psi(\rho)|^2$
which involves generalized Bessel functions 
can be found in \cite{muelld,nikd,blotwue}.
One can see that the rule for constructing $n$-gluon amplitudes is simple.
For every additional gluon with momentum $\kf_j$ 
a factor $(1-\exp(i\kf_j\rho))$
has to be added.
\\
Now we are ready to calculate the partial wave $\Phi_{\omega_1\omega_2}$.
We insert the factorizing momentum space expressions 
for the BFKL (i.\ e.\ the momentum space analogue of 
eq.\ (\ref{solnonzero})) 
amplitudes into eq.\ (\ref{start1}).
The partial wave then decomposes into three parts.
We have two factors which result from integrating the lower 
factor of the BFKL amplitude with the pomeron-proton coupling
\beqn
\Lambda^{(\nu_i,n_i)}(\qf^2)
= \int \frac{d^2 \lf_i}{(2\pi)^3} \phi_P(\lf_i,\qf-\lf_i)
E^{(\nu_i,n_i)\,\ast}(\lf_i,\qf-\lf_i) \;\;\;\;\;\;(i=1,2)
\eeqn
The key element results from integrating the four gluon amplitude 
with the upper factors of the BFKL amplitudes. This defines 
a vertex function $\Theta^{(\nu,\nu_1,\nu_2;n_1,n_2)}(\qf^2)$
\beqn
\Theta^{(\nu,\nu_1,\nu_2;n_1,n_2)}(\qf^2) = 
\int \frac{d^2 \kf_1}{(2 \pi)^3}
\int \frac{d^2 \kf_3}{(2 \pi)^3}
\left[
\sum_{i=1}^4 \left(\frac{\kf_i^2}{Q^2}\right)^{\fez+i\nu}
-\sum_{i=2}^4 \left(\frac{(\kf_1+\kf_i)^2}{Q^2}\right)^{\fez+i\nu}
\right]
\nonumber \\ \cdot \,
E^{(\nu_1,n_1)}(\kf_1,\qf-\kf_1) 
E^{(\nu_2,n_2)}(\kf_3,-\qf-\kf_3)
\phantom{xxx} 
\label{Theta}
\eeqn
where from momentum conservation we have $\kf_2=\qf-\kf_1,\kf_4=-\qf-\kf_3$.
\\
The partial wave amplitude is then given as
\beqn
\frac{1}{Q^4}
\Phi_{\omega_1\omega_2}(\qf^2)=
\sum_{n_1,n_2=-\infty}^{+\infty}
\int \frac{d \nu}{2 \pi}
\int \frac{d \nu_1}{2 \pi} \int \frac{d \nu_2}{2 \pi}
D_{(2,0)}(\nu) \;
\Theta^{(\nu,\nu_1,\nu_2;n_1,n_2)}(\qf^2)
\prod_{i=1}^2
\frac{         
\Lambda^{(\nu_i,n_i)}(\qf^2)
}
{\omega_i-\chi(\nu_i,n_i)}
\label{pwvam}
\eeqn
In the following we drop the index $n$ since we restrict the 
calculation to zero conformal spin.
For the function $\Lambda^{(\nu)}(\qf^2)$ the results from 
the previous section can be used. Performing the steps that led to 
eq.\ (\ref{eq22fin}) we obtain
\beqn
\Lambda^{(\nu)}(\qf^2)= 
-{\cal C}\,\frac{\Delta^2}{\qf^2}\,(\qf^2)^{-\frac{1}{2}+i\nu} 
4^{- 2 i\nu}  \pi \frac{\Gamma(1-i\nu)\Gamma(-\fez+i\nu)}
{\Gamma(i\nu)\Gamma(\ftz-i\nu)}
\phantom{xxxxxxxxxxxxxxxxxxxxxxxxx}
\nonumber \\
\int \frac{d \lambda}{\pi}
\left(\frac{\qf^2}{4\Delta^2}\right)^{\frac{3}{2}+i \lambda}
\frac{\Gamma^2(\frac{3}{2}+i\lambda)
\Gamma(-i\lambda+i\nu)\Gamma(-i\lambda-i\nu)}
{\Gamma(\fez-\fez(i\lambda+i\nu))\Gamma(\fez-\fez(i\lambda-i\nu))
\Gamma(1+\fez(i\lambda+i\nu))\Gamma(1+\fez(i\lambda-i\nu))}
\label{lambda}
\eeqn
The core of this section is the calculation of the vertex function.
The first point to realize is that out of the seven terms in 
the sum in eq.\ (\ref{Theta}) only two give a nonzero 
contribution. This is however true only if we keep $\qf^2 \neq 0$.
If we set $\qf^2=0$ in (\ref{Theta}) it is crucial to keep all seven 
terms to obtain a finite result \cite{markphd}.
For finite $\qf^2$ on the other hand we can go back to the mixed 
representation for $E^{(\nu)}$ (with arguments $\qf,\rho$)
and find that for the cases in which the two gluons from the BFKL amplitude 
are coupled to the same quark line a $\delta^{(2)}(\rho)$-function
is obtained. If then $\mbox{Re}(i\nu)<1/2$ 
is assumed the $\rho$-integration 
of the corresponding term gives zero.
In the end an analytic continuation to 
$\mbox{Re}(i\nu)>1/2$ can be performed.
The same reasoning was used in the previous section when the second term
in eq.\ (\ref{besselk0}) was omitted.
This shows that the contributions in which two gluons couple to the 
same quark line serve as pure subtraction terms which are needed for 
convergence but give no finite contribution. Calculating with finite
$\qf^2$ provides a regularization in which these terms can be omitted from 
the beginning. 
Furthermore by shifting the integration variable it can be shown that 
the two terms which survive are in fact identical.
After all one nontrivial two-loop integration has to be carried out
in (\ref{Theta}). We use the representation (\ref{momspacef}) 
for the $E^{(\nu)}$ functions and get 
the following expression
\beqn
\Theta^{(\nu,\nu_1,\nu_2)}(\qf^2) = 
- 2
\int \frac{d^2 \kf_1}{(2 \pi)^3}
\int \frac{d^2 \kf_3}{(2 \pi)^3}
\left(\frac{(\kf_1+\kf_3)^2}{Q^2}\right)^{\fez+i\nu}
\phantom{xxxxxxxxxxxxxxxxxxxx}\label{theta2} 
\\ 
4 \pi  4^{2i\nu_1} \frac{\Gamma(1+i\nu_1)}{\Gamma(-i\nu_1)}
\frac{\Gamma(\ftz-i\nu_1)\Gamma(-\fez-i\nu_1)}
{\Gamma(\fez+i\nu_1)\Gamma(\fez-i\nu_1)}
\int_0^1 dx[x(1-x)]^{-\fez+i\nu_1}[\qf^2x(1-x)+(\kf_1-x\qf)^2]^{-\ftz-i\nu_1}
\nonumber \\
4 \pi  4^{2i\nu_2} \frac{\Gamma(1+i\nu_2)}{\Gamma(-i\nu_2)}
\frac{\Gamma(\ftz-i\nu_2)\Gamma(-\fez-i\nu_2)}
{\Gamma(\fez+i\nu_2)\Gamma(\fez-i\nu_2)}
\int_0^1 dy[y(1-y)]^{-\fez+i\nu_2}[\qf^2y(1-y)+(\kf_3+y\qf)^2]^{-\ftz-i\nu_2}
\nonumber 
\\
_2F_1\left(\!\!\!\!
\begin{matrix} 
\ftz+i\nu_1\!\!\!\!&\!\!\!\!,\,-\fez+i\nu_1  
\\ \phantom{xxxxxx}1&\phantom{.}
\end{matrix}
;\frac{(\kf_1-x\qf)^2}{\qf^2x(1-x)+(\kf_1-x\qf)^2}\!\!
\right)\,
_2F_1\left(\!\!\!\!
\begin{matrix}
\ftz+i\nu_2\!\!\!\!&\!\!\!\!,\,-\fez+i\nu_2   
\\ \phantom{xxxxxx}1&\phantom{.}
\end{matrix}
;\frac{(\kf_3+y\qf)^2}{\qf^2y(1-y)+(\kf_3+y\qf)^2}\!\!
\right)
\nonumber 
\eeqn
It is now necessary to use the usual 
Mellin-Barnes representation for the 
hypergeometric function \cite{bateman}.
The momentum dependence then appears in terms of the form
$[\qf^2x(1-x)+(\kf-x\qf)^2]^{\gamma}$ with some exponent $\gamma$.
To perform the momentum integration it is then convenient to use 
a Mellin-Barnes representation also for these factors.
Performing these steps we end with the following expression for 
the last three lines in eq.\ (\ref{theta2})
\beqn
\prod_{i=1}^2
4 \pi  \,4^{2i\nu_i} \frac{\Gamma(1+i\nu_i)}{\Gamma(-i\nu_i)}
\frac{\Gamma(\ftz-i\nu_i)\Gamma(-\fez-i\nu_i)}
{\Gamma(\fez+i\nu_i)\Gamma(\fez-i\nu_i)}
\int_0^1 d x[x(1-x)]^{-\fez+i\nu_1}\int_0^1 d y[y(1-y)]^{-\fez+i\nu_2}
\nonumber  \\
\frac{1}{|\Gamma(\ftz+i\nu_1)|^2|\Gamma(\ftz+i\nu_2)|^2}
\int \frac{d s_1}{2 \pi i} \int \frac{d s_2}{2 \pi i}
\prod_{i=1}^2 \frac{\Gamma(-s_i)\Gamma(-s_i-2i\nu_i)\Gamma(s_i+\ftz+i\nu_i)}
{\Gamma(-s_i-\fez-i\nu_i)}
\nonumber \\
\left[\qf^2 x(1-x)\right]^{s_1}
\left[\qf^2 y(1-y)\right]^{s_2}
\left[(\kf_1-x\qf)^2\right]^{-s_1-\ftz-i\nu_1}
\left[(\kf_3+y\qf)^2\right]^{-s_2-\ftz-i\nu_2}
\eeqn
Now we can perform a shift of the integration variable and
the momentum integration takes the simple form
\beqn
(Q^2)^{-\fez-i\nu}
\int \frac{d^2 \kf_1}{(2\pi)^3}
\int \frac{d^2 \kf_3}{(2\pi)^3}
\frac{\left[(\kf_1+\kf_3+(x-y)\qf)^2\right]^{\fez+i\nu}}
{(\kf_1^2)^{s_1+\ftz+i\nu_1}(\kf_3^2)^{s_2+\ftz+i\nu_2}}
\\
=
\frac{(Q^2)^{-\fez-i\nu}}{(2 \pi)^6}
\pi^2 
\left[\qf^2|x-y|^2\right]^{-\fez-s_1-s_2+i\nu-i\nu_1-i\nu_2}
\frac{\Gamma(\ftz+i\nu)}{\Gamma(-\fez-i\nu)}
\nonumber \\
\cdot
\prod_{i=1}^2\frac{\Gamma(-s_i-\fez-i\nu_i)}
{\Gamma(s_i+\ftz+i\nu_i)}
\frac{\Gamma(\fez+s_1+s_2-i\nu+i\nu_1+i\nu_2)}
{\Gamma(\fez-s_1-s_2+i\nu-i\nu_1-i\nu_2)}
\eeqn
This result is now inserted in eq.\ (\ref{theta2})
and it remains to perform the 
integration of the Feynman parameters. Due to the symmetry 
properties of the integrand the integral can be readily
expressed as a $_3F_2$ generalized hypergeometric function 
which can be reduced to $\Gamma$-functions using Whipples theorem 
\cite{bateman} 
\beqn
\int_0^1 d x[x(1-x)]^{-\fez+s_1+i\nu_1}
\int_0^1 d y[y(1-y)]^{-\fez+s_2+i\nu_2}
\left[|x-y|^2\right]^{-\fez-s_1-s_2+i\nu-i\nu_1-i\nu_2}
\nonumber \\
= 
4^{1+s_1+s_2+i\nu_1+i\nu_2-2 i\nu}
\frac{\pi}{\Gamma(\fez+i\nu)}
\prod_{i=1}^2
\frac{\Gamma(\fez+s_i+i\nu_i)}{\Gamma(\fez-s_i-i\nu_i+i\nu)}
\phantom{xxxxxxxxxxxxxx}
\nonumber \\
\cdot
\frac{\Gamma(-2s_1-2s_2-2i\nu_1-2i\nu_2+2i\nu)
\Gamma(-s_1-s_2-i\nu_1-i\nu_2+2i\nu)}
{\Gamma(\fez-s_1-s_2-i\nu_1-i\nu_2+i\nu)}
\label{whipple}
\eeqn
Collecting then the terms from eqs.\ 
(\ref{theta2}) - (\ref{whipple}) and performing 
all possible cancellations we obtain the final result for the 
vertex function 
\beqn
\Theta^{(\nu,\nu_1,\nu_2)}(\qf^2)= - 2
\frac{(\qf^2)^{-\fez+i\nu-i\nu_1-i\nu_2}}{16\pi^3(Q^2)^{\fez+i\nu}} 
4^{i\nu_1+i\nu_2-2 i\nu}
\frac{\Gamma(\ftz+i\nu)}
{\Gamma(-\fez-i\nu)\Gamma(\fez+i\nu)}
\phantom{xxxxxxxxxxxxxxxxxxxxxx}
\label{thetafin}
\\
\prod_{i=1}^2 \left[4 \pi \, 4^{2i\nu_i}
\frac{\Gamma(1+i\nu_i)}{\Gamma(-i \nu_i)}
\frac{\Gamma(-\fez-i\nu_i)}
{\Gamma(\ftz+i\nu_i)\Gamma(\fez+i\nu_i)
\Gamma(\fez-i\nu_i)}\right]
\prod_{i=1}^2 \left[
\int \frac{d s_i}{2 \pi i}
\Gamma(-s_i)\Gamma(-s_i-2i\nu_i)
\frac{4^{s_i}\Gamma(\fez+s_i+i\nu_i)}{\Gamma(\fez-s_i-i\nu_i+i\nu)}
\right]
\nonumber \\
\frac{\Gamma(\fez+s_1+s_2-i\nu+i\nu_1+i\nu_2)}
{\Gamma(\fez-s_1-s_2+i\nu-i\nu_1-i\nu_2)}
\frac{\Gamma(-2s_1-2s_2-2i\nu_1-2i\nu_2+2i\nu)
\Gamma(-s_1-s_2-i\nu_1-i\nu_2+2i\nu)}
{\Gamma(\fez-s_1-s_2-i\nu_1-i\nu_2+i\nu)}
\nonumber
\eeqn
The partial wave amplitude $\Phi_{\omega_1\omega_2}(\qf^2)$ can now 
be determined by inserting the expression for the vertex function 
$\Theta^{(\nu,\nu_1,\nu_2)}$ and the functions $\Lambda^{(\nu_i)}$ 
into eq.\ (\ref{pwvam}). The result is fairly complicated and even a
numerical evaluation does not seem to be a simple task. 
We will investigate in the following the limits $\qf^2 \to 0$ and 
$\qf^2 \to \infty$.
\subsubsection{The limit $\qf^2=0$}
Let us first observe that the $\qf^2$-dependence of the vertex 
function is simply $(\qf^2)^{-\fez+i\nu-i\nu_1-i\nu_2}$
which follows of course from purely dimensional considerations.
From this dependence it can be concluded that a finite limit of the 
vertex function in the limit $\qf^2 \to 0$ implies a constraint 
on the conformal dimensions $\nu,\nu_1,\nu_2$. 
A well-defined, nonzero limit requires $ -\fez+i\nu-i\nu_1-i\nu_2=0$,
otherwise the vertex function would either be zero or infinity. 
To show how this works in detail one has to consider the singularity 
structure of the $\nu$-integration in (\ref{pwvam})
\\
It is important to recall first from the previous section that the 
$\qf^2 \to 0$ limit of the functions $\Lambda^{(\nu_i)}$
is determined by the poles at $i\lambda=\pm i\nu$ in the integrand 
in eq.\ (\ref{lambda}). 
This means that for $\qf^2 \to 0$ we have 
$\Lambda^{(\nu_i)}(\qf^2)=
(\Delta^2)^{-1/2}[a(\nu_i)+b(\nu_i)(\qf^2)^{2i\nu_i}]$.
\\
As to the $\nu$-integration, the $\nu$-contour in (\ref{pwvam})
runs on the real axis and the relevant part of the integrand 
is $(\qf^2)^{-1/2+i\nu}$.
For $\qf^2 \to 0$ one has to shift the contour into the lower
half plane in order to obtain a power series in $\qf^2$.
The integrand has two factors which generate singularities in the 
lower half plane. From combining the coefficient function $D_{(2,0)}(\nu)$
with the three $\Gamma$-functions in the first line of 
eq.\ (\ref{thetafin}) one gets simple poles at $i\nu=1/2,3/2,\, ...$
(these poles are absent in the longitudinal case).
Furthermore one has a string of simple poles from the function 
$\Gamma(1/2+s_1+s_2+i\nu_1+i\nu_2-i\nu)$ in the third line of 
(\ref{thetafin}).
The other two $\Gamma$-functions generate poles in the upper half plane.
One has to consider only the first pole of each of these strings since 
the following ones are $\qf^2$-suppressed.
For the first group taking the residue at $i\nu=1/2$ leads to a 
remaining $\qf$-dependence $(\qf^2)^{-i\nu_1-i\nu_2}$.
If this factor is combined with the results for the functions
$\Lambda^{(\nu_i)}(\qf^2)$ (see above) we end with the result
\beqn
\Phi_{\omega_1\omega_2}(\qf^2) = 
C(\nu_1,\nu_2)
\cdot
\prod_{i=1}^2 
\int_{-\infty}^{+\infty} \frac{d \nu_i}{2 \pi}
\frac{1}{\omega_i-\chi(\nu_i)}
\left[
a(\nu_i)(\qf^2)^{-i\nu_i}+b(\nu_i)(\qf^2)^{i\nu_i}
\right]
\label{coeffpole}
\eeqn
Since $C(\nu_1,\nu_2)$, $a(\nu_i)$ and $b(\nu_i)$ can be shown to be analytic
in a strip around the real $\nu_i$-axis we can shift each $\nu_i$-contour
either into the upper or lower $\nu_i$-plane.
In that way each term in (\ref{coeffpole}) can be shown to vanish 
for $\qf^2=0$. The pole for the coefficient function hence 
gives no contribution for zero momentum transfer. 
\\
As to the leading pole of the second string of singularities we take the 
residue at $i\nu=1/2+i\nu_1+i\nu_2+s_1+s_2$.
This leads to the factor $(\qf^2/Q^2)^{s_1+s_2}$
where $s_1$ and $s_2$ still have to be integrated.
Since we are interested in the limit 
$\qf^2 \to  0$ we have to shift the $s_i$-contours to the right,
past the poles at $s_i=0$, $s_i=-2i\nu_i$,
i.\ e.\ we get a sum of four contributions.
Combining these with the results for the functions $\Lambda{(\nu_i)}$
one gets 16 terms.
Four of these are independent of $\qf^2$ and the other 12 have a 
$\qf^2$-dependence analogous to eq.\ (\ref{coeffpole})
and vanish after deforming the contours of 
the $\nu_i$-integration.
Finally one can show that due to the symmetry properties of the 
integrand the remaining four terms are in fact identical 
and the result for the partial wave amplitude in the limit 
$\qf^2=0$ reads 
\beqn
\Phi_{\omega_1\omega_2}(\qf^2)_{\vert \qf^2=0} 
= \frac{{\cal C}^2}{64\pi^2} 
\frac{Q^2}{\Delta^2}
\prod_{i=1}^2 \int \frac{d \nu_i}{2 \pi} 
\frac{(\Delta^2/Q^2)^{i\nu_i}}{\omega_i-\chi(\nu_i)}
\Gamma(\fez+i\nu_i)\Gamma(\fez-i\nu_i) 
\,\cdot \,
D_{(2,0)}(\fez+i\nu_1+i\nu_2)
\nonumber \\
\cdot \frac{\Gamma(2+i\nu_1+i\nu_2)}{\Gamma(-1-i\nu_1-i\nu_2)}
\frac{\Gamma(-\fez-i\nu_1)\Gamma(-\fez-i\nu_2)}
{\Gamma(\ftz+i\nu_1)\Gamma(\ftz+i\nu_2)}
\label{pwvares}
\eeqn
Here the factors in the second line correspond to the zero 
momentum transfer expression of the vertex function.
We have seen that the constraint on the conformal dimensions 
which was anticipated above is implemented in the vertex function 
through the pole of the $\Gamma$-function
$\Gamma(1/2+i\nu_1+i\nu_2+s_1+s_2-i\nu)$. In the $\qf^2=0$-limit
one is forced to take this pole which expresses the conservation
of conformal dimensions.
The remaining $\nu_i$-integrals in eq.\ (\ref{pwvares}) have to be 
evaluated numerically. For small $x$ we can obtain an approximate 
analytical result by using the saddle point approximation.
It is convenient to use the integration variables
$\nu_{\pm}=\nu_1 \pm \nu_2$. The saddle point of the $\nu_-$-integration
is then found at $\nu_{-\,s}=0$ and the corresponding integration is 
straightforward. The saddle point of the $\nu_+$-integration 
is found at 
$i\nu_{+\,s}= 1/2 \log (\Delta^2/Q^2)/\log (1/x)/|\chi''(0)| \ll 1$.
For the longitudinal part the corresponding integration 
is simple since the non-exponential parts of the integrand are analytic
in a neighbourhood of the saddle point. For the transverse case  
we have a pole at $i\nu_+=0$, i.\ e.\ in the limit of small $x$ 
the saddle point approaches a pole of the integrand. 
In this case the saddle point contour consists of two parts. We have 
the principal value of the $\nu_+$-integration and the contribution
of a small semicircle around the pole at $\nu=0$.
In the limit of small $x$ the semicircle contribution is dominant
and the principal value integral can be neglected.
The results for the photon-proton cross sections in the small-$x$
limit then read
\beqn
\frac{d \sigma_T^{\gamma^*P}}{d t}(x,t)_{\vert t=0}&=&
\frac{1}{Q^2\Delta^2}
{\cal C}^2
\sum_{f}e_f^2\alpha_{\mbox{\tiny em }}\alpha_s^2 \frac{2}{9}  
\frac{e^{2 \frac{N_c\alpha_s}{\pi}\,4 \log 2 \,\log 1/x}}
{\sqrt{7 N_c\alpha_s\zeta(3)\log1/x}}
\\
\frac{d \sigma_L^{\gamma^*P}}{d t}(x,t)_{\vert t=0}&=&
\frac{1}{Q^2\Delta^2}
{\cal C}^2 
\sum_{f}e_f^2\alpha_{\mbox{\tiny em }}\alpha_s^2 \frac{1}{9}
\frac{e^{2 \frac{N_c\alpha_s}{\pi}\,4 \log 2 \,\log 1/x}}
{7 N_c\alpha_s\zeta(3)\log1/x}
e^{-\frac{\log\Delta^2/Q^2}{28 N_c\alpha_s/\pi \zeta(3)\log1/x}}
\eeqn
The ratio of the longitudinal to the transverse cross section 
is approximately (disregarding the exponential factor which is close
to unity)
$1/(2\sqrt{7 N_c\alpha_s\zeta(3)\log(1/x)})$, i.\ e.\ the longitudinal 
part can be neglected. 
\\
The calculation above was based on the assumption that the large photon 
virtuality $Q^2$ justifies the perturbative approach to
diffractive dissociation at $t=0$. At first sight one could think
that the large scale plays a similar role as in inclusive DIS, namely
it pulls the diffusion of transverse momenta 
in the BFKL ladder out of the infrared region. 
In this case one would conclude that for high $Q^2$ a
large part of the phase space is treated correctly and that 
corrections which are important in the infrared region lead to a 
reduction of the exponent of $1/x$ from 1 to $0.4 ... 0.5$, which is 
twice the exponent of the observed rise of $F_2$.   
Numerical investigations of the BFKL evolution in diffractive 
dissociation \cite{mvogt} however show that these expectations are not 
correct. It turns out that $Q^2$ is not the relevant scale 
at the upper end of the BFKL ladders, but that on the contrary
the diffusion is driven to even lower momentum scales than the 
proton scale at the lower end. 
Almost the whole phase space which is covered by the evolution is 
located in the infrared domain where perturbation theory is not applicable.
This shows that in inclusive diffractive dissociation corrections 
to the BFKL pomeron in the infrared region are even more important than
in ordinary DIS and that the deviation of the experimentally 
observed $x$-dependence from the BFKL prediction should be large, in other
words, the BFKL based calculation provides no reasonable approximation 
to inclusive diffractive DIS.  
The findings of \cite{mvogt} thus confirm the physical picture 
of the aligned jet model \cite{ajm}. In this model one of the quarks
into which the virtual photon dissociates carries almost the whole momentum
of the photon whereas the other on carries almost none. Both
quarks have almost zero transverse momentum (they are aligned
to the photon-proton direction).
The slow quark interacts with the proton after it has travelled a long
time from the point where it was produced. During this time it evolves
nonperturbatively and there is no hard scale which can be associated 
with the quark-proton interaction. 
It was also shown in \cite{mvogt} that the situation changes when 
restrictions are imposed on the $q\bar{q}$ final state. An additional 
hard scale in the $q\bar{q}$ final state, e.\ g.\ a large 
transverse momentum or the mass of a formed vector meson acts similarly 
to the scale $Q^2$ in ordinary DIS. It pulls the diffusion out of the 
infrared and increases the contribution of perturbative scales.
Consequently a steeper $x$-dependence is expected for processes of 
this type (for the production of quarks with large 
transverse momenta see the next section).   
\subsubsection{The limit of large momentum transfer}
We now turn to the investigation of the large momentum transfer
limit of the partial wave amplitude. In this region
perturbation theory is applicable and the BFKL prediction has a better
theoretical foundation. 
\\
This consideration is also interesting for another reason.
If we turn the diagram in fig.\ \ref{figincdiff1} upside down and replace 
the virtual photon with a proton we end with a model for diffractive
vector meson production at large $t$, where the BFKL amplitude is now
coupled to the proton in a gauge invariant way. Clearly one has 
to reinterpret the function $D_{(4,0)}$ in this model.
One could think of $D_{(4,0)}$ as the momentum space expression of 
the square of the inclusive $q\bar{q}$ wavefunction of the proton.
With $Q^2$ being replaced by a hadronic scale $\Lambda^2$ one then 
has a consistent model for the coupling of the BFKL amplitude to 
a hadron which does not need the effective Mueller-Tang
\cite{muellertang}
prescription. The investigation of the limit $\qf \to \infty$
allows us to discuss the validity of this prescription.
\\
The starting point is again eq.\ (\ref{pwvam}) with 
$\Theta^{(\nu,\nu_1,\nu_2)}$ and $\Lambda^{(\nu_i)}$
being given by (\ref{lambda}) and (\ref{thetafin}).
Using the results from the previous section we derive that 
$\Lambda^{(\nu_i)}$ behaves for large $\qf^2$ as 
\beqn
\Lambda^{(\nu_i)} = 
(\qf^2)^{-3/2+ i \nu_i} \log\qf^2/(4 \Delta^2) a(\nu_i) 
\eeqn
with some analytic function $a(\nu_i)$.
For the $\qf^2$-dependence of the vertex function we again have 
to consider the singularity structure of the $\nu$-integration.
The essential factor is $(\qf^2/Q^2)^{i\nu}$
and for $\qf^2 \to \infty$ we have to shift the $\nu$-contour
into the upper half plane.      
The Mellin transform $D_{(2,0)}(\nu)$ has a pole at $i\nu=-1/2$.
It seems as if the $\Gamma$-functions in the denominator of eq.\
(\ref{thetafin}) cancel this singularity. 
But closer inspection shows that this 
is not the case. Let us assume that the pole at $i \nu=-1/2$ is the one which 
is nearest to the $\nu$-contour, i.\ e.\ we have to take it first.
Then with $i\nu =-1/2+\epsilon$ and as $\epsilon \to 0$ one finds 
that both the contours of the $s_1$ and $s_2$ integrations are pinched
by the poles of the $\Gamma$-functions inside the $s_i$-integrations 
in eq.\ (\ref{thetafin}). 
Each pinching implies a pole at $\epsilon=0$ or, equivalently, at 
$i\nu =-1/2$. We conclude that the $s_i$-integrals in fact contain
a double pole at $i\nu=-1/2$ which cancels the pole of the 
$\Gamma$-functions in the denominator.
Hence we can calculate the vertex function $\Theta^{(\nu,\nu_1,\nu_2)}$
for $i\nu=-1/2$
\beqn
\Theta^{(\nu=i\fez,\nu_1,\nu_2)}
=
\frac{2}{16\pi^3}(\qf^2)^{-1+i\nu_1+i\nu_2}
\prod_{i=1}^2 \left[
4 \pi 4^{2i\nu_i} \frac{\Gamma(1+i\nu_i)}{\Gamma(-i\nu_i)}
\frac{\Gamma(-\fez-i\nu_i)}{\Gamma(\ftz+i\nu_i)} 
\right]
\frac{1}{2\pi}
\label{thetaqasy}
\eeqn
The precise $\qf$-behaviour of the partial wave amplitude
depends on the nature of the singularity
of $D_{(2,0)}(\nu)$ at $i\nu=-1/2$. If it has a single pole, like in the
longitudinal case, the result is  
\beqn
\Phi_{\omega_1\omega_2\;L}(\qf^2)=
\sum_f e_f^2 \alpha_{\mbox{\tiny em}}\alpha_s^2
\frac{(Q^2\Delta^2)^2}{(\qf^2)^4} \log^2 \frac{\qf^2}{4 \Delta^2}
\frac{64}{9}{\cal C}^2
\prod_{i=1}^2 \int \frac{d \nu_i}{2 \pi}
\frac{\nu_i^2}{\omega_i-\chi(\nu_i)}
\frac{\Gamma(-\fez+i\nu_i)\Gamma(-\fez+i\nu_i)}
{|\Gamma(\frac{5}{4}+i\frac{\nu_i}{2})|^2
|\Gamma(\frac{1}{4}+i\frac{\nu_i}{2})|^2}
\eeqn
If there is a double pole at $i\nu=-1/2$, like in the transverse case
then there is an additional $\log \qf^2/Q^2$-enhancement
\beqn
\Phi_{\omega_1\omega_2\;T}(\qf^2)=
2 \, \log \frac{\qf^2}{Q^2}
\Phi_{\omega_1\omega_2\;L}(\qf^2)
\eeqn
The small-$x$ asymptotics of the cross section can now be calculated 
in the usual way by evaluating the $\nu_i$-integrals in the saddle
point approximation with the saddle point at $\nu_{i \,s}=0$.
Note the factor $\nu_i^2$ in the numerator of the integrand which 
is characteristic for the BFKL pomeron in the non-forward direction.
\\
Now let us comment on the Mueller-Tang prescription.
As anticipated in the last section this prescription states to consider 
only the diagram in which both gluons from the BFKL amplitude 
couple to the same quark line and to subtract from the conformal 
eigenfunctions $E^{(\nu)}$ $\delta$-function like terms which are not 
present in perturbation theory.
In our case this corresponds to keeping only one term in the sum
in eq.\ (\ref{de4}) with momentum dependence $(\qf^2/Q^2)^{1/2+i\nu}$
and to substitute the functions $E^{(\nu)}$ with the 
Mueller-Tang subtracted eigenfunctions in momentum space
\beqn
E^{(\nu)}_{MT}(\kf,\qf-\kf)=
E^{(\nu)}(\kf,\qf-\kf) - 
\left[\delta^{(2)}(\kf)+\delta^{(2)}(\qf-\kf)\right]
4^{2i\nu}\,4\pi^2\,(\qf^2)^{-\fez-i\nu}
\frac{\Gamma(1+i\nu)}{\Gamma(-i\nu)}
\frac{\Gamma(-\fez-i\nu)}{\Gamma(\ftz+i\nu)} 
\eeqn
Using the same arguments as presented below eq.\ (\ref{lambda}) it can be 
shown that only the $\delta$-function terms give a finite contribution 
and the result for the vertex function is exactly the expression 
in eq.\ (\ref{thetaqasy}) multiplied by $(\qf^2/Q^2)^{1/2+i\nu}$.
Taking then the pole of $D_{(2,0)}(\nu)$ at $i\nu=-1/2$ one arrives at 
the large-$\qf^2$ limit of the exact calculation. 
This shows that the Mueller-Tang prescription indeed gives the 
correct large-$\qf^2$ asymptotics. 
The effective parton formfactor which was used in the last section 
to calculate diffractive vector meson production therefore gives a
correct description for large $\qf^2$. For smaller $\qf^2$, however, 
a gauge invariant coupling like the one presented in this 
section should be used.  
\\
In contrast to the open $q\bar{q}$-pair production, the Mueller-Tang
prescription fails when applied to the coupling of the BFKL amplitude
to the photon - vector meson vertex $\phi_V$ 
(eq.\ (\ref{phot-vecmes})).
It was shown in the last section that the large-$\qf^2$ behavior 
in this case contains a $\log (\qf^2)$ which is not obtained 
when the Mueller-Tang subtraction is used. 
This demonstrates that the Mueller-Tang prescription
does not apply universally but has to be checked in each single case.
From the examples which have been studied here one can conjecture
\cite{baliplofo}
that the prescription works in inclusive cases, where configurations
with far separated quarks dominate. In exclusive cases where the 
quarks stay close together the prescription fails.
\\
It remains to discuss the $\delta$-function terms which are
subtracted in the Mueller-Tang prescription.
It is clear from the definition that the momentum space eigenfunctions
$E^{(\nu)}(\kf,\qf-\kf)$ are singular for $\kf \to 0$ and $\kf \to \qf$.
To investigate this singularity it is useful to introduce a small positive
conformal dimension $\lambda$ for the (reggeized) gluon field as a 
regularization parameter \cite{blotwue}. The regularized momentum
space expression $E^{(\nu,\lambda)}$ can be obtained in the same way 
as for $\lambda=0$. Investigation of the limit $\kf\to 0$ of this 
expression shows that the most singular term is
\beqn
E^{(\nu)}(\kf,\qf-\kf) \stackrel{\kf \to 0}{\approx}
\lambda \cdot 
(\kf^2)^{-1+\lambda} 
\; (\qf^2)^{-\fez -i\nu}
4 \pi
\frac{\Gamma(1+i\nu)}{\Gamma(-i\nu)}
\frac{\Gamma(-\fez-i\nu)}
{\Gamma(\ftz+i\nu)}
\left[ 1+O(\lambda)\right]
\label{sylt}
\eeqn    
which can be regarded as a smeared $\delta^{(2)}(\kf)$-function.
The origin of these terms which cannot be associated with Feynman
diagrams is clear. When proving conformal invariance of the BFKL equation 
it was necessary to add logarithms of differences of coordinates
to the solution of the zero order ($\alpha_s=0$) equation to ensure 
conformal symmetry of the inital condition. 
These fictious terms cancel when the BFKL 
amplitude is integrated with color neutral impact factors. The same
is true for the $\delta$-function terms which do not appear
in the physical amplitude as long as the wave functions of the external 
particles obey the color neutrality condition.
One can therfore can conclude that the presence of the $\delta$-functions
is due to the addition of these logarithmic terms to the 
configuration space amplitude.
\subsubsection{Additional gluons in the final state}
In this part we will generalize the above results to final states 
which contain also gluons in addition to the $q\bar{q}$-pair.
The considerations are based on the analysis of photon 
diffractive dissociation in the triple Regge limit 
\cite{bartels,bartels-wue}
in which $s \gg M^2 \gg Q^2$, where $M^2$ is the invariant mass
of the produced hadronic system.
The partial wave representation of the cross section now reads
\beqn
\frac{d \sigma^{\gamma^*P}}{d t d M^2}
=
\frac{1}{M^2 Q^4}
\int \frac{d \omega}{2 \pi i}
\int \frac{d \omega_1}{2 \pi i} 
\int \frac{d \omega_2}{2 \pi i}
\left(\frac{s}{M^2}\right)^{\omega_1+\omega_2} 
\left(\frac{M^2}{Q^2}\right)^{\omega}
\Phi_{\omega\,\omega_1\omega_2}
\eeqn
It was shown in \cite{bartels,bartels-wue}
that the partial wave amplitude $\Phi_{\omega\,\omega_1\omega_2}$
comes as a sum of two terms.
In both terms the additionally produced gluons are described by a BFKL
ladder which is coupled to the quark loop.
In the first term this BFKL amplitude couples to the lower two 
BFKL pomerons through a disconnected vertex. This vertex results from 
the reggeization of pairs and triplets of gluons in the same way as 
described in the discussion of the four gluon ampitude $D_{(4,0)}$.
A typical contribution is displayed in fig. \ref{figincdiff2}.
\begin{figure}[!h]
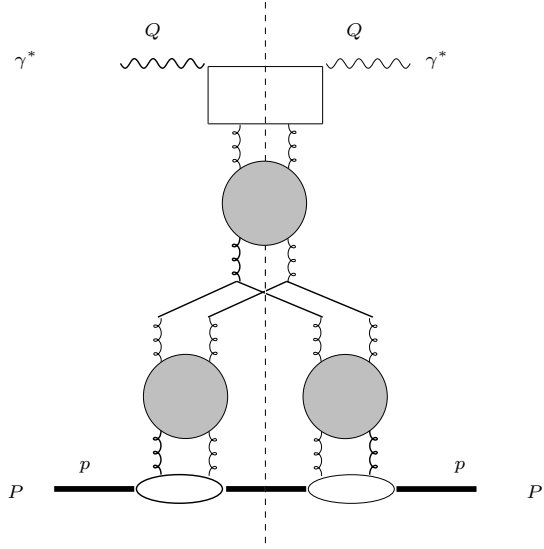

\begin{center}
\input incdiffregg.pstex_t
\end{center}
\caption{
Disconnected contribution to photon diffractive
dissociation into $q\bar{q}$ + $n$ gluons. 
\label{figincdiff2}
}
\end{figure}
In the second term a connected vertex appears which couples the upper 
BFKL ladder to an interacting four gluon state.
This interacting four gluon state splits into two BFKL ladders 
which couple to the proton.
Due to our incomplete understanding of the connected vertex and the 
interacting four gluon state we cannot treat the second term exactly
and restrict our considerations to the first term.
The vertex and the four gluon state will be studied in the last chapter 
of this thesis.
\\
Given the results of the $q\bar{q}$-case it is not difficult to obtain the 
partial wave amplitude for the diagrams in fig.\ \ref{figincdiff2}.
We have to insert a factor $1/(\omega-\chi(\nu))$ into the expression for 
the partial wave amplitude, corresponding to the Mellin transform of the upper
BFKL pomeron. The vertex function $\Theta^{(\nu,\nu_1,\nu_2)}$
which now has the interpretation of a triple pomeron vertex is the 
same as before. 
The constraint on the conformal dimensions which holds for $\qf^2=0$
now leads to a remarkable interplay between the energy and the momentum
transfer dependence in the region of small $\qf^2$.
After performing the $\omega$-integrations the energy dependence 
of the $\gamma^*-P$ cross section 
is encoded in the function
\beqn
\exp \left[y_M \chi(\nu) + y_s (\chi(\nu_1)+\chi(\nu_2)) \right]
\label{exp}
\eeqn
where $y_M=\log M^2/Q^2$ and $y_s=\log s/M^2$ were introduced 
and $\chi(\nu)$ is the BFKL eigenvalue.
In addition to the exponential we 
have also the factor $(\qf^2/Q^2)^{i\nu}$
which becomes important if $\qf^2$ becomes small.
First we keep $\qf^2 \approx Q^2$, i.\ e.\ $\log \qf^2/Q^2 \ll 1$,
and $y_M$ and $y_s$ are large. Then all $\nu$ integrals are determined
by the respective stationary point of the exponent in (\ref{exp}) 
which is found at $\nu_s=\nu_{1\,s}=\nu_{2\,s}=0$.
From this one can conclude that the triple pomeron vertex 
$\Theta$ behaves as $(\qf^2)^{-\fez}$ in the triple Regge limit.
This behavior was also observed in the color dipole formalism 
\cite{muellerpatel}. One can also expect that this result holds 
for the connected vertex since the $\qf^2$-behavior follows from 
purely dimensional arguments. 
Upon saddle point integration the  
following expression for the cross section is finally found 
\beqn
\frac{d \sigma^{\gamma^*P}}{d t d M^2}=
\frac{1}{M^2}\frac{Q^4}{|\qf|^6}\frac{1}{\sqrt{\qf^2Q^2}}
\log^2 \frac{\qf^2}{4 \Delta^2}
\left(\frac{M^2}{Q^2}\right)^{4 \log 2 \frac{N_c\alpha_s}{\pi}}
\left(\frac{s}{M^2}\right)^{8 \log 2 \frac{N_c\alpha_s}{\pi}}
\nonumber \\ \phantom{xxxxxx}
\sqrt{\pi} {\cal S}
\frac{1}{(14 N_c\alpha_s \zeta(3))^3}\frac{1}{\sqrt{y_M y_s^6}}
\frac{2{\cal C}^2}{\Gamma^2(1/4)\Gamma^2(5/4)}
\eeqn
where the functions $\Lambda^{(\nu_i)}$ (eq.\ (\ref{lambda}))
have been evaluated in the limit 
$4\Delta^2 \ll \qf^2$ and ${\cal S}$ is a constant 
resulting from the $s_i$ integration in eq.\ (\ref{thetafin}) when all
conformal dimensions are zero. 
\\
Now if $\qf^2$ becomes small the factor $(\qf^2/Q^2)^{i\nu}$ 
becomes dominant again. In the limit $\qf^2=0$ we are again forced 
to perform the $\nu$-integration by taking the  conservation pole
at $i\nu=1/2+i\nu_1+i\nu_2$ in the same way as described before 
eq. (\ref{pwvares}).
Performing the $\omega$-integrations we end with the energy 
dependence 
\beqn
\exp\left[y_M \chi(\nu_1+\nu_2-i\fez)+
y_s (\chi(\nu_1)+\chi(\nu_2)) +(i\nu_1+i\nu_2)
\log \frac{\Delta^2}{Q^2}\right]
\eeqn  
Depending on the relative magnitude of $y_M$, $y_s$ 
and $\log\Delta^2/Q^2$ very different energy dependences are obtained.  
We will assume that $y_s$ is the largest parameter, i.\ e.\ 
$\nu_1$ and $\nu_2$ are close to zero. Introducing then again $\nu_+$
and $\nu_-$ we find the saddle point of $\nu_-$ at $\nu_{-\,s}=0$.
Imposing the the condition 
$(\log Q^2/\Delta^2)^3 \gg y_My_s^2$ 
we find a saddle point of the 
$\nu_+$-integration at 
\beqn
i\nu_{+\,s}= 
\sqrt{\frac{N_c\alpha_s}{\pi}\frac{y_M}{\log Q^2/\Delta^2}}
\eeqn
which leads to the energy dependence 
\beqn
\frac{d \sigma^{\gamma^*P}}{dt d M^2}_{|t=0}
\sim
\left(\frac{s}{M^2}\right)^{8 \log 2 \frac{N_c\alpha_s}{\pi}}
\exp\left[ 
2 \sqrt{ \frac{N_c\alpha_s}{\pi} \log \frac{Q^2}{\Delta^2}
\cdot y_M} \right]
\label{crappx1}
\eeqn
If we instead impose the condition 
$\log (\log Q^2/\Delta^2)^3 \ll y_My_s^2 $  
the saddle point is located at
\beqn
i\nu_{+\,s}= -\left[\frac{1}{14\zeta(3)} 
\frac{y_M}{y_s}\right]^{\frac{1}{3}} 
\eeqn 
from which follows the energy dependence
\beqn
\frac{d \sigma^{\gamma^*P}}{dt d M^2}_{|t=0}
\sim
\left(\frac{s}{M^2}\right)^{8 \log 2 \frac{N_c\alpha_s}{\pi}}
\exp\left[\frac{3}{2} \frac{N_c\alpha_s}{\pi}
\left(y_M^2y_s 14 \zeta(3)\right)^{\frac{1}{3}}
\right] 
\label{crappx2}
\eeqn
In both cases we have retained only the part which follows from the 
zero order term of the expansion of the exponent around the saddle
point. The contribution of the fluctuations have been left out
since their treatment depends on the singularity structure of 
the coefficient function as already shown in the $q\bar{q}$-case. 
Further saddle points have been discussed in \cite{blotwue}. 
\\
These results demonstrate that the conservation law for the 
conformal dimensions at $\qf^2=0$ has remarkable implications 
on the energy dependence of the cross section. In particular it is 
excluded that in all three BFKL amplitudes the point $\nu=0$
(the BFKL point) dominates.  
In the case (\ref{crappx1}) we find the energy dependence of the 
double-logarithmic (DLA) limit for the upper BFKL-ladder. This corresponds
to a large ratio $Q^2/\Delta_0^2$ where $\Delta_0^2$ is the effective 
scale at the triple pomeron vertex.
This result confirms the assertion that for $\qf^2=0$ the scale at 
the upper end of the BFKL amplitudes coupling to the proton is not 
$Q^2$ but a nonperturbative scale. 
In (\ref{crappx2}) the upper ladder is also near the DLA limit since $\nu$
is close to -1. From the energy dependence one can derive 
an effective scale 
$\Delta_0^2=Q^2\cdot\exp(-\sqrt[3]{y_My_s^2} \; c)$ 
which determines 
the scale at the vertex ($c$ is a constant of order unity).
An increase of the mass hence leads to a lowering of the 
effective scale at the vertex. 
From both results we conclude that   
should expect serious modifications 
of our $\qf^2=0$ results
from nonperturbative corrections.
\newpage
\subsection{Production of 
$q\bar{q}$
pairs in 
DIS diffractive dissociation}
\label{sec24}
This section is devoted to the study of deep inelastic diffractive
reactions with a final state consisting of the diffracted 
proton and two jets with large transverse momenta.
In lowest order perturbative QCD the jets originate from a quark-antiquark
pair which is produced by the virtual photon. 
We calculate 
the cross section for the process
$\gamma^*(q) + P (p) \to q(k) + \bar{q}(q-k+\zeta) + P (p-\zeta)$
(fig.\ \ref{fig231}).
\begin{figure}[hhh]
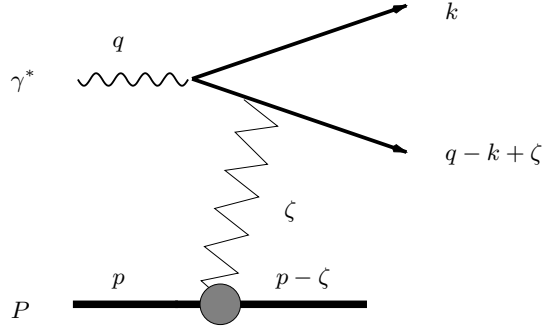

\begin{center}
\input diffqq1.pstex_t
\end{center}
\caption{
Diffractive production of a quark-antiquark pair.
\label{fig231}
}
\end{figure}
We are in interested in the limit of high photon-proton center 
of mass energy $W^2=(p+q)^2$, in which the outgoing proton is well 
separated in rapidity from the $q\bar{q}$-pair.
Since we regard the process as being mediated by the exchange
of a composite object of gluons between the $q\bar{q}$-pair 
and the proton we are forced to conclude that this object
is colorless because otherwise the formation of the rapidity 
gap would be exponentially suppressed, or, stated differently,
a produced gap would be filled by final state radiation.
As the lowest order contribution we take two-gluon exchange
and since we require the transverse momentum $\kf^2$ of the 
(anti)quark to be large ($\kf^2 \gg \Lambda_{QCD}^2$)
we treat both gluons perturbatively.
A similar approach has been reported in \cite{rysjet} 
and \cite{nikjet}, whereas in \cite{diehl} the process was studied 
assuming the exchange of nonperturbative gluons.
A different approach towards jet production in diffractive DIS
has been pursued in
\cite{buchmcdermheb} using a semiclassical picture in which
the proton is treated as a classical color field. 
\subsubsection{Kinematics and observables}
We begin with some kinematical considerations limiting
ourselves to the case of zero momentum transfer $t=-\zeta^2=0$.
For the momenta $k$ and $\zeta$ we use a Sudakov decomposition 
w.\ r.\ t.\ the light cone momenta $p$ and $q'=q+xp \; (x=Q^2/(2 p q))$
\beqn
k&=&\alpha \, q'+\beta \, p+\kf \\
\zeta&=&\alpha_{\zeta} \, q' + x_{\Pam} \, p +\Delta
\eeqn
Using the mass shell conditions and the fact that $W^2$ is the large 
variable one can show that $\alpha_{\zeta}$ and $\Delta^2$ can be
neglected. 
The phase space can be cast into the form
\beqn
d \Gamma =
\frac{\pi}{8 pq} \frac{1}{M^2}
\frac{1}{\sqrt{1-4\frac{\kf^2+m_f^2}{M^2}}}  
d M^2 d t d^2 \kf  
\eeqn
with the quark mass $m_f$
and $M^2$ being the invariant mass of the $q\bar{q}$ 
pair which is related 
to the light cone momentum fraction 
$\alpha$ through the important
relation
\beqn
\alpha(1-\alpha)M^2=\kf^2+m_f^2
\eeqn
Energy-momentum conservation leads to the phase space restriction
$M^2 \geq 4(\kf^2+m_f^2)$.
Furthermore the longitudinal momentum fraction $x_{\Pam}$
transferred from the proton to the $q\bar{q}$ pair is fixed as
\beqn
x_{\Pam}=\frac{M^2+Q^2}{W^2+Q^2}
\eeqn
Another variable often used is $\beta$ defined as
\beqn
\beta=\frac{Q^2}{Q^2+M^2}
\eeqn
and one has $\beta=x/x_{\Pam}$.  
In addition we introduce two angular observables which can be defined from 
scalar products of the quark momentum with the momenta of the incoming 
particles. Going to the $q\bar{q}$ cms in which quark and antiquark are 
produced back-to-back we define the angle $\theta$ between the proton 
direction and the outgoing quark (or antiquark) which points into the 
proton hemisphere. One should remark here that the cross section of the 
process is symmetric w.\ r.\ t.\ to the exchange of quark and antiquark. 
For $\theta$ we find the relation $\sin^2 \theta = 4 \kf^2/M^2$. 
In terms of $\alpha$ one has $\cos \theta = 1-2 \alpha$.
The second angle $\phi$ is defined between the particle which points
into the proton hemisphere and the plane formed by the incoming 
and outgoing electron.
The definition of the variables is summarized in fig.\ \ref{fig232}.
\begin{figure}[hhh]
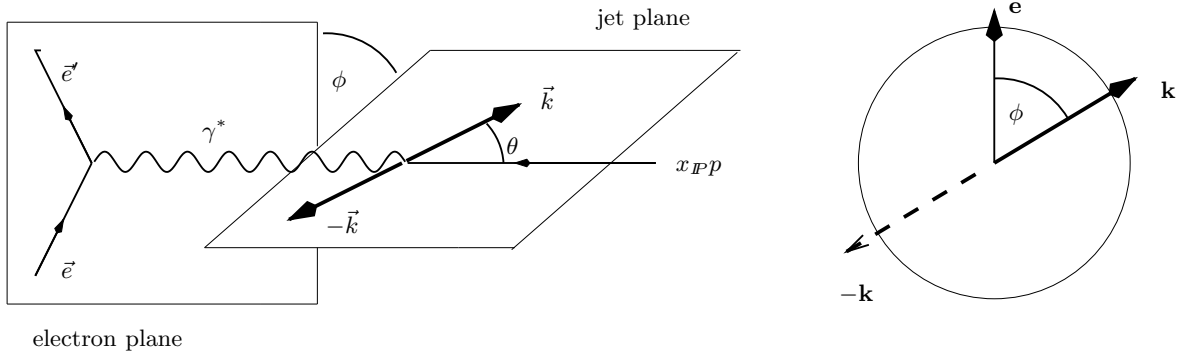

\begin{center}
\input geometry.pstex_t
\end{center}
\caption{
Definition of planes and angles in the $q\bar{q}$ cms.
The right hand side displays the projection onto the transverse plane
perpendicular to the proton direction. 
\label{fig232}
}
\end{figure}
Introducing the angle $\phi$ as an additional observable one 
obtains a more complicated expression for the electron-proton cross 
section than in inclusive (angle integrated) DIS.
Using the Sudakov decomposition of the electron momentum
$e= 1/y q' + x (1-y)/y p + \ef$ one finds for the lepton tensor 
\beqn
L_{\mu \nu}
&=&
\frac{1}{2}
\left[2 e_{\mu}e_{\nu} - g_{\mu \nu} \frac{Q^2}{2}\right]
\nonumber \\
&=&
\left[
4(1-y)\frac{x^2}{y^2} p_{\mu}p_{\nu} 
+\ef_{\mu}\ef_{\nu}- g^{\perp}_{\mu \nu}\frac{Q^2}{4}
+ (2-y)\frac{x}{y}(p_{\mu}\ef_{\nu}+p_{\nu}\ef_{\mu})
\right] 
\nonumber \\
g^{\perp}_{\mu \nu} &=& g_{\mu \nu}-(p_{\mu}q'_{\nu}+p_{\nu}q'_{\mu})/pq'
\eeqn
The different terms in this expression represent the contributions
of photons with different polarization states. The first term correspond
to the scattering of longitudinal photons, the second and third term 
belong to transversely polarized photons and the fourth term is an 
interference term. 
\subsubsection{Formalism and result for the cross section}
As to the hadronic tensor we use the high energy factorization theorem
to express the amplitude of the photon-proton scattering in terms of the 
unintegrated gluon density of the proton, which was discussed in section 
\ref{sec21}.
The relevant diagrams for the amplitude are shown in fig. \ref{fig233}
and the hadronic tensor has the form 
\beqn
H^{\mu \nu} = |\int d \lf^2 \; C(\lf^2;\kf^2,Q^2,M^2) 
\, {\cal F}_G(x_{\Pam},\lf^2) \, |_{\mu\nu}^2
\label{cmunu}
\eeqn
with the coefficient function $C_{\mu}$
\footnote{The notation in eq.\ (\ref{cmunu}) has to be understood in 
the sense that $C$ carries one Lorentz index.}. 
This factorization is valid in 
the leading-log ($1/x_{\Pam}$) approximation, in which the imaginary
part of the diagrams in fig.\ \ref{fig233} contributes. 
\begin{figure}[hhh]
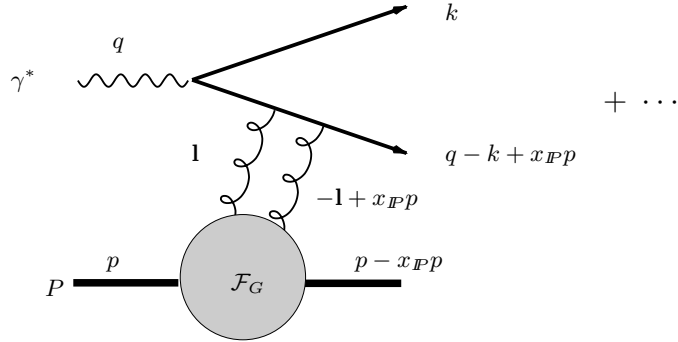

\begin{center}
\input diffqq2.pstex_t
\end{center}
\caption{
Representation of the amplitude in terms of the unintegrated gluon 
structure function. The dots represent three other diagrams which 
are generated by attaching the gluons to the quark lines in all 
possible ways. 
\label{fig233}
}
\end{figure}
In contrast to inclusive DIS where the cross section is given as the imaginary
part of a true forward amplitude, in the present case the longitudinal momenta
of the gluons coupling to the quarks are not symmetric. Their difference 
in longitudinal momentum fraction is $x_{\Pam}$ which is needed to transform
the virtual photon into a real $q\bar{q}$ - state. 
Consequently the longitudinal momentum argument of ${\cal F}_G$ 
is not well-defined.
To leading-log ($1/x_{\Pam}$) accuracy, to which this study is restricted,
this difference does not contribute since in this approximation the 
longitudinal momenta of both gluons are neglected relative to the 
longitudinal momenta of the gluons one rung below (strongly ordered
longitudinal momenta). 
In order to go beyond this approximation 
one has to take into account one gluon rung below the $q\bar{q}$-pair
in a more accurate fashion.
One should remark that in the leading-log ($1/x$) approximation the 
longitudinal momentum argument of the gluon density is not well-defined
as a matter of principle.  
\\ 
The calculation of the diagrams in fig.\ \ref{fig233} is straightforward
as soon as one realizes that the gluon polarization proportional
to $p^{\mu}$ gives the dominant contribution in the small-$x$ limit
\cite{muelld,nikd,levwued}.
Contraction of the lepton tensor and the hadron tensor
then gives the following structure for the electron-proton cross section
\beqn
\frac{d \sigma^{e P}}
{d y d Q^2 d M^2 d \kf^2 d \phi d t}_{|t=0}
= \frac{\alpha_{\mbox{\tiny em}}}{2 y Q^2 \pi^2}
\left[
\frac{1+(1-y)^2}{2}
\frac{d \sigma^{\gamma^* p}_{D,T}}{d M^2 d \kf^2 d t}_{|t=0}
- 2 (1-y) \cos 2 \phi
\frac{d \sigma^{\gamma^* p}_{D,A}}{d M^2 d \kf^2 d t}_{|t=0}\right.
\nonumber \\ 
\left. + (1-y) \frac{d \sigma^{\gamma^* p}_{D,L}}{d M^2 d \kf^2 d t}_{|t=0}
+(2-y)\sqrt{1-y} \cos \phi
\frac{d \sigma^{\gamma^* p}_{D,I}}{d M^2 d \kf^2 d t}_{|t=0}
\right] \;\;\;\;\;\;
\label{eq1} 
\eeqn
where the indices T, L, and I
refer to the contributions of transverse 
and longitudinal photons 
and the interference term, respectively.
The term with index A is an angular asymmetric contribution
which also comes from transverse photons.
The expressions for the photon-proton cross sections read
\beqn
\frac{d \sigma^{\gamma^* P}_{D,T}}{d M^2 d \kf^2 d t}_{|t=0} &=&
\frac{1}{M^4}\frac{1}{\kf^2}
\frac{1}{12}
\sum_f e_f^2 \alpha_{\mbox{\tiny{em}}}\pi^2 \alpha_s^2
\frac{\kf^2+m_f^2}{\sqrt{1-4 \frac{\kf^2+m_f^2}{M^2}}}
\left[
\left(1-2 \frac{\kf^2+m_f^2}{M^2}\right)
\;[I_T(Q^2,M^2,\kf^2,m_f^2)]^2
\nonumber \right. \\
& &\left. + 
\;\; 4 m_f^2 \;\; 
\frac{\kf^2 M^4}{(\kf^2+m_f^2)^2Q^4}
\;[I_L(Q^2,M^2,\kf^2,m_f^2)]^2
\right] 
\label{tra}
\\
\frac{d \sigma^{\gamma^* P}_{D,L}}{d M^2 d \kf^2 d t}_{|t=0} &=&
\frac{1}{M^4}\frac{1}{Q^2}
\frac{4}{3}
\sum_f e_f^2 \alpha_{\mbox{\tiny{em}}}\pi^2 \alpha_s^2
\frac{\kf^2+m_f^2}{\sqrt{1-4 \frac{\kf^2+m_f^2}{M^2}}}
\;[I_L(Q^2,M^2,\kf^2,m_f^2)]^2
\label{long}
\\
\frac{d \sigma^{\gamma^* P}_{D,I}}{d M^2 d \kf^2 d t}_{|t=0} &=&
\frac{1}{M^4}\frac{1}{\sqrt{\kf^2Q^2}}
\frac{1}{3}
\sum_f e_f^2 \alpha_{\mbox{\tiny{em}}}\pi^2 \alpha_s^2
(\kf^2\!+\!m_f^2) 
I_T(Q^2,M^2,\kf^2,m_f^2)\!I_L(Q^2,M^2,\kf^2,m_f^2)
\label{interf}
\\
\frac{d \sigma^{\gamma^* P}_{D,A}}{d M^2 d \kf^2 d t}_{|t=0}
&=&
\frac{1}{M^6}\frac{1}{\kf^2}
\frac{1}{12}
\sum_f e_f^2 \alpha_{\mbox{\tiny{em}}}\pi^2 \alpha_s^2
\frac{(\kf^2+m_f^2)^2}{\sqrt{1-4 \frac{\kf^2+m_f^2}{M^2}}}
\; [I_T(Q^2,M^2,\kf^2,m_f^2)]^2
\label{asy}
\eeqn
The essential dynamics is contained in the universal functions 
$I_T,I_L$ for which we have the following expressions
\beqn
I_L(Q^2,M^2,\kf^2,m_f^2) \!\!\!
&=& \!\!\!
-\int \frac{d \lf^2}{\lf^2}
{\cal F}_G(x_{\Pam},\lf^2)
\left[
\frac{Q^2}{M^2+Q^2}-\frac{(\kf^2+m_f^2)Q^2}{M^2 \sqrt{P_{\kf,\lf}}}
\right]
\label{il}
\\
I_T(Q^2,M^2,\kf^2,m_f^2) \!\!\!
&=& \!\!\!
-\int \frac{d \lf^2}{\lf^2}
{\cal F}_G(x_{\Pam},\lf^2)
\left[
\frac{2 M^2\kf^2}{(\kf^2+m_f^2)(Q^2+M^2)}-1
\right. \nonumber \\ 
& &\left.
            \phantom{xxxxx\frac{2 M^2\kf^2}{(\kf^2+m_f^2)(Q^2+M^2)}}
+\frac{\lf^2+\frac{\kf^2}{M^2}(Q^2-M^2)+m_f^2(1+\frac{Q^2}{M^2})}
{\sqrt{P_{\kf,\lf}}}
\right]
\label{it}
\eeqn
with $P_{\kf,\lf}$ being defined as
\beqn
P_{\kf,\lf}=(\lf^2+\frac{\kf^2}{M^2}(Q^2-M^2)
+\frac{m_f^2}{M^2}(Q^2+M^2))^2
+4 \kf^2( \frac{\kf^2}{M^2} Q^2+\frac{m_f^2}{M^2}(Q^2+M^2))
\eeqn
These expressions simplify considerably in the case of massless flavors 
($m_f^2=0$) \cite{baj}.
To proceed we need an explicit representation for the unintegrated
gluon structure function ${\cal F}_G$.
One could in principle use the relation in eq.\ (\ref{kt-dla})
to calculate ${\cal F}_G$ from the usual gluon density
by differentiation and evaluate the $\lf$-integral numerically
\cite{opcharm}.
For this one has to introduce an infrared cutoff $\lf_0^2$
since the gluon density is not known for momenta $ \lf^2 \lesssim 1 \,
\mbox{GeV}^2$.
In this section we prefer to work consistently in the 
leading-log $(1/x_{\Pam})$ approximation in which ${\cal F}_G$
is determined by the BFKL resummation. Using the results of 
section \ref{sec11} we have the following representation for the 
unintegrated structure function
\beqn
{\cal F}_G (x_{\Pam},\lf^2) = 
\frac{1}{\Lambda_0^2}
\int_{-\infty}^{+\infty} 
\frac{d \nu}{2 \pi}
\left(\frac{\lf^2}{\Lambda_0^2}\right)^{-\frac{1}{2}-i\nu}
\, \phi(\nu) \, 
\exp\left[ \chi(\nu) \log \frac{1}{x_{\Pam}}\right]
\eeqn
with $\Lambda_0^2$ being a nonperturbative scale, $\phi(\nu)$ 
an integrable function of $\nu$ which is analytic in the strip
$-1/2 < \mbox{Im}(\nu) < 1/2  $ and $\chi(\nu)$ 
the eigenvalue of the BFKL kernel.
This representation is inserted into the $\lf$-integrals in eqs. 
(\ref{il}), (\ref{it}).
It is useful to define the scaling variable
\beqn
\xi = \frac{1}{M^2}\left(Q^2+\frac{m_f^2}{\kf^2}(Q^2+M^2)\right)
\eeqn
The results for $I_L,I_T$ then read
\beqn
I_L(Q^2,M^2,\kf^2,m_f^2)
&=&
\frac{1}{\Lambda_0^2} 
\frac{(\kf^2+m_f^2)Q^2}{\kf^2M^2}
\frac{1}{1+\xi}
\int_{-\infty}^{+\infty} \frac{d \nu}{2 \pi} \phi(\nu)\,e^{\chi(\nu) 
\log 1/ x_{\Pam}}
\, \Gamma(\frac{1}{2}+i\nu)\Gamma(\frac{1}{2}-i\nu)
\nonumber \\
& &
\left[\frac{\kf^2}{\Lambda_0^2}(1+\xi)\right]^{-\frac{1}{2}-i\nu}
\, _2F_1\left(\frac{3}{2}+i\nu,-\frac{1}{2}-i\nu
,1;\frac{1}{1+\xi}\right)
\label{il1}
\\
I_T(Q^2,M^2,\kf^2,m_f^2)
&=&
\frac{1}{\Lambda_0^2}  \frac{2}{1+\xi}
\int_{-\infty}^{+\infty} \frac{d \nu}{2 \pi} \phi(\nu)\,e^{\chi(\nu) 
\log 1/ x_{\Pam}} 
\, \Gamma(\frac{1}{2}+i\nu)\Gamma(\frac{1}{2}-i\nu)(\frac{3}{2}+i\nu)
\nonumber \\
& &
\left[\frac{\kf^2}{\Lambda_0^2}(1+\xi)\right]^{-\frac{1}{2}-i\nu}
\, _2F_1\left(\frac{3}{2}+i\nu,-\frac{1}{2}-i\nu
,2; \frac{1}{1+\xi}\right)
\label{it1}
\eeqn
Note that since $\xi > 0$ (for $Q^2 \neq 0$) the argument of the 
hypergeometric function is in the unit circle, which means that this 
part of the integrand is an analytic function of $\nu$, even for imaginary
$\nu$. The important variable which emerges from the calculation is
the dimensionless ratio 
\beqn
\Delta =\frac{\kf^2}{\Lambda_0^2}(1+\xi)
=\frac{Q^2+M^2}{M^2} \cdot \frac{\kf^2+m_f^2}{\Lambda_0^2}
\eeqn
For the perturbative QCD calculation to be reliable one should 
postulate $\Delta \gg 1$. If we set $m_f^2=m_c^2$ (charm quarks)
this relation is fulfilled for all $\kf^2$, but for the three light 
flavours the result has to be restricted to large transverse
momenta $\kf^2 \geq 1 \, \mbox{GeV}^2$. 
For DIS ($Q^2 > 0$) the 'hardness' of the process is enhanced 
due to multiplication with the ratio $(Q^2+M^2)/M^2$.
\\
The $\nu$-integration in eqs.\ (\ref{il1}), (\ref{it1}) can be performed
using the saddle point approximation.
In the case $\log 1/x_{\Pam} \gg \log \Delta$ the saddle point is located
at $i\nu=0$ and one obtains the usual BFKL results with the power 
rise in $1/x_{\Pam}$.
We give them here for completeness, although this is not the main objective
of the section. 
\beqn
I_L
&=&
\frac{(\kf^2+m_f^2)Q^2}{M^2\sqrt{\Lambda_0^2 [\kf^2(1+\xi)]^3}}
\;\pi\; _2F_1(\frac{3}{2},-\frac{1}{2},1;\frac{1}{1+\xi})
\left(\frac{1}{x_{\Pam}}\right)^{\frac{N_c\alpha_s}{\pi}4 \log2}
\frac{e^{
- \frac{\log^2 \Delta}{N_c\alpha_s/\pi \log1/x_{\Pam}56 \zeta(3)}
}}{\sqrt{N_c\alpha_s \log1/x_{\Pam}56 \zeta(3)}}
\\
I_T
&=&
\frac{1}{\sqrt{\Lambda_0^2 [\kf^2(1+\xi)]^3}}
 \;3 \pi\; _2F_1(\frac{3}{2},-\frac{1}{2},2;\frac{1}{1+\xi})
\left(\frac{1}{x_{\Pam}}\right)^{\frac{N_c\alpha_s}{\pi}4 \log2}
\frac{e^{
- \frac{\log^2 \Delta}{N_c\alpha_s/\pi \log1/x_{\Pam}56 \zeta(3)}
}}{\sqrt{N_c\alpha_s \log1/x_{\Pam}56 \zeta(3)}}
\eeqn
In the opposite case, $\log 1/x_{\Pam} \ll \log \Delta $,
the exponent becomes 
stationary at 
\beqn
i\nu_s=1/2-\sqrt{\frac{N_c\alpha_s}{\pi} 
\frac{\log(1/x_{\Pam})}{\log \Delta}}
\eeqn
which leads to the double-logarithmic approximation for $I_L,I_T$
\beqn
I_L
&=& 
\frac{(\kf^2+m_f^2)Q^2}{\kf^4 M^2}\frac{\xi-1}{(1+\xi)^3}
x_{\Pam}
f_G(x_{\Pam},\kf^2(1+\xi))
\label{ildla}
\\
I_T
&=&
\frac{4}{\kf^2}\frac{\xi}{(1+\xi)^3}
x_{\Pam}
f_G(x_{\Pam},\kf^2(1+\xi))
\label{itdla}
\eeqn
where we have used the identity (cf.\ eq.\ (\ref{dla-gluon})) 
\beqn
x_{\Pam}
f_G(x_{\Pam},\kf^2(1+\xi))
=
\sqrt{\frac{\pi}{4}}\left[\frac{N_c \alpha_s}{\pi}
\frac{\log (1+\xi)\kf^2/\Lambda_0^2}
{\log1/x_{\Pam}}\right]^{\frac{1}{4}}
\exp{\sqrt{4 \frac{N_c \alpha_s}{\pi} \log (1+\xi)
\frac{\kf^2}{\Lambda_0^2}
\log \frac{1}{x_{\Pam}}}} \phi(\nu_s)
\label{qqdla}
\eeqn  
This double logarithmic approximation (in $1/x_{\Pam}$ and $\kf^2(1+\xi)$)
could have been obtained directly from eqs.\ (\ref{il}),
(\ref{it}) by assuming dominance of the phase space region 
$\lf^2 \ll \kf^2(1+\xi)$ and corresponding expansion of the integrand.
Retaining the leading term of this expansion and using the basic 
relation (\ref{kt-dla}) then leads to the above results.
The scale of the gluon structure function emerges as the upper
limit of the $\lf^2$-integration which in turn is determined by 
the consistency of the expansion.
\\
We will first concentrate on the case of massless flavors ($m_f^2=0$)
in which we have $\xi=Q^2/M^2$.
Then we find that in the photoproduction limit ($Q^2=0$) or, 
equivalently, in the large mass limit ($M^2 \to \infty$)
both $I_L$ and $I_T$ vanish.
For $I_L$ this has to be the case, of course, since longitudinally 
polarized real photons do not exist. Since, on the other hand, we expect
a contribution from transversely polarized real photons, 
we have to conclude that the double logarithmic approximation is a poor 
one for $Q^2 \to 0$ ($M^2\to\infty$). In the extreme case 
$Q^2 = 0$ the term in squared brackets in eq.\ (\ref{it}) equals
$2 \theta(\lf^2-\kf^2)$, i.\ e.\ the DLA phase space does not 
contribute. Since ${\cal F}_G(x_{\Pam},\lf^2)$ decreases as $1/\lf^2$
(mod.\ logarithms) for large $\lf^2$ the integrand in   
eq.\ (\ref{it}) is $2 \theta(\lf^2-\kf^2)$ times a rapidly decreasing
function of $\lf^2$ and consequently is dominated by the region
$\lf^2 \simeq \kf^2$. It follows that one gets the result
$I_T \sim {\cal F}_G(x_{\Pam},\kf^2) \sim \partial/\partial \kf^2 
x_{\Pam} f_G(x_{\Pam},\kf^2)$. 
Such subleading corrections (in the DLA sense) can be obtained from the 
exact expressions (\ref{il1}), (\ref{it1}) if we, for $i\nu=1/2-\delta$, 
do not only retain the pole term 
(proportional to $1/\delta$) of the coefficient 
function, but also the constant.
Using the fact that 
the constant can be associated with the unintegrated 
structure function
we find the following correction terms for 
$I_L$ and $I_T$
\beqn
I_L^{(c)}
&=&
\frac{(\kf^2+m_f^2)Q^2}{\kf^2M^2}
\frac{1}{(1+\xi)^3}\left(2+(1-\xi) \, \log\frac{\xi}{1+\xi}\right)
\frac{\partial}{\partial \kf^2}  x_{\Pam} f_G(x_{\Pam},\kf^2(1+\xi))
\\
I_T^{(c)}
&=&
2
\frac{1}{(1+\xi)^3}\left(1 - \xi -2\xi \log\frac{\xi}{1+\xi}\right)
\frac{\partial}{\partial \kf^2}  x_{\Pam} f_G(x_{\Pam},\kf^2(1+\xi))
\eeqn
where we have restored the mass dependence for the moment.
Indeed for $I_T$ the correction is finite in the limit $Q^2=0$
($M^2 \to \infty$). These 
correction terms represent only a small subset 
of the complete next-to-leading order corrections.
Their numerical significance will be studied below.
\\
If we now insert the above results for $I_L$ and $I_T$ 
into the cross section formulae (\ref{tra})-(\ref{asy})
we obtain a parameter free prediction for $q\bar{q}$-production
in diffractive DIS. It is on the level of 
double leading logarithmic accuracy with a subset of 
next-to-leading order corrections included, which means 
that an uncertainty in the absolute normalization
cannot be excluded.
The essential feature of the result is the dependence of 
the cross section on the square of the gluon density, with 
the scale of the latter being given by a specific combination 
of the kinematical parameters. The perturbative approach 
should be justified as long as either $\kf^2$ or $m_f^2$ (or both)
are larger than $\simeq 1 \,\mbox{GeV}^2$.
\\
In the perturbative QCD approach used here 
Regge factorization is explicitly violated. Regge factorization
was introduced in \cite{ing-sch} to describe diffractive processes
with a hard subprocess. Regge factorization states that the diffractive 
cross section can be written as the product of a 
$x_{\Pam}$-dependent flux factor (pomeron flux) and a partonic 
structure function of the pomeron which depends only on $M^2$ and $Q^2$.
In our model the $x_{\Pam}$-dependence enters through the function
$x_{\Pam}f_G(x_{\Pam},\kf^2(1+\xi))$ which depends also on $\kf^2,Q^2,M^2$
and $m_f^2$, i.\ e.\ the flux is controlled by the 
hard scales of the process and factorization is broken. 
Our analysis is of course limited to specific final states 
(jets or heavy flavors) and we cannot make a statement on the 
fully inclusive diffractive cross section which might well be
described approximately by Regge factorization. 
\\ \\
We now turn to the numerical evaluation of the cross section formulae
(\ref{tra})-(\ref{asy}). In all subsequent calculations we have used
running $\alpha_s$ with the scale being given by the scale of the 
gluon structure function. The results above have 
been given for $t=0$ where 
the cross section has its maximum. Experiments, however, cover a finite 
$t$-range and to obtain more realistic event rates we should 
integrate over $t$. The continuation of the results to finite 
$t$ can be done by using the non-forward BFKL pomeron
for the unintegrated gluon structure function.
This leads to quite complicated expressions which do not allow for 
a straightforward evaluation.
In order to keep the analysis simple we continue to finite $t$ 
by simply multiplying our expressions for $t=0$ with the elastic 
proton form factor which has been given by Donnachie and Landshoff 
\cite{dlformf} and a factor which takes into account the 
$t$-dependence of the soft pomeron trajectory \cite{donnla} 
\beqn
F_P(x_{\Pam},t)=
\frac{4-2.8 \frac{t}{\mbox{{\small GeV}}^2}}{4 - 
\frac{t}{\mbox{{\small GeV}}^2}}
\left(1-\frac{t}{0.7 \, \mbox{{\small GeV}}^2}\right)
x_{\Pam}^{-0.25 \frac{t}{\mbox{{\small GeV}}^2}}
\eeqn
With this continuation the $t$-integration can even be done 
analytically.
\\
Since our results depend on the gluon density, for a numerical evaluation
we have to decide which parametrization to use.
We choose the gluon density of GRV \cite{grv} because it allows 
for a simple implementation using the explicit formula which covers the 
whole $x,Q^2$ range. 
A more fundamental question is whether the leading or the next-to-leading
order gluon density should be used. Since we have included at least some
subleading corrections the use of the next-to-leading order
parameterization is not totally inconsistent. 
For comparison we have performed a calculation of $F_2$ with the same 
type of approximations as done above, i.e. we have calculated the 
DLA-contribution and the first subleading corrections given by the 
constant parts of the coefficient function.
\begin{figure}[h]
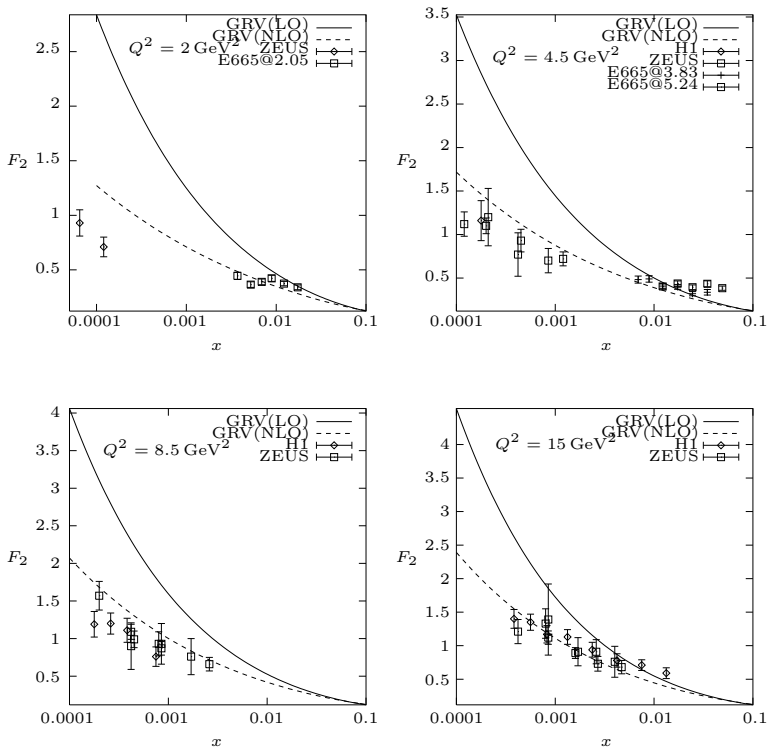

\begin{center}
\input diffqqf2comp.pstex_t
\end{center}
\caption{
Comparison of $F_2$ calculated with the leading order 
(solid line) and next-to-leading 
order (dashed line)
GRV gluon density with experimental data from 
H1 \protect\cite{h1}, ZEUS \protect\cite{zeus} and 
E665 \protect\cite{e665}.
\label{fig234}
}
\end{figure}
The results are shown in fig.\ \ref{fig234} where comparison is made 
with experimental data. For $F_2$ we get a much better description by 
using the next-to-leading order parameterization and we take this as
a justification to use it in the calculation of diffractive DIS 
as well, admitting of course that we do not have a solid theoretical
foundation for this prescription. 
\subsubsection{Numerical results}
We start our numerical analysis with giving integrated 
electron-proton cross sections for $q\bar{q}$-pair production in 
DIS diffractive dissociation for three massless flavors.
All phase space variables are integrated using a Monte-Carlo
integration algorithm \cite{vegas}. 
The available phase space space is constrained
by the cuts $Q^2 > 10 \, \mbox{GeV}^2$, $x_{\Pam}<10^{-2}$ and 
$ 50 \,\mbox{GeV} < W < 220 \,\mbox{GeV}$ ($W^2=(p+q)^2$), corresponding
to typical HERA values. For $\kf^2$ we have chosen three different
lower cut-offs $2 \,\mbox{GeV}^2$, 
$4 \,\mbox{GeV}^2$ and  $8 \,\mbox{GeV}^2$.      
In table \ref{tab1} we show the results both for the GRV leading
and the next-to-leading order gluon density. To demonstrate the 
magnitude of the corrections we compare the result containing the 
corrections with the pure DLA result.  
\begin{table}[!h]
\begin{center}
\begin{tabular}{|c|c|c|c|c|c|c|}   \hline
 & $\kf^2_0 = 2 \mbox{GeV}^2$& & $\kf^2_0 = 4 \mbox{GeV}^2$ & 
 & $\kf^2_0 = 8 \mbox{GeV}^2$ &\\
 \hline \hline
   &  GRV(LO)  &  GRV(NLO) & GRV(LO)   & GRV(NLO) &GRV(LO)  &  GRV(NLO) \\
 \hline \hline
 \multicolumn{7}{|c|} {DLA + corrections}  \\ \hline \hline
 $\sigma^{eP}_T$    & 193     & 108     & 49    & 30 & 8 & 6\\
  \hline
 $\sigma^{eP}_L$    & 15     &  9    &  3   & 2 & 1 &  0.6\\
  \hline
 $\sum_{i=T,L}\sigma^{eP}_i$
                    &  208    &  117    & 52    & 32  & 9 & 6.6\\
\hline \hline 
\multicolumn{7}{|c|} {DLA}   \\ \hline \hline
 $\sigma^{eP}_T$    & 113     & 56     & 28    & 16  & 5 & 3 \\
  \hline
 $\sigma^{eP}_L$    & 10     &  5    &  4   & 2  & 1.5 & 0.9 \\
  \hline
 $\sum_{i=T,L}\sigma^{eP}_i$
                    &  123    &  61    & 32    & 18  & 6.5 &  3.9 \\
  \hline
\end{tabular}
\end{center}
\caption{Results for total $eP$-cross sections (in pbarn) of 
diffractive dijet production for two 
different parameterizations of the gluon density and 
three different cuts on the transverse momentum of the jets.
\label{tab1}
}
\end{table}
The numbers show that the jet cross section is strongly suppressed 
at large $\kf^2$. The ratio of the longitudinal cross section to the 
transverse one is of the order of 1:10. The other conclusion to be drawn 
is that the corrections are substantial, especially for the transverse
cross section, where they are almost as large as the leading term.
It becomes clear from the numbers that an experimental analysis of the 
process requires a very good transverse momentum resolution.
\\
Next we turn to the most prominent feature of the cross section,
namely its $x_{\Pam}$-distribution.
Being proportional to the gluon density squared the cross section rises
steeply at small $x_{\Pam}$. Mainly due to this enhancement there might
be a chance to observe this type of event at HERA, although it 
is strongly momentum suppressed (it is a higher twist effect).
Due to the strong sensitivity to the gluon density the process might 
in the long run serve as a means to discriminate between different
sets of gluon distributions. Compared with diffractive vector meson
production \cite{vecmes} it has the advantage that the cross section does 
not depend on a theoretically poorly determined quantity like the 
vector meson wave function. 
The $x_{\Pam}$-dependence of the $\gamma^*$-proton cross section is 
displayed in fig.\ \ref{fig235}, separately for the transverse and the 
longitudinal part. The kinematical parameters are chosen as 
$Q^2=50 \, \mbox{GeV}^2, \beta =2/3$ and 
$\kf^2$ is integrated from $2\,\mbox{GeV}^2$
to the phase space boundary.
We show both the results based on the leading order and next-to-leading
order gluon densities to demonstrate the strong sensitivity to this 
quantity.
\begin{figure}[!h]
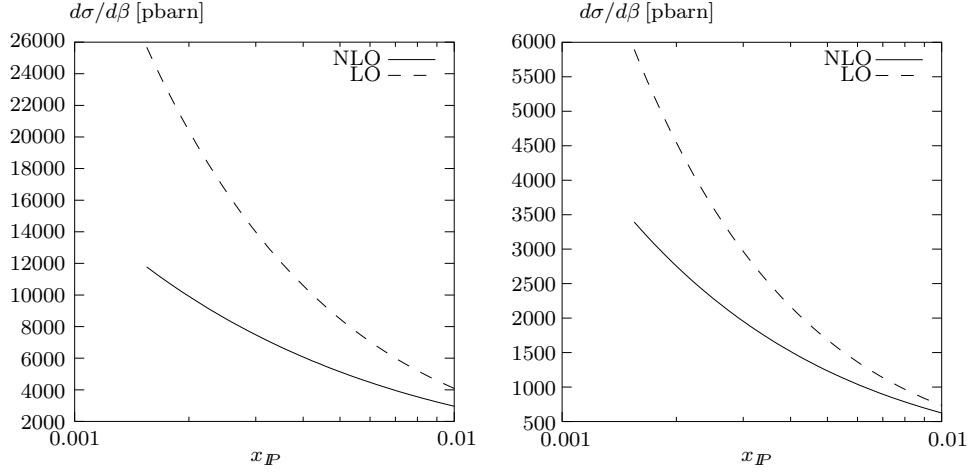

\begin{center}
\input diffqqxpom.pstex_t
\end{center}
\caption{
$x_{\Pam}$-distribution of $d \sigma_{T,L}^{\gamma^*P}/d \beta$ for fixed 
$Q^2=50 \, \mbox{GeV}^2, \beta =2/3$ and 
$\kf^2$ integrated from $2\,\mbox{GeV}^2$ to $M^2/4$.
The transverse cross section is on the left and the longitudinal one on the 
right hand side.
\label{fig235}
}
\end{figure}
\begin{figure}[!h]
\begin{center}
\input diffqqxpomfit.pstex_t
\end{center}
\caption{
$x_{\Pam}$-distribution of $1/N \, d \sigma_{T}^{\gamma^*P}/d \beta$  
($N$ is the integral of the cross section over the $x_{\Pam}$-range 
displayed).
The kinematical parameters are:
$Q^2=80 \,\mbox{GeV}^2, \beta =2/3$, $\kf^2$ integrated between 2 and 
$4 \,\mbox{GeV}^2$;
$Q^2=80 \,\mbox{GeV}^2, \beta =2/3$, $\kf^2$ integrated between 4 and 
$8 \,\mbox{GeV}^2$ and 
$Q^2=80 \,\mbox{GeV}^2, \beta =5/6$, $\kf^2$ integrated between 2 and 
$4 \,\mbox{GeV}^2$.
The variable $\alpha$ denotes the slope of each curve. 
\label{fig236}
}
\end{figure}
In fig.\ \ref{fig236} we take a closer look on the slope of the cross 
section and its dependence on the parameters $\kf^2$ and $\beta$.
For massless flavors the scale of the gluon density is
$\kf^2/(1-\beta)$ and it is known that an increase of the 
momentum scale leads to an increase of the slope.
We choose three different combinations of $\kf^2$ and $\beta$ and display
the normalized transverse cross section, calculated with the GRV
next-to-leading order gluon density.
The results from a fit to the function $(1/x_{\Pam})^{\alpha}$
confirm the expectation. The slope increases with increasing 
$\kf^2/(1-\beta)$, demonstrating the breaking of Regge factorization.
\\ \\
Next we want to present a more detailed discussion of the 
$\kf^2$-dependence of the transverse and longitudinal cross section,
again for massless flavors.
From the expressions for $I_L$ and $I_T$ and the cross section
formulae (\ref{tra}), (\ref{long}) we see that for large $\kf^2$
the transverse cross section falls roughly as $1/\kf^4$
whereas the longitudinal one decreases like $1/\kf^2$.
This power behavior is modified by the $\kf^2$-dependent scaling
violations of the gluon density which cause a logarithmic 
enhancement for larger $\kf^2$ and a flattening of the power
behavior for small $\kf^2$.
In addition we have the (integrable) Jacobi singularity
$1/\sqrt{1-4\kf^2/M^2}$ which changes the power fall-off, 
although in a region where the cross section is rather small.
The situation for $d \sigma_T^{\gamma^*P}$
and $d \sigma_L^{\gamma^*P}$
is summarized in figs.\ \ref{fig237a} - \ref{fig237d} where we 
display $\kf$-spectra for different $Q^2$ at fixed $\beta$ and for 
different $\beta$ at fixed $Q^2$.
In the figures we give results of a fit to a behavior 
$(\kf^2)^{-\delta}$ for each curve.
From fig.\ \ref{fig237c} it can be seen that a variation in $\beta$
indeed has a substantial effect on the slope. Due to the 
scaling violations an increase in $1/(1-\beta)$ leads to a drastic 
modification of the global power behavior $(\kf^2)^{-2}$.
What is seen in fig.\ \ref{fig237d} is a zero of the longitudinal
cross section in $(\kf^2,Q^2,M^2)$-space.
In the DLA the longitudinal cross section has a zero at $Q^2=M^2$ ($\xi=1$),
but when the corrections are taken into account, this zero becomes 
$\kf^2$-dependent, due to the nontrivial 
$\kf^2$-dependence of the ratio 
$f_G(x_{\Pam},\kf^2(1+\xi))/ \partial_{\kf^2}f_G(x_{\Pam},\kf^2(1+\xi))$. 
\setlength{\unitlength}{1cm}
\begin{figure}[!h] 
\begin{minipage}[h]{6cm}
\input fig237a.pstex_t
\caption{
$\kf^2$-spectra 
for transverse photons: $x_{\Pam}=5\cdot 10^{-3}$, fixed $\beta=2/3$.
\label{fig237a}
}
\end{minipage}
\hspace{1.5cm}
\begin{minipage}[h]{6cm}
\input fig237b.pstex_t
\caption{
$\kf^2$-spectra 
for longitudinal photons: $x_{\Pam}=5\cdot 10^{-3}$, fixed $ \beta=2/3$.
\label{fig237b}
}
\end{minipage}
\end{figure}
\begin{figure}[!h] 
\begin{minipage}[h]{6cm}
\input fig237c.pstex_t
\caption{
$\kf^2$-spectra 
for transverse photons: $x_{\Pam}=5\cdot 10^{-3}$, fixed $Q^2=10
\,\mbox{GeV}^2$.
\label{fig237c}
}
\end{minipage}
\hspace{1.5cm}
\begin{minipage}[h]{6cm}
\input fig237d.pstex_t
\caption{
$\kf^2$-spectra 
for longitudinal photons: $x_{\Pam}=5\cdot 10^{-3}$, fixed $Q^2=10
\,\mbox{GeV}^2$.
\label{fig237d}
}
\end{minipage}
\end{figure}
The angular asymmetric term $d \sigma_A^{\gamma^*P}$
which will be discussed in detail later 
has the same $\kf^2$-dependence as the longitudinal cross section
(for $m_f^2=0$). This means that the asymmetry is enhanced at larger 
$\kf^2$. 
\\
It is also worthwile to have a closer look on the small 
$\kf^2$-behavior of the cross section, although this does not 
correspond to the production of jets anymore. First, for $m_f^2=0$,
we observe from eqs.\ (\ref{il1}) and (\ref{it1}) that 
$I_L$ and $I_T$ behave the same for $\kf^2 \to 0$, up to constants.
For $\kf^2 \ll \Lambda_0^2$ one can expand around the saddle point 
at $i\nu_s=-1/2+\sqrt{N_c \alpha_s/\pi \, \log(1/x_{\Pam})/\Delta}$
and finds $I_L,I_T \sim \mbox{const.}$ up to logarithms and the 
usual double logarithmic exponent(cf.\ eq.\ (\ref{qqdla})). 
Inserting this into the cross section formulae (\ref{tra}), (\ref{long})
one finds that $d \sigma_L^{\gamma^*P}$ vanishes for $\kf^2 \to 0$
and that $d \sigma_T^{\gamma^*P}$ approaches a constant.
This can be traced back to the fact that the photon wave function has 
a different end-point behavior in the longitudinal and transverse case.
The wave function of a $\gamma_L^*$ is proportional to 
$\alpha(1-\alpha) \sim \kf^2$ and vanishes at the end-points 
$\alpha=0,\alpha=1$. The wave function of a $\gamma_T^*$, on the other hand
is proportional to $\alpha^2+(1-\alpha)^2 \sim 1 - 2\kf^2/M^2$ and does not 
vanish at the end points. Nonvanishing end-point contributions can be 
associated with large non-perturbative effects because the 
cross section is then dominated by the region where $\kf^2$ is small.
One can conclude that in cases where one would like to integrate
over $\kf^2$, e.\ g.\ in vector meson production or in inclusive 
diffractive scattering, the transverse cross section receives large 
nonperturbative contributions whereas the perturbative result for the 
longitudinal cross section is rather stable.  
\\ 
We finish the numerical calculations with the $\beta$-spectra 
for the longitudinal and transverse case, shown in 
fig.\ \ref{fig238}. We have kept $x_{\Pam}$ fixed and have integrated 
$\kf^2$ from $2 \,\mbox{GeV}^2$ to the phase space boundary. The minimal 
$\kf^2$ leads to a maximal $\beta$. The transverse cross section 
has a maximum at $\beta \simeq 0.4$ and tends to zero for $\beta \to 0$ 
and $\beta \to 1$. The longitudinal cross section has a zero 
at $\beta \simeq 0.4$ and also goes to zero for $\beta \to 0$
and $\beta \to 1$. The zero at $\beta \simeq 0.4$ is the zero of the 
DLA at $\beta=1/2$ shifted due to the subleading corrections. 
Only or $\beta \to 1$ the longitudinal cross section becomes comparable in 
magnitude to the transverse one. 
\\ \\
\begin{figure}[!h]
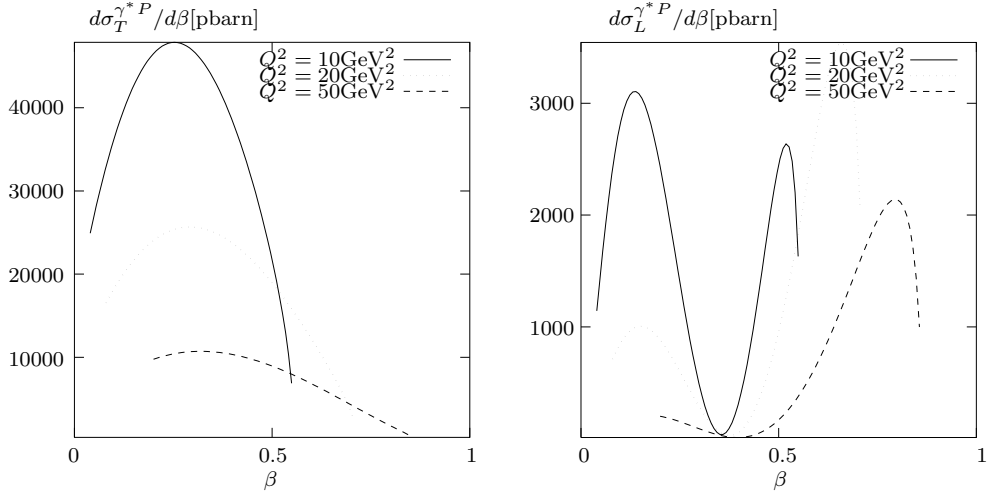

\begin{center}
\input diffqqbeta.pstex_t
\end{center}
\caption{
The $\beta$-dependence of the transverse and longitudinal cross section
for $x_{\Pam}=5\cdot 10^{-3}$, $\kf^2$ integrated from 
$2 \,\mbox{GeV}^2$ to the phase space boundary and for different values of 
$Q^2$. 
\label{fig238}
}
\end{figure}
\subsubsection{The angular asymmetry}
We now come to the discussion of the angular dependence of the 
electron-proton cross section. The dependence upon the angle 
$\theta$ is not so interesting since this can be derived
easily from the dependence upon $\kf^2$ due to the simple 
relation $\sin^2 \theta = 4 \kf^2/M^2$. It is clear
that the jets are concentrated near the 
incoming proton direction and large angle 
scattering is strongly suppressed. 
\\
The dependence upon $\phi$ is more interesting, not alone 
since a measurement of the $\phi$-distribution allows to disentangle
the transverse and the longitudinal contribution. 
We have two asymmetric terms, the term $d \sigma_A^{\gamma^*P}
\sim \cos 2 \phi$ which is symmetric w.\ r.\ t.\ $\phi=\pi$
and the interference term $d \sigma_I^{\gamma^*P}
\sim \cos \phi$ which is antisymmetric w.\ r.\ t.\ $\phi=\pi$.
It follows that the interference contribution can be eliminated
by adding the contributions at $\phi$ and $\phi+\pi$.
For $d \sigma_A^{\gamma^*P}$ we note that this term has an additional 
factor $(\kf^2+m_f^2)/M^2$ compared to the angular symmetric 
transverse term $d \sigma_T^{\gamma^*P}$. 
This offers the possibility to enhance the $\phi$-asymmetry by restricting 
measurements to larger $\kf^2$. 
It is remarkable that the $d \sigma_A^{\gamma^*P}$ 
contribution enters the electron-proton cross section 
with a negative sign, i.\ e.\ the jets which are produced in 
DIS diffractive dissociation prefer to lie in a plane 
perpendicular to the plane defined by the incoming and outgoing 
electron. These features of the $\cos 2 \phi$ asymmetry are 
illustrated in fig.\ \ref{fig239}.
\begin{figure}[!h]
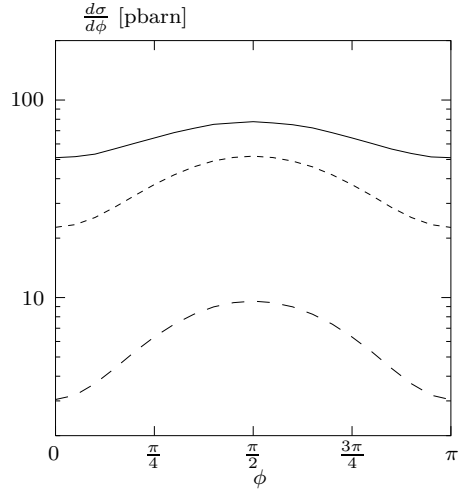

\begin{center}
\input diffqqphi.pstex_t
\end{center}
\caption{
The total $eP$-cross section as a function of $\phi$ with 
$\kf^2$ integrated from $1 \,\mbox{GeV}^2$ to $2 \,\mbox{GeV}^2$
(solid curve), from  $2 \,\mbox{GeV}^2$ to the phase space boundary
(dotted curve)
and from $5 \,\mbox{GeV}^2$ to the phase space boundary (dashed curve).
\label{fig239}
}
\end{figure}
We have again calculated the integrated electron proton cross section
with cuts $Q^2 > 10 \,\mbox{GeV}^2, x_{\Pam}<10^{-2}, 
50 \,\mbox{GeV} < W < 220 \,\mbox{GeV}$ and three different bins in $\kf^2$.
The contributions at $\phi$ and $\phi+\pi$ are added, hence the interference
term drops out and the total cross section is recovered by integrating 
$\phi$ from 0 to $\pi$.
As expected for larger $\kf$2 the asymmetry becomes enhanced. 
We find a factor of $8/5$ between $\phi=\pi/2$ and 
$\phi=0$  for the lowest $\kf^2$-bin and a factor of 3 for the highest 
$\kf^2$-bin.
\\
The point which deserves further discussion is the peak of the 
diffractive cross section at $\phi=\pi/2$, i.\ e.\ the negative sign 
of the $\cos 2 \phi$-term.
Inspection of $q\bar{q}$-pair production in the photon-gluon 
fusion process (i.\ e.\ one-gluon exchange between the $q\bar{q}$-pair
and the proton) shows \cite{bgf}
that in this case the $\cos 2 \phi$-term has a positive sign, i.\ e.\
the cross section has a minimum at $\phi=\pi/2$.
Adding a further gluon in the $t$-channel hence completely changes 
the orientation of the final state. Clearly this striking phenomenon
makes the azimuthal distribution a very interesting
experimental signal which could be used to test the two-gluon nature 
of the hard diffractive interaction.
We sketch below the derivation of the photon-gluon fusion result 
to elucidate the origin of the sign change, although a simple 
physical interpretation, based on angular momentum arguments for example,
is not at our hand at the moment.
\\
We use the same formalism as in the two-gluon case, namely 
$\kf$-factorization, to calculate the process 
$\gamma^*(q) +P(p) \to q(k) + \bar{q}(q-k+\xi)+X(p-\xi)$
and take $\xi=\eta p +\lf$ 
($\eta = x(1+M^2/Q^2$)). Here in the 
leading-log($1/x$)-approximation the transverse momentum $\lf$
of the exchanged gluon goes into the $q\bar{q}$-state, i.\ e.\ when
we integrate over $\lf$ we do not describe back-to-back
$q\bar{q}$-production, but inclusive one-jet production
(the other jet is integrated). The analogous situation to 
the diffractive case is only recovered in the collinear (or DLA) limit,
when $\lf \to 0$. In the photon-gluon case the cross section,
not the amplitude, is proportional to the unintegrated gluon 
structure function. 
Now, for photon-gluon fusion the contraction of $\ef_{\mu}\ef_{\nu}$
with the hadronic tensor results in
\beqn
\ef^2 \left[\Phi(\kf)-\Phi(\kf-\lf)\right]^2-4\alpha(1-\alpha)
\left[\ef\cdot \Phi(\kf)-\ef\cdot\Phi(\kf-\lf)\right]^2
\label{bgfwv}
\\
\mbox{with}\;\;\;\;\;
\Phi(\kf)=\frac{\kf}{\alpha(1-\alpha)Q^2+\kf^2}
\phantom{XXXXX}
\eeqn
The corresponding term in the diffractive case reads
\beqn
\ef^2 \left[2\Phi(\kf)-\Phi(\kf-\lf)-\Phi(\kf+\lf) \right]
      \left[2\Phi(\kf)-\Phi(\kf-\lf')-\Phi(\kf+\lf') \right]
\phantom{XXXXXXXX}
\\ \nonumber 
-4\alpha(1-\alpha)
\left[\ef\cdot \Phi(\kf)-\ef\cdot\Phi(\kf-\lf)\ef\cdot\Phi(\kf+\lf) \right]
\left[\ef\cdot \Phi(\kf)-\ef\cdot\Phi(\kf-\lf')\ef\cdot\Phi(\kf+\lf')\right]
\label{diffwv}
\eeqn
The structures in brackets can be related to the quark-antiquark-gluon
components of the light cone wave function of a transversely 
polarized virtual photon \cite{baj}.
Now we concentrate on the term proprtional to $\alpha(1-\alpha)$ since 
this one will lead to the $\cos 2 \phi$-dependence.
In the collinear limit we have to extract the $O(\lf^2)$
contribution in eq. (\ref{bgfwv}) and the $O(\lf^2\cdot \lf{'}^2)$
contribution in eq.\ (\ref{diffwv}). 
In the latter case we can expand in each bracket separately and get 
from each bracket an identical coefficient $\sim \cos\phi$.
The result is 
\beqn
-4 \alpha(1-\alpha)\, 16 \pi^2 \ef^2 \frac{(\kf^2)^3}
{(\alpha(1-\alpha)Q^2+\kf^2)^6}\cos^2 \phi
\\
=
-4 \frac{\kf^2}{M^2}\, 16 \pi^2 \ef^2 \left(\frac{M^2}{M^2+Q^2}
\right)^6 \frac{1}{(\kf^2)^3}(\cos 2 \phi+1)
\eeqn
where we have used the relation $\kf^2 =M^2 \alpha(1-\alpha)$.
The coefficient which contributes in the 
collinear approximation
is a square and as such positive
definite. In the photon-gluon fusion situation we have
\beqn
4 \alpha(1-\alpha)\, \pi \ef^2 
\frac{1}
{(\alpha(1-\alpha)Q^2+\kf^2)^4}
( \kf^4+\alpha^2(1-\alpha)^2Q^4-2 \cos 2\phi \alpha(1-\alpha)Q^2\kf^2) 
\\
=-4 \frac{\kf^2}{M^2} \, \pi \ef^2 
\left(\frac{M^2}{M^2+Q^2}
\right)^2
\frac{1}{(\kf^2)^2}
\left(1+\frac{Q^4}{M^4}-2 \frac{Q^2}{M^2}\cos 2 \phi\right)
\phantom{xxxxxxxxxx} 
\eeqn
Here the coefficient of the collinear approximation has a more complicated 
structure in which the $\cos 2 \phi$-term happens to have a negative sign.
This leads to the relative sign compared to the diffractive case.
\\
We compare the 
(normalized)
two-gluon and the one-gluon cross section in 
fig.\ \ref{fig2310}. 
Several interesting features can be read off from these graphs.
The difference between the $\cos 2 \phi$ and the $-\cos 2 \phi$ 
behavior is clearly visible.
Since we have, in contrast to fig.\ \ref{fig239}, not added
the contributions at $\phi$ and $\phi+\pi$ the effect of the interference
term can be seen (we display only the interval of $\phi$ between  
0 and $\pi$ since the cross section for $\phi > \pi$ can be obtained 
by reflection w.\ r.\ t.\ the line $\phi=\pi$).
We have $\beta =1/3$ on the left hand side and $\beta=2/3$ on the right hand
side. It is obvious that the $\cos \phi$-term has a different sign in these 
two situations which is due to the zero of $I_L$ located at 
$\beta \simeq 0.4$.
It is furthermore encouraging that the asymmetry is stronger for 
the diffractive situation than for photon-gluon fusion 
(corresponding to normal DIS). 
Since we have no color rearrangement between the jets and the 
proton in the diffractive case (in contrast to normal DIS),
one can expect that the signal is rather stable against hadronization
effects. 
For the normal DIS situation it is known \cite{ingelman}
that hadronization effects
considerably modify the asymmetry created on the partonic level.
\begin{figure}[!h]
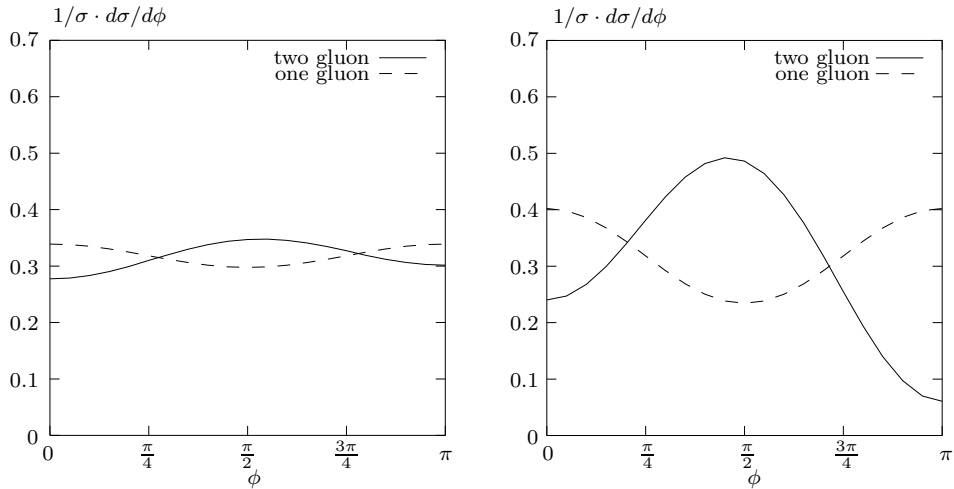

\begin{center}
\input diffqqcomp.pstex_t
\end{center}
\caption{
The $\phi$-distribution of $q\bar{q}$-pair production based on one-gluon
and two-gluon exchange. The parameters are $\beta=1/3$ (left hand side),
$\beta=2/3$ (right hand side), $Q^2 = 100 \,\mbox{GeV}^2$
and $\kf^2$ is integrated above $5 \, \mbox{GeV}^2$. 
\label{fig2310}
The normalization is the integral of the cross section from 0 to $\pi$.
}
\end{figure}
\subsubsection{Production of charm quarks}
We now generalize our analysis to finite quark mass $m_f$. 
All previous numerical calculations have been done for three massless
flavors but now we will calculate the contribution that can be 
expected from charm quarks with mass $m_c = 1.5 \, \mbox{GeV}$.
For this mass the characteristic scale $(\kf^2+m_f^2)(Q^2+M^2)/M^2$
is always large, even for small $\kf^2$. This offers the possibility
to integrate the double-logarithmic results 
(including the corrections) over $\kf^2$ and to obtain in this way
the charm contribution to the diffractive structure function.
Note that for $m_f^2 \neq 0$ the DLA expression for $I_L$ (\ref{ildla})
becomes a constant 
and the DLA expression for 
$I_T$ (\ref{itdla}) vanishes in the limit $\kf^2 \to 0$. 
The $\kf^2$-integration is therefore infrared finite.
\begin{table}[!h]
\begin{center}
\begin{tabular}{|c|c|c|c|c|c|c|}   \hline
 & $\kf^2_0 = 2 \mbox{GeV}^2$& & $\kf^2_0 = 4 \mbox{GeV}^2$ & 
 & $\kf^2_0 = 8 \mbox{GeV}^2$ &\\
 \hline \hline
   &  GRV(LO)  &  GRV(NLO) & GRV(LO)   & GRV(NLO) &GRV(LO)  &  GRV(NLO) \\
 \hline \hline
 \multicolumn{7}{|c|} {DLA + corrections}  \\ \hline \hline
 $\sigma^{eP}_T$    & 39     & 27     & 16    & 11 & 5 & 3.4    \\
  \hline
 $\sigma^{eP}_L$    & 3.4    &  2.3   &  0.6  & 0.4 & 0.2 & 0.1   \\
  \hline
 $\sum_{i=T,L}\sigma^{eP}_i$
                    &  42.4    &  29.3    & 16.6    & 11.4  & 5.2 & 3.5\\
\hline \hline 
\multicolumn{7}{|c|} {DLA}   \\ \hline \hline
 $\sigma^{eP}_T$    & 34     & 22     & 14    & 9  & 3.8 & 2.6 \\
  \hline
 $\sigma^{eP}_L$    &2.5&1.7&0.5&0.4&0.2&0.1 \\
  \hline
 $\sum_{i=T,L}\sigma^{eP}_i$
                    &36.5&23.7&14.5&9.4& 4 &  2.7 \\
  \hline
\end{tabular}
\end{center}
\caption{Results for total $eP$-cross sections (in pbarn) of 
diffractive dijet production for charm quarks.
The cuts are the same as in table \protect\ref{tab1} 
\label{tab2}.
}
\end{table}
Before we turn to the calculation of the structure function
we give in table \ref{tab2} the charm contribution to the jet
cross section. The total $eP$ cross section was calculated with the 
same cuts as in table \ref{tab1}.
Depending on the momentum cut,
charm gives a contribution between 25 and 50 percent  
relative to the three massless flavors.
It is obvious that the suppression for large transverse momenta 
is not as strong as in the massless case.
This is due to the nontrivial $\kf^2$-dependence of $\xi$
in the results for $I_L$ and $I_T$ which damps the 
$1/\kf^2$ decrease.
It is also noticeable that the relative importance of the
correction terms is much smaller than in the massless case.
\\
We now turn to the charm contribution to the diffractive 
structure function $F_2^D$. 
The latter has been introduced in analogy to the inclusive 
structure function $F_2$ to describe diffractive events
(events with a rapidity gap) in electron-proton collisions.
In terms of $F_2^D$ the diffractive cross section reads
\beqn
\frac{d \sigma_{\mbox{\tiny{DIFF}}}^{eP}}{d \beta d Q^2 d x_{\Pam}}
= \frac{2 \pi \alpha_{\mbox{\tiny{em}}}^2}{\beta Q^4}
[1+(1-y)^2]\; F_2^D(\beta,Q^2,x_{\Pam})
\eeqn
where the longitudinal contribution has been neglected.
With this definition of $d \sigma_{\mbox{\tiny{DIFF}}}^{eP}$, 
$F_2^D$ can be obtained 
from the photon-proton cross section
in the following way
\beqn
F_2^D(\beta,Q^2,x_{\Pam}) = 
\frac{Q^2}{4 \pi^2 \alpha_{em}}\int_0^{\infty}d t
\int_0^{M^2/4-m_c^2}
d \kf^2 
\left[
\frac{d \sigma_T^{\gamma^*P}}{d x_{\Pam}d \kf^2 d t}
+ 
\frac{d \sigma_L^{\gamma^*P}}{d x_{\Pam}d \kf^2 d t}
\right]
\eeqn  
In fig.\ \ref{fig2311} the $x_{\Pam}$-dependence of the diffractive 
structure function $F_2^{D\,\mbox{\tiny{(charm)}}}$ is shown.
What is seen here is again the steep rise in $x_{\Pam}$ which 
is determined by the square of the gluon distribution. 
Compared with experimental data on $F_2^D$ this rise is too strong
but our analysis is of course only valid for heavy quarks. 
For the light quarks which make up the larger part of the cross 
section a slower rise is expected since nonperturbative 
contributions dominate. 
The ZEUS collaboration \cite{zeusf2d} 
quotes a value of $\sim 30$ for $F_2^D$
at $Q^2= 16 \,\mbox{GeV}^2, \beta=0.65$, and $x_{\Pam}\simeq 10^{-3}$.
From the right hand side of fig.\ \ref{fig2311} we read off a value of
$\sim 2.3$, i.\ e.\ 
at first sight
we predict a 10 \% contribution of charm 
in $F_2^D$.
It would of course be interesting
to isolate the charm contribution to $F_2^D$ experimentally
to confirm the perturbative character of this type of events.    
It can be seen from the graphs
that the $Q^2$-dependence is rather weak, i.\ e.\ 
$F_2^D$ shows a leading-twist behavior
\footnote{It is understood that all statements are referring to the 
case of charm production. The index $^{\tiny{(charm)}}$ will be omitted in the
following.}.
This is true, however, only for the transverse part of $F_2^D$.
As can be seen from eq.\ (\ref{long}) the longitudinal cross section has 
a $1/Q^2$ suppression. The most striking effect of the variation of 
$Q^2$ is again the change in the slope. These curves thus
demonstrate again the breaking of Regge-factorization.
\begin{figure}[!h]
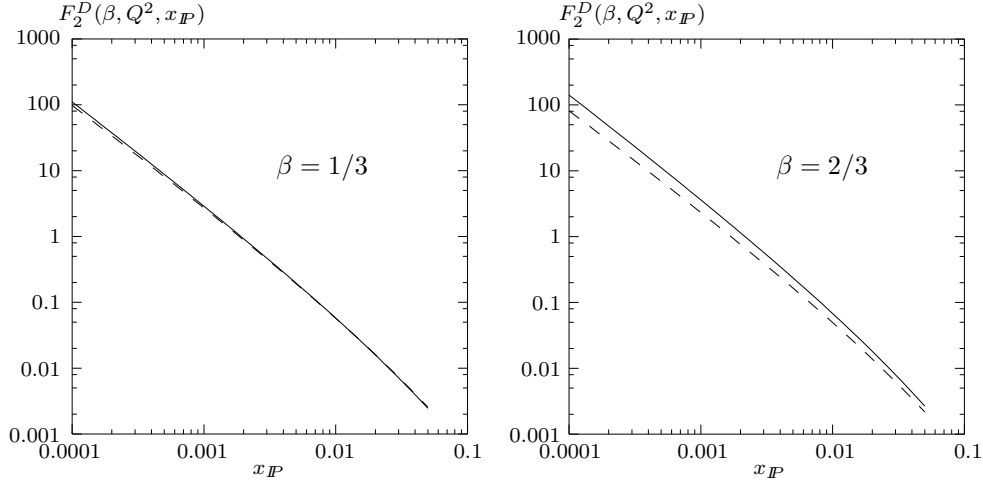

\begin{center}
\input diqqf2dxpom.pstex_t
\end{center}
\caption{
The $x_{\Pam}$-dependence of the diffractive structure function for 
$\beta=1/3$, $\beta=2/3$ and $Q^2=50 \,\mbox{GeV}^2$ (solid line)
and $Q^2=20\,\mbox{GeV}^2$ (dashed line).
\label{fig2311}
}
\end{figure}
In fig.\ \ref{fig2312} we display the $\beta$-dependence
of $F_2^D$ for $x_{\Pam}=10^{-3}$ and two values of $Q^2$, separating the 
transverse and the longitudinal part. On the right hand side of each 
figure the charm-threshold can be seen, 
i.\ e.\ there is no phase space left if $M^2$ becomes equal 
to the threshold mass $M_{th}^2=4 m_c^2$.
Increasing the mass from the threshold, i.\ e.\ going from  
higher to lower $\beta$, one sees first an increase in 
$F_2^D$ due to the increasing phase space, but ultimately 
with growing masses the cross section is mass-suppressed 
and $F_2^D$ goes to zero for $\beta \to 0$. Again we see that for 
large $\beta$ there is a region where the longitudinal contribution 
becomes larger than the transverse one.
When one compares the $\beta$-dependence of fig.\ \ref{fig2312}
with experimental data one realizes one major shortcoming 
of our approach. The measured $F_2^D$ does not vanish for 
$\beta \to 0$. For large masses quark-antiquark-gluon final states 
give the dominant (and for $\beta=0$ non-vanishing) contribution.
These contributions are beyond the reach of the present analysis.    
It should therefore be realized that all calculations  
presented here are limited to the quark-antiquark component 
of the of $F_2^{D\,(\mbox{\tiny{charm}})}$. Including higher order 
components will above all modify the $\beta$-spectrum and the overall
normalization. The $x_{\Pam}$-dependence is not expected to change.
\begin{figure}[!h]
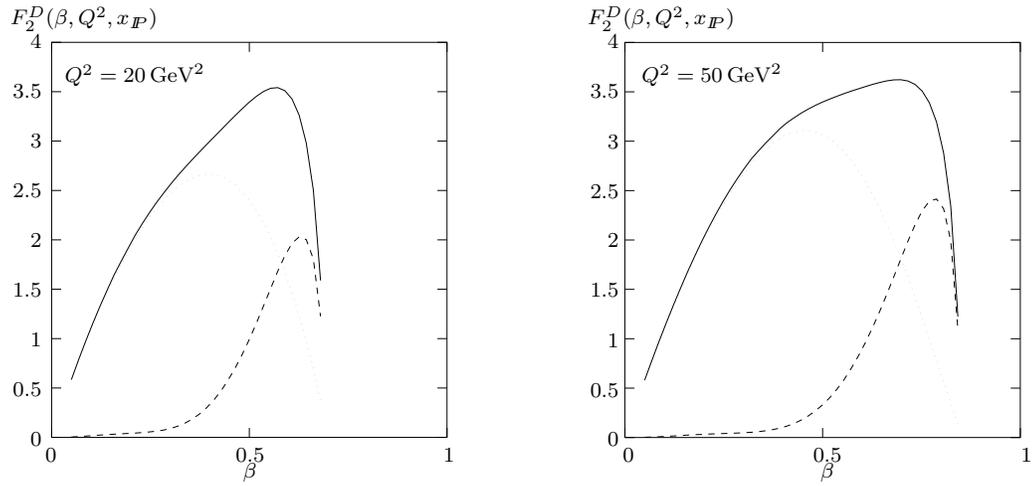

\begin{center}
\input charmbeta2.pstex_t
\end{center}
\caption{
The $\beta$-dependence of the diffractive structure function 
for $Q^2 = 20\,\mbox{GeV}^2$ and $Q^2 = 50\,\mbox{GeV}^2$ and 
$x_{\Pam}=10^{-3}$. The dotted line is the transverse contribution 
the dashed line is the longitudinal part and the solid lines 
represent the sum.
\label{fig2312}
}
\end{figure}
\clearpage
\newpage
\setcounter{equation}{0}
\setcounter{figure}{0}
\setcounter{table}{0}
\section{Unitarity Corrections}
\label{chap3}
In the preceding chapter we have discussed phenomenological implications of 
the resummation of large logarithms of $1/x$.
The basic principle of these calculations was the insertion 
of the BFKL amplitude which resums the leading logarithms into a specific 
environment corresponding to respective initial and final states of a 
deep inelastic electron proton scattering process.
The common feature of the results is the steep increase of the cross section
as $x$ decreases, either according to a power behavior with a power of order 
unity or - when only double logarithms are taken into account - according to 
the $\exp(\sqrt{\log1/x})$ behavior. 
This steep increase cannot persist down to arbitrarily low values of $x$ 
since it violates a fundamental principle of quantum theory, i.\ e.\ 
unitarity. In the context of relativistic quantum field theory 
of the strong interaction unitarity implies that the cross section of a 
hadronic scattering reaction
cannot increase with increasing energy $s$ stronger than $\log^2(s)$.
This statement is usually referred to as Froissarts theorem \cite{froissart}.
Applied to deep inelastic scattering 
\footnote{Although it is not proven that the theorem is applicable to DIS
this is generally believed to be the case. Compare footnote 
\protect\ref{foot} in chapter 2.} 
this theorem states that at small 
$x$ the total cross section (the structure function) cannot increase faster 
than $\log^2 1/x$.    
This shows that the results of the preceding chapter can be valid only in 
a limited range of $x$ since for arbitrarily small $x$ they are in conflict 
with unitarity.
This is one of the major shortcomings of the leading logarithmic approximation.
Although consistent from the point of view of a perturbative expansion
this approximation scheme violates a fundamental principle of quantum 
field theory. 
To fulfill the unitarity requirement one inescapably has to go beyond
the leading logarithmic approximation.
\\
In this chapter we introduce and investigate contributions that arise 
when subleading logarithms are taken into account in the 
perturbative expansion.
\\
One should remark that the program which is pursued
in the following constitutes only one of the many recent
(effective actions \cite{lipeffac,verlinde}, 
eikonal approximation in soft gluon background \cite{nachtmann},
operator expansion \cite{balitskii}, semiclassical approach
\cite{buchmcdermheb,buchheb}) 
approaches to the unitarization problem in QCD.
It is however the only approach which starts from the well-established
and consistent framework of perturbative QCD which has been highly 
successful in describing strong interaction physics in deep inelastic 
scattering and elsewhere.
On the other hand, it is not expected that a solely perturbative approach 
will be sufficient to restore unitarity.
Ultimately one will certainly need some nonperturbative information
regarding the structure of the colliding hadrons. Perturbation theory 
might nevertheless be a sensible starting point since it is known that a 
connection to nonperturbative physics might be drawn from ambiguities
associated with the QCD perturbation series \cite{renormalon}.
\\
So far the most complete approach to subleading perturbative logarithms 
has been formulated by Bartels \cite{bart-veryold,bart-old,bartels}.
It is based on using unitarity and dispersion relations from the start
as a tool to construct higher order amplitudes.
As the main outcome of this approach one finds that it is necessary
to take into account contributions with higher numbers of reggeized gluons
in the $t$-channel, compared to the BFKL amplitude with two reggeized gluons.
Ultimately it will be required to consider general
$n$-reggeized gluon states in the $t$-channel and to sum over $n$.  
To this end it will be sensible to decompose the $n$-gluon problem
into two major sectors. First one identifies transition vertices which 
mediate the transition between states with a different number of 
reggeized gluons. Second one has to solve the $n$-gluon problem.
There is hope that one can find an effective field theory which 
incorporates all these elements and which might be solvable.
If one recalls the symmetry properties of the BFKL amplitude which
constitutes the first step in this program, one might even conjecture
that this effective field theory is conformally symmetric. 
\\
In order to gain insight into the structure of a potential field theory
one has in a first step to investigate the simplest new elements which 
arise beyond the leading logarithmic approximation. This is the objective
of the present chapter.
\\
We start with an introduction and review the basic concepts underlying
the approach of \cite{bartels}. In this introductory part we stress the 
importance of the notion of a reggeon as a collective excitation
of elementary degrees of freedom in QCD. 
The reggeon which contains an infinite number of Feynman diagrams
might be the right 
candidate for the basic field in an effective theory. 
At the present stage it is at least convenient to interpret the amplitudes
which emerge from the calculation as reggeon amplitudes.
To illustrate this we start with an interpretation of the BFKL equation in 
this framework. Since $t$-channel unitarity will ultimately be an important 
requirement to construct amplitudes of physical processes from reggeon
amplitudes we demonstrate as an example the $t$-channel unitarity
relations which follow from the BFKL equation. 
\\
We then give the coupled system of equations for reggeon amplitudes
with up to $n=4$ reggeized gluons in the $t$-channel. 
The principle how
to build equations with higher numbers of reggeons will become clear 
then. The step which was anticipated before, namely the identification
of transition vertices has for this case been accomplished by 
Bartels and W\"usthoff \cite{bartels,bartels-wue} and we describe the outcome
of their analysis. Having isolated the two most important elements of 
the first subleading corrections, namely the transition vertex and the 
interacting four reggeized gluon state we then turn to the detailed 
investigation of these objects.  
We first derive a compact symbolic representation for the transition 
vertex. This representation is in the spirit of Lipatovs 
Hamilton operator formulation of the BFKL kernel \cite{holsep}.
The symbolic representation allows on the one hand a quite 
straightforward proof of conformal invariance of the vertex \cite{balipwue}
and on the other hand the adressation of the property of holomorphic
separability which has been shown to hold for the BFKL kernel \cite{holsep}.
We then show that after projection with the elementary conformal three
point functions which were introduced in chapter \ref{chap1}
to diagonalize the BFKL kernel, the transition vertex can be consistently
interpreted as the conformal three point function of a composite operator. 
This demonstrates that the interpretation of the amplitudes
in the framework of a conformal field theory which was sketched for the 
BFKL amplitude in chapter \ref{chap1} can be extended 
to the first corrections. 
The coordinate representation of the vertex is then completely fixed.
The nontrivial part of the vertex in this representation is a 
dimensionless function of three conformal dimension
which is given in terms of two-dimensional integrals. 
\\
The remaining sections of this chapter are devoted to the investigation 
of the four reggeized gluon state. We state the defining equation, namely the
four particle BKP equation \cite{bart-veryold,bkp-k} 
and discuss its properties. 
The phenomenological 
and theoretical significance of the solution of this equation is then 
discussed in some detail. We stress that a fully unitary amplitude requires
the resummation of all contributions with arbitrarily large number of 
reggeized gluons in the $t$-channel.
From the point of view of a potential effective conformal field theory
the four gluon amplitude already contains interesting information.
Performing a short-distance expansion of the amplitude it should
be possible to derive the operator algebra of the composite operators
associated with the BFKL pomeron. We discuss the concept of short-distance
expansion and show using the BFKL amplitude as an example the relation
between the anomalous dimensions and the spectrum of the system.
This then leads us to our approach to the spectrum which is based
on the twist expansion, i.\ e.\ the short-distance expansion
of the amplitude in momentum space. The basic idea is to reconstruct the 
spectrum from the anomalous dimensions.
\\
In the concluding section we then show how in principle the twist expansion
of the four gluon state can be derived. We reformulate the defining equation
using a method which goes back to Faddeev \cite{faddeev}.
Using this method one can reformulate the problem as an
effective two particle problem with highly complicated propagators and 
interaction vertices. 
We show how the singularities that are needed for the twist expansion can be
extracted from the propagators and the vertices.
For the simplest case we demonstrate how these singularities can be iterated
to obtain the singularity of the four reggeized gluon amplitude.
These singularities can be interpreted as the anomalous dimension of a
corresponding operator associated with the four gluon state. 
We discuss its relation to the spectrum of the 
four reggeized gluon state. The complications which arise for the 
subleading singularities are finally indicated. 
\\
One final remark is in order here. In our investigations we keep the full 
color structure of the four gluon problem. In this case only conformal
symmetry of the elementary interaction kernel can be and is used to simplify 
the problem. It has been shown that a remarkable simplification 
occurs in the large-$N_c$ approximation \cite{largenc}.
The key observation is that in this approximation the $n$-gluon system
becomes holomorphic separable and in turn reduces to the product 
of two one-dimensional problems.
After Lipatov had proven \cite{quant}
the existence of 
nontrivial conserved quantities of the $n$-gluon system,  
Faddeev and Korchemsky \cite{fad-kor,kor}
have shown that the system can be identified with the 
XXX Heisenberg model for spin $s=0$. 
From this follows that the system is completely solvable
in the sense that there exists a sufficiently 
large number of integrals of motion.
Using these results one can try to employ the 
quantum inverse scattering method and solve 
the system with the Bethe ansatz
\cite{kor,wall}.
We do not follow this direction in the present work.
\newpage
\subsection{Generalized leading-log approximation}
Let us first show where the amplitudes which are 
investigated in the following parts of this chapter have their 
origin and sketch the role they play in the program towards 
an unitary effective high energy theory derived from QCD.
The term 'generalized leading-log approximation' which appears here
in the title is not really well defined. It is intended to denote the 
approximation scheme in which a minimal subset of next-to-leading 
logarithmic terms is taken into account which are needed to 
construct a unitary amplitude. 
\\
The fundamental idea is to obtain an effective description of QCD in 
the Regge limit in terms of a reggeon field theory \cite{gribov,abarbanel}
in which 
unitarity is built in from the start.
The reggeon is a collective excitation of the underlying field 
theory (QCD) that can be associated with singularities of 
partial wave amplitudes. It can be regarded as a nonrelativistic field
in two transverse space dimensions which carries signature and belongs to 
an irreducible representation of the gauge group of the underlying
theory. The immediate example is the reggeized gluon which emerges from an 
infinite order resummation of elementary gluon diagrams.
It can be associated with a pole of the partial wave amplitude in the 
color octet channel and it has negative signature.
\\
The basic objects of the effective theory are the $n$-reggeon amplitudes.
To be specific we consider two-particle to $n$-reggeon amplitudes 
$D_n(\omega)$ where the reggeons 
\footnote{Unlike otherwise stated the term 'reggeons' in the following
always refers to reggeized gluons.} 
are coupled  
to a species of external elementary particles of the theory.
Here $\omega$ is a complex angular momentum variable which admits the 
interpretation as the energy of the $n$-reggeon system.
These amplitudes arise from multiple discontinuities of multigluon 
amplitudes and can be shown to obey a set of coupled integral equations.
The solutions of these equations are then plugged into the 
reggeon unitarity equations - which are unitarity equations in the 
$t$-channel - to calculate partial wave amplitudes \cite{white,bart-old}.
In the intermediate step in which the $t$-channel unitarity is applied
one has to go to the physical region in the $t$-channel, i.\ e.\ 
$t>0$. Afterwards an analytical continuation to $t<0$ has to be performed.
In this way the partial wave amplitude is obtained as a sum of 
contributions with different numbers of reggeons in the $t$-channel.
It is important to emphasize that the dynamical equations for the reggeon
amplitudes are coupled, i.\ e.\ there are transition vertices from 
$n$ to $m$ reggeons and the reggeon number in the $t$-channel is not 
conserved.
\\
The starting point in this formalism is the two reggeon amplitude
$D_2(\omega)$ determined by the BFKL equation which 
can be written in the form \cite{klf2}
\beqn
\left[\omega+\beta(\kf)+\beta(\qf-\kf)\right]
D_2^I(\omega;\kf,\qf-\kf) = D_{2,0}^I
+ 
\left[K_{(2,2)}^I \otimes D_2^I(\omega)\right](\kf,\qf-\kf)
\label{bfklregg}
\eeqn
Here $D_{2,\,0}^I$ is an elementary particle-gluon vertex,
$K_{(2,2)}^I$ and $\beta(\kf)$ are the kernel and the trajectory 
function (\ref{traj})
of the BFKL equation
and the 
index $I$ labels the irreducible representation of the gauge group.
The function $\omega+\beta(\kf)+\beta(\qf-\kf)$
can be interpreted as the inverse reggeon propagator. 
The equation (\ref{bfklregg}) then admits the interpretation as 
a two reggeon integral equation where the kernel 
$K_{(2,2)}^I$ represents the reggeon interaction.
The kernel has the form
\beqn
K_{(2,2)}^I(\qf;\kf,\kf') =  
\frac{N_c}{2} g^2
\frac{c_I}{\kf{'}^2(\qf-\kf')^2}
\left[- \qf^2 +  
\frac{\kf^2(\qf-\kf')^2+\kf{'}^2(\qf-\kf)^2}
{(\kf-\kf')^2}
\right]
\label{kernregg}
\eeqn
and the symbol $\otimes$ in eq.\ (\ref{bfklregg}) represents integration 
w.\ r.\ t.\ the measure $d^2 \kf'/(2 \pi)^3 $. 
The weight $c_I$ is a real number depending on the 
irreducible representation ($g$ is the gauge coupling). 
In terms of $D_2(\omega)$ the partial wave 
amplitude $A(\omega,t)$ can be calculated as
\beqn
A^I(\omega,t) =
\int \frac{d^2 \kf}{(2\pi)^3} \frac{1}{\kf^2(\qf-\kf)^2}
D_{2,0}^I\,\cdot \,D_{2}^I(\kf-\qf-\kf)
\phantom{xx},\phantom{x}-\qf^2=t
\label{factregg}
\eeqn
When $I$ corresponds to 
the color octet representation in QCD we have $c_I=1$
and the solution of eq.\ (\ref{bfklregg}) is \cite{klf1}
\beqn
D_{2}^{8_A}(\kf,\qf-\kf) = \frac{D_{2,0}^{8_A}}{\omega+\beta(\qf^2)}
\label{pole}
\eeqn
from which follows
\beqn
A^{8_A}(\omega,t)= 
\frac{2}{3g^2}
(D_{2,0}^{8_A})^2\frac{\beta(\qf^2)}{\omega+\beta(\qf^2)}
\label{follo}
\eeqn
This is the manifestation of the reggeization of the gluon.
The partial wave amplitude has a Regge pole in the complex angular momentum
plane.
Of course the solution (\ref{pole}) is infrared divergent.
In a nonabelian gauge theory with a massless gauge boson, 
amplitudes with nonsinglet 
exchange always contain infrared divergencies. Since the reggeization is such 
a fundamental property in the present approach it is desirable to give a 
meaning to the solution in eq. (\ref{pole}). To this end one has to perform
a regularization. In the following we assume a mass regularization, i.\ e.\
the gluon has acquired a mass $\lambda$. To do this consistently one has to 
add scalar particles to the theory which give additional contributions 
to the kernel in eq.\ (\ref{kernregg}).   
We will not consider these additional terms in detail since they do not 
play any role in the following.
\subsubsection{Unitarity relations from the BFKL equation}
In this short interlude we want to show that a set of unitarity 
equations can be derived from the integral equation (\ref{bfklregg})
which could in principle serve to calculate the partial wave amplitude.
The unitarity corrections which are the objective of this chapter will 
provide generalizations of the reggeon integral equations to 
higher numbers of reggeons.   
This formalism does however not give 
a generalization to eq.\ (\ref{factregg}).
It is therefore not clear in which way the $n$-reggeon amplitudes
contribute to the partial wave amplitude. The idea is to use $t$-channel 
unitarity to construct the partial wave from the reggeon amplitudes.
The following considerations should serve as an illustration of 
$t$-channel unitarity relations. 
For the use of unitarity in the $t$-channel it is first of all 
necessary to continue the amplitudes to the 
physical region of the $t$-channel, i.\ e.\ the region of 
positive $t$.
In this region the trajectory function $\beta(\qf^2)$
has a cut starting at $t=-\qf^2 > 4 \lambda^2$. For the discontinuity 
along this cut one finds
\beqn
\disc_t \;\beta(\qf^2) &=&
\disc_t \; \frac{3}{2} g^2 \int \frac{d^2 \kf}{(2 \pi)^3} 
\frac{\qf^2+\lambda^2}
{[\kf^2+\lambda^2][(\qf-\kf)^2+\lambda^2]}
\nonumber \\
&=&
\frac{3}{2} \frac{g^2}{(2\pi)^3}  2 \pi^2 \,i\, 
\frac{t-\lambda^2}{\sqrt{t^2-4 t \lambda^2}}   
\\
&=& 
\frac{3}{2} \frac{g^2}{(2\pi)^3}  (t-\lambda^2)
\int \prod_{i=1}^2 [d^2 \kf_i 2 \pi \delta(\kf_i^2-\lambda^2)]
\delta^{(2)}(\qf-\sum_{i=1}^2\kf_i)
\eeqn
The last expression shows that this discontinuity can be associated 
with two $t$-channel particles going on the mass-shell. 
It is then possible to derive from eqs.\ (\ref{bfklregg}), (\ref{factregg})
the following equation for the discontinuity of the partial wave 
amplitude (continued to the region of positive $t$)
\beqn
\disc_t\;A^I(\omega,t) = \frac{\omega}{(2 \pi)^3}
\int  \prod_{i=1}^2 
\left[d^2 \kf_i 2 \pi \delta(\kf_i^2-\lambda^2)\right]  
\delta^{(2)}(\qf-\sum_{i=1}^2\kf_i)
D^I_2(\omega;\kf_1,\kf_2)D^{I*}_2(\omega;\kf_1,\kf_2) 
\eeqn
The right hand side of this equation can be regarded as an unitarity integral
for the two reggeon amplitude $D_2(\omega)$.
The partial wave amplitude has further discontinuities corresponding
to the thresholds for the production of $n$ particles in the 
$t$-channel ($n\geq3$). The three particle cut starts at $-\qf^2=9\lambda^2$
and the corresponding unitarity condition can be expressed in terms of 
the two reggeon amplitude as
\beqn
\disc_t\;A^I(\omega,t) &=& 
\frac{3}{2}
\frac{g^2}{(2 \pi)^6} 
\int  \prod_{i=1}^3 
\left[d^2 \kf_i 2 \pi \delta(\kf_i^2-\lambda^2)\right]
\delta^{(2)}(\qf-\sum_{i=1}^3\kf_i)
\nonumber \\  
& & \phantom{x}
\cdot 2
\left[
 D^I_2(\kf_1+\kf_2,\kf_3) D^{I*}_2(\kf_1+\kf_2,\kf_3)
-
 c_I D^I_2(\kf_1+\kf_2,\kf_3) D^{I*}_2(\kf_1,\kf_2+\kf_3)
\right]
\label{3cut}
\eeqn 
The first term comes from cutting two virtual lines
from a 
trajectory function whereas the 
second term comes from cutting a real line from the interaction kernel.
The two and three particle unitarity relations have been first discussed
in \cite{klf2}.
Similarly one finds for the four particle unitarity relation 
for $t=-\qf^2 > 16 \lambda^2$
\beqn
\disc_t\;A^I(\omega,t) &=& 
\frac{9}{4}\frac{g^2}{(2\pi)^9}
\int  \prod_{i=1}^4 
\left[d^2 \kf_i \delta(\kf_i^2-\lambda^2)\right]
\delta^{(2)}(\qf-\sum_{i=1}^4\kf_i)
\nonumber \\ & &
\left[
- 2 D^I_2(\kf_1+\kf_2,\kf_3+\kf_4) 
D^{I*}_2(\kf_1+\kf_2,\kf_3+\kf_4)
\right. \nonumber \\ & &\left. 
- 2 c_I^2  D^I_2(\kf_1,\kf_2+\kf_3+\kf_4) 
D^{I*}_2(\kf_1+\kf_2+\kf_3,\kf_4)
\right. \nonumber \\ & &\left. 
+ 2 c_I D^I_2(\kf_1+\kf_2,\kf_3+\kf_4) 
D^{I*}_2(\kf_1+\kf_2+\kf_3,\kf_4)
\right. \nonumber \\  & &\left. 
+ 2 c_I D^I_2(\kf_1+\kf_2+\kf_3,\kf^4) 
D^{I*}_2(\kf_1+\kf_2,\kf_3+\kf_4)
\right]
\label{4cut}
\eeqn
Due to the reggeization the
$n$-particle cuts of the  
partial wave amplitude in the color octet channel have
to vanish for $n>2$. 
This can easily be shown to be true from 
eqs.\ (\ref{3cut}), (\ref{4cut})
since the solution (\ref{follo})
of the integral equation for the color octet 
representation depends only on $\qf^2$ and we have $c_{8_A}=1$.
\subsubsection{Higher order equations}
In this part we turn to the higher order
reggeon amplitudes.
For the generalization of the two reggeon integral equation (\ref{bfklregg})
particle number nonconserving vertices $K_{(2,n)}$ 
of order $g^n$ are needed.
These kernels generalize the BFKL kernel and they are calculated 
in perturbation theory. They have been obtained in \cite{bart-old}
for the massive gauge theory and they 
are given in \cite{bartels-wue} for QCD.
With these kernels the integral equations for the three 
and four reggeon amplitudes read
\beqn
\left[\omega+\sum_{i=1}^3\beta(\kf_i)\right]
D_3(\omega;\{\kf_i\})
\!\!&=&\!\!
D_{3,0}
\!+\!
\left[K_{(2,3)}\otimes D_2(\omega)\right](\{\kf_i\})
\!+\!
\sum_{ 1 \leq i < j \leq 3} 
\!
\left[K^{(i,j)}_{(2,2)}\otimes D_3(\omega)\right](\{\kf_i\})
\label{eqd3}
\\
\left[\omega+\sum_{i=1}^4\beta(\kf_i)\right]
D_4(\omega;\{\kf_i\})
\!\!&=&\!\!
D_{4,0}
\!+\!
\left[K_{(2,4)}\otimes D_2(\omega)\right](\{\kf_i\})
\!+\!
\sum_{1 \leq i < j \leq 3} 
\!
\left[K^{(i,j)}_{(2,3)}\otimes D_3(\omega)\right](\{\kf_i\})
\nonumber \\
& &\phantom{xxxx}
+
\sum_{1 \leq i < j \leq 4} 
\left[K^{(i,j)}_{(2,2)}\otimes D_4(\omega)\right](\{\kf_i\})
\label{eqd4}
\eeqn
The summation indicates that all pairwise interactions have to be 
summed up. The relation $\sum_{i=1}^n \kf_i=\qf$ is implicitly assumed. 
The construction principle for these equations is 
easy to understand. On the left hand side we have the inverse 
$n$-reggeon propagator and on the right hand side the kernels $K_{(2,n)}$
are used to construct all contributions with $n$ reggeons in the 
$t$-channel. It is clear from the construction that the reggeon number
in the $t$-channel never decreases.
In the above equations we have completely suppressed the color structure.
Each reggeon amplitude $D_n(\omega)$ carries $n$ color indices and 
each interaction kernel carries $n+2$ color indices which are not 
displayed. All following considerations are restricted to total color zero
in the $t$-channel, i.\ e.\ the four gluon state is in the color singlet
representation. For this case it has been shown \cite{bartels}
that eqs.\ (\ref{eqd3}) and (\ref{eqd4}) 
are infrared finite, i.\ e.\ all divergencies 
cancel mutually.
\\
It turned out that solutions to the above integral equations can 
be found \cite{bart-old,bartels,bartels-wue}. 
The three reggeon amplitude can be determined exactly
\beqn
D^{a_1a_2a_3}_3(\omega;\{\kf_i\})
=
g \,c_3 \,f^{a_1a_2a_3}
\left[
 D_2(\omega;\kf_1+\kf_2,\kf_3)
-D_2(\omega;\kf_1+\kf_3,\kf_2)
+D_2(\omega;\kf_1,\kf_2+\kf_3) 
\right]
\phantom{x}
\eeqn
Here $g$ is the gauge coupling, $c_3$ is a normalization and 
$f^{a_1a_2a_3}$ are the structure constants of $SU(3)$.
What is observed here is again the reggeization of the gluon in QCD.
Each single reggeon line is in a color octet state with negative 
signature. But since the total system is in a color singlet state 
the remaining pair also has to be in the negative signature color
octet state. Due to the bootstrap property of the BFKL equation
each two reggeon state in the color octet collapses into one reggeon
and the three reggeon system can be reduced to the two reggeon system.    
\\
The four reggeon equation can be solved only partially
\cite{bartels,bartels-wue}. 
The amplitude 
$D_4(\omega)$ can be decomposed into two terms the first of which 
can be reduced again to the two reggeon amplitude.
The second one can be written as a convolution of a two reggeon system
and a four reggeon system with a transition vertex $V_{(2,4)}$. 
We have
\beqn
D_4(\omega;\{\kf_i\})
=D_4^R(\om;\{\kf_i\})+ D_4^I(\om;\{\kf_i\}) 
\label{comp}
\eeqn
with
\beqn
D_4^{R\,,\,a_1a_2a_3a_4}(\om;\{\kf_i\})&=&
g^2 c_4  
\left[-d^{a_2a_1a_3a_4}
\left(D_2(\om;\kf_{12},\kf_{34})+D_2(\om;\kf_{13},\kf_{24})
\right)
-d^{a_1a_2a_3a_4}D_2(\om;\kf_{14},\kf_{23})
\right. \nonumber \\ & & \left. \phantom{xxx} 
+ d^{a_1a_2a_3a_4}\left(D_2(\om;\kf_1,\kf_{234})
+D_2(\om;\kf_4,\kf_{123})\right)
\right. \nonumber \\ & & \left. \phantom{xxx}
+d^{a_2a_1a_3a_4}\left(D_2(\om;\kf_2,\kf_{134})
+D_2(\om;\kf_3,\kf_{124})\right)
\right]
\label{d4r}
\\
D_4^{I\,,\,a_1a_2a_3a_4}(\om;\{\kf_i\})&=&  
\left[G_4(\om)\otimes V_{(2,4)} 
\otimes D_2(\om)\right]^{a_1a_2a_3a_4}(\{\kf_i\})
\label{d4irr}
\eeqn
Here $c_4$ is again a normalization constant and
we have defined $\kf_{ij}=\kf_i+\kf_j$. 
The convolution $\otimes$ now also contains
a summation over color indices. The color tensor $d^{a_1a_2a_3a_4}$ 
is defined as
\beqn
d^{a_1a_2a_3a_4} = Tr[T^{a_1}T^{a_2}T^{a_3}T^{a_4}]
+ Tr[T^{a_4}T^{a_3}T^{a_2}T^{a_1}]
\eeqn
where the $T^a$ are the generators of $SU(3)$ in the fundamental 
(three-dimensional) representation.
There are two new elements which appear in the above solution of the 
four reggeon integral equation. The first is the  
two-to-four reggeon vertex $V_{(2,4)}$. It is an effective vertex
which emerges after summation of different elementary interactions.
It must not be confused with the elementary transition kernels $K_{(2,n)}$. 
An explicit representation and an investigation of some of its properties
will be given in the next section.
The second new element which deserves interest is the fully interacting 
four reggeon amplitude $G_4$. This function is a solution of the 
BKP equation \cite{bart-veryold,bkp-k} for four reggeized gluons.
The BKP equation is of the Bethe Salpeter type with pairwise interactions
specified by the BFKL kernel $K_{(2,2)}$. 
It resums the pairwise 
interactions of four reggeized gluons in the $t$-channel. 
This equation will be formulated and investigated in section
\ref{bkp}.
The important feature of the decomposition in eq.\ (\ref{d4irr})
is the fact that all particle number nonconserving contributions
have been absorbed in the function $V_{(2,4)}$. The problem which remains
to be solved is one with fixed particle number.   
\\
Let us close this introductory part with a remark on the reggeized
contributions $D_3(\om)$ and $D_4^R(\om)$.
One notices that these amplitudes are expressed through terms 
which appear in the unitarity relations (\ref{3cut}) and (\ref{4cut})
for the BFKL amplitude. This means that
if one plugs in $D_3(\om)$ and $D_4^R(\om)$
into $t$-channel unitarity equations then certain parts of the 
result with the correct color quantum numbers should in fact be identified
with the results that follow from the BFKL equation.
Contributions with correct color quantum numbers in this sense are those
for which the group of gluons which momenta appear summed in the 
respective $D_2$-amplitude is in the antisymmetric color octet 
representation, i.\ e.\ their color quantum number coincides with that 
of the gluon.
It follows that one should associate the complete $D_3$-amplitude
with contributions that are already contained in the BFKL equation.
To show this consistently one has to formulate rules to construct
$t$-channel unitarity integrals from the three reggeon amplitude.
These rules have to reproduce the correct weight factors in eq.\
(\ref{3cut}).
In the case of $D_4^R(\om)$, the construction of the $t$-channel 
unitarity integrals with the appropriate rules should yield    
one contribution with the correct color quantum numbers
that corresponds to eq.\ (\ref{4cut}).
The remaining part could be associated with a part of the next-to-leading
order corrections to the BFKL kernel as done by White and Coriano
\cite{whit-cor}. 
From our point of view, however, a consistent set of $t$-channel
unitarity rules has not yet been found.                        
It should be remarked, in any case, that a complete
interpretation of the $n$-reggeon amplitudes 
requires still some work to be done. 
In particular it is important to separate the true corrections that 
arise from the $n$-reggeon amplitudes from terms which are already contained 
in the BFKL equation.
\subsection{The transition vertex}
\label{vertex}
The starting point is the expression for the irreducible part 
of the four reggeon amplitude (\ref{d4irr}). We assume that we 
convolute this amplitude from below with some kind of impact 
factor. This gives then rise to the amplitude
\beqn
{\cal A}_4 =
\prod_{i=1}^2 d^2 \qf_i \prod_{i=1}^4 d^2 \kf_i 
\Phi_{\omega,2}(\qf_1,\qf_2) 
\phantom{
\Phi_{\omega,4}^{a_1a_2a_3a_4}
(\kf_1,\kf_2,\kf_3,\kf_4)
\delta^{(2)}(\sum_{i=1}^2\qf_i-\sum_{i=1}^4 \kf_i)
}
\nonumber \\
V_{(2,4)}^{a_1a_2a_3a_4}(\{\qf_i\};\{\kf_i\}) 
\:
\Phi_{\omega,4}^{a_1a_2a_3a_4}
(\kf_1,\kf_2,\kf_3,\kf_4)
\;\delta^{(2)}(\sum_{i=1}^2\qf_i-\sum_{i=1}^4 \kf_i)
\label{eq41}
\eeqn 
where we have slightly changed our notation and introduced the 
two reggeon amplitude $\Phi_{\omega,2}$ which is a solution of the 
BFKL equation and the four reggeon amplitude $\Phi_{\omega,4}$ 
which is a solution of the 
four-particle BKP equation.
The vertex has the structure \cite{bartels-wue}
\beqn 
V_{(2,4)}^{a_1a_2a_3a_4}(\{\qf_i\};\{\kf_i\})
&=&\phantom{+}
\delta^{a_1a_2}\delta^{a_3a_4} V(\{\qf_i\};\kf_1,\kf_2,\kf_3,\kf_4)
\nonumber \\
& &+
\delta^{a_1a_3}\delta^{a_2a_4} V(\{\qf_i\};\kf_1,\kf_3,\kf_2,\kf_4)
\nonumber \\
& &+
\delta^{a_1a_4}\delta^{a_2a_3} V(\{\qf_i\};\kf_1,\kf_4,\kf_2,\kf_3)
\eeqn
From this representation it is clear that the vertex $V_{(2,4)}$ is
completely symmetric under the interchange of the outgoing gluons
since $V$ will be shown to be symmetric w.\ r.\ t. the exchange
of the first and the second momentum argument, respectively
the third and the fourth momentum argument.   
The momentum space representation of the function $V$ is 
given explicitly in \cite{bartels-wue} and can be represented graphically 
as in fig.\ \ref{fig41}.
Here for each line with momentum $\kf_i$ we have a propagator 
$1/\kf_i^2$ 
and for each vertex we have a factor 
$\qf^2$ with $\qf^2$ denoting the sum of momenta above or below the 
vertex. 
All momenta are integrated according to eq.\ (\ref{eq41})
and for each momentum integration we have a factor $1/(2\pi)^3$.
As indicated all permutations have to be summed finally and one 
has to multiply with the global factor $g^4 \sqrt{2}/8$.
The grouping in fig.\ (\ref{fig41}) is organized in such a way 
that for each group the $\qf_1,\qf_2$-integration in eq.\ (\ref{eq41})
is infrared finite separately.
It should be noted that the two gluon state $\Phi_{\omega,2}$ 
vanishes if either $\qf_1=0$ or $\qf_2=0$.
Besides this infrared finiteness the function $V$ fulfills another 
important consistency condition. If we absorb the propagators of the 
four outgoing gluons in the four gluon state $\Phi_{\omega,4}$ 
the function $V$
vanishes whenever one outgoing momentum or a group of outgoing 
momenta is set equal to zero. In this case cancellation takes 
place between terms of different groups and it is not possible
to find a subgroup for which this property holds separately.  
\begin{figure}[!h]
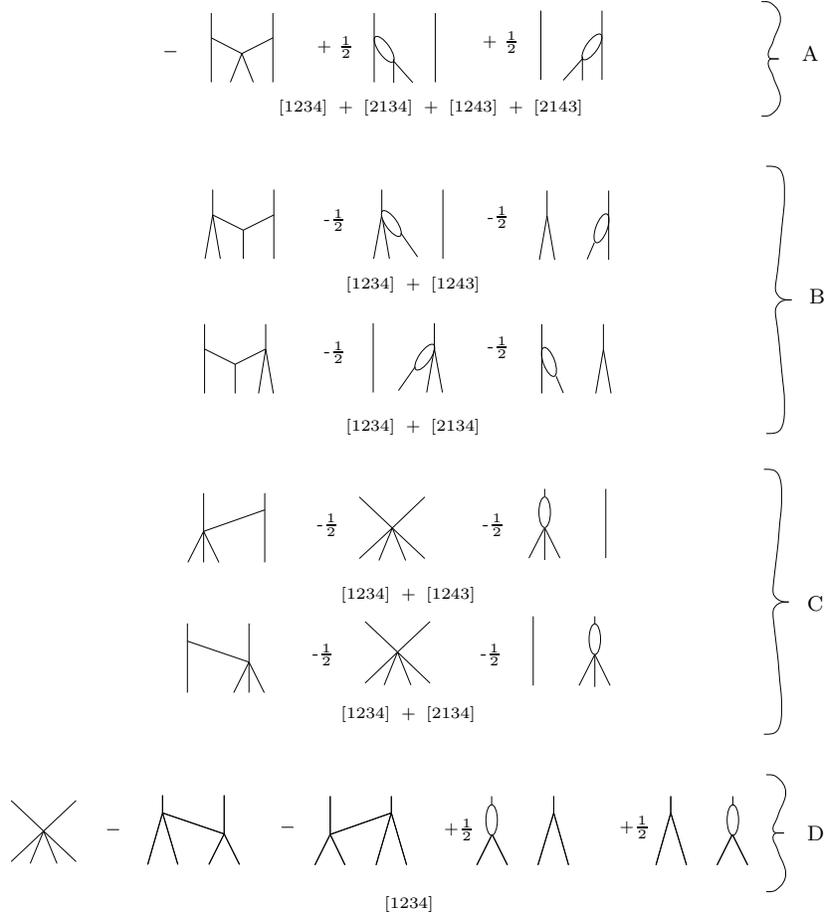

\begin{center}
\input conform4.pstex_t
\end{center}
\caption{
Graphical representation of the transition vertex
function $V$. The notation is explained in the text.
The brackets indicate the permutations.
\label{fig41}
}
\end{figure}
\subsubsection{The operator representation}
In \cite{balipwue} the configuration space representation of the 
above introduced function $V$ was derived and it was proven that this 
function is conformally  invariant, i.\ e.\ it has the same 
symmetry properties as the BFKL kernel.
The techniques employed in \cite{balipwue} 
are identical to the ones that were used in chapter 1 of this work to 
prove the conformal invariance of the BFKL kernel.
In this section we derive an alternative representation for the
transition vertex $V$ which permits a more direct proof of conformal 
symmetry. From this representation we can furthermore address the 
important question of holomorphic separability of the vertex.
\\ \\
Lipatov has shown \cite{holsep} that beyond conformal symmetry the BFKL 
kernel possesses the further property of being holomorphic separable.
He developed a representation of the BFKL kernel in terms of 
pseudodifferential operators. In this representation the kernel
decomposes into a sum of two terms which are complex 
conjugates of each other.  
The first one acts only on the coordinates $\rho_1,\rho_2$, whereas the
second one acts only on the complex conjugate coordinates 
$\rho^*_1,\rho^*_2$. This property is termed holomorphic separability.
Consequently the eigenfunctions of the kernel factorize
into a part depending on $\rho_1,\rho_2$ and a second part depending
on $\rho^*_1,\rho^*_2$. The real eigenvalues come as a sum of 
two mutually 
complex conjugate contributions.
The explicit representation of the kernel reads
\beqn
{\cal K} = \frac{N_c g^2}{8 \pi^2} \left( K + K^* \right)
\eeqn
with 
\beqn
K &=& 
\log\left(\rho^2_{12}\partial_1\right)
+
\log\left(\rho^2_{12}\partial_2\right)
-2 \log \rho_{12} - 2 \psi(1)
\label{holsep1}
\\ 
&=&
\frac{1}{2}\sum_{l=0}^{\infty}
\left(\frac{2 l+1}{l(l+1)+\rho^2_{12}\partial_1\partial_2}
-\frac{2}{l+1}\right)
\label{holsep2}
\eeqn
The second representation is of particular interest since it expresses
the BFKL kernel in terms of the Casimir operators 
$L^2=\rho^2_{12}\partial_1\partial_2,
L^{*\,2}=\rho^{*\,2}_{12}\partial^{*}_{1}\partial^{*}_{2}$ 
(cf.\ eq.\ (\ref{casimir})) of the conformal 
group. 
In this representation the conformal symmetry becomes manifest.
The equivalence of the representation (\ref{holsep2}) 
with the BFKL kernel can be proven 
using eq.\ (\ref{casimir}) and the series representation
(\ref{app3eigen}) of the BFKL eigenvalue $\chi(\nu,n)$.
The way in which the representation (\ref{holsep1}) is obtained will become
clear below.  
In the operator representation the BFKL kernel becomes formally local
\footnote{ ${\cal K}$ is of course only formally local. 
In representation (\ref{holsep1}) appears the 
pseudodifferential operator $\log \rho^2 \partial$ which is defined 
through Fourier transformation. In representation (\ref{holsep2})
an infinite summation over a differential operator is implied. 
This means that mathematically ${\cal K}$ is a nonlocal operator.} 
and can be interpreted as a Hamilton operator of a two-particle system.
The corresponding stationary Schr\"odinger equation
with the energy $\omega$ reads
\beqn
-\omega \;\Phi_{\omega}(\rho_1\rho_2)={\cal K} \Phi_{\omega}(\rho_1\rho_2)
\eeqn
The maximal eigenvalue of ${\cal K}$ corresponds to the ground state 
energy of the two particle system. In section \ref{bkp} we will meet the 
four-particle generalization of this Schr\"odinger equation. 
\\ \\
In the following we derive a representation similar to 
(\ref{holsep1}) for the transition vertex $V$. 
We start with the group D in fig.\ \ref{fig41} which has the form of the 
BFKL kernel. 
Using complex notation and mass regularization 
and omitting the global factors
we have the
following expression for this group
\beqn
{\cal A}_4^D = -
\int d^2 \qf_1 |\qf_1|^2 |\qf_2|^2 
\hat{\Phi}_{\omega,2}(\qf_1,\qf_2)
\prod_{i=1}^4 d^2 \kf_i
\left[ \frac{1}{|\qf_1-\kf_1-\kf_2|^2+\lambda^2}
\left(\frac{(k_1+k_2)(k_3+k_4)^*}{q_1^*q_2}
+\mbox{h.c.} \right) 
\right. \nonumber \\ \left.
- \pi 
\left( 
\log\frac{|\qf_1|^2}{\lambda^2}\delta^{(2)}(\qf_1-\kf_1-\kf_2)
+
\log\frac{|\qf_1|^2}{\lambda^2}\delta^{(2)}(\qf_1-\kf_3-\kf_4)
\right)\right] 
\hat{\Phi}_{\omega,4}(\kf_1,\kf_2,\kf_3,\kf_4)_{
\vert \qf_2=\sum_{i=1}^4\kf_i-\qf_1}
\eeqn
The propagators of the incoming and outgoing gluons are now absorbed 
in the functions $\hat{\Phi}_{\omega,2}$ and $\hat{\Phi}_{\omega,4}$.
The momentum integration of the bubbles has already been performed 
leading to the logarithms.
Now we switch to the configuration space representation by inserting
\beqn
\hat{\Phi}_{\omega,2}(\qf_1,\qf_2)
&=&
\prod_{i=1}^2[d^2\rho_{i'}e^{-i\qf_i\rho_{i'}}]
\Phi_{\omega,2}(\rho_{1'},\rho_{2'})
\\
\hat{\Phi}_{\omega,4}(\kf_1,\kf_2,\kf_3,\kf_4)
&=&
\prod_{i=1}^4[d^2\rho_{i}e^{-i\kf_i\rho_{i}}]
\Phi_{\omega,4}(\rho_{1},\rho_{2},\rho_{3},\rho_{4})
\eeqn
and obtain 
\footnote{A global factor $(4 \pi^2)^4$ from the fourfold $\kf$-integration
which appears here and in the following contributions
has been omitted. 
Furthermore we have omitted a global factor which results 
from the color contraction of the vertex and the four gluon function.}  
upon integration of the momentum variables
\beqn
{\cal A}_4^D=
-\int d^2 \rho_1 d^2 \rho_2 \left[\Delta_1 \Delta_2 
\Phi_{\omega,2}(\rho_{1},\rho_{2})\right]
\left[ \frac{1}{\partial_1\partial_2^*}
\int d^2 \qf \frac{e^{-i\qf\rho_{12}}}{|\qf|^2+\lambda^2}
\partial_1^*\partial_2 \;+\; \mbox{h.c.}
\right. \nonumber \\ \left.
-\pi \left(\log (4|\partial_1|^2)+\log (4|\partial_2|^2) \right)
+ 2 \pi \log \lambda^2
\right]
\Phi_{\omega,4}(\rho_{1},\rho_{1},\rho_{2},\rho_{2})
\eeqn
Here $\Delta$ is the 2-d Laplace operator 
and the product $\Delta_1\Delta_2$ acts only on 
$ \Phi_{\omega,2}$ but not further to the right.
The pseudodifferential operators $\partial^{-1}$ and
$\log |\partial|^2$ have been introduced which are defined 
by the relations
\beqn
\frac{1}{\partial} \, \phi(\rho,\rho^*) &=&
\int d^2 \qf \frac{2}{i\qf^*}
e^{i\qf\rho}\hat{\phi}(\qf)
\\
\log( 4|\partial|^2) \, \phi(\rho,\rho^*) &=&
\int d^2 \qf \log |\qf|^2
e^{i\qf\rho}\hat{\phi}(\qf)
\\
\mbox{where :}\phantom{xxxxxx} 
\phi(\rho,\rho^*) &=&
\int d^2 \qf e^{i\qf\rho}\hat{\phi}(\qf)
\eeqn
An explicit integral operator representation can be found  
in eq.\ (\ref{gammaint}) in the appendix. The $\qf$-integral is also 
calculated in the appendix. It yields
\beqn
\int d^2 \qf \frac{e^{-i\qf\rho_{12}}}{|\qf^2|+\lambda^2}
=  \pi [-\log|\rho_{12}|^2+\log 4 -\log \lambda^2 +2\psi(1)]
\eeqn
and we end with the result 
\beqn
{\cal A}_4^D =  \pi
\int d^2 \rho_1 d^2 \rho_2 \Delta_1 \Delta_2 
\Phi_{\omega,2}(\rho_{1},\rho_{2})
\left[ \frac{1}{\partial_1\partial_2^*}
\log |\rho_{12}|^2 \,\partial_1^*\partial_2 \;+ \;\mbox{h.c.}
\right. \nonumber \\ \left.
+\log \partial_1\partial_1^*+ \log\partial_2\partial_2^* -4 \psi(1)\right]
\Phi_{\omega,4}(\rho_{1},\rho_{1},\rho_2,\rho_2)
\eeqn 
From this representation on can already conclude that
the operator ${\cal K}_D$ acting on $\Phi_{\omega,4}$ 
is holomorphic separable. Simple manipulations give
\beqn
{\cal K} &=& K_D+K_D^*
\\
K_D &=& \partial_1^{-1}\log\rho_{12}\,\partial_1 +
     \partial_2^{-1}\log\rho_{12}\,\partial_2
     +\log \partial_1\partial_2 - 2\psi(1)
\\
    &=&   
\rho_{12}\log (\partial_1\partial_2)\rho_{12}^{-1}
+2\log \rho_{12} -2 \psi(1)
\eeqn
where the last line follows from the identity
$[\rho_{12},\log \partial_1]=-1/\partial_1$.
\\
To arrive at a representation which is manifest conformally invariant
additional steps are required. It is useful to collect the 
noncommuting operators $\partial_1$ and $\rho_{12}^2$ in the form
$\rho_{12}^2\partial_1$ in the argument of the logarithm.
This can be achieved by using the commutator identity given above
and the operator identities \cite{private}
\beqn
\log \partial +\log \rho &=& 
\frac{1}{2}\left[\psi(-\rho\partial)
+\psi(1+\rho\partial)\right]
\label{op1}
\\
\log \rho^2\partial -\log \rho &=& 
\frac{1}{2}\left[\psi(\rho\partial)
+\psi(1-\rho\partial)\right]
\label{op2}
\eeqn
The holomorphic part of $K_D$ can then finally be represented
as
\beqn
K_D =
\log\left(\rho_{12}\partial_1\right)
+
\log\left(\rho_{12}\partial_2\right)
-2 \log \rho_{12} - 2 \psi(1)
\eeqn
Comparison with eq.\ (\ref{holsep1}) shows that $K_D$ is identical
with the BFKL kernel. It follows immediately that the group D 
of the transition vertex $V$ is conformally invariant and 
holomorphic separable.
\\
Now we turn to the group C in fig.\ \ref{fig41}
and start with the permutation $[1234]$
in the first line. 
For the moment we omit the second term (the contact term) 
which will be considered separately below.
The corresponding momentum space expression reads
\beqn
{\cal A}_4^{C_1} =  
\int d^2 \qf_1 d^2 \qf_2 |\qf_1|^2 |\qf_2|^2 
\hat{\Phi}_{\omega,2}(\qf_1,\qf_2)
\prod_{i=1}^4 d^2 \kf_i
\left[ \frac{(\kf_1+\kf_2+\kf_3)^2}
{[|\qf_1-\kf_1-\kf_2-\kf_3|^2+\lambda^2]}
\frac{1}{|\qf_1|^2} 
\right. \nonumber \\ \left.
- \pi 
\log\frac{|\qf_1|^2}{\lambda^2}\delta^{(2)}(\qf_1-\kf_1-\kf_2-\kf_3)
\right] 
\hat{\Phi}_{\omega,4}(\kf_1,\kf_2,\kf_3,\kf_4)_{
\vert \qf_2=\sum_{i=1}^4\kf_i-\qf_1}
\eeqn
Fourier transformation along the lines above then leads to 
\beqn
{\cal A}_4^{C_1} = - \pi 
\int d^2 \rho_1 d^2 \rho_2
\left[\Delta_1 \Delta_2 
\Phi_{\omega,2}(\rho_1,\rho_2)\right]
\left[
\frac{1}{\partial_1\partial_1^*}
\log(|\rho_{12}|^2) \, \partial_1\partial_1^*
+\log \partial_1\partial_1^*
-2\psi(1)
\right]
\Phi_{\omega,4}(\rho_1,\rho_1,\rho_1,\rho_2)
\eeqn
If we associate an operator ${\cal K}_{C_1}$ with these terms 
we find that we can decompose it again into a holomorphic 
and an antiholomorphic part
\beqn
-{\cal K}_{C_1} &=&  K_{C_1} + K^*_{C_1}  
\\
K_{C_1} &=& \partial_1^{-1}\log (\rho_{12}) \,\partial_1 + \log \partial_1
         -\psi(1)
\\
        &=& \log (\rho^2_{12} \partial_1) -  \log \rho_{12}
         -\psi(1)
\eeqn
where the last line follows again after application of the operator
identities (\ref{op1}) and (\ref{op2}).
This representation proves to be useful when investigating the 
conformal properties of the group C.
For the second line of group C and the permutation $[1234]$
we get a similar contribution the holomorphic part of 
which reads (again we omit the 
contact term)
\beqn
K_{C_2}=\log (\rho^2_{12} \partial_2)  -\log \rho_{12}
         -\psi(1)
\eeqn
and this operator and its conjugate act on the function 
$\Phi_{\omega,4}(\rho_1,\rho_2,\rho_2,\rho_2)$. 
\\
Next we consider the contact term which appears with the weight
$-2$ in group C. In group D it was possible to combine this term in a very 
compact way with the remaining contributions. In group C it is not possible
to obtain such a combination and the term has to be treated separately.
In momentum space we have 
\beqn
{\cal A}_4^0 = - 2  
\int d^2 \qf_1 d^2 \qf_2 |\qf_1|^2 |\qf_2|^2 
\hat{\Phi}_{\omega,2}(\qf_1,\qf_2)
\frac{(\qf_1+\qf_2)^2}{|\qf_1|^2|\qf_2|^2}
\prod_{i=1}^4 d^2 \kf_i 
\hat{\Phi}_{\omega,4}(\kf_1,\kf_2,\kf_3,\kf_4)_{
\vert \qf_2=\sum_{i=1}^4\kf_i-\qf_1}
\eeqn
Explicit Fourier transformation leads to 
\beqn
{\cal A}_4^0 = - 2
\int d^2 \rho_1 d^2 \rho_2 \left[\Delta_1 \Delta_2 \Phi_{\omega,2}
(\rho_1,\rho_2)\right]
\int d^2 \rho_0
\left[ - \pi\left(\delta^{(2)}(\rho_{10})+\delta^{(2)}(\rho_{20})\right)
\left( \log|\rho_{12}|^2  \right)
\right. \nonumber \\ \left.
-2 \,\nabla_1 \cdot \nabla_2 \log |\rho_{10}| \log|\rho_{20}|
\right]
\Phi_{\omega,4}(\rho_0,\rho_0,\rho_0,\rho_0)
\eeqn
which can be rewritten as 
\beqn
{\cal A}_4^0 &=&  2 \pi
\int d^2 \rho_1 d^2 \rho_2 \left[\Delta_1 \Delta_2 \Phi_{\omega,2}
(\rho_1,\rho_2)\right]
\log\frac{|\rho_{12}|^2}{\epsilon^2}
\left(
\Phi_{\omega,4}(\rho_1,\rho_1,\rho_1,\rho_1)
+
\Phi_{\omega,4}(\rho_2,\rho_2,\rho_2,\rho_2)
\right)
\nonumber \\
& &
-2 
\int d^2 \rho_1 d^2 \rho_2 \left[\Delta_1 \Delta_2 \Phi_{\omega,2}
(\rho_1,\rho_2)\right]
\int d^2 \rho_0
\frac{|\rho_{12}|^2}{(|\rho_{10}|^2+\epsilon^2)(|\rho_{20}|^2+\epsilon^2)}
\Phi_{\omega,4}(\rho_0,\rho_0,\rho_0,\rho_0)
\label{contact}
\eeqn
Here we have used that 
$\int d^2 \rho_1 \Delta_1 \Phi_{\omega,2}(\rho_1,\rho_2)=0$ 
(and equivalently for $\rho_2$).
This property is valid since we have
required $\Phi_{\omega,2}(\qf_1,\qf_2)$ - 
which is obtained from $\hat{\Phi}_{\omega,2}$ by amputation of the 
propagators - 
to vanish if either $\qf_1=0$ or $\qf_2=0$. 
The fictious parameter $\epsilon$ has 
been introduced to regularize 
the singularity of the $\rho_0$-integration in the second line. 
It is not possible to decompose the second line in eq.\ (\ref{contact})
into a holomorphic and an antiholomorphic part. 
Group C thus contains a contribution which is not holomorphic 
separable.
Now we use the identity 
\beqn
\frac{1}{2} \nabla_0^2 \log^2 \frac{|\rho_{10}|}{|\rho_{20}|}
=
2 \pi \log \frac{\epsilon}{|\rho_{12}|}
\left(
\delta^{(2)}(\rho_{10})+\delta^{(2)}(\rho_{20})
\right)
+
\frac{|\rho_{12}|^2}
{(|\rho_{10}|^2+\epsilon^2)(|\rho_{20}|^2+\epsilon^2)}
\label{distident}
\eeqn 
to combine the two terms in eq.\ (\ref{contact}) in the 
compact expression. 
\beqn
{\cal A}_4^0 &=& -
\int d^2 \rho_1 d^2 \rho_2 \Delta_1 \Delta_2 \Phi_{\omega,2}
(\rho_1,\rho_2)
\int d^2 \rho_0 \Delta_0 \log^2 \frac{|\rho_{10}|}{|\rho_{20}|}
\Phi_{\omega,4}(\rho_0,\rho_0,\rho_0,\rho_0)
\eeqn
This finishes the discussion of the contact term.
\\
Finally we come to the terms in groups A and B.
Consider first group A with the permutation $[1234]$.
We omit the momentum space representation and turn directly to configuration
space where the sum of the three terms can be written as
\beqn
{\cal A}_4^A = 
\int d^2 \rho_1 d^2 \rho_2 \left[\Delta_1 \Delta_2 \Phi_{\omega,2}
(\rho_1,\rho_2)\right]
\int d^2 \rho_0 
\left[ \pi \log\frac{|\rho_{12}|^2}{\epsilon^2}
\left(
\delta^{(2)}(\rho_{10})+\delta^{(2)}(\rho_{20})
\right)
\right. \nonumber \\ \left.
-\frac{|\rho_{12}|^2}
{(|\rho_{10}|^2+\epsilon^2)(|\rho_{20}|^2+\epsilon^2)}
\right]
\Phi_{\omega,4}(\rho_1,\rho_0,\rho_0,\rho_2)
\label{aterm}
\eeqn
We make again use of the identity (\ref{distident}) 
to bring this in the final form
\beqn
{\cal A}_4^A &=& - 
\frac{1}{2}
\int d^2 \rho_1 d^2 \rho_2 \left[\Delta_1 \Delta_2 \Phi_{\omega,2}
(\rho_1,\rho_2)\right]
\int d^2 \rho_0 \log^2 \frac{|\rho_{10}|}{|\rho_{20}|}
\Delta_0
\Phi_{\omega,4}(\rho_1,\rho_0,\rho_0,\rho_2)
\eeqn
where we have used integration by parts
to shift the $\Delta_0$-operator 
to the right. 
All permutations in group A and B can be cast into this form.
Hence we can represent all terms in group A and B  
as well as the remaining contact term from group C through an 
integral operator with the kernel 
\beqn
{\cal K}_A = \log^2 \frac{|\rho_{10}|}{|\rho_{20}|}\Delta_0
\eeqn
As to the holomorphic separability we find that 
this integral operator 
can not be decomposed into a holomorphic and an antiholomorphic part.
Consequently we have to conclude from our analysis that 
the transition vertex $V$ is not holomorphic separable. 
\subsubsection{Conformal invariance}
With the expressions which have been found above 
the conformal invariance of the transition vertex can be shown 
in a rather straightforward way. 
We collect all terms and find for the operator representation
of $V$ acting on the four-gluon function $\Phi_{\omega,4}$
\beqn
&-& 
\frac{1}{2}
\int d^2 \rho_0 \log^2 \frac{|\rho_{10}|}{|\rho_{20}|}
\Delta_0
\left[
 \Phi_{\omega,4}(1,0,0,2)
+\Phi_{\omega,4}(1,0,2,0) 
+\Phi_{\omega,4}(0,1,0,2)
+\Phi_{\omega,4}(0,1,2,0)
\right]
\nonumber \\
&+&
\frac{1}{2}
\int d^2 \rho_0 \log^2 \frac{|\rho_{10}|}{|\rho_{20}|}
\Delta_0
\left[
 \Phi_{\omega,4}(1,1,0,2)
+\Phi_{\omega,4}(1,1,2,0) 
+\Phi_{\omega,4}(0,1,2,2)
+\Phi_{\omega,4}(1,0,2,2)
 \right]
\nonumber \\
&-&
\int d^2 \rho_0 \log^2 \frac{|\rho_{10}|}{|\rho_{20}|}
\Delta_0
\Phi_{\omega,4}(0,0,0,0)
\nonumber \\
&-&\pi
\left[
\log (\rho_{12}^2\partial_1) -\log \rho_{12} - \psi(1) + \mbox{h.c.}
\right] 
\left[
\Phi_{\omega,4}(1,1,1,2) + \Phi_{\omega,4}(1,1,2,1)
\right]
\nonumber \\
&-&\pi
\left[
\log (\rho_{12}^2\partial_2) -\log \rho_{12} - \psi(1) + \mbox{h.c.}
\right] 
\left[
\Phi_{\omega,4}(1,2,2,2) + \Phi_{\omega,4}(2,1,2,2)
\right]
\nonumber \\
&+&  \pi 
\left[
\log (\rho_{12}^2\partial_1)  + \log (\rho_{12}^2\partial_2) 
-2 \log \rho_{12} - 2 \psi(1)
+ \mbox{h.c.}
\right]
\Phi_{\omega,4}(1,1,2,2) 
\label{vertexop}
\eeqn
The amplitude ${\cal A}_4$ is obtained by multiplying with
$ \int d^2\rho_1 d^2\rho_2 [\Delta_1\Delta_2 \Phi_{\omega,2}]$.
The invariance of the above structure under rotation, translation 
and dilatation is obvious.
It remains to investigate the behavior under the inversion 
transformation $\rho_i \to 1/\rho_i, \rho_i^* \to 1/\rho_i^*$.
The last line corresponds to the BFKL kernel and we already know that 
it is invariant. 
For the terms in the fourth and fifth line we find under inversion 
\beqn
\log \rho_{12}^2\partial_1 -\log \rho_{12}\;
\longrightarrow \; \log \rho_{12}^2\partial_1 -\log \rho_{12} 
+ \log \frac{\rho_1}{\rho_2}
\eeqn
and conclude that under inversion we 
produce the original structure and
get the additional terms
\beqn
-  \pi \log \frac{|\rho_1|^2}{|\rho_2|^2}
\left[
\Phi_{\omega,4}(1,1,1,2) + \Phi_{\omega,4}(1,1,2,1) 
-\Phi_{\omega,4}(1,2,2,2) - \Phi_{\omega,4}(2,1,2,2) 
\right]
\label{vertinv1}
\eeqn
For the first three lines the 
transformation properties of the 
operator ${\cal K}_A$
have to be considered
\beqn
{\cal K}_A = \Delta_0 \log^2 \frac{|\rho_{10}|}{|\rho_{20}|}
&\to& |\rho_0|^4 \Delta_0 
\log^2 \frac{|\rho_{10}||\rho_2|}{|\rho_{20}||\rho_1|}
\nonumber \\
&=&\Delta_0 \log^2 \frac{|\rho_{10}|}{|\rho_{20}|}
+ 2 \pi \log \frac{|\rho_2|^2}{|\rho_1|^2}
\left( \delta^{(2)}(\rho_{10}) - \delta^{(2)}(\rho_{20}) \right)
\eeqn
Upon inversion we therefore obtain from the first and second
line after some cancellations 
besides the original structure
the additional terms 
\beqn
\pi \log \frac{|\rho_1|^2}{|\rho_2|^2}
\left[
\Phi_{\omega,4}(1,1,1,2)
-\Phi_{\omega,4}(1,2,2,2)
+\Phi_{\omega,4}(1,1,2,1)
-\Phi_{\omega,4}(2,1,2,2)
\right]
\eeqn
These terms cancel exactly against the additional terms in 
the expression (\ref{vertinv1}). The contact term in the third line 
of expression (\ref{vertexop}) is invariant in itself. The additional terms
vanish here since they depend only on either $\rho_1$ or $\rho_2$
and one can use the property discussed 
after eq.\ (\ref{contact}).
This finishes the proof of conformal invariance of the 
transition vertex function $V$ and in turn of the two-to-four
vertex $V_{2,4}$. 
\\
To summarize, the transition vertex $V$ has three important properties.
It is infrared finite, it vanishes whenever the momentum of 
one of the outgoing gluons vanishes 
(zeropoint property)
and it is symmetric under
conformal transformations. The question arises wether these properties are 
independent or if conformal symmetry follows from the other ones.
The presence of the isolated contact term which is invariant in itself
shows that conformal symmetry is independent from the zeropoint property.
The contact term is not necessary for conformal invariance but it is 
needed to have the zeropoint property.   
It is remarkable that the first term in the first line of 
fig.\ \ref{fig41} representing the new element that appears 
when one considers four gluons can be written in configuration space
- up to regularization -
as $\int d^2 \rho_0 |\rho_{12}|^2/(|\rho_{10}|^2|\rho_{20}|^2)$
(cf.\ eq.\ (\ref{aterm})). 
From this one might have concluded that this term together with its
infrared regulators is conformally invariant separately. The above 
results however show that this conjecture is not true. To render group 
A conformally invariant one inevitably has to add group C which is 
independent from A as far as infrared finiteness is concerned.
This shows that when one starts from momentum space 
the conformal symmetry of the vertex
is a quite nontrivial property which emerges from a very specific
combination of momentum space structures.
It is not possible 
to see from the present analysis if this property persists in higher orders,
i.\ e.\ when the five and six reggeon system     
\cite{ewerz} is considered.
\subsubsection{Projection on conformal three-point functions}
In this part we show that the two-to-four vertex $V_{(2,4)}$ has 
a nice interpretation in terms of the three-point function of the conformal 
field $O_{h\bar{h}}$ which was introduced in eq.\ (\ref{desy1}) 
to describe the bound state of two reggeized gluons. 
To this end we perform the projection of the vertex on three 
conformal eigenfunctions $E^{(\nu)}(\rho_{10},\rho_{20})$. 
This means that we calculate the amplitude ${\cal A}_4$ and use for 
the configuration space representation of the functions $\Phi_{2,\omega}$
and $\Phi_{4,\omega}$ the expressions
\beqn
\Phi_{2,\omega}(\rho_{1'},\rho_{2'}) &=&  
E^{(\nu_c)}(\rho_{1'c},\rho_{2'c})
\nonumber \\
\Phi_{4,\omega}^{a_1a_2a_3a_4}(\rho_1,\rho_2,\rho_3,\rho_4)
&=&  \delta^{a_1a_2}\delta^{a_3a_4}
E^{(\nu_a)\,\ast}(\rho_{1a},\rho_{2a})
\cdot E^{(\nu_b)\,\ast}(\rho_{3b},\rho_{4b})
\eeqn
The color projection corresponds to color singlet states in the 
$(12)$ and $(34)$ subsystems of the outgoing gluons.
The amplitude ${\cal A}_4$ then depends on the three conformal dimensions
and the respective 
bound-state coordinates $\rho_a,\rho_b,\rho_c$.
Let us consider again the transition vertex function $V$ which is displayed
in fig.\ \ref{fig41}. In configuration space each vertex corresponds 
to a configuration space point $\rho$. Consequently for each pair of 
outgoing gluon lines $i,j$ which merge into the same vertex
we have a $\delta$-function $\delta^{(2)}(\rho_{ij})$. If we 
assume $\mbox{Re}(1/2+i\nu_a),\mbox{Re}(1/2+i\nu_a) > 0$
then all contributions vanish in which either the gluons $(12)$ or the 
gluons $(34)$ merge into the same vertex. This is exactly the same 
argument that was used in the discussion of
diffractive vector meson production
in section \ref{sec22} and diffractive $q\bar{q}$ production 
in section \ref{sec23}.
It follows that after projection only the group A of the 
vertex function $V$ in fig.\ \ref{fig41} gives a finite contribution.   
The function $V$ as displayed in fig.\ \ref{fig41} corresponds to the 
color structure $\delta^{a_1a_2}\delta^{a_3a_4}$ of the vertex. 
For the other two color structures, however,  
also B and D give contributions.
\\
Let us first have a closer look at group A. 
The configuration space representation of the permutation $[1234]$
is given in eq.\ (\ref{aterm}).
With the same argument as given just before the logarithmic terms  
corresponding to the bubble diagrams can be shown to vanish. 
In the remaining term the regulator $\epsilon^2$ can be omitted 
since with the above constraint on the conformal dimensions
the $\rho_0$-integration is finite. 
All what is left from the vertex after projection is the first 
term in the first line of group A.     
The integral associated with this term reads
\beqn
{\cal A}_4^A  \!=\! - \!
\int d^2 \!\rho_1 d^2 \rho_2 
\!
\left[\Delta_1 \Delta_2 \!
\left(
\frac{|\rho_{12}|^2}{|\rho_{1c}|^2|\rho_{2c}|^2}
\right)^{\fez-i\nu_c} 
\right]\!
\int \! d^2 \!\rho_0 
\frac{|\rho_{12}|^2}{|\rho_{10}|^2|\rho_{20}|^2}
\!
\left(
\frac{|\rho_{10}|^2}{|\rho_{1a}|^2|\rho_{0a}|^2}
\right)^{\fez+i\nu_a}
\!\!\!\!
\left(
\frac{|\rho_{20}|^2}{|\rho_{0b}|^2|\rho_{2b}|^2}
\right)^{\fez+i\nu_b}
\!\!\!
\label{threeint}
\eeqn
When acting on the conformal eigenfunction the twofold Laplace operator
yields a factor $|\rho_{12}|^{-4}$. 
It is then possible 
with the repeated use of Feynman parameter techniques and the Mellin Barnes
representation for the hypergeometric function  
to perform the integral in (\ref{threeint}). 
An important condition for this to be possible is that 
for every $i \in \{0,1,2\}$
the sum of the 
powers of all monomials $|\rho_{ik}|^2$ which contain the index $i$ is -2.
The result then reads 
\beqn
{\cal A}_4^A =
(|\rho_{ab}|^2)^{-\fez-i\nu_c-i\nu_a-i\nu_b}
(|\rho_{bc}|^2)^{-\fez+i\nu_c+i\nu_a-i\nu_b}
(|\rho_{ac}|^2)^{-\fez+i\nu_c-i\nu_a+i\nu_b}
\cdot
\Omega(\nu_a,\nu_b,\nu_c)
\label{threeex}
\eeqn
The function $\Omega(\nu_a,\nu_b,\nu_c)$ is too complicated to 
display it explicitly here. It can be written as a sum of 
strings of iterated contour integrals over fractions of $\Gamma$-functions.
What is important is that the dependence of ${\cal A}_4$ upon the 
coordinates is completely explicit and turns out to be very simple.
It can be checked easily that (\ref{threeint}) and (\ref{threeex})
transform identically under conformal transformations.  
By comparison with the general form of a three-point function of 
conformal fields (\ref{three}) one finds that ${\cal A}_4^A$ as given 
above can be associated with the three point function of the field 
$O_{h,\bar{h}}$ 
\beqn
<O_{h_a,\bar{h}_a}^*(\rho_a)
 O_{h_b,\bar{h}_b}^*(\rho_b)
 O_{h_c,\bar{h}_c}(\rho_c)> = {\cal A}_4^A + \cdots
\label{susi}
\eeqn
where $h_k=\bar{h}_k$ is given as $h_k=1/2+i\nu_k$ 
for $k=a,b$ and $h_k=1/2-i\nu_k$ for $k=c$
and
the dots indicate further contributions.
We have $h=\bar{h}$ here since we have considered zero conformal spin only.
Without doubt the above relation also generalizes to nonzero 
conformal spin.
\\
It would of course be desirable to obtain more information on the 
function $\Omega(\nu_a,\nu_b,\nu_c)$. It would e.\ g.\ be interesting
to isolate the singularity which implements the conservation of the 
conformal dimensions as it was done for the disconnected vertex 
(which results from the reggeized part of $D_4$)
in section \ref{sec23}.
Another point of interest is the triple pomeron point 
$\nu_a=\nu_b=\nu_c=0$.
Since the coordinate dependence of ${\cal A}_4$ is explicitly known
one could use the points $\rho_a=0,\rho_b=1,\rho_c=\infty$ to calculate 
the function $\Omega$. But even with this simplification applied 
$\Omega(\nu_a,\nu_b,\nu_c)$ presently cannot be expressed in a transparent 
way. As to the point $\nu_a=\nu_b=\nu_c=0$ one could 
of course think of a numerical 
Monte Carlo integration to determine the value $\Omega(0,0,0)$.
\\
So far we have only considered one part of the vertex $V_{(2,4)}$ 
which belongs to  
the color structure $\delta^{a_1a_2}\delta^{a_3a_4}$.
If we turn to the other two contributions then different terms 
give a nonzero result after projection. The momentum space structure
which accompanies the color structure $\delta^{a_1a_3}\delta^{a_2a_4}$
is obtained from the function $V$ displayed in fig.\ \ref{fig41} 
by interchanging the momenta of 
gluons $(2)$ and $(3)$. 
Projecting then in the same 
way as above on the $(12)$ and $(34)$ subsystems one finds that from
group A the permutations $[1234]$ and $[2143]$ give 
a nonzero result which 
corresponds to ${\cal A}_4^A$ given in eq.\ (\ref{threeint}).
As to the group B the situation is more complicated. With the same 
arguments as given above the second term in each line can be shown to 
vanish for all four permutations. The other two terms, however, have 
to be kept. We consider all four permutations together and find 
after projection the following expression 
\beqn
{\cal A}_4^B &=& 
\int \frac{d^2 \rho_1 d^2 \rho_2}{|\rho_{12}|^4}   
\left(
\frac{|\rho_{12}|^2}{|\rho_{1c}|^2|\rho_{2c}|^2}
\right)^{\fez-i\nu_c}
\left(
\frac{|\rho_{12}|^2}{|\rho_{1b}|^2|\rho_{2b}|^2}
\right)^{\fez+i\nu_b}
\nonumber \\
& &
\int d^2 \rho_0 
\left[
\frac{|\rho_{12}|^2}{|\rho_{10}|^2|\rho_{20}|^2}
\theta
\left(\frac{|\rho_{20}|}{|\rho_{12}|}-\epsilon\right)
+\pi \delta^{(2)}(\rho_{20})\log \epsilon^2
\right]
\left(
\frac{|\rho_{10}|^2}{|\rho_{1a}|^2|\rho_{0a}|^2}
\right)^{\fez+i\nu_a}
\nonumber \\
& & + [ (\nu_a,\rho_a) \longleftrightarrow (\nu_b,\rho_b)]
\nonumber \\
& & + [ \rho_2 \longleftrightarrow \rho_1]
\nonumber \\
& & + [ (\nu_a,\rho_a) \longleftrightarrow (\nu_b,\rho_b) \wedge 
\rho_2 \longleftrightarrow \rho_1]
\label{grunz}
\eeqn
Note that here the regularization is needed since the 
conformal dimension regulates only one singularity of the 
$\rho_0$-integration.
We have chosen the $\theta$-function regularization
since this is convenient for the following steps.
One can show that the $\rho_0$-integral in the second line together with 
the one obtained after interchange of 
the coordinates $\rho_1$ and $\rho_2$ (fourth line)
is conformally invariant. The logarithms obtained after 
transformation of the argument of the $\theta$-function cancel between
the two terms due to antisymmetry. This works in the same way as 
in the proof of conformal invariance of the BFKL kernel.
From this we conclude that the conformal three-point function
which appears in the $\rho_0$-integral is an eigenfunction of the 
integral operator associated with this integral. 
For the remaining terms in (\ref{grunz}) the same is true with 
$(\nu_a,\rho_a)$ and $(\nu_b,\rho_b)$ interchanged.
We thus find the interesting feature that after 
interchange of the outgoing gluons $(2)$ and $(3)$ the operator
which is associated with group B decomposes into two terms which are
diagonalized by conformal eigenfunctions in the $(12)$ subsystem 
and $(34)$ subsystem, respectively.     
${\cal A}_4^B$ can therefore be expressed as
\beqn
{\cal A}_4^B \!= \!  
\int \frac{d^2 \rho_1 d^2 \rho_2}{|\rho_{12}|^4}   
\left(
\frac{|\rho_{12}|^2}{|\rho_{1c}|^2|\rho_{2c}|^2}
\right)^{\fez-i\nu_c}\!\!
\left(
\frac{|\rho_{12}|^2}{|\rho_{1b}|^2|\rho_{2b}|^2}
\right)^{\fez+i\nu_b}\!\!
\left(
\frac{|\rho_{12}|^2}{|\rho_{1a}|^2|\rho_{2a}|^2}
\right)^{\fez+i\nu_a}
\left[ \xi(\nu_a)+\xi(\nu_b) \right]
\label{gutzi}
\eeqn
An explicit calculation yields for the function $\xi$
\beqn
\xi(\nu) = 2 \pi \left[2 \psi(1)-\psi(\fez+i\nu)-\psi(\fez-i\nu)\right]
\label{haiti}
\eeqn
which coincides with the eigenvalue of the BFKL kernel
\footnote{We do not use the symbol $\chi$ here 
since this was defined to 
include a coefficient $N_c\alpha_s/(2\pi^2)$.}
.
We have thus found the striking result that
after a permutation of the outgoing gluons the group B from the 
vertex function V acts as the BFKL kernel on 
pairs of the outgoing gluons.
The integration over $\rho_1$ and $\rho_2$ can now be performed with the
result
\beqn
{\cal A}_4^B = 
(|\rho_{ab}|^2)^{-\fez-i\nu_c-i\nu_a-i\nu_b}
(|\rho_{bc}|^2)^{-\fez+i\nu_c+i\nu_a-i\nu_b}
(|\rho_{ac}|^2)^{-\fez+i\nu_c-i\nu_a+i\nu_b}
\nonumber \\
\cdot
\Lambda(\nu_a,\nu_b,\nu_c)
\left[ \xi(\nu_a)+\xi(\nu_b) \right]
\eeqn
We conclude that also this term has the general form of a 
conformal three-point function of the field 
$O_{h,\bar{h}}$. 
It gives a contribution to the terms which are indicated as dots 
in eq.\ (\ref{susi}).
Before we have a closer look at the function $\Lambda(\nu_a,\nu_b,\nu_c)$
we discuss the contribution of the group D.  
This group has the structure of the BFKL kernel and it is clear that
this kernel is diagonal in the conformal eigenfunction 
used for projection from above. 
The $\rho_0$-integration hence yields the familiar BFKL eigenvalue
and the remaining $\rho_1,\rho_2$-integration is exactly the same as 
in eq.\ (\ref{gutzi}) above.  
The result for the group D thus reads 
\beqn
{\cal A}_4^D = -
(|\rho_{ab}|^2)^{-\fez-i\nu_c-i\nu_a-i\nu_b}
(|\rho_{bc}|^2)^{-\fez+i\nu_c+i\nu_a-i\nu_b}
(|\rho_{ac}|^2)^{-\fez+i\nu_c-i\nu_a+i\nu_b}
\;
\Lambda(\nu_a,\nu_b,\nu_c) \, \xi(\nu_c)
\eeqn 
with $\xi(\nu)$ as given above.
As to the function $\Lambda(\nu_a,\nu_b,\nu_c)$ we can make use of the fact
that the coordinate dependence 
of the integral in eq.\ (\ref{gutzi})
is known and therefore use 
the coordinates $\rho_a=0,\rho_b=1,\rho_c=\infty$ to determine 
$\Lambda$. In this way we are able to extract the singularity which expresses
the conservation of conformal dimensions from the function $\Lambda$.
We obtain
\beqn
\Lambda(\nu_a,\nu_b,\nu_c) = 
\frac{\Gamma(\fez+i\nu_a+i\nu_b-i\nu_c)} 
{\Gamma(\fez-i\nu_a-i\nu_b+i\nu_c)} \lambda(\nu_a,\nu_b,\nu_c)
\eeqn
The function $\lambda$ can be expressed as a twofold 
Mellin Barnes integral 
over a rational function of $\Gamma$-functions. The $\Gamma$-function which 
we have extracted expresses in the same way as discussed in section 
\ref{sec23} the conservation of conformal dimensions at the 
two-to-four transition vertex. 
It is easy to see 
that this singularity is associated with 
the region $|\rho_{12}|\sim 0$ of the integral in eq.\ (\ref{gutzi}). 
\\
Finally one has to consider the third part of the vertex $V_{(2,4)}$
which belongs to the color structure $\delta^{a_1a_4}\delta^{a_2a_3}$.
One immediately realizes that 
here the situation is the same as in the case discussed just before,
i.\ e.\ we have to multiply the above results 
for the color structure 
$\delta^{a_1a_3}\delta^{a_2a_4}$ with a factor of 
two.
\\
If we collect all terms we find the following result 
for the $V_{(2,4)}$ vertex projected on conformal 
eigenfunctions in the subsystems of the gluon pairs $(1'2')$, $(12)$
and $(34)$
\beqn
{\cal A}_4 &=&
\frac{1}{64}
\int \prod_{i=1}^2 d^2 \rho_{i'} 
E^{(\nu_c)}(\rho_{1'c}\rho_{2'c})
\int \prod_{i=1}^4 d^2 \rho_i 
V^{a_1a_2a_3a_4}_{(2,4)}
(\{\rho_{i'}\};\{\rho_i\}) \delta^{a_1a_2}\delta^{a_3a_4}
E^{(\nu_a)\,\ast}(\rho_{1a}\rho_{2a})
E^{(\nu_b)\,\ast}(\rho_{3b}\rho_{4b})
\nonumber \\
&=&
(|\rho_{ab}|^2)^{-\fez-i\nu_c-i\nu_a-i\nu_b}
(|\rho_{bc}|^2)^{-\fez+i\nu_c+i\nu_a-i\nu_b}
(|\rho_{ac}|^2)^{-\fez+i\nu_c-i\nu_a+i\nu_b}
\nonumber \\
& &\left[
4 \Omega(\nu_a,\nu_b,\nu_c)
+\frac{1}{4}\left[2 \Omega(\nu_a,\nu_b,\nu_c) +  
\Lambda(\nu_a,\nu_b,\nu_c)
\left( \xi(\nu_a)+\xi(\nu_b)-\xi(\nu_c) \right)
\right]
\right]
\label{gutz2}
\eeqn
The factor $1/4$ results from the two identical contributions
which have a color factor of $1/8$ relative to the first term whose 
color structure matches the color structure of the projection operator. 
The factor $1/64$ is a convenient normalization of the color projector.
\\
The above equation states that one can consistently interpret 
the amplitude ${\cal A}_4$ as a three point function 
of three 
fields of a conformal field theory
with conformal dimensions $\nu_a,\nu_b,\nu_c$  
\beqn
{\cal A}_4 = 
<O_{h_a,\bar{h}_a}^*(\rho_a)
 O_{h_b,\bar{h}_b}^*(\rho_b)
 O_{h_c,\bar{h}_c}(\rho_c)> 
\label{susi2}
\eeqn
In chapter \ref{chap1} we have seen that the results of the BFKL theory 
fit into the framework of a general conformal field theory.
Here we have demonstrated that the first nontrivial unitarity corrections
- the effective two-to-four - vertex also has a natural interpretation
in this setup. This strongly supports the expectation that the concept
of an effective conformal field theory might serve as a powerful guideline in
the investigation of the Regge limit of QCD.
\\
${\cal A}_4$ is not yet the complete three point function 
of the conformal field $O_{h,\bar{h}}$ 
since we have not taken into account
the four-reggeon state which follows below the vertex.
Since this state is constructed from conformally invariant elements
it is however clear that the coordinate dependence 
of the full three point function will be identical with the one in 
eq.\ (\ref{gutz2}). The four reggeon state will therefore only 
modify the coefficient which depends on the conformal dimensions.
We turn to this problem in the next section. 
\subsection{The four gluon state I\,: Introduction, Motivation and Examples}
\label{bkp}
In this section we investigate the second new element which arises 
from the study of subleading unitarity corrections to the 
BFKL pomeron, the fully interacting four reggeon amplitude 
$G_4(\omega)$. This amplitude is determined through the following 
generalized Bethe-Salpeter equation
\beqn
\omega G_4^{\{a_i\},\{a'_{j}\}}
(\om;\{\kf_i\},\{\kf'_j\})
=
\prod_{i'=1}^4 \delta^{a_{i'} a'_{i'}} \delta^{(2)}(\kf_{i'}-\kf'_{i'})
+ \sum_{1\leq i' < j' \leq 4}  
\left({\cal K}_{i'j'}
\otimes
G_4(\omega)
\right)^{\{a_i\},\{a'_{j}\}}(\{\kf_i\},\{\kf'_j\})
\label{bumm0}
\eeqn
where the operator ${\cal K}_{ij}$ describes the two reggeon
interaction, i.\ e.\ it acts on the reggeons with the indices $i$ and 
$j$ and leaves the others unchanged. The interaction kernel
acts on the reggeons $i$ and $j$ according to 
\beqn
({\cal K}_{ij}\otimes \phi)^{a_1a_2}(\kf_1,\kf_2)
= 
\frac{1}{3}
f^{a'_{1}ca_1}f^{ca_2a'_{2}}
\left[
\left(
K^0_{(2,2)}\otimes 
\phi^{a_{1'}a_{2'}}
\right)(\kf_1,\kf_2)
-  \left(\beta(\kf_1)+\beta(\kf_2)\right)
\phi^{a_{1'}a_{2'}}(\kf_1,\kf_2)
\right]
\label{bumm}
\eeqn
where $K^0_{(2,2)}$ is the operator which is
given in eq.\ (\ref{kernregg}) ($c_0=2$), $\beta(\kf)$
is the trajectory function of the gluon and the 
$f^{ijk}$ are the generators 
of $SU(3)$ in the adjoint representation, i.\ e.\ the structure constants.
It is assumed that the total color of the four reggeon system is zero.
The momentum structure of the two reggeon interaction corresponds 
to the one of the BFKL kernel in the color singlet channel. This has the 
immediate consequence that $G_4(\om)$ contains no infrared singularities.  
The coefficient of the interaction kernel depends on the color state
of the two reggeon system. For the moment we keep the color structure 
in the explicit tensor form. The coefficient of the respective color 
state can be obtained by applying
the corresponding projection operators.
\\
If one compares the equations above with the initial equation 
(\ref{eqd4}) for the four reggeon amplitude $D_4(\omega)$
one should note that a nontrivial step has been done to arrive at 
the form (\ref{bumm}) of the interaction kernel. 
The original equation contains one trajectory function for each 
reggeon. 
Above, each trajectory function appears three times, multiplied 
with a nontrivial color structure which is normally associated with 
an $s$-channel gluon-production diagram.  
That eqs.\ (\ref{bumm0}) and (\ref{eqd4})
are indeed equivalent follows from the consideration of the infrared
behavior. It has been shown in \cite{bartels} that eq.\ (\ref{eqd4})
is free from infrared singularities, i.\ e.\ all divergencies cancel
mutually, provided that the four reggeon system is in a color singlet state.
The kernel $K_{(2,2)}^0$ in (\ref{bumm}) including the color structure
is identical with the one that appears in (\ref{eqd4}).
Since the kernel (\ref{bumm}) and in turn eq.\ (\ref{bumm0})
are manifestly infrared finite it follows that the distribution of the 
trajectory functions into two-particle interaction kernels 
\footnote{
This interaction is not a reggeon interaction any more,
since the corresponding equation contains no reggeon propagator.
Nevertheless we will use the term 'four-reggeon system' in the 
following to denote the system which is 
described by eq.\ (\ref{bumm0}).
} 
in (\ref{bumm0}) is equivalent to the representation in terms of 
a reggeon propagator in eq.\ (\ref{eqd4}). If it were not equivalent 
there would appear isolated infrared singularities in (\ref{bumm}).
The advantage of the 
representation above is its manifest infrared finiteness and the 
appearance of the BFKL kernel. The symmetry properties of the latter
facilitate to a large extent
the analysis of the four reggeon system. 
\subsubsection{The onset of unitarity}
The ultimate aim is to solve eq.\ (\ref{bumm0}) or, equivalently,
to diagonalize the four-particle operator which appears on the 
right hand side.
Let us first sketch schematically the significance of such a solution.
Assume that a complete set of eigenfunctions of the four-particle 
operator is known. We denote these eigenfunctions by
$\psi_{\{\alpha\}}(\{\kf_i\})$ where the $\kf_i\,(i=1,2,3,4)$
are the momentum space arguments and the set $\{\alpha\}$
represents a collection of quantum numbers that parameterize the solutions. 
These quantum numbers will be related to the different symmetries of the 
problem. These symmetries include the color symmetry and the conformal
symmetry. This means that there will be at least one quantum number 
related to the irreducible representations of the color group 
$SU(3)$ and quantum numbers related to the representation of the conformal
group, i.\ e.\ conformal dimension and conformal spin.
In addition, quantum numbers belonging to extra symmetries are expected.
The functions $\psi_{\{\alpha\}}(\{\kf_i\})$ diagonalize the four particle
operator
\beqn
\sum_{1\leq i' < j' \leq 4}  
\left(
{\cal K}_{i'j'}
\otimes \psi_{\{\alpha\}}
\right)
(\{\kf_i\}) 
=
\chi_4(\{\alpha\}) \psi_{\{\alpha\}}(\{\kf_i\})
\eeqn
with eigenvalues $\chi_4(\{\alpha\})$. Now the formal solution 
of eq.\ (\ref{bumm0}) reads
\beqn
G_{4}(\omega,\{\kf_i\},\{\kf'_j\}) =
\sum_{\{\alpha\}} 
\frac{\psi_{\{\alpha\}}(\{\kf_i\})\psi^*_{\{\alpha\}}(\{\kf'_j\})}
{\omega-\chi_4(\{\alpha\})}
\eeqn
where the summation symbol stands for summation of discrete and 
integration of continuous quantum numbers. We have implicitly assumed that 
the functions $\psi_{\{\alpha\}}$ satisfy suitable 
completeness and orthonormalization conditions. The color index of 
$G_4(\om)$ is suppressed here and in the following. 
We have assumed that $\psi_{\{\alpha\}}$ is a tensor in color space.
\\
This formal solution can now be inserted into eq.\ (\ref{d4irr})
to obtain the result for the irreducible part of the four reggeon amplitude
$D_4^I(\omega;\{\kf_i\})$. We use the fact that the solution 
for the two reggeon amplitude $D_2(\om)$ is known to be given by the BFKL 
pomeron and represent $D_4^I(\om)$ in the form
\beqn
D_4^I(\omega;\{\kf_i\}) 
=
\int_{-\infty}^{+\infty}\frac{d \nu}{2 \pi}
\frac{D_{2,0}(\nu)}{\omega-\chi_2(\nu)} \sum_{\{\alpha\}}
V_{(2,4)}(\nu,\{\alpha\}) \frac{1}{\omega-\chi_4(\{\alpha\})}
\psi_{\{\alpha\}}(\{\kf_i\})
\label{supp}
\eeqn
The function $\chi_2(\nu)$ is the BFKL eigenvalue, formerly denoted 
by $\chi(\nu)$. 
$D_{2,0}(\nu)$ is an impact factor in the Mellin representation
which couples the two reggeon state
to colorless external particles. 
The function $V_{(2,4)}(\nu,\{\alpha\})$
arises from integrating the two to four transition vertex 
studied in the last section with the eigenfunctions 
$E^{(\nu)}$ of the BFKL equation from above 
and the eigenfunctions $\psi_{\{\alpha\}}$ of the four reggeon
system from below. 
The momentum space integration which is involved here is expected to 
lead to nontrivial correlations between the quantum number 
$\nu$ associated with the two reggeon state and the quantum numbers 
of the four reggeon system which are related to the conformal symmetry.
As an example of such a correlation we have already discussed the 
conservation of conformal dimensions at the triple pomeron vertex at an
earlier stage of this work \ref{sec23}.
\\
Now, for simplicity, we construct the partial wave amplitude
$A(\om,t)$ 
from $D_4^I(\om)$ by convoluting eq.\ (\ref{supp}) from below 
with a phenomenological impact factor $B$ that depends on the four 
gluon momenta and contracts the color indices
\footnote{In this way one would proceed when this formalism is applied
to inclusive deep inelastic scattering.}.
Having constructed the partial wave  
amplitude one gets the energy dependence of the amplitude by performing 
the $\omega$-integration. Taking the residues at $\om=\chi_2(\nu)$ and 
$\om=\chi_4(\{\alpha\})$ one obtains a sum of two terms
\footnote
{The energy $s$ is measured in units of a mass scale of the process.}
\beqn
A(s,t)=i s 
\int \frac{d\nu}{2\pi}\sum_{\{\alpha\}}
D_{2,0}(\nu) V_{(2,4)}(\nu,\{\alpha\}) B(\{\alpha\})
\left[
\frac{s^{\chi_2(\nu)}}
{\chi_2(\nu)-\chi_4(\{\alpha\})}
+\frac{s^{\chi_4(\{\alpha\})}}
{\chi_4(\{\alpha\})-\chi_2(\nu)}
\right]
\label{supp2}
\eeqn
For large energy $s$ the first term will be dominated by the region 
near $\nu=0$ and one obtains the usual energy behavior of the BFKL 
amplitude $A(s,t) \sim s^{N_c\alpha_s/\pi \,4 \log2}$ with a 
coefficient proportional to $\alpha_s^2$. Recall that  
the transition vertex $V_{(2,4)}$ is proportional to $\alpha_s^2$.
With regard to the enrgy dependence
the first contribution is therefore on the same level as the part of 
the amplitude that arises from the reggeizing part 
$D_4^R$ of the four reggeon amplitude. 
The latter can be expressed in terms of the 
two reggeon amplitude $D_2$ with a global coefficient of order $\alpha_s$
(cf.\ eq.\ (\ref{d4r})).
Consequently it has the same energy
dependence as the first term 
in (\ref{supp2}). Both terms therefore lead to a higher order
renormalization of the coefficient function of the 
BFKL amplitude. 
The second contribution in eq.\ (\ref{supp})
is of greater significance. If we assume that 
$\hat{\chi}_4=\chi_4(\{\hat{\alpha}\})=\max_{\{\alpha\}}\chi_4(\{\alpha\})$
exists then this term introduces a new energy dependence
in next-to-leading order in $\alpha_s$
\beqn
A(s,t) = i s^{1+\hat{\chi}_4}
\int \frac{d\nu}{2\pi} 
\frac{D_{2,0}(\nu) V_{(2,4)}(\nu,\{\hat{\alpha}\}) B(\{\hat{\alpha}\})}
{\chi_4(\{\hat{\alpha}\})-\chi_2(\nu))}
\label{suff}
\eeqn
This contribution now defines the first unitarity correction to the 
BFKL pomeron. Let us assume that the relation 
$ \hat{\chi}_4 > \hat{\chi}_2 =N_c\alpha_s/\pi \,4 \log2\,$ holds and 
furthermore that the coefficient of the term (\ref{suff})
turns out to be negative.  
By taking into account the contribution (\ref{suff}) and the contribution
that is obtained from the two reggeon amplitude $D_2(\om)$
we find the following structure of the scattering amplitude
\beqn
A(s,t) = i s \left[c_0 s^{\hat{\chi}_2} - \alpha_s^2 c_1 s^{\hat{\chi}_4}\right] 
\eeqn
This demonstrates the effect of the unitarity corrections which are 
encoded in the second term. If the energy $s$ starts to increase 
the first term dominates and the amplitude will rise as a function of $s$
with the power given by the BFKL exponent. With a further increase of the 
energy the $O(\alpha_s^2)$-suppressed second term will gradually become
important. 
This term will damp the rise of the amplitude and ultimately stop it.
Clearly, for even larger energies also this term violates unitarity
since it induces a power dependence on $s$.
This shows the necessity to take into account
all higher order reggeon amplitudes.
For each $n$-reggeon amplitude a reasoning along the lines 
above leads to a new contribution to the amplitude which has
a coefficient 
\footnote{For some $n$ the coefficient might be zero. See e.\ g.\ $n=3$
above and the discussion in \cite{ewerz}.} 
of order $\alpha_s^{\frac{n}{2}}$ and depends on the 
energy according to the power behavior $s^{\hat{\chi}_n}$.
Each of these terms individually violates unitarity. Since, however, 
the formalism from which these results are obtained was designed from the 
beginning to provide a unitary scattering amplitude, unitarity 
is expected to be restored when all terms are summed up.
As an example \cite{korchemsky1} of a series with such properties consider   
\beqn
\sum_{n=0}^{\infty} \frac{(-\alpha_s)^n}{n+1} s^{(n+1)\lambda}
=\frac{1}{\alpha_s}\log(1+\alpha_s s^{\lambda})
\approx \frac{\lambda}{\alpha_s}\log s 
\eeqn
where the last expression is obtained in the limit of asymptotically 
large $s$.
This example is intended to demonstrate that from a summation of powers 
a logarithm can be obtained.
It is clear from this discussion that the solution of the 
four reggeon problem constitutes only one step on the path 
towards an unitary amplitude. From a phenomenological point of 
view, however, it is of great significance. Without it, 
it is impossible to decide at which energies the first 
modifications of the leading-log behavior become important.
With the knowledge of at least the four reggeon correction
a much more consistent comparison than presently can be made
of experimental data with 
the results of the BFKL theory.
It would e.\ g.\ be highly interesting to see if a 
satisfactory description of the 
behavior of the deep inelastic structure function $F_2$ as measured in 
the small $x$ regime of the HERA collider is possible by adding the 
four reggeon contribution to the BFKL pomeron.
From eq.\ (\ref{suff}) we see that the transition vertex $V_{(2,4)}$
plays an important role in this context. It contributes to the 
coefficient function which ultimately determines the sign 
of the correction. 
\subsubsection{Correlation functions, short distance limit 
and the BFKL amplitude as an example}
In the last subsection the new energy dependence 
which is obtained from the four reggeon state was considered.
The focus of the present section is the transverse momentum or,
equivalently, the coordinate dependence of the reggeon amplitudes.
This will ultimately lead us to an approach towards the four reggeon
spectrum $\chi_4(\{\alpha\})$ which is pursued in the subsequent sections
of this chapter.
\\
In eq.\ (\ref{susi2}) of the preceding section we have already discussed 
the three point correlation function of the field $O_{h\bar{h}}(\rho)$
that has been introduced to describe a compound state of two 
reggeons (reggeized gluons). At that point the calculation of the 
coefficient of this correlation function remained incomplete,
but with the knowledge of the four reggeon amplitude $G_4(\om)$
this function can be completely determined within the approximation 
scheme that takes into account up to four reggeons.
It is clear in which way the three point function is calculated.
First one has to convolute the function $G_4(\om)$ with the 
transition vertex $V_{(2,4)}$. Then the vertex is convoluted 
from above with the conformal eigenfunction $E^{(\nu_c)}$.
Simultaneously the four gluons at the lower end of $G_4(\om)$ 
are grouped into two pairs, e.\ g.\ $(12)$ and $(34)$, and the corresponding 
legs are convoluted with conformal eigenfunctions 
$E^{(\nu_a)*}$ and $E^{(\nu_b)*}$. In the configuration space 
representation the result 
\footnote{
Note that we do not include a two reggeon state above the vertex
in our definition of of the three point function. Likewise we do not include
a full two reggeon amplitude in the subsystems $(12)$ and $(34)$ at the 
lower end.}
reads
\beqn
<\!O_{h_c\bar{h}_c}(\rho_c)O^*_{h_a\bar{h}_a}(\rho_a)
O^*_{h_b\bar{h}_b}(\rho_b)\!>
= \!\left[
\prod_{{i,j \in\{a,b,c\}\choose i \neq j}} 
(|\rho_{ij}|^2)^{-h_i-h_j+2\sum_k h_k}
\right]
\sum_{\{\alpha\}}
\frac{V_{(2,4)}(\nu_c,\{\alpha\}) g_4(\{\alpha\},\nu_a,\nu_b)}
{\omega-\chi_4(\{\alpha\})}
\eeqn
The relation between the conformal weights and the conformal 
dimension of the field 
$O_{h\bar{h}}(\rho)$
is in each case $h_i=\bar{h}_i=1/2-i\nu_i$ since we have
restricted ourselves to zero conformal spin. 
For the conjugate field $O^*_{h\bar{h}}(\rho)$
we have $h_i=\bar{h}_i=1/2+i\nu_i$.
The residue functions $V(\nu_c,\{\alpha\})$ and 
$g_4(\{\alpha\},\nu_a,\nu_b)$ arise from the convolution of the 
conformal eigenfunctions with the momentum space dependent 
part of the vertex function $V_{(2,4)}$ and the amplitude $G_4(\om)$.
\\
From the function $G_4(\om)$ also the four point function of the 
field $O_{h\bar{h}}(\rho)$ can be constructed. This can be done by grouping
the four gluons at the upper end of $G_4(\om)$ into two pairs
$(1'2')$ and $(3'4')$ and convoluting these pairs with conformal 
eigenfunctions $E^{(\nu_c)}$ and $E^{(\nu_d)}$.
The lower end of $G_4(\om)$ is convoluted in the same way as for the 
three point function.
This defines the four point correlation function
\beqn
<O_{h_c\bar{h}_c}(\rho_c)O_{h_d\bar{h}_d}(\rho_d)
O^*_{h_a\bar{h}_a}(\rho_a)O^*_{h_b\bar{h}_b}(\rho_b)>
= 
\left[
\prod_{{i,j \in\{a,b,c,d\}\choose i\neq j}} 
(|\rho_{ij}|^2)^{-h_i-h_j+\frac{1}{3}\sum_k h_k}
\right]
\nonumber \\
\phantom{
O_{h_c\bar{h}_c}(\rho_c)O_{h_d\bar{h}_d}(\rho_d)
}
\cdot
\sum_{\{\alpha\}}
\frac{g^*_4(\{\alpha\},\nu_c,\nu_d)g_4(\{\alpha\},\nu_a,\nu_b)}
{\omega-\chi_4(\{\alpha\})}
\,\Psi(\{\nu\},\{\alpha\};\eta,\eta^*)
\eeqn
Here $\eta$ denotes the anharmonic ratio 
$(\rho_{ab}\rho_{cd})/(\rho_{ac}\rho_{bd})$
and $\Psi$ is an unknown function. 
The construction of the four point and the three point correlation
function is illustrated in fig.\ \ref{fig42}.
For both cases the coordinate dependence 
follows entirely from 
conformal covariance. 
\begin{figure}[!h]
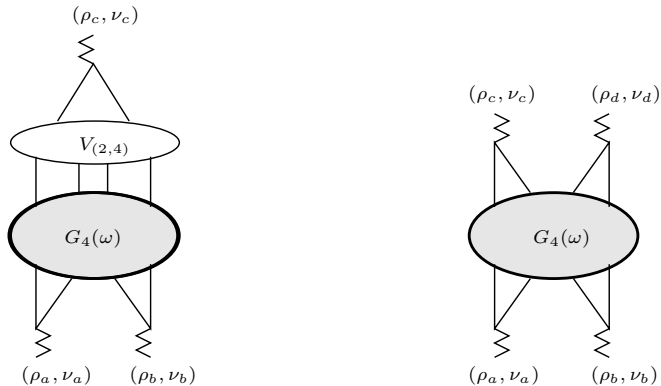

\begin{center}
\input corr1.pstex_t
\end{center}
\caption{Construction of the three point and the four point function
of the field $O_{h\bar{h}}(\rho)$ from the four reggeon amplitude
$G_4(\om)$ and the vertex $V_{(2,4)}$. 
\label{fig42}
}
\end{figure}
\\
Having found the four point function of the field $O_{h\bar{h}}$
one is able to derive the operator algebra of this field. In general 
the operator algebra \cite{dot,bpz,wilson} is contained in expansions 
of the form
\footnote{
This is a shorthand notation. The operator $O_k$ represents a 
whole series of operators, belonging to a conformal family \cite{bpz}.
The term which is denoted in eq.\ (\ref{hipp}) represents the 
most singular term (in $|\rho_{ij}|$) of each conformal family.} 
\beqn
O_i(\rho_i)O_j(\rho_j) = \sum_{k}\frac{C^{ij}_k}
{|\rho_{ij}|^{\Delta_i+\Delta_j-\Delta_k}}O_k(\rho_i)
\label{hipp}
\eeqn
which holds when the operators are inserted into some correlation function.
The quantities $\Delta_i$ are usually called the scaling dimensions
\footnote{We assume here that the conformal spin of the operators
is zero. For the general case compare \cite{bpz}.}
of the fields $O_i$. It is clear from the transformation properties of 
(\ref{hipp}) that they coincide with the conformal dimensions, i.\ e.\
they are related to the conformal weights $h,\bar{h}$ through
$\Delta_i=h_i+\bar{h}_i$.
The summation on the rhs extends over an infinite set of operators.  
From the point of view of conformal field theory \cite{bpz,dot}, 
the scaling dimensions
$\Delta$ and the expansion coefficients (fusion coefficients) $C^{ij}_k$
contain the whole information on the theory. The knowledge of all anomalous
dimensions and all fusion coefficients in (\ref{hipp}) is hence equivalent
to having the explicit representation of the four point function. 
In particular the operator algebra contains the information on the 
spectrum $\chi_4({\alpha})$
\\
The operator algebra in (\ref{hipp}) can be determined from the 
four point function by performing short distance limits w.\ r.\ t.\ 
the coordinates. 
\\
Since the relation between the operator algebra and the spectrum 
is the central idea of what follows below we will illustrate 
it using the BFKL four point function as an example.
\\
Let us consider the four point function of the field
$\phi_{00}$ associated with the 
reggeized gluon. This function was calculated explicitly in 
section \ref{secconf} and we found the result (cf.\ eq.\ (\ref{new}))
\beqn
<\phi_{00}(\rho_1)\phi_{00}(\rho_2)\phi_{00}(\rho_{1'})\phi_{00}(\rho_{2'})>
= \int_{-\infty}^{+\infty} d \nu \frac{1}{\omega-\chi_2(\nu)}
\phantom{xxxxxxxxxxxxxxxxxxxxxxxxxxxx}
\nonumber \\
\cdot
\left[c(\nu) \eta^{\fez+i\nu}\eta^{* \,\fez+i\nu}
F(\nu,\eta)F(\nu,\eta^*)
+c(-\nu)\eta^{\fez-i\nu}\eta^{* \,\fez-i\nu}
F(-\nu,\eta)F(-\nu,\eta^*)
\right] 
\label{zupp}
\eeqn 
where $\eta=(\rho_{12} \rho_{1'2'})/(\rho_{11'}\rho_{22'})$ and 
$F(\nu,\eta)$ is a shorthand notation for 
$_2F_1(1/2+i\nu,1/2+i\nu,1+2i\nu;\eta)$.
We examine the short-distance
limit $|\rho_{12}| \ll |\rho_{1'2'}| \sim |\rho_{12'}|
\sim |\rho_{1'2'}|$. In this limit the hypergeometric function is 
regular and we replace it with unity. More interesting are
the factors $|\rho_{12}|^{i\nu},|\rho_{12}|^{-i\nu}$.
From these factors we obtain an expansion in $|\rho_{12}|$ 
by shifting the contour of the $\nu$-integration
past the singularities of the $\nu$-dependent coefficient.
We restrict our considerations to the singularities  
generated by the factor $1/(\om-\chi_2(\nu))$.
The singularities of this term are obtained from the solution of the equation
\beqn
\omega =\chi_2(\nu)
\label{sipp}
\eeqn
If we assume $\alpha_s \ll \omega$, i.\ e.\ we are not in the limit of 
asymptotically large energies, solutions of this equation can be found
at
\beqn
\nu_k = \pm \frac{1}{i}
\left(\frac{1}{2}+k-\frac{N_c\alpha_s}{\pi\omega}\right) 
+ O(\alpha_s^2) \;,\; k \in \Bbb{N}
\label{ano}
\eeqn
To obtain the series representation of the four point function we
close the $\nu$-contour in the lower half plane for the first term in 
eq.\ (\ref{zupp}) and in the upper half plane for the second term.
Retaining only terms of order $\alpha_s$ in the 
solution (\ref{ano}) the resulting series reads
\beqn
<\phi(\rho_1)\phi(\rho_2)\phi(\rho_{1'})\phi(\rho_{2'})>
\;\overset{|\rho_{12}|\to 0}{=}\;
\sum_{k=0}^{\infty}
C_k(\alpha_s) |\rho_{12}|^{2+2k-2\frac{N_c\alpha_s}{\pi\omega}}
\left(\frac{|\rho_{1'2'}|}{|\rho_{11'}||\rho_{22'}|}\right)^{
2+2k-2\frac{N_c\alpha_s}{\pi\omega}}
\eeqn 
where the constants $C_k(\alpha_s)$ are calculated from the coefficients 
$c(\nu)$ and the function $F$.
Note that in the limit
$|\rho_{12}| \to 0$ 
the coefficients of the expansion have the approximate form
\beqn
\left(\frac{|\rho_{1'2'}|}{|\rho_{11'}||\rho_{22'}|}\right)^{
2+2k-2\frac{N_c\alpha_s}{\pi\omega}}
= <\phi(\rho_{1'})\phi(\rho_{2'})O_{h_k\bar{h}_k}(\rho_1)>
\left[1+O(|\rho_{12}|)\right]
\label{identi}
\eeqn
with $h_k=\bar{h}_k=1+k-N_c\alpha_s/(\pi\om)$.
We want to interpret the result in the following way. We have 
realized an expansion of 
the product $\phi(\rho_1)\phi(\rho_2)$ in a series 
\footnote{We have made two approximations. In eq.\ (\ref{identi}) we have
neglected higher orders and we have not considered the contributions 
that arise from higher orders of
the expansion of the $F$-function in eq.\ (\ref{zupp}).
Both neglected groups give contributions with an integer power
of $|\rho_{12}|$. One might identify these terms with contributions 
from secondary fields \cite{bpz}.}
over the 
operators $O_{h_k\bar{h}_k}$ where the the scaling dimension $\nu_k$
is determined by eq.\ (\ref{sipp}). 
Note that the $O(\alpha_s)$ deviation of the scaling dimensions
from a half integer value is degenerate, i.\ e.\ identical
for all $k$. This degeneracy is lifted in higher orders.  
As discussed in section \ref{secconf} the scaling dimensions 
of the field $\phi$ are zero.  
\\
It is clear that the scaling dimensions of the operators 
$O_{h\bar{h}}$ which appear in the expansion of the operator product of
the fields $\phi$
are intimately related to the spectrum $\chi_2(\nu)$ of the 
bound state of two $\phi$ fields. The scaling dimensions 
in order $O(\alpha_s)$ correspond to the residues of the singularities
of $\chi_2(\nu)$. In other words, if we knew all scaling dimensions
we could determine the spectrum $\chi_2(\nu)$ by means of a dispersion 
relation.  
In the case discussed here, one subtraction constant is needed
to formulate the dispersion relation. To determine the subtraction constant
one could use e.\ g.\ an higher order (in $\alpha_s$) coefficient of a 
scaling dimension, which corresponds mathematically to a nonsingular 
contribution in the Laurent expansion of the function $\chi_2(\nu)$ around 
one of its singularities.
\\
We have considered here one particular short distance limit.
Other limits such as e.\ g.\ $|\rho_{12'}| \to 0$ can be easily performed 
after using analytic continuations for the hypergeometric function 
$F$ in eq.\ (\ref{zupp}) \cite{private}. In the BPZ \cite{bpz} approach to 
two dimensional conformal field theory associativity of the
operator algebra is an important principle.
It leads to relations between the coefficients $C^{ij}_k$
of different expansions. It would be an interesting subject for future 
work to study these relations for the BFKL amplitude.
This could lead to a better understanding of the connection between 
the two dimensional BFKL theory and the general theory with 
infinite dimensional conformal symmetry of \cite{bpz}.
\\
Using the BFKL amplitude as an example we have demonstrated that the 
knowledge of the operator algebra suffices to determine the spectrum
of the compound state of two reggeized gluons. 
The transfer of this connection
to the four reggeon system defines the basis 
of our approach to the spectrum of the four reggeon problem.
We will develop a method with which 
the operator algebra of this system can be constructed.
More precisely, we show how to calculate the scaling dimensions 
of the relevant operators up to 
order $\alpha_s$. Equipped with this information one could in the next step 
try to determine the function $\chi_4$ and in turn the energy dependence
of the four reggeon system. 
\subsubsection{The twist expansion}
In the preceding section we have discussed short-distance expansions in 
configuration space where conformal covariance dictates to a large extent
the coordinate dependence of the amplitudes. For our actual computation we 
find it easier to use the momentum space representation.
The short-distance expansion discussed above then transforms into 
the twist expansion.
\\
Consider again the BFKL amplitude, which we now convolute with
impact factors $T(Q^2)$, $B(\Lambda^2)$ of colorless particles which 
are characterized by mass scales $Q^2$ and $\Lambda^2$.
For simplicity we consider the forward direction ($t=0$) and obtain 
for the partial wave amplitude
\beqn
\Phi_{\om}(Q^2,\Lambda^2)
=\frac{1}{\sqrt{Q^2\Lambda^2}}\int \frac{d \nu}{2 \pi}  
\left(\frac{Q^2}{\Lambda^2}\right)^{i\nu}
\frac{T(\nu)B(\nu)}{\om-\chi_2(\nu)}
\eeqn
with the Mellin transformed impact factors $T(\nu)$ and $B(\nu)$.
We assume $Q^2 \gg \Lambda^2$ and expand the amplitude in
powers of the ratio $\Lambda^2/Q^2$. This works exactly as described above
by closing the $\nu$-contour in the upper half plane and taking the residues
of the poles of $(\om-\chi_2(\nu))^{-1}$. This leads to
\beqn
\Phi_{\om}(Q^2,\Lambda^2)
= \frac{1}{\om Q^2} 
\sum_{k=1}^{\infty}T_kB_k \,\gamma_k(\omega)\cdot 
\left(\frac{Q^2}{\Lambda^2}\right)^{1-k+\gamma_k(\om)}
\label{twist}
\eeqn
where the $\gamma_k(\om)$ are obtained from the equation $\om=\chi_2(\nu)$
as the solutions in $\nu$.
To leading order we have again $\gamma_k(\om)=N_c\alpha_s/(\pi\om)$
for all $k$. 
Eq.\ (\ref{twist}) constitutes the twist expansion of the 
partial wave amplitude, i.\ e.\ the expansion in inverse powers of 
a large momentum scale. The quantities $\gamma_k$ correspond
to the anomalous dimensions of operators 
constructed from the quark and gluon fields
with twist $1+k$. 
In the present 
case only gluonic operators appear since quark operators do not 
contribute in the leading logarithmic approximation. 
Comparison with the results from the last subsection shows that the 
anomalous dimensions are related to the scaling dimensions of 
the fields $O_{h_k\bar{h}_k}$. The expansion (\ref{twist}) reveals that 
the BFKL equation contains contributions from an infinite number of 
operators with 
different twist. In leading order all anomalous dimensions 
which exactly correspond to the residues of the $\chi_2$ function
are degenerate.
\\
So far we have constructed the twist (or short-distance) expansion by 
making use of our complete knowledge of the BFKL four point function.
In the four reggeon case the latter is of course not known  
and we have to develop an alternative method to obtain the expansion and 
especially the anomalous dimensions. 
The basic idea is the following.
The solutions of the reggeon integral equations can be represented
as an infinite series of nested momentum loop integrations.
The latter arise from iteratively applying interaction kernels to some initial 
condition.
For each individual loop integration a twist expansion can be performed 
by expanding the kernel of the corresponding loop in the ratio of the inner 
and the outer momentum scale (or alternatively the inverse ratio).
The assertion is then that the $O(\alpha_s)$-contribution to the anomalous 
dimension of the operator with twist $l$ can be found by retaining in each 
single loop the term with the corresponding twist and iterating this 
procedure through the whole succession of nested loops.
Let us demonstrate this for the example of the partial
wave amplitude $\Phi_{\om}(Q^2,\Lambda^2)$ calculated before. 
\\ 
We can represent $\Phi_{\om}$ in the form
\beqn
\Phi_{\om}(Q^2,\Lambda^2) 
=
B(\Lambda^2)\otimes \sum_{n=0}^{\infty}
\left(\frac{{\cal K}_{\mbox{\tiny BFKL}}}{\omega}\right)^n
\otimes \frac{1}{\om}T(Q^2)
\eeqn
where the symbol $\otimes$ denotes momentum integration and 
${\cal K}_{\mbox{\tiny BFKL}}$ is the forward BFKL kernel including the 
gluon trajectory function $\beta(\kf^2)$.
Using a Mellin representation for $T(Q^2)$ the first momentum loop
has the form
\beqn
\int_0^{\infty}d^2\kf{'}
\left[
\frac{N_c\alpha_s}{\pi}
\frac{\kf^2}{\kf{'}^2|\kf^2-\kf{'}^2|}
-\delta(\kf{'}^2-\kf^2)\beta(\kf^2)
\right]\left(\frac{\kf{'}^2}{Q^2}\right)^{\fez-i\nu}
T(\nu)
\eeqn
Now we expand the kernel in the ratio  $\kf^2/\kf{'}^2$, assuming 
$\kf^2 \ll \kf{'}^2$ (strong ordering).
The trajectory function gives no contribution since in the strong 
ordering limit the $\delta$-function yields zero.
From the first term we get
\beqn
\frac{N_c\alpha_s}{\pi}\sum_{k=1}^{\infty}
\int_{\kf^2}^{\infty} \frac{d\kf{'}^2}{\kf{'}^2}
\left(\frac{\kf^2}{\kf{'}^2}\right)^k
\left(\frac{\kf{'}^2}{Q^2}\right)^{\fez-i\nu}
T(\nu)
=
\frac{N_c\alpha_s}{\pi}
\left(\frac{\kf^2}{Q^2}\right)^{\fez-i\nu}
\sum_{k=1}^{\infty}
\frac{1}{i\nu+k-1/2}
T(\nu)
\eeqn
We take the $l$-th term in the sum and repeat this in each following loop.
The last convolution yields 
$\int d\kf^2/\kf^4 B(\Lambda^2)(\kf^2/Q^2)^{\fez-i\nu}
=(Q^2\Lambda^2)^{-\fez}(Q^2/\Lambda^2)^{i\nu}B(\nu)$ 
and the corresponding
contribution to the partial wave amplitude reads 
\beqn
\Phi_{\om}^{(l)}&=& 
\frac{1}{\sqrt{Q^2\Lambda^2}}
\int \frac{d\nu}{2\pi}B(\nu)T(\nu)\frac{1}{\om}
\left(\frac{Q^2}{\Lambda^2}\right)^{i\nu}
\sum_{n=0}^{\infty}
\left(\frac{\frac{N_c\alpha_s}{\pi\om}}
{i\nu+l-1/2}\right)^n
\\
&=&\frac{1}{\sqrt{Q^2\Lambda^2}}
\int \frac{d\nu}{2\pi}B(\nu)T(\nu)\frac{1}{\om}
\left(\frac{Q^2}{\Lambda^2}\right)^{i\nu}
\frac{1}{1-\frac{N_c\alpha_s}{\pi\om}\frac{1}{i\nu+l-1/2}}
\label{helgoland}
\eeqn
By taking the residue at $i\nu=1/2+N_c\alpha_s/(\pi\om)-l$ 
we obtain exactly the $l$-th term 
in the series in eq.\ (\ref{twist}) with the anomalous dimension 
determined to $O(\alpha_s)$.
Repeating this procedure for each $l$ we generate the whole twist 
expansion. In fact, we were able to obtain the whole set of anomalous 
dimensions to order $\alpha_s$
without knowledge of the explicit form of the BFKL eigenvalue 
$\chi_2(\nu)$.  
\\
The same procedure is in the next section applied to the four gluon state.
\subsection{The four gluon state II\,: Beginning of the Twist Expansion}
\label{sec33}
We now turn back to the Bethe Salpeter equation (\ref{bumm0})
which defines the central problem of the present chapter.
As anticipated in the preceding section our aim is the twist 
expansion of the four reggeon amplitude 
$G_4^{\{a_i\},\{a'_j\}}(\om;\{\kf_i\},\{\kf'_j\})$.
In order to analyze the singularity structure of the Mellin transformed
amplitude we follow the approach of Bartels \cite{bartels}
by reordering the interaction kernels ${\cal K}_{ij}$ according to a method
developed by Faddeev \cite{faddeev}.
The idea is to rearrange the infinite summation which is obtained 
by iteratively solving eq.\ (\ref{bumm0}). One combines the four 
reggeons into two pairs, e.\ g.\ $(12)$ and $(34)$.
Then the interactions of each pair are resummed separately. 
After this infinite resummation one puts the first cross pair interaction,
e.\ g.\ of reggeons $(13)$. Then one performs the infinite resummation
of interactions in the corresponding coupling scheme $(13)$ and $(24)$.
This procedure is infinitely iterated.
The virtue of this method lies in the fact that the infinite resummation
of interactions in one fixed coupling scheme can be compactly expressed
by the BFKL amplitude.
\\
Before we realize this idea in terms of mathematical expressions 
we change the initial condition of the equation (\ref{bumm0}).
To get rid of the $\delta$-functions in momentum and color space it is 
convenient to convolute the primed arguments of $G_4(\om)$ with an impact
factor. The natural impact factor for the four reggeon state is of course
the transition vertex $V_{(2,4)}^{\{a'_j\}}(\{\qf_i\},\{\kf'_j\})$ and from 
now on we take $G_4^{\{a_i\}}(\om;\{\kf_i\})$
to be the four reggeon amplitude with the vertex as the initial condition.
\\
Next we have to discuss the color structure.
Since the color structure of the interaction kernel (\ref{bumm})
can be expressed through projection operators 
of irreducible representations of $SU(3)$
according to
\cite{bartels,lev-rysk-sh}
\beqn
f^{a'_1ca_1}f^{ca_2a'_2} = 3P_1 + \frac{3}{2}P_{8_A}+\frac{3}{2}P_{8_S}
-P_{27}
\eeqn 
it is sensible to decompose the amplitude $G_4^{\{a_i\}}(\om;\{\kf_i\})$
w.\ r.\ t.\ the color states of the two pairs of reggeized gluons.
The total color of the system is zero, hence the color states of the two 
pairs are identical. After the color decomposition 
has been performed the iteration 
of pairwise interactions within a fixed coupling scheme is simple.
The infinite resummation leads to the BFKL amplitude with the eigenvalue
$\chi_2^{(i)}(\nu,n)$ where the index $i=1,2,3,4$ labels the respective 
color state $(1,8_A,8_S,27)$ and $\chi_2^{(i)}$ is obtained  
by multiplying $\chi_2^{(i)}$ in eq.\ (\ref{chi}) with 
$a_i=1/N_c\cdot(N_c,N_c/2,N_c/2,-1)$.  
It is clear that when the coupling scheme is changed, i.\ e.\ when the first
cross-pair interaction occurs, the basis in color space has to be changed.
This basis change is implemented by a 
$4\times4$ matrix $\Lambda$ (cf.\ eq.\ (\ref{elba})).
\\ 
When the Faddeev method is applied the four reggeon amplitude 
decomposes into three parts
\beqn
G_4^{(i)}(\om;\{\kf_j\})\!=\!
G_{4\,(12)}^{(i)}(\om;(\kf_1,\kf_2);(\kf_3,\kf_4))
+
G_{4\,(13)}^{(i)}(\om;(\kf_1,\kf_3);(\kf_2,\kf_4))
+
G_{4\,(14)}^{(i)}(\om;(\kf_1,\kf_4);(\kf_2,\kf_3))
\label{foehr}
\eeqn
where the index $i$ indicates that from now on 
$G_4(\om;\{\kf_i\})$ is a vector in the four dimensional color state space.
The function $G_{4\,(jk)}^{(i)}(\om;(\kf_j,\kf_k),(\kf_l,\kf_m))$ 
contains all contributions in which the last pairwise interaction was 
in the $(jk),(lm)$ coupling scheme.
Now the 12 amplitudes on the rhs of eq.\ (\ref{foehr})
can be collected into a 12 dimensional vector ${\bf G}_4$
with elements
\beqn
{\bf G}_4=\mbox{diag}
\left(G_{4\,(12)}^{(i)}((1,2);(3,4)),G_{4\,(13)}^{(i)}((1,3);(2,4)),
G_{4\,(14)}^{(i)}((1,4);(2,3))\right)
\eeqn
The solution of the Bethe Salpeter equation can now be expressed in a 
very compact form as
\beqn
{\bf G}_4= {\boldsymbol \Phi}_{\om} \otimes \sum_{n=0}^{\infty}
\left({\bf S}(\Lambda){\boldsymbol \Phi}_{\om}\right)^n \otimes V
+ {\boldsymbol 1} V
\label{elba}
\eeqn
On the right hand side ${\boldsymbol \Phi}_{\om}$ and ${\bf S}(\Lambda)$
are $12\times 12$ matrices and $V$ is a 12-dimensional vector
(${\boldsymbol 1}$ is the unit matrix).
The first four elements of $V$ are obtained by applying the 
projection operators $P_1,P_{8_A},P_{8_S},P_{27}$
in the coupling scheme $(12),(34)$ to the transition vertex 
$V_{(2,4)}^{\{a_i\}}$, the next four elements by applying the projection 
operators in the coupling scheme $(13),(24)$ and the last four by projecting
in coupling scheme $(14),(23)$.
The explicit expressions for the projection operators can be 
found in \cite{bartels}.
The matrix ${\bf S}$ describes the transition between 
different coupling schemes. 
It has the form \cite{bartels-rysk}
\beqn
{\bf S} = 
\begin{pmatrix}
 0           & \Lambda               & P \Lambda P \\
 \Lambda^T   &   0                   & \Lambda^T P \Lambda P \\
 P \Lambda P^T & P \Lambda^T P \Lambda & 0 
\end{pmatrix}
\,,\,
\Lambda=
\begin{pmatrix}
\frac{1}{8} & \frac{1}{\sqrt{8}}&\frac{1}{\sqrt{8}}& \sqrt{3}\frac{3}{8} \\
\frac{1}{\sqrt{8}} & 1/2 & 1/2 & -\frac{1}{2}\sqrt{\frac{3}{2}} \\
\frac{1}{\sqrt{8}}&\frac{1}{2}&-\frac{3}{10}&\frac{3}{10}\sqrt{\frac{3}{2}} \\
\sqrt{3}\frac{3}{8}& -\frac{1}{2}\sqrt{\frac{3}{2}}& 
\frac{3}{10}\sqrt{\frac{3}{2}}& \frac{7}{40}
\end{pmatrix}
\,,\, P = \mbox{diag}(1,-1,1,1)
\eeqn
The matrix ${\boldsymbol \Phi}_{\om}$ represents the infinite summation 
of interactions within a fixed coupling scheme.
It can be written in the form
\beqn
{\boldsymbol \Phi}_{\om}
=\mbox{diag}\left(
\Phi^{(i)}_{\om}(\kf_1,\kf_2;\kf_3,\kf_4),
\Phi^{(i)}_{\om}(\kf_1,\kf_3;\kf_2,\kf_4),
\Phi^{(i)}_{\om}(\kf_1,\kf_4;\kf_2,\kf_3)
\right)
\label{cypern}
\eeqn
with
\beqn
\Phi^{(i)}_{\om}(\kf_1,\kf_2;\kf_3,\kf_4)&=&
\sum_{n_1,n_2\!=\!-\infty}^{+\infty}
\int_{-\infty}^{+\infty}\frac{d\nu_1}{2 \pi}
\int_{-\infty}^{+\infty}\frac{d\nu_2}{2 \pi}
\left[
\frac{\om}{\om-\chi^{(i)}_2(\nu_1,n_1)-\chi^{(i)}_2(\nu_2,n_2)}
-1\right]
\nonumber \\& &
E^{(\nu_1,n_1)\ast}(\kf'_1,\kf'_2)E^{(\nu_1,n_1)}(\kf_1,\kf_2)
E^{(\nu_2,n_2)\ast}(\kf'_3,\kf'_4)E^{(\nu_2,n_2)}(\kf_3,\kf_4)
\prod_{i=1}^4 \kf_i^2
\label{aruba1}
\eeqn
This expression corresponds to the product of two 
BFKL amplitudes with the respective color coefficient $a_i$
in nonforward direction in the momentum space 
representation. The functions $E^{(\nu_j,n_j)}$ are the momentum 
space eigenfunctions which have been calculated in section \ref{sec13}.
One has to subtract unity in the first line of (\ref{aruba})
since at least one interaction has to take place for the 
coupling scheme to be defined. The no-interaction contribution is 
restored by the extra term in (\ref{elba}).
The symbol $\otimes$ in eq.\ (\ref{elba}) represents momentum 
integration w.\ r.\ t.\ the measure $d^2\kf/(2\pi)^3$.
\\
The first line of eq.\ (\ref{aruba1}) can be regarded as a two reggeon
propagator with the trajectory function of the reggeon given by 
$\chi^{(i)}_2(\nu_j,n_j)$. 
When the Faddeev procedure is applied to the four reggeized gluon system,
new reggeons 
\footnote{Note that by now we generalize the terminology.
Up to now the term 'reggeon' was exclusively reserved for reggeized gluons.
From now on the term refers to bound states of two reggeized gluons 
with different color states.}
appear naturally as bound states of pairs of reggeized gluons.
From this point of view the right hand side of 
(\ref{elba}) can be regarded as 
the iteration of two intermediate reggeon states.
There appear four different types of reggeons corresponding to the four 
color projections and each state is threefold degenerate corresponding to 
the three different coupling schemes. In total we therefore have 
a 12 dimensional reggeon state space. Each intermediate state
is described by a matrix ${\boldsymbol \Phi}_{\om}$ as given in 
eq.\ (\ref{cypern}).  
\\
It is clear that the momentum space integration associated with 
the change of the coupling scheme constitutes the essential complication
in the iteration of the two reggeon state. This momentum space integration
can be cast into an effective reggeon-reggeon interaction vertex.
To this end we contract the lower part of the two reggeon function
$\Phi_{\om}^{(i)}(\kf_1,\kf_2;\kf_3,\kf_4)$ 
(these are by definition the factors with the unprimed momentum arguments 
in eq.\ (\ref{aruba1})) with the upper part of the two reggeon
function $\Phi_{\om}^{(i)}(\lf_1,\lf_3;\lf_2,\lf_4)$ belonging to a different 
coupling scheme. This defines the vertex function
\beqn
{\bs \Theta}^{(\nu_1,n_1;\nu_2,n_2)}_{(\nu'_1,n'_1;\nu'_2,n'_2)} (\qf,\qf') = 
\int \frac{d^2 \kf}{(2\pi)^3}
E^{(\nu_1,n_1)}(\kf,\qf-\kf) E^{(\nu_2,n_2)}(\qf'-\kf,-\qf-\qf'+\kf) 
\phantom{xxxxxxxxxxxxxxxxxx}
\nonumber \\
E^{(\nu'_1,n'_1)*}(\kf,\qf'-\kf) E^{(\nu'_2,n'_2)*}(\qf-\kf,-\qf-\qf'+\kf) 
\kf^2(\qf-\kf)^2(\qf'-\kf)^2(-\qf-\qf'+\kf)^2 \;\;
\label{juist}
\eeqn 
We have introduced momenta $\qf=\kf_1+\kf_2,\qf'=\lf'_1+\lf'_2$.
Note that the total system is in the forward direction 
$\sum_{i=1}^4\kf_i=0$. A graphical representation for the vertex function is 
given in fig.\ \ref{fig51}.
\begin{figure}[!h]
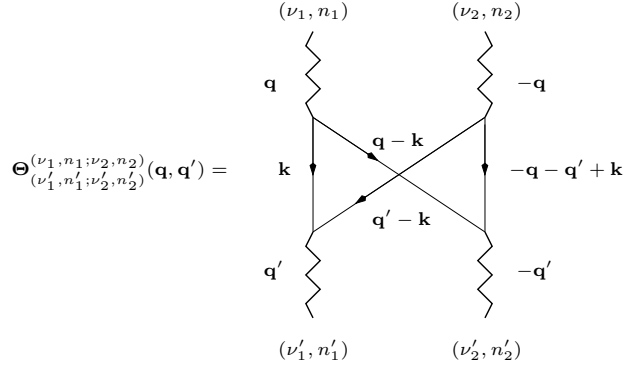

\begin{center}
\input pom_vert.pstex_t
\end{center}
\caption{
Definition of the effective reggeon interaction vertex
${\bs \Theta}$.
\label{fig51}
}
\end{figure}
Having defined the vertex function ${\bs \Theta}$ we can 
simplify the structure of our above expressions essentially.
The matrix ${\bs \Phi}_{\om}$ is replaced by
\beqn
{\bs \Phi}(\om)= \mbox{diag}
\left(\Phi^{(i)}(\om), \Phi^{(i)}(\om),\Phi^{(i)}(\om)\right)
\eeqn
with 
\beqn
\Phi^{(i)}(\om) = 
\sum_{n_1,n_2=-\infty}^{+\infty}
\int_{-\infty}^{+\infty}\frac{d\nu_1}{2 \pi}
\int_{-\infty}^{+\infty}\frac{d\nu_2}{2 \pi}
\left[
\frac{\om}{\om-\chi^{(i)}_2(\nu_1,n_1)-\chi^{(i)}_2(\nu_2,n_2)}
-1\right]
\eeqn
and equation (\ref{elba}) can be expressed as
\beqn
{\bf G}_4= {\bf E}\,{\boldsymbol \Phi}(\om) \otimes \sum_{n=0}^{\infty}
\left({\bf S}(\Lambda)
\,{\bs \Theta}\,{\boldsymbol \Phi}(\om)\right)^n \otimes \hat{V}
+ {\boldsymbol 1} V
\label{elba2}
\eeqn
Note that ${\bs \Theta}$ is only a coefficient, not a matrix. 
The virtue of eq.\ (\ref{elba2}) is that the matrix ${\boldsymbol \Phi}(\om)$
is simplified and the momentum space integrations have been plugged into
the coefficient ${\bs \Theta}$.
The integration over the momenta $\qf$ which can be regarded as 
loop momenta of the intermediate two reggeon states are still contained 
in the convolution $\otimes$. The diagonal matrix ${\bf E}$ at the end of the 
iteration in (\ref{elba2}) restores the momentum dependence of 
the lowest two reggeon states. It can be written as
\beqn
{\bf E} \!= \!\mbox{diag}\!\left(\!
\hat{E}^{(\!\nu_1\!,n_1\!)}(\kf_1\!,\!\kf_2) 
\hat{E}^{(\!\nu_2\!,n_2\!)}(\kf_3\!,\!\kf_4),
\!\hat{E}^{(\!\nu_1\!,n_1\!)}(\kf_1\!,\!\kf_3) 
\hat{E}^{(\!\nu_2\!,n_2\!)}(\kf_2\!,\!\kf_4),
\!\hat{E}^{(\!\nu_1\!,n_1\!)}(\kf_1\!,\!\kf_4) 
\hat{E}^{(\!\nu_2\!,n_2\!)}(\kf_2\!,\!\kf_3)
\!
\right)
\eeqn
where each element represents a $4\times4$ diagonal matrix 
and the hat on $E^{(\nu,n)}$ means that the lower gluon propagators 
have been amputated, i.\ e.\ 
$\hat{E}^{(\nu,n)}(\kf_1,\kf_2) = \kf_1^2\kf_2^2 E^{(\nu,n)}(\kf_1,\kf_2)$.
The vertex $V$ has been changed to $\hat{V}$ in (\ref{elba2}) to indicate 
that it is now not only color but also momentum projected, i.\ e.\ 
convoluted with a product of eigenfunctions 
$E^{(\nu'_1,n'_1)*}E^{(\nu'_2,n'_2)*}$ according to the respective 
coupling scheme.
A graphical representation of the structure of the four reggeized 
gluon amplitude can be found in fig.\ \ref{fig52}.
\\
Having applied the Faddeev procedure we arrive at a ladder-like 
structure (fig.\ \ref{fig52})
for the four reggeized gluon state. This is quite welcome since
ladder cells are easy to iterate. The elements of the ladder,
however, are quite complicated.
Each cell consists of a two reggeon intermediate state and an 
effective reggeon interaction vertex. This effective vertex is defined 
through an integral over four conformal eigenfunctions $E^{(\nu,n)}$. 
In order to obtain the twist expansion of $G_4(\omega)$ with the method 
introduced for the BFKL amplitude in the preceding section, we have to 
isolate all sources of singularities of the amplitude.
For each ladder cell we have an integration over a loop momentum $\qf$.
This loop momentum can be identified with the momentum that is carried
by the two reggeons in that cell. Furthermore we have for each 
two reggeon intermediate state the integration over the conformal 
dimensions $\nu_j$ associated with the two reggeons.
As a first important result it turns out that the integration 
over the loop momentum $\qf$ generates a singularity which implies 
the conservation of the sum of the conformal dimensions of a ladder cell
according to the conservation law $\fez+i\nu_1+\fez+i\nu_2=\fez+i\nu$.
This is the first conserved quantity which characterizes the 
four gluon system. It is clear that this quantity
is related to the conformal dimension of the operator which 
will be associated with the four gluon state. In particular, we expect the 
singularities of $G_4(\om)$ in the $\nu$-plane to be associated with 
the residues of the spectrum $\chi_4(\nu,\{\alpha'\})$. 
Keeping $\nu$ fixed, we still have one conformal dimension for each 
loop which has to be integrated. This integration leads to nontrivial 
singularities of $\Phi^{(i)}(\om)$ in the $\nu$-plane. 
The main complication arises from the 
effective interaction vertex ${\bs \Theta}$. 
Since this vertex has an inner momentum loop
we will also get a singularity from the vertex which has to be combined
with the singularities from the two reggeon intermediate state.
\\
In the next subsection we will first discuss the singularities arising
from the integration of the conformal dimension of a single ladder cell.
The singularities of the vertex will be considered in the next but one
subsection.   
\begin{figure}[!h]
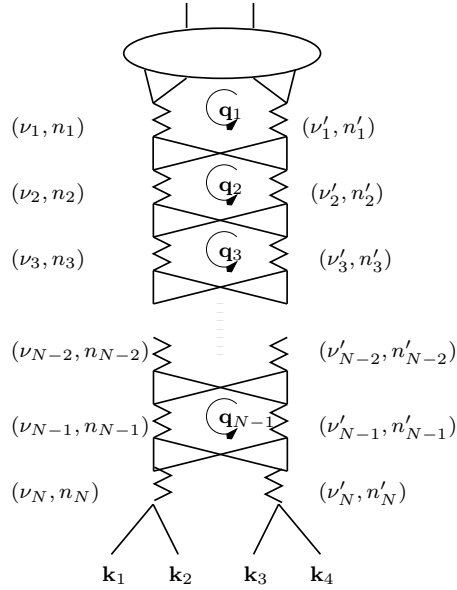

\begin{center}
\input iteration.pstex_t
\end{center}
\caption{
\label{fig52}
Schematic representation of the four reggeized gluon amplitude 
after the Faddeev procedure has been applied. The bubble on top 
denotes the vertex, the vertical curly lines represent the 
reggeons and the crossed box stands for the effective vertex.
Indicated are the conformal dimensions and spins of the reggeons 
as well as the loop momenta.
}
\end{figure}
\subsubsection{The two reggeon intermediate state}
For each two reggeon intermediate state we have a function
\beqn
\Phi^{(i)}(\omega,\nu)
=
\sum_{n_1,n_2=-\infty}^{+\infty}
\int_{-\infty}^{+\infty} \frac{d\nu_1}{2\pi}
\frac{\om}{\omega
-\chi_2^{(i)}(\nu_1,n_1)
-\chi_2^{(i)}(\nu_2,n_2)}
\label{blub}
\eeqn
where the index $i$ labels the color of the exchanged reggeon 
and $\nu_1,\nu_2$ are related to the conserved conformal weight 
$1/2+i\nu$ of the four (reggeized) gluon
system by $1/2+i\nu_1+1/2+i\nu_2=1/2+i\nu$. 
The additive constant $(-1)$ has been omitted 
from the two reggeon propagator
since it contains no singularities.
We aim at extracting the singularities of $\Phi^{(i)}(\omega,\nu)$
(in $\nu$ at fixed $\omega$) from the integral.
Since the contour is closed (on $S^2$), end-point singularities
are absent and one only expects pinch singularities.
Let us illustrate the origin of these pinch singularities.
For simplicity we take $n_1=n_2=0$.
It is convenient to introduce the new variables $\mu_i=1/2+i\nu_i$  
because the conservation law for the conformal weights then reduces 
to the simple form $\mu_1+\mu_2=\mu$.
The $\mu_1$-contour then runs along the imaginary axis and intersects
the real axis to the right of $\mu_1=0$ and to the left of 
$\mu_1=\mu$. We require from the beginning $0 < \mbox{Re} (\mu) < 1$
and we are interested in the singularities in $\mu$ which are located in  
the left half plane $\mbox{Re} (\mu) \leq 0$. It is clear how the pinching 
occurs.
If $\mu$ approaches zero, the $\mu_1$-contour is pinched between the 
singularities of $\chi_2^{(i)}(\mu_1)$ at $\mu_1=0$ and 
 $\chi_2^{(i)}(\mu-\mu_1)$ at $\mu_1=\mu$.
Expanding around these singularities and taking the pole in $\mu_1$ one finds
\beqn
\Phi^{(i)}(\omega,\mu)
=
\int_{{\cal C}} \frac{d\mu_1}{2\pi i}
\frac{\om}{\omega
-a_i\al/(\pi\mu_1)
-a_i\al/(\pi(\mu-\mu_1))}
=
\frac{a_i\al}{\pi\om}\frac{1}{\sqrt{1-\frac{4 a_i\al}{\pi\om\mu}}}
\label{quad}
\eeqn
It follows that for each two reggeon intermediate state we have a 
square-root singularity (a cut) starting at $\mu=a_i\al/(\pi\om)$.
For the case of color zero ($i=1$) this singularity is known as
the two-pomeron cut.
If the effective reggeon-reggeon vertex were absent, this cut would 
already determine the anomalous dimension. 
This was the case in the analysis of Gribov, Levin and Ryskin \cite{glr}
in which only noninteracting multi-reggeon states (fan diagrams)
appeared as the corrections to the BFKL amplitude. 
\\
If we shift $\mu$ further to the left additional singularities 
appear due to the pinching of different poles of the functions
$\chi_2^{(i)}(\mu_1)$ and $\chi_2^{(i)}(\mu-\mu_1)$.
In the following we want to classify these singularities. 
To this end we make use 
of the Landau equations which are usually exploited to extract singularities
of contour integrals (see e.\ g.\ \cite{itzykson}).
Keeping still $n_1=n_2=0$ we define the function
\beqn
\phi_{(0,0)}^{(i)}(\mu_1,\mu)
=
\om -\chi_2^{(i)}(\mu_1,0)-\chi_2^{(i)}(\mu-\mu_1,0)
\eeqn
The Landau equations then read
\beqn
\phi_{(0,0)}^{(i)}(\mu_1,\mu) &=& 0 
\\
\frac{\partial}{\partial \mu_1}
\phi_{(0,0)}^{(i)}(\mu_1,\mu) &=& 0
\eeqn
We use the familiar form of $\chi_2^{(i)}$ and write the 
second equation explicitly
\beqn
\psi'(\mu_1)-\psi'(1-\mu_1)-\psi'(\mu-\mu_1)+\psi'(1-\mu+\mu_1) = 0
\eeqn
where the prime denotes differentiation w.\ r.\ t.\ $\mu_1$.
This equation is solved if we set $\mu_1=\mu/2$. When inserting this 
solution into the first Landau equation we obtain
\beqn
\omega=2\chi_2^{(i)}(\frac{\mu}{2},0)
\label{landau}
\eeqn 
and through this equation we define a function 
$\omega_{(0,0)}^{(i)}= 2\chi_2^{(i)}(\frac{\mu}{2},0)$
where the upper index of $\om$ refers to the color state and the lower
ones to the conformal spins of the two intermediate reggeons.
Now we express $\Phi^{(i)}(\om,\mu)$ in term of this function.
We expand the function $\phi^{(i)}_{(0,0)}$ around $\mu_1=\mu/2$
\beqn
\phi^{(i)}_{(0,0)}(\mu,\mu_1)=\om -2 \chi_2^{(i)}(\frac{\mu}{2})
+\frac{1}{2}(\mu_1-\frac{\mu}{2})^2\xi^{(i)}(\mu)\;\;,\;\;
\xi^{(i)}(\mu)= \frac{\partial^2}{\partial \mu_1^2}
[-\chi_2^{(i)}(\mu_1)-\chi_2^{(i)}(\mu-\mu_1)]_{|\mu_1=\frac{\mu}{2}}
\eeqn
This is inserted into the $\mu_1$-integral in eq.\ (\ref{blub})
and the integration is performed by taking one of the poles.
The result is
\beqn
\Phi^{(i)}(\om,\mu) = \frac{\om}
{\sqrt{2 \xi^{(i)}(\mu)\left(2\chi_2^{(i)}(\frac{\mu}{2})-\omega\right)}}
\label{crash}
\eeqn
In order to investigate the singularity in the vicinity of 
$\mu=0$ we use $\chi_2^{(i)}(\frac{\mu}{2}) \simeq 2a_i\al/(\pi\mu)$ and 
$\xi(\mu) \simeq -32a_i\al/(\pi\mu^3)$ and find
\beqn
\Phi^{(i)}(\om,\mu) \simeq 
\sqrt{\frac{\mu^{3}\omega\pi}{64 a_i\al}}
\frac{1}{\sqrt{1-\frac{4a_i\al}{\pi\om\mu}}}
\label{lork}
\eeqn
In the vicinity of the tip of the cut we have $\mu=4a_i\al/(\pi\om)$ 
and one recognizes that (\ref{lork}) behaves exactly like (\ref{quad}).
This demonstrates that we can correctly reproduce the pinching from the 
solution of the Landau equations.
One can easily read off further singularities from eq.\ (\ref{crash}).
These are determined by the poles of $\chi_2^{(i)}(\frac{\mu}{2})$
which are located at (we restrict ourselves to the left halfplane)
$\mu=0,-2,-4, \cdots$. 
These poles lead to cuts of $\Phi^{(i)}(\om,\mu)$ in the 
$\mu$-plane, starting at $\mu=-k+4a_i\al/(\pi\om), \,k=0,2,4,\cdots$.
This means that 
from the solution (\ref{landau}) of the 
Landau equations we have obtained a whole 
string of singularities in $\mu$. 
\\
Let us now have a closer look at the vicinity of $\mu=-2$.
From the reasoning above we find a pole of $\chi_2^{(i)}(\frac{\mu}{2})$
which transforms into a cut of $\Phi^{(i)}(\om,\mu)$. We can associate this 
singularity with the pinching of the poles at $\mu_1=-1$ of 
$\chi_2^{(i)}(\mu_1)$ and at $\mu-\mu_1=-1$ of $\chi_2^{(i)}(\mu-\mu_1)$.
But at $\mu=-2$ also other types of pinching are possible,
namely $\mu_1=-2$ with $\mu-\mu_1=0$ and conversely
 $\mu_1=0$ with $\mu-\mu_1=-2$.
By using the same method as in eq.\ (\ref{blub}) one can show that all 
three pinchings lead to the same type of 
square-root singularity.
We conclude that the singularity at $\mu=-2$ is degenerate. 
This can be understood as soon as one recognizes that due to the 
quasiperiodicity 
\footnote{
Quasiperiodicity here means that $\psi$ has an infinite periodic series
of poles.} 
of $\psi(x)$ the Landau equations have infinitely many solutions.
We start again from the second Landau equation
\beqn
\partial_{\mu_1}
[\psi(\mu_1)+\psi(1-\mu_1)+\psi(\mu-\mu_1)+\psi(1-\mu+\mu_1)]=0
\eeqn
By using the functional equation $\psi(x+1)=\psi(x)+1/x$ one can 
shift the arguments of the first two terms on the left hand side
\beqn
\partial_{\mu_1}
[\psi(\mu_1+k)+\psi(1-\mu_1-k) -2\sum_{l=0}^{k-1}(\mu_1+l)^{-1}
+\psi(\mu-\mu_1)+\psi(1-\mu+\mu_1)] \nonumber \\
=
[\psi'(\mu_1+k)-\psi'(1-\mu_1-k) +2\sum_{l=0}^{k-1}(\mu_1+l)^{-2}
-\psi'(\mu-\mu_1)+\psi'(1-\mu+\mu_1)]
\label{lan1}
\eeqn
Equivalently one can shift the arguments of the last two terms 
\beqn
\partial_{\mu_1}
[\psi(\mu_1)+\psi(1-\mu_1)
+\psi(\mu-\mu_1+k)+\psi(1-\mu+\mu_1-k)
-2\sum_{l=0}^{k-1}(\mu-\mu_1+l)^{-1}
] \nonumber \\
=
[\psi'(\mu_1)-\psi'(1-\mu_1) 
-\psi'(\mu-\mu_1+k)+\psi'(1-\mu+\mu_1-k)
+2\sum_{l=0}^{k-1}(\mu-\mu_1+l)^{-2}
]
\label{lan2}
\eeqn
We have already discussed that the pinching is due to the poles of 
the $\psi$-functions . Then, neglecting the nonsingular double
poles in the sum, the following solution of eq.\ (\ref{lan1}) can
be found
\beqn
\mu_1=\frac{\mu-k}{2}+\Delta_{-k}(\mu)
\eeqn
Similarly a solution of eq.\ (\ref{lan2}) reads 
\beqn
\mu_1=\frac{\mu+k}{2}+\Delta_{k}(\mu)
\eeqn 
By inserting these solutions into 
eqs.\ (\ref{lan1}) and (\ref{lan2})
resp., one finds that near the integer points $\mu=-k-2m\,,\,m=0,2,4,\cdots$
the functions $ \Delta_{-k}(\mu)$ and $\Delta_{k}(\mu)$
both behave as $\Delta(\mu)\sim \mbox{const.}(\mu+k+2m)^3$.
Now we can again insert the above solutions into the first Landau equation
and define in analogy to the previously discussed case ($k=0$)
functions $\omega_{(0,0)\,-k}^{(i)}$ and $\omega_{(0,0)\,k}^{(i)}$
through the equations
\beqn
\omega_{(0,0)\,-k}^{(i)}(\mu)&=&
\chi_2^{(i)}(\frac{\mu-k}{2}+\Delta_{-k}(\mu),0) 
+\chi_2^{(i)}(\frac{\mu+k}{2}-\Delta_{-k}(\mu),0) 
\\
\omega_{(0,0)\,k}^{(i)}(\mu)&=&
\chi_2^{(i)}(\frac{\mu+k}{2}+\Delta_{k}(\mu),0) 
+\chi_2^{(i)}(\frac{\mu-k}{2}-\Delta_{k}(\mu),0)
\eeqn
These functions have a quite simple singularity structure in the 
left half plane. Both 
$\omega_{(0,0)\,-k}^{(i)}$ and $\omega_{(0,0)\,k}^{(i)}$ have 
simple poles at $\mu=-k-2m\,,\,m=0,2,4,\cdots$ with identical residues
$4a_i\al/\pi$. All these poles have their origin in the pinching of the 
$\mu_1$-contour by the poles of the two $\chi_2$-functions in the 
denominator in eq.\ (\ref{blub}).
Introducing the functions $\omega_{(0,0)\,\pm k}^{(i)}$ 
($k \in {\Bbb{N}}$)
we have thus obtained a classification of the singularities of the two 
reggeon intermediate state. 
Each function $\omega_{(0,0)\, k}^{(i)}$ introduces at most one 
square root singularity of $\Phi(\om,\mu)$ in the vicinity of an
integer point $\mu=-m$. For each integer point $\mu=-m$ we find a 
$m+1$-fold degeneracy, i.\ e.\ $\Phi(\om,\mu)$ has
$m+1$ cuts starting at 
$\mu=-m+4a_i\al/(\pi\om)$. 
\\
The functions $\omega_{(0,0)\, k}^{(i)}$ are degenerate on the 
poles, i.\ e.\ they have identical residues. This degeneracy is lifted
when the finite parts are taken into account.    
The finite parts can easily be calculated by applying the functional equation
of the $\psi$-function.
Since $\Delta_{\pm k}(\mu)$ vanishes in the vicinity of the integer
points $\mu=-m$ proportional to $(\mu+m)^3$ it can be neglected  
in the calculation of the finite part. Now at a fixed integer point 
$\mu=-m$ a pole appears in the functions $\omega_{(0,0)\, \pm l}^{(i)}(\mu)$  
with $l=m,m-2,m-4,\cdots,0(1)$ for $m$ even (odd). The expansion
of $\omega_{(0,0)\, \pm l}^{(i)}(\mu)$ around the integer point up to the 
finite part then reads
\beqn
\omega_{(0,0)\, \pm l}^{(i)}(\mu)
= \frac{4a_i\al}{\pi(\mu+m)} + c_l 
\;\;,\;\;
c_l = -4 \frac{a_i\al}{\pi}
\left[
\sum_{i=0}^{(m+l)/2-1}\frac{2}{(m+l)/2-i}
+
\sum_{i=0}^{(m-l)/2-1}\frac{2}{(m-l)/2-i} 
\right]
\eeqn
which shows that the finite parts are different for different $k$.
\\
So far we have only considered the contribution with zero 
conformal spin ($n_1=n_2=0$) for both reggeons.  
The same analysis can be repeated for nonzero spin. Without going 
into the details we only give the results.
For each pair $(n_1,n_2)$ functions 
$\omega_{(n_1,n_2)\, \pm k}^{(i)}(\mu)$ can be defined from the
solutions of the Landau equations. 
In the left half of the $\mu$-plane
each function has an infinite
string of simple poles located at 
$\mu=-2k-(|n_1|+|n_2|)/2,\,k=0,1,2,\cdots$
with residues $4 a_i\al/\pi$. 
The degeneracy of the residues is lifted by the finite parts.
These poles can again be related to the pinching of 
the $\mu_1$-contour by poles from the two $\chi_2$-functions in the 
denominator in eq.\ (\ref{blub}).
In the vicinity of (half)integer points $\mu=-l-(|n_1|+|n_2|)/2$
the poles of $\omega_{(n_1,n_2)\, \pm k}^{(i)}(\mu)$ 
for $k=l,l-2,l-4,\cdots,0(1)$
lead to  
square root singularities of 
$\Phi(\om,\mu)$ at $\mu=-l-(|n_1|+|n_2|)/2+4a_i\al/(\pi\om)$. 
\\
It is obvious that by taking nonzero $n_1,n_2$ into account,
the degeneracy of the singularities increases strongly.
Let us consider e.\ g.\ a fixed even integer point $\mu=-m < 0$.
Simple counting then shows that there are 
$m(m+2)$ functions $\omega_{(n_1,n_2)\, \pm k}^{(i)}(\mu)$
having a pole with residue $4a_i\al/\pi$ at that point.  
\\
To summarize the results of this subsection, we have derived the 
following representation for the two reggeon intermediate state
\beqn
\Phi^{(i)}(\mu,\omega)=
\sum_{n_1,n_2=-\infty}^{+\infty}
\sum_{k=-\infty}^{+\infty}
\frac{\om}{\left[\left(\omega-\om_{(n_1,n_2)\,k}^{(i)}(\mu)\right)
|2\xi_{(n_1,n_2)\,k}^{(i)}|\right]^{\fez}}
\label{aruba}
\eeqn 
where $\xi_{(n_1,n_2)\,k}^{(i)}$ is related to the second derivative
of $\om_{(n_1,n_2)\,k}^{(i)}$.
We expect the representation (\ref{aruba}) 
to be correct up to nonsingular terms. In particular it reproduces 
the behavior of $\Phi^{(i)}(\mu,\omega)$ in the vicinity 
of integer and half-integer points in the left half of the 
$\mu$-plane.
The most interesting result is that we could relate the singularities 
of $\Phi^{(i)}(\mu,\omega)$ to strings of poles of an infinite 
number of different functions $\om_k(\mu)$.
It is of course tempting to identify the label $k$ as a discrete quantum
number of the noninteracting two reggeon system. It is, however,
not clear from the present analysis if this quantum number can be associated
with a symmetry.
In addition to this new quantum number $k$ we have to sum in 
(\ref{aruba}) over the quantum numbers $n_1,n_2$ which are 
associated with the conformal symmetry.
\\
Now the two reggeon state has to be iterated. 
Let us first assume that the 
effective reggeon interaction vertex ${\boldsymbol \Theta}$ is diagonal in
all quantum numbers $k,n_1,n_2$ and its only effect is the conservation
of the conformal dimension parameter $\mu$. In this case $k$ could even be 
interpreted as a quantum number of the interacting two reggeon system.
Let us furthermore assume for the sake of the argument that the 
matrix structure 
of the iteration in (\ref{elba2}) is completely
ignored, i.\ e.\ we consider only one color state 
\footnote{Consequently we drop the index $i$ referring to the color state
in the following.}
and one coupling 
scheme and assume the vertex to be diagonal both in color and in the 
coupling scheme
\footnote{Of course this simplification is paradoxical since if we had only 
one coupling scheme we wowuld not have a vertex.
The toy model which we use here for illustration
is intended to capture all aspects
which arise from the iteration of $\mu$-plane singularities while 
avoiding the complications associated with the matrix structure.}
. 
Then one can iterate each term in the sum in eq.\ (\ref{aruba}) separately
with the result
\beqn
\sum_{j=0}^{\infty}
\left(\frac{\om\theta_{(n_1,n_2)\,k}(\mu)}
{\left[\left(\omega-\om_{(n_1,n_2)\,k}(\mu)\right)
|2\xi_{(n_1,n_2)\,k}(\mu)|\right]^{\fez}}
\right)^j \frac{\om}
{\left[\left(\omega-\om_{(n_1,n_2)\,k}(\mu)\right)
|2\xi_{(n_1,n_2)\,k}(\mu)|\right]^{\fez}}
\nonumber \\
=
\frac{\om}{\left[\left(\omega-\om_{(n_1,n_2)\,k}(\mu)\right)
|2\xi_{(n_1,n_2)\,k}(\mu)|\right]^{\fez}-\om\theta_{(n_1,n_2)\,k}(\mu)}
\label{korsika}
\eeqn
The function $\theta_{(n_1,n_2)\,k}(\mu)$ is the coefficient 
which comes from the vertex. 
Now we use the fact that in the vicinity of 
the (half)integer point $\mu=-m$ the function   
$\om_{(n_1,n_2)\,k}(\mu)$ behaves as 
$\om_{(n_1,n_2)\,k} \sim 4a_i\al/(\pi(\mu+m))$.
Expanding around that point we find from eq.\ (\ref{korsika}) the analogue 
of eq.\ (\ref{helgoland}) which was derived for the BFKL amplitude.
In the present case 
in addition to a cut starting at $\mu=-m+4a_i\al/(\pi\om)$ we find 
a pole 
\footnote{If we had retained the constant $(-1)$ in the definition
of the two reggeon propagator the location of the pole had been slightly
different. For the present qualitative discussion, however, 
this does not play any role.}
at 
$\mu =- m+4a_i\al/(\pi\om) \cdot 
1/(1-\om\theta_{(n_1n_2)\,k}^2/(2\xi_{(n_1n_2)\,k}))$ 
where $\theta_{(n_1n_2)\,k}$
is obtained from the expansion of $\theta_{(n_1,n_2)\,k}(\mu)$
around $\mu=-m$
and $\xi_{(n_1n_2)\,k}$ is a remainder from 
$\xi_{(n_1,n_2)\,k}(\mu)$.
One finds that the singularity structure in the $\mu$-plane, 
even in this highly simplified case, is quite complicated.
Near the (half)integer point there are a pole and a cut the distance
of which is determined by the parameter $\theta_{(n_1n_2)\,k}$.
In particular the cut might be difficult to interpret when the twist 
expansion is considered. 
It is, however, encouraging that a pole in the $\mu$-plane appears.
One can conjecture that this pole belongs to a meromorphic function
$\chi_4(\mu;n_1,n_2,k)$ which can be identified with an eigenvalue 
of the four (reggeized) gluon system.
The further poles and residues of the function $\chi_4(\mu;n_1,n_2,k)$
could also be easily calculated from the known properties of the 
function $\om_{(n_1,n_2)\,k}^{(i)}(\mu)$.
\\ 
Unfortunately things are not so easy as considered here since the
interaction vertex is neither diagonal in color space,
nor in the quantum numbers $k,n_1,n_2$.
The nontrivial color structure 
is not too problematic since the color state space is finite 
dimensional, which leads to reformulating 
eq.\ (\ref{korsika}) in matrix form.  
This is not the case 
for the other quantum numbers which are unbounded.
\subsubsection{The effective interaction vertex}
This section is devoted to the effective reggeon interaction vertex defined 
in eq.\ (\ref{juist}) and in particular to an attempt to extract its 
singularities.
For the following discussion it is helpful to recall the definition of the 
vertex here
\beqn
{\bs \Theta}^{(\nu_1,n_1;\nu_2,n_2)}_{(\nu'_1,n'_1;\nu'_2,n'_2)} (\qf,\qf') = 
\int \frac{d^2 \kf}{(2\pi)^3}
E^{(\nu_1,n_1)}(\kf,\qf-\kf) E^{(\nu_2,n_2)}(\qf'-\kf,-\qf-\qf'+\kf) 
\phantom{xxxxxxxxxxxxxxxxxx}
\nonumber \\
E^{(\nu'_1,n'_1)\ast}(\kf,  \qf'-\kf) E^{(\nu'_2,n'_2)\ast}
(\qf-\kf,-\qf-\qf'+\kf) \cdot
\kf^2(\qf-\kf)^2(\qf'-\kf)^2(-\qf-\qf'+\kf)^2
\label{juist2}
\eeqn  
From which region of the momentum integration can we expect singularities?
One possibility is that a singularity arises from the upper limit 
of the integration, i.\ e.\ the region $\kf^2 \to \infty$.
The other basic possibility is that a singularity appears when one of the 
momenta of the propagators in the crossed box diagram in fig.\ \ref{fig51} 
vanishes, i.\ e.\ one of the arguments of the $E^{(\nu,n)}$-functions
in (\ref{juist2}) goes to zero. 
\\
Let us investigate the first possibility. In the case $\kf \to \infty$ we can 
neglect the momenta $\qf,\qf'$ of the external reggeons, i.\ e.\ we can take
the zero momentum transfer limit of $E^{(\nu,n)}$ which was discussed 
in section \ref{sec13}. In this limit $E^{(\nu,n)}(\kf,\qf-\kf)$ 
behaves as $(\kf^2)^{-\ftz-i\nu}$. From this we find the following 
behavior of the integrand $I(\kf^2)$ in (\ref{juist2})
\beqn
I(\kf^2) \stackrel{\kf \to \infty}{\simeq} 
(\kf^2)^{-2-i\nu_1-i\nu_2+i\nu'_1+i\nu'_2}
\eeqn
In the last but one subsection we have anticipated that from the 
$\qf$-integration we obtain a conservation law for the conformal dimensions in 
the form $i\nu_1+i\nu_2=i\nu'_1+i\nu'_2$. Using, this we
see that $I(\kf^2)$ effectively behaves as $(\kf^2)^{-2}$
and the upper limit of the $\kf$-integration is convergent, i.\ e.\ 
does not produce a singularity.
\\
We therefore turn to the second possibility, the vanishing of one propagator
momentum.
This implies a technical difficulty since we have to investigate 
the behavior of $E^{(\nu,n)}(\kf,\qf-\kf)$ for small argument.
Starting from the explicit expression (\ref{momspacef}) 
one can derive the following expansion for the simplest case $n=0$
\beqn
E^{(\nu,0)}(\kf,\qf-\kf)&=& c(\nu,0) (\qf^2)^{-\fez-i\nu}
\frac{1}{\kf^2}
\left[b_1 \frac{|\kf|}{|\qf|}
+ b_2 \frac{|\kf|^2}{|\qf|^2}
+ b_3 \frac{|\kf|^3}{|\qf|^3}
+O\left(\frac{|\kf|^4}{|\qf|^4}\right)
\right]
\label{ruegen}
\\
\mbox{with}\;\;\; & & 
b_1 = b(\nu,0)\cdot  2 \cos \phi
\nonumber 
\\
\phantom{\mbox{with}\;\;\;}& &
b_2 = b(\nu,0)\cdot \left[
                    (\nu^2+\frac{1}{4})\left( 2 
                    \left(\log \frac{|\kf|}{|\qf|}
                    -\xi(\nu)\right) -3\right) +\fez(4\nu^2+9) \cos^2\phi
                    \right]
\nonumber 
\\
\phantom{\mbox{with}\;\;\;}& &
b_3 = b(\nu,0)\cdot \left[
                    \left( 2\left(\log \frac{|\kf|}{|\qf|}-\xi(\nu)\right)
                    (\nu^2+\frac{1}{4})(\nu^2+\frac{9}{4}) 
                    -4(\nu^2+\frac{3}{4})(\nu^2+\frac{5}{4}) \right)\cos\phi
\right. \nonumber \\  & & \left. 
\phantom{xxxxxxxxxxxxxxxxxxxxxxxxxxxxxxxxxxxx}
         +\frac{2}{3}(\nu^2+\frac{9}{4})(\nu^2+\frac{25}{4}) \cos^3\phi
                    \right]
\nonumber
\eeqn
The coefficient $c(\nu,0)$ is given in eq.\ (\ref{bali}),
$b(\nu,0)$ is defined as 
$b(\nu,0)=8\pi^24^{i\nu}\Gamma(3/2+i\nu)\Gamma(3/2-i\nu)/\Gamma^2(1/2-i\nu)$ 
and $\xi(\nu)=2\psi(1)-\psi(1/2+i\nu)-\psi(1/2-i\nu)$ 
is the same as in eq.\ (\ref{haiti}).  
The angle $\phi$ is defined through $\cos \phi= (\kf\cdot\qf)/(|\kf||\qf|)$.
Due to symmetry the same expansion is obtained for the argument $(\qf-\kf)$.
The derivation of the coefficients in (\ref{ruegen})
requires considerable effort the basic steps
of which are given in appendix \ref{app4}.
Since there are two 
$E^{(\nu)}$-functions in (\ref{juist}) which have the momentum $\kf$ as 
an argument we find from the leading term in (\ref{ruegen}) for the 
small-$\kf$ behavior of the integrand $I(\kf^2) \simeq \mbox{const}$.
One concludes that the vertex function ${\bs \Theta}$ is also regular
when one of the propagator momenta vanishes.
\\
This shows that we cannot expect a singularity from the effective vertex 
if both external momenta $\qf,\qf'$ are different from zero. 
\\
The situation changes if one of the the momentum transfers e.\ g.\ 
$\qf'$ is set equal to zero.
Inserting the expression (\ref{eigenzero})
for the eigenfunctions in forward direction we then obtain the 
following simplified form of the vertex
\beqn
{\bs \Theta}^{(\nu_1,n_1;\nu_2,n_2)}_{(\nu'_1,n'_1;\nu'_2,n'_2)} (\qf,0) = 
\int \frac{d^2 \kf}{(2\pi)^3}
(2 \pi\sqrt{2})^2 (\kf^2)^{\fez+i\nu'_1}
                  ((\qf-\kf)^2)^{\fez+i\nu'_2}
\left(\frac{k^*}{k}\right)^{\frac{n'_1}{2}}
\left(\frac{(q-k)^*}{(q-k)}\right)^{\frac{n'_2}{2}}
\nonumber \\
\cdot E^{(\nu_1,n_1)}(\kf,\qf-\kf) E^{(\nu_2,n_2)}(-\kf,-\qf+\kf) 
\label{helena}
\eeqn  
Here the complex notation (cf.\ eq.\ (\ref{coord}))
was used to display the angular dependence
of the eigenfunctions.
Considering again the limit $\kf \to 0$ we find, using the leading 
term of the expansion (\ref{ruegen}), the following behavior of the 
integrand (for $n_1=n_2=0$)
\beqn
I(\kf^2) \stackrel{\kf \to 0}{\simeq} 
(\kf^2)^{-\fez+i\nu'_1} \cdot\mbox{const.}
\cdot
(\qf^2)^{\ftz+i\nu'_2-i\nu_1-i\nu_2}
\eeqn 
The $\kf$-integration now leads to a pole in the $\nu'_1$ plane
\beqn
(\qf^2)^{-\ftz+i\nu'_2-i\nu_1-i\nu_2}
\int_0^{\qf^2} \frac{d\kf^2}{\kf^2} (\kf^2)^{\fez+i\nu'_1} 
= \frac{1}{\fez+i\nu'_1}
(\qf^2)^{-1+i\nu'_1+i\nu'_2-i\nu_1-i\nu_2}
\label{mon}
\eeqn  
where the upper limit of the integration is required from consistency 
with the expansion in powers of $|\kf|/|\qf|$.
Equivalently a pole in the $\nu'_2$ plane can be obtained when the 
limit $\kf \to \qf$ is considered.
\\
By setting $\qf'=0$ we have extracted a singularity of the vertex 
function in the lower half of the $\nu'_1$ and the $\nu'_2$ plane.
It is easily seen that further singularities in the 
respective $\nu'_j$-planes
can be obtained by expanding the integrand in (\ref{helena}) to higher 
orders in $|\kf|/|\qf|$.
These are exactly the singularities we look for since they are of the same
type as the singularities which appear in the 
$\chi_2^{(i)}(\nu'_j)$-functions belonging to the two reggeons below the 
vertex. 
\\
From the above observations we formulate the following conjecture.
The relevant singularities of the vertex function 
${\bs \Theta}(\qf,\qf')$ which can be consistently combined with 
the singularities of the reggeon states below the vertex are obtained by 
setting the momentum transfer $\qf'$ of the lower reggeons to zero 
and expanding the integrand in the vertex function around the singular 
points $\kf=0$ and $\kf=\qf$, respectively, where $\qf$ is the momentum 
transfer of the upper reggeons. If the momentum transfer $\qf'$ of the 
lower reggeons is different from zero it acts as an effective regularization 
which prevents the integration from becoming singular.
\\
It remains to discuss the integration of the momentum transfer. 
After $\qf'$ has been set to zero the $\qf$-dependence of the vertex function
follows from scaling arguments and is as given in eq.\ (\ref{mon}).
Obviously, after integration of $\qf^2$ a function 
$\delta(i\nu'_1+i\nu'_2-i\nu_1-i\nu_2)$ is obtained which expresses 
the conservation of conformal dimensions.    
\\
Now let us consider in more detail
in which way the singularities from the lower two reggeon
state and the singularities from the vertex function are combined.
We start with the leading poles which were discussed above.
Adding the poles at $i\nu'_1=-1/2$ and $i\nu'_2=-1/2$ 
and restoring all coefficients correctly we obtain for eq.\ (\ref{helena})
integrated over $\qf$
\beqn
{\bs \Theta}^{(\nu_1,0;\nu_2,0)}_{(\frac{i}{2},0;\frac{i}{2},0)}
= \frac{\mu}{\mu'_1(\mu-\mu'_1)} \frac{1}{4\pi^2}
c(\nu_1,0)b(\nu_1,0)c(\nu_2,0)b(\nu_2,0)
\label{maui}
\eeqn   
We have considered the simplest case $n_1=n_2=n'_1=n'_2=0$. Furthermore we 
have again introduced the variables $\mu'_1=1/2+i\nu'_1$
and $\mu-\mu'_1=1/2+i\nu'_2$.
Multiplying with the reggeon propagator $\Phi^{(i)}(\om,\mu)$
which has been expanded around the corresponding singularities 
(cf.\ eq.\ (\ref{quad})) 
from below and performing the $\mu'_1$-integration we obtain
\beqn
\frac{1}{\sqrt{1-4\frac{a_i\al}{\pi\om\mu}}}
\frac{1}{4\pi^2}
c(\nu_1,0)b(\nu_1,0)c(\nu_2,0)b(\nu_2,0)
\eeqn
One recognizes that due to the singularities arising 
from the vertex the result 
on the right hand side of eq.\ (\ref{quad}) has been slightly modified.
At this point all singularities associated with the two reggeon state 
below the vertex have been treated and we turn to the upper reggeons
with quantum numbers $\nu_1,\nu_2$. 
\\
From the lower reggeon state the singularity at $\mu \simeq 0$
has been extracted. Conservation of conformal dimensions hence 
implies $1/2+i\nu_1+1/2+i\nu_2=\mu_1+\mu_2 \simeq 0$. 
There is only one possibility to fulfill this equation and to 
simultaneously generate singularities in the upper reggeon state,
namely $i\nu_1=i\nu_2 \simeq -1/2$, 
corresponding to $\mu_1 = \mu_2 \simeq 0$.  
Conservation of conformal dimensions near the point $\mu \simeq 0$ thus 
implies that one has to expand the upper two reggeon state around 
$i\nu_1=-1/2$ and $i\nu_2=-1/2$, respectively. 
In particular the coefficients $c(\nu_1,0)$ and $b(\nu_i,0)$ have to be 
evaluated at these points.
Using the explicit expressions for these functions
we then obtain for eq.\ (\ref{maui})
\beqn
{\bs \Theta}^{(\frac{i}{2},0;\frac{i}{2},0)}_{(\frac{i}{2},0;\frac{i}{2},0)}
= \frac{1}{2}\,\frac{\mu}{\mu'_1(\mu-\mu'_1)} 
\eeqn
We find that evaluated near the singular points of the upper two
reggeon state the vertex function ${\bs \Theta}$
integrated over the momentum transfer of the upper two reggeon state
reduces to the factor $1/2$.
This is the famous factor $1/2$ which has been found with a different
method in \cite{bartels}.
Consequently if one is interested in the singularity  
of the four reggeized gluon state near the point $\mu=0$ the following 
structure has to be iterated
\beqn
\frac{1}{2}\left[\frac{1}{\sqrt{1-\frac{4a_1}{\om\mu}}}-1\right]
\eeqn 
In \cite{bartels} it was shown that when the full structure of the 
recoupling matrix ${\bs S}$ is taken into account the iteration leads 
to a pole in the $\mu$-plane. This pole which was found to the right of the 
two pomeron cut was identified as the anomalous dimension of the 
twist four operator associated with the four gluon state.
\\
So far we have extracted the rightmost poles in the left half of the 
$\mu'_1$ and $\mu'_2$ plane. It is clear how one has to proceed 
further in our 
setup. By expanding the integrand in (\ref{helena}) to higher orders 
in $|\kf|/|\qf|$ one generates further singularities at 
$\mu'_1,\mu'_2=-2,-3,-4, \cdots$ (we still keep $n_i,n'_i=0$).
These singularities are combined in the way described above with 
the poles from the two reggeon propagator. Finally one applies the 
conservation of conformal dimensions and evaluates the coefficients
$c(\nu_i),b(\nu_i)$ at the respective integer points.
\\
There appears to arise a problem when the integrand is expanded to 
higher orders since the next-to-leading terms in the expansion 
(\ref{ruegen}) contain logarithms in $|\kf|$. This seems to indicate
the appearance of double poles. 
Closer inspection however shows that these logarithms disappear
when the coefficients $c(\nu_i),b(\nu_i)$ are evaluated on the integer
points. The same is true for the additional poles which one could expect
from the $\xi$-functions in ({\ref{ruegen}).
In fact it can be shown that the functions $E^{(\nu,n)}(\kf,\qf-\kf)$
can be evaluated explicitly on the integer points $i\nu=-(1+|n|)/2-k$
and turn out to be rational functions of $|\kf|$ and $|\qf|$. This then
proves that only single poles appear in the vertex as long as the 
conformal dimensions of the upper reggeon state are near the integer 
points. 
\\
It is nevertheless obvious that for the poles which are found to the left
of the ones discussed above a new complication arises due to 
the degeneracy of singularities at a fixed integer point.
We have seen in the preceding subsection that at an integer point $\mu=-m$
several functions $\om^{(i)}_{\pm k}(\mu)$ 
have a singularity corresponding 
to the different types of possible pinchings. For all these contributions
the respective poles have to be extracted from the vertex. The same 
degeneracy is of course also present in the upper two reggeon state.
Conservation of conformal dimensions tells us that $\mu$ is fixed but 
it does not determine which combination of poles builds up the singularity in 
$\mu$ in the upper reggeon state given a particular type of pinching 
in the lower two reggeon state. 
Different choices of combinations
will clearly lead to different results for the coefficients 
$c(\nu_i),b(\nu_i)$. Consequently, the vertex function ${\bs \Theta}$ is not 
diagonal in the quantum number $k$ which was introduced to label the different
types of possible pinchings. W.\ r.\ t.\ this quantum number the vertex 
has a matrix structure.   
It is clear that the dimension of this matrix becomes even higher if 
nonzero conformal spins $n_i,n'_i$ are taken into account. 
\\ 
Let us consider an example which displays the full complexity. 
We are interested in the behavior near the singular point 
$\mu=-2$. The following table 
shows the different combinations of quantum 
numbers which lead to a singularity at $\mu=-2$ in the 
lower two reggeon state.
\begin{center}
\begin{tabular}{|c|c|c|c|}   
\hline
 $\;\;|n'_1|\;\;$ & $\;\;|n'_2|\;\;$ & $\;\;\mu'_1\;\;$ & $\;\;\mu'_2\;\;$  
 \\ \hline 
  \hline 
    0  &  0   &   -2   &  0       \\
 \hline 
    0  &  0   &   -1   &  -1       \\
 \hline 
    0  &  0   &    0   &  -2       \\
 \hline
    1  &  1   &    -3/2   &  -1/2       \\
 \hline
    1  &  1   &    -1/2   &  -3/2       \\
 \hline
    2  &  0   &    -1   &  -1       \\
 \hline
    2  &  0   &    -2   &   0       \\
 \hline
    0  &  2   &    -1   &  -1       \\
 \hline
    0  &  2   &     0   &  -2       \\
 \hline
    3  &  1   &    -3/2   &  -1/2       \\
 \hline
    1  &  3   &    -1/2   &  -3/2       \\
 \hline
    2  &  2   &    -1   &  -1       \\
 \hline
    4  &  0   &     -2   &  0       \\
 \hline
    0  &  4   &     0   &  -2       \\
 \hline
\end{tabular}
\end{center}
We have 56 combinations of quantum numbers which lead to a 
singularity at $\mu=-2$. Corresponding to this 56-fold degeneracy
we have a $56\times56$ matrix associated with the vertex function
${\bs \Theta}$.
This matrix can be obtained from the representation $\ref{helena}$
with the following steps.
\begin{itemize}
\item
Fix $n'_1$ and $n'_2$ according to the first two columns in the table.
\item
Find the poles in $\mu'_1$ and $\mu'_2$ by expanding the integrand in 
(\ref{helena}) to the respective order.
\item
With $\mu_1=1/2+i\nu_1,\mu_2=1/2+i\nu_2$, calculate the coefficients
$c(\nu_1,n_1),c(\nu_2,n_2)$ and $b(\nu_1,n_1),b(\nu_2,n_2)$
for all combinations of $n_1,n_2,\mu_1,\mu_2$ which appear in the 
table above.  
\end{itemize}
We see that in order to calculate the complete matrix we need in principle
the explicit momentum space representation of the functions 
$E^{(\nu,n)}(\kf,\qf-\kf)$ for general $n$ and moreover their expansion
in powers of $|\kf|/|\qf|$.
In section \ref{sec13} we have considered these functions for nonzero 
$n$ and we have recognized that their calculation is an extremely arduous
task. What is really needed, however, is not the complete dependence
of the function $E^{(\nu,n)}(\kf,\qf-\kf)$ on $\nu$ but only the value 
at specific points $i\nu=-(1+|n|)/2-k$.
It turns out that these quantities, the eigenfunctions evaluated on the
integer points, can be obtained in a much simpler way. 
To see this, we have to recall the result for the coefficients 
$c(\nu,n)$ which have been calculated for general $n$ in section \ref{sec13}
\beqn
c(\nu,n) = i^{n}\frac{\sqrt{2}}{2 \pi}4^{i \nu}
\frac{\Gamma(1+i\nu+\frac{|n|}{2})}{\Gamma(-i\nu+\frac{|n|}{2})}
\frac{\Gamma(-\fez-i\nu+\frac{|n|}{2})\Gamma(\fez-i\nu+\frac{|n|}{2})}
{\Gamma(\fez+i\nu+\frac{|n|}{2})\Gamma(\ftz+i\nu+\frac{|n|}{2})}
\label{koln}
\eeqn 
Inspection shows that for $|n| \neq 0$ this coefficient has a zero
for $i\nu=-1/2-|n|/2$ and a double zero 
for $i\nu=-1/2-|n|/2-k, k=1,2,3,\cdots$.
Consequently the function which multiplies these coefficients must have 
a single, respectively a double pole at the corresponding points, otherwise
the value of the function $E^{(\nu,n)}(\kf,\qf-\kf)$ on the integer point
$i\nu=-1/2-|n|/2-k$ 
would be zero.  
It follows that to calculate $E^{(\nu,n)}(\kf,\qf-\kf)$ on the integer points
we just have to extract the residues from the function multiplying the 
coefficient $c(\nu,n)$ in eq.\ (\ref{koln}).
Returning to the basic equation (\ref{mom1}) we find that this function is 
just the ordinary Fourier integral of the conformal three point function in 
configuration space
\beqn
\int d^2 \rho_1 d^2 \rho_2 
e^{i \kf \rho_1+i(\qf-\kf)\rho_2}\left(
\frac{\rho_{12}}{\rho_1\rho_2}\right)^{\frac{1+n}{2}-i\nu}
\left(\frac{\rho_{12}^{\ast}}{\rho_1^{\ast}
\rho_2^{\ast}}\right)^{\frac{1-n}{2}-i\nu}
\eeqn 
Now to extract the respective poles which cancel the zeroes of $c(\nu,n)$ 
from this integral is far simpler than performing the Fourier transformation 
explicitly. The origin of these poles is clear. They arise when
either $\rho_1$ or $\rho_2$ (or both) go to zero. They can be extracted 
from the integral by isolating the factors 
$1/(|\rho_1|^2)^{\fez-i\nu}$ and $1/(|\rho_2|^2)^{\fez-i\nu}$
and successively performing integration by parts.
Without going into the details we give here a list of results 
\footnote{For $n=0$ and the first integer point $i\nu=-1/2$ there is no 
zero in the coefficient $c(\nu,0)$ in eq.\ (\ref{koln}).
For this case the result can be obtained directly from the momentum
space representation by putting $i\nu=-1/2$.}
for the first integer points and the cases $|n|=0,1,2$.
\beqn
E^{(i\fez,0)}(\kf,\qf-\kf) &=& \sqrt{2}\pi
\left[
\frac{1}{\kf^2}+\frac{1}{(\qf-\kf)^2}-\frac{\qf^2}{\kf^2(\qf-\kf)^2}
\right]
\nonumber
\\
E^{(i\ftz,0)}(\kf,\qf-\kf) &=&
2\sqrt{2}\pi
\nonumber
\\
E^{(i\frac{5}{2},0)}(\kf,\qf-\kf) &=&
\frac{1}{2}
2\sqrt{2}\pi\left(4\kf^2-4|\kf||\qf|\cos\phi+\qf^2\right)
\\
E^{(i 1,\pm 1)}(\kf,\qf-\kf) &=&
\sqrt{2}\pi
e^{\mp i \alpha}\frac{(2|\kf|-|\qf|e^{\pm i\phi})}
                     {|\kf|(|\kf|-|\qf|e^{\pm i\phi})}
\nonumber
\\
E^{(i 2, \pm 1)}(\kf,\qf-\kf) &=&
\sqrt{2}\pi e^{\mp i \alpha}(2|\kf|-|\qf|e^{\mp i\phi})
\\
E^{(i\ftz,\pm 2)}(\kf,\qf-\kf) &=&
\frac{1}{2}
\sqrt{2}\pi e^{\mp 2 i \alpha}
\frac{(2|\kf|-|\qf|e^{\pm i \phi})(2|\kf|-|\qf|e^{\mp i\phi})}
                     {|\kf|(|\kf|-|\qf|e^{\pm i\phi})}
\nonumber
\\
E^{(i\frac{5}{2},\pm 2)}(\kf,\qf-\kf) &=&
\frac{2}{5}\sqrt{2}\pi
e^{\mp 2 i \alpha} 
\left(5\kf^2-5|\kf||\qf|e^{\mp i \phi} + \qf^2 e^{\mp 2 i \phi} \right)
\eeqn
Here $\phi$ is defined as in eq.\ (\ref{ruegen}) and 
$\exp(i\alpha)=\sqrt{k/k^*}$. A check of these results is the 
forward direction limit $\qf=0$. In this case the above results are 
consistent with eq.\ (\ref{eigenzero}). 
\\
Starting form these expressions, 
the values of the expansion coefficients of $E^{(\nu,n)}(\kf,\qf-\kf)$
in the $|\kf|/|\qf|$-expansion can be obtained without recurrence 
to the explicit representation of the momentum space functions.
This offers the possibility to calculate the elements of the matrix 
${\bs \Theta}$ for nonzero $n_1,n_2$. 
In this respect the calculation of the coefficients $c(\nu,n)$ in 
section \ref{sec13} has been very important. 
\subsubsection{Concluding remarks}
Since the section \ref{sec33} has become increasingly technical 
it appears advisable to recall our starting point, the results that have
been obtained and the difficulties we have encountered. 
\\
The aim was to formulate the twist expansion of the four (reggeized)
gluon amplitude $G_4(\om)$. The background motivation, inspired 
by the analysis of the BFKL amplitude, is to construct from the series
of anomalous dimensions, supplemented by a number of subtraction constants,
the spectrum $\chi_4(\{\alpha\})$ of the four gluon state by means 
of a dispersion relation.
The function $\chi_4$ is the fundamental new quantity governing the 
energy and the scaling behavior of $G_4(\om)$.
We have argued that in terms of $\chi_4$, $G_4(\om)$ can be expressed as
\beqn
G_4(\om) = \sum_{\{\alpha\}} 
\frac{\psi_{\{\alpha\}}\psi^*_{\{\alpha\}}}
{\omega-\chi_4(\{\alpha\})}
\eeqn
with quantum numbers $\{\alpha\}$ and eigenfunctions $\psi_{\{\alpha\}}$.
The fundamental observation is that the anomalous dimensions which 
appear in the twist expansion of $G_4(\om)$ are related to the 
residues of the singularities of $\chi_4$ in the 
vicinity of the integer points in the $\nu$-plane, with $\nu$ being a 
conformal dimension parameter.  
Guided by the experience from the BFKL amplitude we have argued that the twist 
expansion can be accomplished by a loopwise extraction of logarithms
(corresponding to poles in the conformal dimension plane) and iteration
of these singular terms.
To this end we have used the Faddeev method to construct a solution of the 
defining equation for $G_4(\om)$ in terms of an infinite iteration
of two basic elements, namely a two reggeon intermediate state and an
effective reggeon interaction vertex.
It is important to emphasize that the two reggeons which make up the 
intermediate state are bound states of two reggeized gluons.  
In the present context it is essential to distinguish between these reggeons
and the reggeized gluons. In our approach the four reggeized gluons are 
grouped into intermediate two reggeon states. These reggeons can be 
expressed through the BFKL amplitude with a generalized color coefficient.
Consequently we associate a two dimensional momentum transfer $\qf$
as well as three quantum numbers, namely 
color and conformal dimension and spin with them. 
Within the Faddeev formalism both the intermediate state and the vertex
are $12 \times 12$-matrices which substantially complicates the analysis
of the analytic structure. The momentum dependence, however, can be collected
into a single loop integral ${\bs \Theta}$ which is extracted from the
vertex matrix. The further analysis concentrated on extracting the
singularities from the two reggeon intermediate state and the vertex function 
${\bs \Theta}$. 
\\
The singularities of the two reggeon intermediate
state were obtained and classified by 
employing the Landau equations. In addition to the quantum numbers
$i$ (color), $n_1,n_2$ (conformal spins of the two 
intermediate reggeons) and $\mu$ 
(the sum $\mu=1/2+i\nu_1+1/2+i\nu_2$ of conformal dimensions)
a new quantum number $k$ was identified.
With each $k$ an analytic function $\om^{(i)}_{(n_1,n_2)\,k}(\mu)$ 
could be associated which contains
an infinite number of single poles with degenerate residues at the 
integer points of the left half of the $\mu$ plane.
The outcome of these considerations was that the singularities of the two 
reggeon intermediate state in the $\mu$-plane are degenerate.
In the vicinity of an integer point $\mu=-m$ a definite number 
(depending on $m$) of different 
cuts at $\mu=-m+4a_i\al/(\pi\om)$ have their origin. These different 
cuts can be associated with different combinations of the quantum
numbers $n_1,n_2$ and $k$.
The number of degenerate cuts increases with increasing $m$.
\\
Associated with each iteration are two momentum integrals, the one 
contained in the vertex function ${\bs \Theta}$
and the integration over the momentum 
transfer of the intermediate reggeons. 
The latter can be shown to lead to a conservation 
of the sum of conformal dimensions $\mu$ associated 
with the two reggeon state.
The former can be demonstrated to generate an infinite string of 
poles in the $\mu'_1$ and $\mu'_2$ planes 
where $\mu'_1$ and $\mu'_2$ are related to the conformal dimensions
of the two intermediate reggeons below the vertex
by $\mu'_i=1/2+i\nu'_i$. 
These singularities can be obtained by
expanding the integrand of the vertex in powers of the ratio of the 
loop momentum and the momentum transfer of the upper two reggeon state. 
They are combined in a straightforward way with the corresponding 
singularities of the lower reggeon state.
For the leading singularity which is found near $\mu=0$ the vertex 
function ${\bs \Theta}$ then reduces to the coefficient $1/2$.
The iteration of singularities near $\mu=0$ is rather easy and finally leads 
to a pole of the four gluon amplitude in the $\mu$-plane to the 
right of the two pomeron cut.
For the singularities located to the left of $\mu=0$ the 
situation becomes more complicated. Here the vertex function becomes 
a matrix in the space which is spanned by the set of the quantum 
numbers $n_1,n_2,k$ which lead to the same singularity in the 
$\mu$-plane. This matrix can in principle be calculated. An important 
information which is needed then is the value of the conformal momentum space 
eigenfunctions $E^{(\nu,n)}$ on the integer points $i\nu=(1+|n|)/2-k$. 
The iteration of singularities near the integer points to the left 
of $\mu=0$ has to take into account the matrix structure.
The appearance of this matrix means that the quantum 
numbers $n_1,n_2$ and 
$k$ which were identified for the two reggeon state do not 
correspond to the true quantum numbers of the four reggeized gluon system.
One can not identify $k$ as a label of the eigenvalues $\chi_4$.
It is in fact questionable if the quantum numbers of the 
four reggeized gluon system can be
identified with the method presented here. 
The appearance of the quantum number $k$ in the two reggeon 
intermediate state however indicates that $\chi_4$ has extra
discrete quantum numbers in addition to the quantum numbers related to the 
conformal symmetry.
The investigation of the singularity structure 
is of course complicated by the additional matrix 
structure but it is at least in principle clear how to proceed. 
For each integer point $\mu=-m$ 
which is considered one could diagonalize
the matrix and iterate the eigenvalues in analogy to the factor $1/2$
for the nondegenerate point $\mu=0$. 
For each eigenvalue this will lead after 
iteration to a different singularity of the function $G_4(\om)$ in the 
$\mu$-plane. 
These singularities can be identified as contributions to the twist 
expansion of $G_4(\om)$.
The technical problem is the rather rapid increase of the dimension of the 
matrix when going farther to the left in the $\mu$-plane.
A principle problem is the classification of the resulting singularities.
One would like to collect them into groups corresponding
to different quantum numbers of the potential function $\chi_4(\{\alpha\})$.
From the present point of view it is not yet clear according to which 
principles this grouping should be performed.  
The quite regular pattern of singularities which appeared in both the 
reggeon intermediate state through the functions 
$\om^{(i)}_{(n_1,n_2)\,k}(\mu)$ and in the vertex function 
${\boldsymbol \Theta}$ gives rise to the hope that there exists some 
underlying structure which eventually could be identified 
using an additional information.   
\newpage
\section{Conclusions}
In the introduction we raised two questions 
concerning the phenomenological and theoretical 
significance of the BFKL pomeron. We want to summarize the 
results of this thesis with reference to these issues.
\\
The first question points to the phenomenological relevance 
of the BFKL pomeron, i.\ e.\ the perturbative resummation of
leading logarithms in $1/x$. 
To address this question we investigated in chapter 2 four different 
processes in deep inelastic scattering on the basis of BFKL pomeron
exchange. The aim was to work out specific properties of the cross 
sections and to check the theoretical consistence.
DIS is well-suited for this purpose since the large virtuality of the 
photon allows to start from perturbation theory.
\\
First we performed a calculation of the inclusive structure function $F_2$
based on a numerical solution of the BFKL equation and the high 
energy factorization formalism. Special attention was paid to 
the distribution of transverse momenta in the evolution. The point was 
made that due to the diffusion mechanism inherent in the
BFKL evolution the contribution of transverse momenta from the infrared 
region where perturbation theory cannot be consistently applied
is large. We have discussed a modification of the equation which 
suppresses the region of low transverse momenta. With such a modification 
it is possible to obtain good agreement with the experimental data
but due to the arbitrariness inherent in the modification
it is not conclusive.
In summary one can say that the strong rise of $F_2$ at low $x$ indicates 
the importance of the large logarithms of $1/x$ but a consistent 
application of the BFKL pomeron   
to this observable has to take into account higher order corrections.
\\
Since it is difficult to trace the BFKL pomeron in the $x$-dependence 
of the inclusive structure function, exclusive observables like e.\ g.\ the 
transverse energy distribution in the final state have been proposed for this 
purpose. We have studied some characteristic features of the transverse 
energy distribution of the gluons in the BFKL evolution.
The idea behind this is that the broad distribution of transverse momenta 
typical for the diffusion behavior transforms into some broad 
distribution of the transverse energy of the final state.
The problem with this observable, however, is that it is strongly 
affected by hadronization effects \cite{kuhlen}.
\\
To overcome the infrared problem of the BFKL equation 
one can consider a process in which the infrared momentum region 
is naturally suppressed by the kinematical requirements.
As an example of such a process we have discussed diffractive production 
of vector mesons with a large momentum transfer in DIS.
This process provides an interesting application of the BFKL 
pomeron in the nonforward direction. 
We have derived an explicit formula for the 
cross section which has been evaluated numerically and analytically in some
limiting cases. Based on this evaluation we have
obtained estimates for the cross 
section of the process in the HERA region. The result is of the order 
of $10^2$ nbarn, i.\ e.\ the prospects to observe this process 
at HERA are not bad. Being $t$-dependent this process also allows to 
study the slope of the Regge trajectory associated with the BFKL pomeron.
We have defined an effective slope parameter $\alpha'_{\mbox{\tiny eff}}$
which turned out to be quite small compared to the slope of 
the soft pomeron. This smallness can be traced back to the conformal 
symmetry of the BFKL pomeron which is only weakly broken by mass scales 
associated with the impact factors of the physical particles.
\\
BFKL pomeron exchange has then been applied to 
inclusive photon diffractive dissociation.
First, the situation has been studied in which the hadronic state is 
made out of a quark-antiquark pair. For this case a complete expression
for the cross section has been derived.
Simple expressions have been found for the $t=0$ and the large-$t$ limit.
The large $t$-limit is interesting 
since it allows a discussion of the Mueller-Tang effective prescription 
for the coupling of the BFKL pomeron to quarks.
It was demonstrated that for the inclusive case in which the final state 
of the quarks is integrated the Mueller-Tang prescription stating to 
couple the two gluons from the BFKL pomeron to a single quark is correct.    
Using the effective photon-meson transition vertex from the preceding 
calculation it has been shown that this prescription does not hold in general.
In this case where the quark-antiquark pair forms a specific
final state (exclusive situation) the effective prescription fails.
In any case, for the consistent application of the BFKL pomeron 
one should try to work out a gauge invariant coupling of the gluons.
\\
When the invariant mass of the diffractively produced hadronic system
increases final states with additional gluons have to be taken into account.
In the triple Regge limit these corrections decompose into two parts.
A reggeizing part which resembles a triple pomeron situation and an
irreducible part in which a four gluon state appears as a new element.
\\
The first part of these corrections has been investigated in this thesis by
generalizing the calculation for the quark-antiquark case.
It was shown that the zero momentum transfer limit is finite in 
contradiction to a result by Mueller and Patel. 
The calculation reveals a conservation law for the conformal dimensions
of the three BFKL pomerons at the effective triple ladder vertex in the 
limit $t=0$. This conservation law has important consequences for the 
energy dependence. In particular it is not possible to 
couple three ladders which are in the BFKL limit with the 
anomalous dimension $-1/2$. Instead, if the two lower
ladders are in the BFKL limit the upper ladder is forced into the
double logarithmic limit of strongly ordered transverse momenta.
As a consequence the transverse momentum decreases rapidly from the 
photon virtuality $Q^2$ to the effective scale at the vertex.
The latter is shifted deep into the infrared.
\\
The same is true if the BFKL pomeron couples to the quark-antiquark
pair directly. The photon virtuality $Q^2$ does not act as an effective hard 
scale. On the contrary the scale which determines the gluon coupling
to the quarks is soft. This shows that corrections to the BFKL pomeron 
which operate in the infrared region are very important in diffractive
dissociation.
\\
The scale at the effective photon-pomeron vertex can be increased
by imposing restrictions on the final state of the 
quark-antiquark pair.
As an example we have considered in this thesis a large transverse 
momentum or a large mass of the (anti)quark. Indeed the effective scale 
was found to be $(\kf^2+m^2)/(1-\beta)$ where $\kf^2$ and $m^2$ are the 
transverse momentum and the mass of the (anti)quark and 
$\beta$ is the familiar variable used in diffraction. This scale is
perturbative if either $\kf^2$ or/and $m^2$ are larger than 
$1 \,\mbox{GeV}^2$. The cross section of the process has been 
obtained in the framework of high energy factorization
and expressed through the unintegrated gluon structure function.
Performing the 
double-logarithmic approximation in the result we have been able to 
express the cross section in terms of the square of the gluon density
evaluated at
the momentum scale  $(\kf^2+m^2)/(1-\beta)$. This result displays
the explicit violation of Regge factorization. 
Extensive numerical studies
on the dependences on the 
different kinematical parameters have been performed. 
Characteristic features are the steep rise at small $x$ and the rapid decrease
with increasing transverse momentum (higher twist behavior).
For kinematical cuts corresponding to the HERA characteristics 
a total cross section of the order of $10^2$ pbarn has been estimated.
One should however keep in mind that our calculations are based on a
leading logarithmic approximation and are therefore subject to 
normalization uncertainties.
An interesting observable turned out to be the azimuthal angle 
$\phi$ between the quark-antiquark plane and the scattering plane of the 
electron. The $\phi$-spectrum peaks at $\phi=\pi/2$ quite in contrast 
to one gluon exchange models which peak at $\phi=0$ and $\phi=\pi$.
This might serve as a signal to reveal the two gluon exchange nature
of the process. 
\\
For charm quarks the contribution to the diffractive structure
function has been obtained in this model by integration over 
the transverse momentum. In the $\beta$-spectrum a zero at $\beta=0$ 
is found both for the transverse and the longitudinal cross section.
This behavior is in contrast to the data which is flat in the low
$\beta$ region. This highlights the necessity to   
extend the model by taking into account
production of additional gluons.
With one additional gluon being produced in the final state the cross section
becomes constant for $\beta=0$.
\newline
\newline
The theoretical aspects of the BFKL pomeron have been the focus of the 
second part of this thesis. Starting from the observation that both 
the violation of unitarity and the infrared consistency problems
require to take into account higher order corrections beyond
the leading logarithmic approximation we have investigated the first 
unitarity corrections. The basis of our study was the approach of Bartels 
\cite{bart-veryold,bart-old,bartels} in which an effective two-to-four 
gluon vertex and the interacting four (reggeized) gluon state 
appear as the important elements of the first unitarity corrections.
The analysis of both elements is urgently needed:
from the phenomenological
point of view to estimate in which region of $x$ subleading corrections 
leading to a saturation of parton densities become important,
and from the theoretical point of view to obtain indications of how
unitarization in perturbation theory could work.
\\
As to the transition vertex we have concentrated our investigations
on the properties of this element under conformal transformations. 
We have obtained a symbolic operator representation of the vertex
which was then used to give a simplified proof of conformal invariance 
of the vertex. The question of holomorphic separability has been addressed and
it turned out that in our representation the vertex could not be 
decomposed into a holomorphic and an antiholomorphic part.
We have then projected the vertex on conformal eigenfunctions.
These eigenfunctions have been interpreted before as conformal 
three point functions of two fields associated with the reggeized 
gluon and one composite field associated with the BFKL pomeron. 
We denote this field composite since it appears as a bound
state of the two reggeized gluons.
What we have obtained after projection of the vertex could be 
interpreted as the three point function of this composite field.
This is a very encouraging result. It indicates that it could be fruitful
to associate fields of a conformal field theory with the reggeons
that appear after resummation of elementary Feynman diagrams. 
The elements that appear as unitarity corrections could then be 
identified as correlation function of this conformal field theory.
\\
This approach has also been pursued with regard to the four gluon state.
We have demonstrated in which way the corresponding amplitude can be 
identified as the four point function of the composite field mentioned
above. Important information on the structure of the theory can be 
expected from the operator algebra of these fields. The operator 
algebra, i.\ e.\ the operator product expansion contains encoded in the 
fusion coefficients and the anomalous dimensions the whole information
on a conformal field theory. This initiated our interest in the operator 
product expansion of the four (reggeized) gluon amplitude.
\\
The aim of our approach is the calculation of the anomalous dimensions.
From these the spectrum of the integral operator of the 
Bethe Salpeter equation of the four gluon system can eventually be obtained.
The spectrum of the system is encoded in the new function 
$\chi_4(\{\alpha\})$
which depends on a set of quantum numbers $\{\alpha\}$ associated with the 
symmetries of the system. 
We have derived a close relation between the anomalous dimensions
and the spectrum of the respective integral operator of the 
BFKL equation encoded in the well-known function $\chi_2(\nu,n)$. 
Inspired by this relation we have initiated a similar 
approach for the four gluon system. 
This means that 
in our analysis we focus on the behaviour of this function near the 
integer points in the conformal dimension plane. 
The virtue of the method lies in the fact that this behavior 
can be interpreted physically in terms of the anomalous dimensions of the 
four gluon state. 
\\
Following Bartels \cite{bartels} we have used the Faddeev method to 
transform the equation for the four gluon amplitude into an effective two 
reggeon equation defined in momentum space. 
We have argued that in order to obtain the short distance expansion
corresponding to the twist expansion in momentum space the 
singularities of the elements (reggeon propagators and vertices) 
of this equation have to be isolated and iterated.
We succeeded in classifying the singularities which arise from the 
reggeon intermediate states and the effective reggeon interaction vertex.
With this classification at hand it is possible to see in which 
way these elementary singularities have to be iterated.
The singularities that are generated upon iteration of the elementary
singularities can be identified with the anomalous dimensions of 
new composite fields associated with the four (reggeized) gluon state.
As a result of the iteration
a very complex singularity structure of the four gluon amplitude
is found containing poles as well as cuts.
So far we have gained insight into the structural properties 
of the four gluon state. In our setup we could understand the 
origin of the singularity structure of the corresponding amplitude. 
\\
With our method we were able to reproduce the result of 
\cite{bartels} for the anomalous dimension of the four gluon operator 
belonging to the twist four.
With this anomalous dimension we associate a pole of the function 
$\chi_4$ in the complex conformal dimension plane. We expect further 
poles being associated with the anomalous dimension 
of higher twist operators.
The calculation of the anomalous dimensions belonging to the higher 
twists faces the complication
that the extraction of singularities from the reggeon vertices
turns out to be quite difficult.
This is a problem of technical nature.
As a possible way out one could investigate 
whether in configuration space the relevant information,
namely the singularities of the effective reggeon-reggeon interaction,
can be obtained more easily.
In the last chapter we have demonstrated that the singularities 
belonging to the higher twists appear with a certain multiplicity.
This multiplicity is associated with the quantum numbers of the 
function $\chi_4$.  
Ultimately with each complete set of quantum numbers one single 
singularity should be related.
A conceptual problem arising here is that
it is not easy to see in which way one should classify the resulting 
singularities with regard to the quantum numbers of the system.
To classify the singularities according to the quantum numbers
it is probably necessary to obtain more information on 
the mathematical structure of the problem.
The identification of additional symmetries would of course be 
very helpful. 
Possibly, one has to restrict oneself to the 
large $N_c$-approximation in which the set of commuting operators 
can be identified in configuration space using quantum inverse scattering 
methods. 
\\
The specified problems constitute major challenges for future work.
The analysis of the four gluon state certainly deserves further 
interest since it constitutes the next important step to be taken 
towards a unitary scattering amplitude in perturbative QCD.
\newpage
\section*{Acknowledgments}
I would like to thank my supervisor Professor Jochen Bartels
for constant guidance, encouragement and interest.
I am very much
indebted to him for countless helpful suggestions and discussions
during all the years.
\\
I thank Professor Wilfried Buchm\"uller for delivering his opinion
on this thesis
and Professor Gustav Kramer for the support of my applications
for scholarships.  
\\
The discussions with Mark W\"usthoff were vital for my understanding
of the Regge limit in QCD. I want to express my deep gratitude for his 
assistance and collaboration.
\\
It was a great experience to meet Professor Lev
Lipatov who generously explained 
us his ideas and shared some of his deep insights 
into QCD and field theory.
\\
Special thanks to Jeff Forshaw for
extraordinarily pleasant discussions and collaboration.  
\\
For fruitful collaboration I am indebted to  
Albert De Roeck,
Markus Diehl,
Carlo Ewerz, 
Professor Mikhail Ryskin and 
Matthias Vogt. 
\\
For the very careful reading of the manuscript
and numerous helpful suggestions I would like to 
thank Carlo Ewerz.
Helpful  comments on the manuscript were also delivered by Claas Bontus.
\\
Furthermore I would like to thank all participants 
of the seminar on perturbative 
aspects of QCD for creating such a nice atmosphere in the last four years.  
\\
The financial support of the Deutsche Forschungsgemeinschaft and the 
Studienstiftung des Deutschen Volkes is gratefully 
acknowledged. 
\\
This thesis would not have been possible without 
constant support from my parents.
Finally thanks to Tanja for patience.

\newpage
\setcounter{equation}{0}
\setcounter{figure}{0}
\setcounter{table}{0}
\begin{appendix}
\section{Appendix}
\subsection{Fourier transformation of the Bethe-Salpeter equation}
\label{appft}
The non-trivial part is the transformation of the squared 
production vertex and the gluon trajectory. Regularizing 
the infrared singularities by a fictious gluon mass 
$\lambda$ and multiplying the whole equation (\ref{bfklmom})
with $\kf^2(\qf-\kf)^2$ the homogeneous part reads
\footnote{For simplicity we set $\frac{N_c\alpha_s}{2 \pi^2}=1$
in this and the following section.}
\beqn
\left( {\cal K} \Phi_{\omega} \right)(\kf,\qf-\kf) =
\int d^2 \kf' 
\left(
\frac{k(q-k)^{\ast}k{'}^{\ast}(q-k')}
{|\kf-\kf'|^2+\lambda^2}  + \mbox{h.c.} 
\right)
\Phi_{\omega}(\kf',\qf-\kf') 
\nonumber \\ 
- \pi \left( \log \frac{|\kf|^2}{\lambda^2}
+ \log \frac{|\qf-\kf|^2}{\lambda^2} \right)
\Phi_{\omega}(\kf,\qf-\kf)
\eeqn
Fourier transformation is defined as 
\beqn
\Phi_{\om}(\rho_1,\rho_2)=\frac{1}{(2 \pi)^4}
\int d^2 \kf \,d^2 \qf e^{i\kf \rho_1+
i(\qf-\kf)\rho_2}\Phi_{\om}(\kf,\qf-\kf)
\eeqn
Let us begin with the gluon trajectory. Performing the Fourier
transformation
and inserting
\begin{equation}
1= \frac{1}{(2 \pi)^2} \int d^2 \frho_0  \int d^2 \lf \,
e^{ \left[i\frho_0(\lf-\kf)\right]}
\end{equation}
we obtain
\beqn
- 16 |\partial_1|^2 |\partial_2|^2 
 \int d^2 \frho_0    \left[ \frac{1}{(2 \pi)^2}\int d^2 \lf \,
e^{i\lf(\frho_1-\frho_0)} \pi \log \frac{|\lf|^2}
{\lambda^2} \right]
\Phi_{\omega}(\frho_{0},\frho_{2})
\eeqn
For the integral in brackets we make the ansatz
\beqn
- \left[ \frac{1}{(2 \pi)^2}\int d^2 \lf
e^{i\lf(\frho_1-\frho_0)} \pi \log \frac{|\lf|^2}
{\lambda^2} \right] = \Gamma(\frho_1-\frho_0)  \\
\Gamma(\frho_1-\frho_0) = \frac{1}{|\frho_{10}|^2}
\theta(|\frho_{10}|-\epsilon) +c(\epsilon,\lambda)
\delta^{(2)}(\frho_{10})
\label{gammaint}
\eeqn
and obtain
\begin{equation}
c(\epsilon,\lambda)= 2\pi [  \log \lambda + \log
\frac{\epsilon}{2} - 2 \psi(1)]
\label{ftconst}
\end{equation}
Here we have used 
\beqn
\int d^2 \rho \frac{1}{|\rho|^2} \theta ({|\rho|-\epsilon})
e^{-i\lf\rho}&=& 2 \pi \int_{\epsilon}^{\infty}\frac{ d|\rho|}
{|\rho|}J_0(|\lf||\rho|) \nonumber \\
&=& 2 \pi [\log \frac{1}{\epsilon} + |\lf|
 \int_0^{\infty} d |\rho| \log |\rho|
J_1(|\lf||\rho|)] + O(\epsilon) \nonumber \\& =&
2 \pi[\log \frac{1}{\epsilon} -\log
\frac{|\lf|}{2} +\psi(1)] + O(\epsilon)
\eeqn
Now we turn to the squared production vertex. 
Applying Fourier transformation leads to
\beqn
& &\int d^2 \kf \,d^2 \qf
\frac{1}{(2 \pi)^4}
e^{i\kf_1\frho_1 + i(\qf-\kf)\frho_2}
\int d^2 \kf'
\left(
\frac{k(q-k)^{\ast}k{'}^{\ast}(q-k')}
{|\kf-\kf'{|}^2+\lambda^2}  + \mbox{h.c.} \right)
 \Phi_{\omega}(\kf,\qf-\kf)
\nonumber \\ &=& 
\int d^2 \frho_{1'} d^2 \frho_{2'} \int d^2 \kf \,d^2 \qf
\int d^2 \hat{\kf}\;
\frac{16}
{|\hat{\kf}|^2+\lambda^2}
\partial_1^{\ast} \partial_2
e^{-i\hat{\kf}(\frho_{1'}-\frho_{2'})}
\hspace{2.4cm}
\nonumber \\
& &\cdot \frac{1}{(2 \pi)^4}
e^{i\kf(\frho_1-\frho_{1'})+i(\qf-\kf)(\frho_2-\frho_{2'})}
\partial_{1'}\partial_{2'}^{\ast}
\Phi_{\omega}(\frho_{1'},\frho_{2'})
\hspace{1cm}
+\mbox{h.c.}
\nonumber \\
&=& 16 \;
\partial_1^{\ast} \partial_2
\int d^2 \hat{\kf}
\frac{1}{|\hat{\kf}|^2+\lambda^2}
e^{-i\hat{\kf}(\frho_{1}-\frho_{2})}
\partial_{1} \partial_{2}^{\ast}
\Phi_{\omega}(\frho_{1},\frho_{2})
+ \mbox{h.c.} 
\eeqn
The $\hat{\kf}$ - integral gives
\beqn
\int d^2 \hat{\kf} \frac{e^{i\hat{\kf}\rho}}{\hat{\kf}^2+\lambda^2} &=&
2 \pi \int_0^{\infty} d |\hat{\kf}| \, \frac{|\hat{\kf}| J_0(
|\hat{\kf}||\rho|)}{\hat{\kf}^2+\lambda^2} \hspace{2cm} \nonumber \\
&=& 2 \pi K_0(|\rho| \lambda) \nonumber \\& =&
2 \pi [ \log \frac{2}{|\rho|} - \log \lambda
+\psi(1)] +O(\lambda)
\eeqn
Comparing with eq. (\ref{ftconst}) we find that the combination 
$\psi(1)-\log \lambda +\log 2 $ cancels between the gluon
production vertex and the trajectory function. The final 
form in configuration space reads
\beqn
|\partial_1|^2|\partial_2|^2
({\cal K}\Phi_{\om})(\frho_1,\frho_2) &=&
|\partial_1|^2|\partial_2|^2 \int \frac{d^2 \frho_0}
{|\frho_{10}|^2} \theta(|\frho_{10}| - \epsilon)
\Phi_{\omega}(\frho_{0},\frho_{2}) \nonumber \\& &+
|\partial_1|^2|\partial_2|^2 \int \frac{d^2 \frho_0}
{|\frho_{20}|^2} \theta(|\frho_{20}| - \epsilon)
\Phi_{\omega}(\frho_{1},\frho_{0}) \nonumber \\& &+
2 \pi \log \epsilon^2
|\partial_1|^2|\partial_2|^2 \Phi_{\omega}
(\frho_{1},\frho_{2}) \nonumber \\ & &
- \left[\pi \partial_1^{\ast}\partial_2 \log |\frho_{12}|^2
\partial_1 \partial_2^{\ast}+ \mbox{h.c.} \right]
\Phi_{\omega}(\frho_{1},\frho_{2})
\label{final1}
\eeqn
We rearrange the differential operators in the first two terms
using integration by parts ($g$ is a test function)
 \beqn
\partial_1 \partial_1^{\ast} \int d^2 \frho_0 \,g(|\frho_{10}|^2)
\Phi_{\omega}(\frho_{0},\frho_{2})
= \partial_1 \int d^2 \frho_0 \,\partial_1^{\ast}
g(\rho_{10}\rho_{10}^{\ast})
\Phi_{\omega}(\frho_{0},\frho_{2})
\hspace{1.2cm} \nonumber \\
=\partial_1 \int d^2 \frho_0 \,(-1) \partial_0^{\ast}
g(\rho_{10}\rho_{10}^{\ast})
\Phi_{\omega}(\frho_{0},\frho_{2})
=\partial_1 \int d^2 \frho_0 \,g(|\frho_{10}|^2) \partial_0^{\ast}
\Phi_{\omega}(\frho_{0},\frho_{2})
\eeqn
Distributing the derivatives in a symmetric way we end up with
\beqn
2 \; |\partial_1|^2|\partial_2|^2
({\cal K}\Phi_{\om})(\frho_1,\frho_2) =
\hspace{8cm} \nonumber \\
\partial_1 \partial_2^{\ast}
\int d^2 \frho_0 \left[ \frac{1}
{|\frho_{10}|^2} \theta(|\frho_{10}| - \epsilon)
-2 \pi \delta^{(2)}(\rho_{10}) \log|\rho_{12}| \right]
\partial_0^{\ast} \partial_2
\Phi_{\omega}(\frho_{0},\frho_{2}) \nonumber \\
+
\partial_1^{\ast} \partial_2
\int d^2 \frho_0 \left[ \frac{1}
{|\frho_{10}|^2} \theta(|\frho_{10}| - \epsilon)
-2 \pi \delta^{(2)}(\rho_{10})  \log|\rho_{12}| \right]
\partial_0 \partial_2^{\ast}
\Phi_{\omega}(\frho_{0},\frho_{2})
\nonumber \\
+ \left[ 1 \leftrightarrow 2 \right]
+  \pi \log \epsilon^2
|\partial_1|^2|\partial_2|^2 \Phi_{\omega}
\hspace{2.1cm}
\nonumber \\
= \partial_1 \partial_2^{\ast}
\int d^2 \frho_0 \left[ \frac{1}
{|\frho_{10}|^2} \theta(|\frho_{10}| - \epsilon)
-2 \pi \delta^{(2)}(\rho_{10})   \log|\rho_{12}| \right]
\partial_0^{\ast} \partial_2
\Phi_{\omega}(\frho_{0},\frho_{2}) \nonumber \\
+\left[ \mbox{h.c.} \right] \hspace{6cm} \nonumber \\
+ \left[ 1 \leftrightarrow 2 \right]
+  \pi \log \epsilon^2
|\partial_1|^2|\partial_2|^2 \Phi_{\omega}(\frho_1,\frho_2)
\hspace{0.9cm}
\label{abs}
\eeqn
The final form is obtained by absorbing the logarithm into
the argument of the $\theta$-function
\beqn
 2 \; |\partial_1|^2|\partial_2|^2
({\cal K}\Phi_{\om})(\frho_1,\frho_2) =
\hspace{6cm} \nonumber \\
\partial_1 \partial_2^{\ast}
\int \frac{d^2 \frho_0}
{|\frho_{10}|^2} \theta(\frac{|\frho_{10}|}
{|\rho_{12}|} - \epsilon)
\partial_0^{\ast} \partial_2
\Phi_{\omega}(\frho_{0},\frho_{2}) +\left[ \mbox{h.c.} \right]
\hspace{0.8cm}  \nonumber \\
+ \left[ 1 \leftrightarrow 2 \right]
+  \pi \log \epsilon^2
|\partial_1|^2|\partial_2|^2 \Phi_{\omega}(\frho_1,\frho_2)
\label{con}
\eeqn
\subsection{The eigenvalue of the configuration space kernel}
\label{appeigen}
In this section we want to sketch how the eigenvalue of the 
BFKL-kernel can be obtained in configuration space. For 
our calculation we use the
form of eq. (\ref{final1}) of the preceding section.  
We insert 
\beqn
\Phi_{\om}(\rho_1,\rho_2) = 
\left(\frac{\rho_{12}}{\rho_{10'}\rho_{20'}}\right)^{
\frac{1+n}{2}-i\nu}
\left(\frac{\rho_{12}^{\ast}}
{\rho_{10'}^{\ast}\rho_{20'}^{\ast}}\right)^{
\frac{1-n}{2}-i\nu}
\eeqn
and use the fact that the kernel is conformally invariant. This 
allows to use as a special set of coordinates for which we 
calculate the eigenvalue the points ${\rho_1,\rho_2,\rho_0'=\infty}$. 
It remains to calculate the integral
\beqn
\int \frac{d^2 \rho_0}{|\rho_{10}|^2}
\theta(|\rho_{10}|-\epsilon)
(\rho_{20})^{\frac{1+n}{2}-i\nu}
(\rho_{20})^{\frac{1-n}{2}-i\nu}
\label{integral}
\eeqn
We use the following representation of the 
$\theta$-function ($\delta \rightarrow +0$)
\beqn
\theta(|\rho_{10}|-\epsilon)
=\frac{1}{2 \pi i} \int_{-i \infty +\delta}^{i \infty +\delta}
\frac{d \xi}{\xi} \left(\frac{|\rho_{10}|^2}{\epsilon^2}\right)^{\xi}
\eeqn
and end with the configuration space integral
\beqn
I(\nu,n;\xi)=\int d^2 \rho_0 (|\rho_{10}|^2)^{\xi-1}
(\rho_{20})^{\frac{1+n}{2}-i\nu}
(\rho_{20}^{\ast})^{\frac{1-n}{2}-i\nu}
\nonumber \\
=
(|\rho_{12}|^2)^{\frac{1}{2}-i\nu+\xi} 
\left(\frac{\rho_{12}}{\rho_{12}^{\ast}}\right)^{\frac{n}{2}}
\int d^2 x \,x^{\xi-1}x^{\ast\,\xi-1}(x-1)^{\frac{1+n}{2}-i\nu}
(x-1)^{\ast \, \frac{1-n}{2}-i\nu}
\eeqn
The two-dimensional integral 
which obviously depends only 
on $|n|$ is calculated with a method
which is described in detail in \cite{dot}. After a Wick-rotation,
introduction of light-cone coordinates and deformation of contours 
we get    
\beqn
I(\nu,n;\xi) =
(|\rho_{12}|^2)^{\frac{1}{2}-i\nu+\xi}
\left(\frac{\rho_{12}}{\rho_{12}^{\ast}}\right)^{\frac{|n|}{2}}
\sin \pi(\frac{1-n}{2}-i\nu) 
\hspace{5cm}
\nonumber \\
  \int_0^1 du \,u^{\xi-1}(1-u)^{
\frac{1+|n|}{2}-i\nu}
\int_1^{\infty} d v \,v^{\xi-1}(v-1)^{\frac{1-|n|}{2}-i\nu}
\nonumber \\
=\pi (|\rho_{12}|^2)^{\frac{1}{2}-i\nu+\xi}
\left(\frac{\rho_{12}}{\rho_{12}^{\ast}}\right)^{\frac{|n|}{2}}
\frac{\Gamma(\xi)}{\Gamma(1-\xi)} 
\frac{\Gamma(-\xi-\frac{1-|n|}{2}+i\nu)}{\Gamma(\xi+\frac{3+|n|}{2}-i\nu)}
\frac{\Gamma(\frac{3+|n|}{2}-i\nu)\Gamma(\frac{3-|n|}{2}-i\nu)}
{\Gamma(\frac{1+|n|}{2}+i\nu)\Gamma(\frac{1-|n|}{2}-i\nu)}
\eeqn
Now we can perform the $\xi$-integration by shifting the
contour to the left. The first non-vanishing contribution 
comes from the double-pole at $\xi=0$. All subsequent poles
in the left half-plane vanish in the limit $\epsilon \rightarrow 0$.
As the result we get for eq. (\ref{integral})
\beqn
\pi
(|\rho_{12}|^2)^{\frac{1}{2}-i\nu}
\left(\frac{\rho_{12}}{\rho_{12}^{\ast}}\right)^{\frac{|n|}{2}}
\left[2 \psi(1) -
\psi(\frac{3+|n|}{2}-i\nu)-
\psi(-\frac{1-|n|}{2}+i\nu)
 - \log \frac{\epsilon^2}{|\rho_{12}|^2}
\right] \nonumber \\=
(|\rho_{12}|^2)^{\frac{1}{2}-i\nu}
\left(\frac{\rho_{12}}{\rho_{12}^{\ast}}\right)^{\frac{|n|}{2}}
\left[2 \psi(1) -
\psi(\frac{1+|n|}{2}+i\nu)-
\psi(\frac{1+|n|}{2}-i\nu)
\right. \nonumber \\ \left.
-\frac{2i\nu-1}{(\frac{1+|n|}{2}-i\nu)(\frac{1-|n|}{2}-i\nu)}
- \log \frac{\epsilon^2}{|\rho_{12}|^2}
\right]
\eeqn 
After adding the term with $\rho_1$ and $\rho_2$ interchanged  
the $\log \epsilon^2$-contribution cancels in eq. (\ref{final1}).
The fourth term in the bracket above and the 
$\log |\rho_{12}|^2$ cancel against the remaining terms 
\beqn
-\pi[\partial_1^{\ast}\partial_2 \log|\rho_{12}|^2
\partial_1\partial_2^{\ast}+\mbox{h.c.}]
(|\rho_{12}|^2)^{\frac{1}{2}-i\nu}
\left(\frac{\rho_{12}}{\rho_{12}^{\ast}}\right)^{\frac{|n|}{2}}
\eeqn
after the derivatives are taken. 
After all we obtain for the action of the BFKL-kernel
on the conformal three-point functions 
\beqn
\pi (|\rho_{12}|^2)^{\frac{1}{2}-i\nu}
\left(\frac{\rho_{12}}{\rho_{12}^{\ast}}\right)^{\frac{|n|}{2}}
\left[2 \psi(1) -
\psi(\frac{1+|n|}{2}+i\nu)-
\psi(\frac{1+|n|}{2}-i\nu) \right] 
\eeqn
what was to be proved.
\subsection{Properties of the BFKL eigenvalue $\chi(\nu,n)$}
\label{app3}
For completeness some properties of the function $\chi(\nu,n)$
used at different stages of this work are presented here.
The eigenvalue of the BFKL kernel reads ($\psi(x)$ is the logarithmic
derivative of the $\Gamma$-function $\psi(x)=\Gamma'(x)/\Gamma(x)$)
\beqn
\chi(\nu,n)=
\frac{N_c \alpha_s}{\pi}
\left[2 \psi(1) -
\psi(\frac{1+|n|}{2}+i\nu)-
\psi(\frac{1+|n|}{2}-i\nu) \right]
\eeqn
The function $\chi(\nu,n)$ is a real function of $\nu$ which is anlytic 
in the strip $ -(1+|n|)/2< Im(\nu) < (1+|n|)/2$. 
As a function of $\nu \in \Bbb{R}$, $\chi(\nu,n)$ 
is symmetric w.\ r.\  t.\ $\nu=0$,
has a global
maximum at $\nu=0$ and tends to $-\infty$ for $\nu \to \pm \infty$.
For the high-energy limit the expansion around $\nu=0$ up to 
quadratic order
(harmonic approximation) is important
\beqn
\chi(\nu,0)&=& \frac{N_c\alpha_s}{\pi}
\left[
4 \log 2 -14 \zeta(3) \nu^2
+O(\nu^6)
\right]
\\
\chi(\nu,\pm 1)&=& \frac{N_c\alpha_s}{\pi}
\left[
-2 \zeta(3) \nu^2
+O(\nu^6)
\right]
\\ 
\chi(\nu,\pm 2)&=& \frac{N_c\alpha_s}{\pi}
\left[
-4 +4 \log2 +(16-14\zeta(3))\nu^2
+O(\nu^6)
\right]
\eeqn
In this approximation the BFKL equation becomes equivalent to a diffusion
equation, where the coefficient of the quadratic term determines the diffusion
constant.
Comparing the expansions with zero and nonzero 
conformal spin $n$ 
we see that in the high-energy limit the contribution
with zero conformal spin always dominates because it is the only one with
a strictly positive intercept.
\\
The singularities of $\chi(\nu,n)$ are located on the imaginary 
$\nu$-axis. For each $n$ there are integer spaced simple poles
with identical residues. The singularity structure is most easily read 
off from the representation
\beqn
\chi(\nu,n)= \frac{N_c\alpha_s}{\pi}
\sum_{k=0}^{\infty}
\left[
\frac{1}{k+\frac{1+|n|}{2}+i\nu}
+
\frac{1}{k+\frac{1+|n|}{2}-i\nu}
-
\frac{2}{k+1}
\right]
\label{app3eigen}
\eeqn   
Likewise important is the Laurent expansion of $\chi(\nu,0)$ 
around the pole at $i\nu=-1/2$. 
It reads
\beqn
 \chi(\nu,0)= \frac{N_c\alpha_s}{\pi}  
\;
\frac{1}{i\nu+\fez} 
\left[
1+
2 \sum_{k=1}^{\infty} \zeta(2k+1) \left(i\nu+\fez\right)^{2k+1}
\right]
\eeqn
This expansion is used to calculate higher order contributions 
to the anomalous dimension matrix from the BFKL equation \cite{ktfac}.
From the fact that the first subleading term in this expansion is of 
order $(i\nu+\fez)^2$ one concludes 
that the first subleading contribution 
to the gluon anomalous dimension which follows from the BFKL equation 
is of order $\alpha_s^4$.  
\subsection{Expansion of the momentum space eigenfunction for small argument}
\label{app4}
In this appendix we show how the small argument expansion 
the beginning of which is given in eq.\ (\ref{ruegen}) can be derived 
starting from the momentum space expression (\ref{momspacef})
\beqn
E^{(\nu,0)}(\kf,\qf-\kf)=b(\nu,0)
\int_0^1 dx [x(1-x)]^{-\fez+i\nu}[\kf^2+x((\qf-\kf)^2-\kf^2)]^{-\ftz-i\nu}
\nonumber \\
\,_2F_1\left(\ftz+i\nu,-\fez+i\nu,1;1-\frac{\qf^2x(1-x)}
{\kf^2+x((\qf-\kf)^2-\kf^2)}\right)
\eeqn
where $b(\nu,0)$ collects some unimportant (for the moment) coefficients.
It is necessary to perform first the $x$-integration.
To this end an analytic transformation of the $_2F_1$-function is applied
which changes the argument from $z$ to $1-z$. In the two terms which are 
generated by this procedure the series representation 
for the hypergeometric function is used. This leads to 
\beqn
E^{(\nu,0)}(\kf,\qf-\kf)=b(\nu,0)(\qf^2)^{-i\nu} 
\left[\frac{(\qf^2)^{i\nu}}{(\kf^2)^{\ftz+i\nu}}
\frac{\Gamma(-2i\nu)}{\Gamma(-\fez-i\nu)\Gamma(\ftz-i\nu)} 
\sum_{n=0}^{\infty}\frac{(\ftz+i\nu)_n(-\fez+i\nu)_n}{n!(1+2i\nu)_n}
\left(\frac{\qf^2}{\kf^2}\right)^n
\right. \nonumber \\ \left.
\cdot 
\int_0^1 d x [x(1-x)]^{-\fez+i\nu+n}\left(1+x\frac{(\kf-\qf)^2-\kf^2)}{\kf^2}
\right)^{-n-\ftz-i\nu}
+\,(\mbox{h.c.})\,
\right]
\eeqn
where the Pochhammer symbol $(a)_n=\Gamma(a+n)/\Gamma(a)$ has been used.
The $x$-integral in the second line can now be performed
yielding
\beqn
B(\fez+i\nu+n,\fez+i\nu+n)
\,_2F_1\left(\ftz+i\nu+n,\fez+i\nu+n,1+2i\nu+2n;
\frac{\kf^2-(\kf-\qf)^2}{\kf^2}\right)
\eeqn
and analogously for the complex conjugate. Now, unfortunately the argument
of the hypergeometric function does not allow an expansion 
in powers of $|\kf|$. To invert the argument one has to use another 
analytic continuation for the hypergeometric function which - since 
the first to arguments differ by unity - is logarithmic \cite{bateman}.
In this way one obtains  
\beqn
E^{(\nu,0)}(\kf,\qf-\kf)\!=\!b(\nu,0)(\qf^2)^{-i\nu}
\left[
\frac{\Gamma(-2i\nu)}{\Gamma(-\fez-i\nu)\Gamma(\ftz-i\nu)} 
\sum_{n=0}^{\infty}\frac{(\ftz+i\nu)_n(-\fez+i\nu)_n}{n!(1+2i\nu)_n}
\frac{(\qf^2)^{n+i\nu}}{((\kf\!-\!\qf)^2\!-\!\kf^2)^{n+\ftz+i\nu}}
\right. \nonumber \\ \left. 
(\fez\!+\!i\nu\!+\!n)\!
\left(
\sum_{l=0}^{\infty}
\frac{(\fez\!+\!i\nu\!+\!n)_{l+1}(\fez\!-\!i\nu\!-\!n)_{l+1}}{l!(l+1)!}
\left(\frac{\kf^2}{(\qf\!-\!\kf)^2\!-\!\kf^2}\right)^l
\left[\gamma_l\!-\!\log \frac{\kf^2}{(\qf\!-\!\kf)^2 \!-\!\kf^2)} \right]
\!+\! \frac{(\qf\!-\!\kf)^2\!-\!\kf^2}{\kf^2}
\right)
\right. \nonumber \\ \left.
\!\!\!\!\!\!\!\!\!\!\!\!\!\!\!\!\!\!
\!\!\!\!\!\!\!\!\!\!\!\!\!\!\!\!\!\!
\!\!\!\!\!\!\!\!\!\!\!\!\!\!\!\!\!\!
\!\!\!\!\!\!\!\!\!\!\!\!\!\!\!\!\!\!
\!\!\!\!\!\!\!\!\!\!\!\!\!\!\!\!\!\!
\!\!\!\!\!\!\!\!\!\!\!\!\!\!\!\!\!\!
+\,(\mbox{h.c.})\,
\right] \phantom{xxxx}
\label{tobago}
\eeqn
with 
\beqn
\gamma_l= \psi(2+l)+\psi(1+l)
-\psi(\ftz+i\nu+n+l)-\psi(\ftz-i\nu-n+l)
\eeqn
Starting from this expression the expansion in powers of $|\kf|/||\qf|$
or alternatively $|\qf-\kf|/|\qf|$ is in principle straightforward.
The main problem is to collect all terms which contribute to 
a given order. It is now clear where the logarithms and $\psi$-functions
in eq.\ (\ref{ruegen}) have their origin.
Note that the leading term in the expansion (\ref{ruegen})
comes from the very last term in the second line of (\ref{tobago}).
\end{appendix}
 
\end{document}